\documentclass[12pt]{iopart}
\makeatletter
\@addtoreset{equation}{section}
\renewcommand{\theequation}{\thesection.\arabic{equation}}
\makeatother
\usepackage{iopams}  
\usepackage{tcolorbox}
\usepackage{multirow}
\usepackage{bm}
\usepackage{graphicx}
\usepackage{latexsym}
\usepackage{mathrsfs}
\usepackage{amssymb,bm} 
\usepackage{enumitem}

\newcommand{\mathbfcal}[1]{\boldsymbol{\mathcal{#1}}}
\newcommand{\vi}{{\boldsymbol{i}}}
\newcommand{\vj}{{\boldsymbol{j}}}
\newcommand{\pdagger}{{\phantom{\dagger}}}
\usepackage[colorlinks=true]{hyperref} 
\hypersetup{
    bookmarks=true,         % show bookmarks bar?
    unicode=false,          % non-Latin characters 
    pdftoolbar=true,        % show Acrobat
    pdfmenubar=true,        % show Acrobat 
    pdffitwindow=false,     % window fit to page when opened
    pdfstartview={FitH},    % fits the width of the page to the window
    pdftitle={My title},    % title
    pdfauthor={Author},     % author
    pdfsubject={Subject},   % subject of the document
    pdfcreator={Creator},   % creator of the document
    pdfproducer={Producer}, % producer of the document
    pdfkeywords={keyword1} {key2} {key3}, % list of keywords
    pdfnewwindow=true,      % links in new window
    colorlinks=true,       % false: boxed links; true: colored links
    linkcolor=magenta, %red,          % color of internal links (change box color with linkbordercolor)
    citecolor=blue,        % color of links to bibliography
    filecolor=magenta,      % color of file links
    urlcolor=blue       % color of external links
} 
\usepackage{xcolor}
\definecolor{darkgreen}{RGB}{0,100,0}
\begin{document}
\begin{flushright}
\href{https://arxiv.org/abs/2508.20164}{arXiv:2508.20164}
\end{flushright}
\title[Fractionalized Fermi liquids and the cuprate phase diagram]{Fractionalized Fermi liquids\\ and the cuprate phase diagram}

\author{Pietro M. Bonetti$^{1}$, Maine Christos$^{1,2}$, Alexander Nikolaenko$^1$, Aavishkar A. Patel$^{3,4}$, and Subir Sachdev$^{1,3,5}$}

\address{$^1$Department of Physics, Harvard University, Cambridge MA 02138, USA.}
\address{$^2$Walter Burke Institute for Theoretical Physics and Institute for Quantum Information and Matter, California Institute of Technology, Pasadena, CA 91125, USA.}
\address{$^3$Center for Computational Quantum Physics, Flatiron Institute, 162 5th Avenue, New York, NY 10010, USA.}
\address{$^4$International Centre for Theoretical Sciences, Tata Institute of Fundamental Research, Bengaluru 560089, India}
\address{$^5$The Abdus Salam International Centre for Theoretical Physics, Strada Costiera 11, I-34151, Trieste, Italy.}
%\ead{sachdev@g.harvard.edu}
\vspace{10pt}
\begin{indented}
\item[]\today
\end{indented}

\begin{abstract}
We review a theoretical framework for the cuprate superconductors, rooted in a fractionalized Fermi liquid (FL*) description of the intermediate-temperature pseudogap phase at low doping. The FL* theory predicted hole pockets each of fractional area $p/8$ at hole doping $p$,
in contrast to the area $p/4$ in spin density wave theory. Magnetotransport measurements, including observation of the Yamaji angle, show clear evidence of hole pocket quasiparticles which can tunnel coherently between square lattice layers, and are consistent with the FL* description.
 
The FL* phase of a single-band model is described using a layer construction with a pair of ancilla qubits on each site: the Ancilla Layer Model (ALM). Its mean field theory yields hole pockets of area $p/8$, and matches the gapped photoemission spectrum in the anti-nodal region of the Brillouin zone. Fluctuations are described by the SU(2) gauge theory of a background spin liquid with critical Dirac spinons. A Monte Carlo study of the thermal SU(2) gauge theory transforms the hole pockets into Fermi arcs in photoemission. One route to confinement of FL* upon lowering temperature yields a $d$-wave superconductor via a Kosterlitz-Thouless transition of $h/(2e)$ vortices, with nodal Bogoliubov quasiparticles featuring anisotropic velocities and vortices surrounded by charge order halos. An alternative route yields a charge-ordered metallic state that has quantum oscillations consistent with observations. These confinement transitions are driven by the condensation of a SU(2) fundamental Higgs field, which also provides a fractionalized description of intertwined orders. 
 
Increasing doping from the FL* phase in the ALM drives a transition to a conventional Fermi liquid (FL) at large doping, passing through an intermediate strange metal regime. We formulate a theory of the FL*-FL metal-metal transition without a symmetry-breaking order parameter, using a critical quantum `charge' liquid of mobile electrons in the presence of disorder, developed via an extension of the Sachdev-Ye-Kitaev model to two spatial dimensions.
 
 At low temperatures, and across optimal and over doping, we address the regimes of extended non-Fermi liquid behavior by Griffiths effects near quantum phase transitions in disordered metals.

%This perspective connects the pseudogap, superconducting, charge-ordered, and strange metal phases within a single theoretical landscape.
\end{abstract}

%
% Uncomment for keywords
%\newpage

\noindent
{\tt %\underline{Preliminary incomplete draft.}\\~\\~\\
Partly based on lectures by S.S. at
\begin{itemize}
\item \href{https://boulderschool.yale.edu/2025/boulder-school-2025}{Boulder School 2025, Dynamics of Strongly Correlated Electrons}, July 14-18. \href{https://www.youtube.com/playlist?list=PLcD25rnTeV9i2c4s3C9I8RQjy9Js_mY-h}{Lecture videos}.
\item \href{https://indico.ictp.it/event/10859}{Joint ICTP-WE Heraeus School and Workshop on Advances in Quantum Matter: Pushing the Boundaries},  ICTP, Trieste, August 4,6, 2025.
\href{https://www.youtube.com/playlist?list=PLcD25rnTeV9gAvvrZugfCF1qZh8fZorj1}{Lecture videos}.
\item \href{https://indico.ictp.it/event/10860}{School on Quantum Dynamics of Matter, Light and Information}, ICTP, Trieste, August 18,19, 2025. \href{https://www.youtube.com/playlist?list=PLcD25rnTeV9gAvvrZugfCF1qZh8fZorj1}{Lecture videos}.
\item \href{https://projects.croucher.org.hk/advanced-study-institutes/fractional-chern-insulators-theory-numerics-and-experiment}{Croucher Advanced Study Institute for Fractional Chern Insulators}, University of Hong Kong, September 4,5, 2025. \href{https://doi.org/10.5281/zenodo.17072646}{Lecture slides}.
\item \href{https://indico.ictp.it/event/10879}{Advanced School and Conference on Quantum Matter},  ICTP Trieste, Dec 1-12, 2025. \href{https://www.dropbox.com/scl/fi/0keg5yvkdgwcxr4mwdmu6/lectures_ictp.pdf?rlkey=fjvtv2st56pijm9k1f5pmrm9u&st=x7ikergd&dl=0}{Lecture Notes}.
\href{https://www.youtube.com/playlist?list=PLcD25rnTeV9gAvvrZugfCF1qZh8fZorj1}{Lecture videos}.
\end{itemize}
} 
%{\it Preliminary draft with missing references.}\\

%\noindent{\it Keywords}: superconductivity, pseudogap metal, strange metal, quantum criticality\\
%
% Uncomment for Submitted to journal title message
%\submitto{\RPP}
%
% Uncomment if a separate title page is required
%\maketitle
% 
% For two-column output uncomment the next line and choose [10pt] rather than [12pt] in the \documentclass declaration
%\ioptwocol
%
%\newpage
\vspace{1pc}

\tableofcontents
\title[Fractionalized Fermi liquids and the cuprate phase diagram]{}

\maketitle

\section{Introduction}
\label{sec:intro}

\begin{figure}
\centering
\includegraphics[width=4.5in]{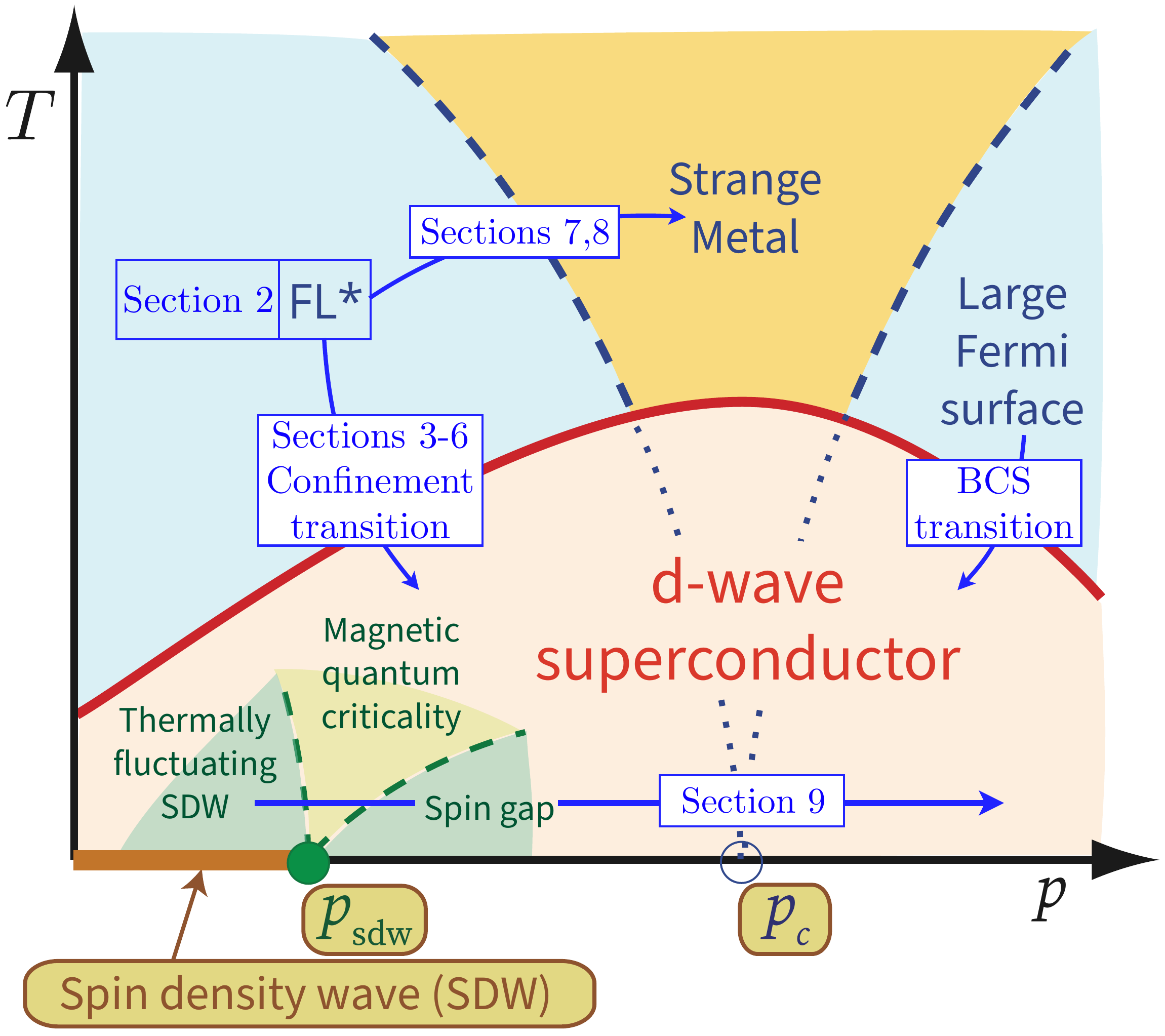}
\caption{Cuprate phase diagram from Fig. 4 of Ref.~\cite{SSwhere}. (The only changes from the 2010 figure are that the pseudogap metal is now labeled FL*, dopings are denoted by $p$ rather than $x$, arrows have been removed, and section number annotations have been added; but all phase boundaries, crossovers, and critical points are the same.)
The incipient FL*-FL transition at $p_c$ \cite{Qi10,SSMetlitskiPunk12,YaHui-ancilla1} is that at $s_c$ in the normal state in Fig.~\ref{fig:flsfan}; there is no transition at $p_c$ in the presence of superconductivity. The SDW transition at $p_{\rm sdw}$ is discussed in Section~\ref{sec:griffiths} in the presence of disorder, where the ``spin gap'' region above becomes a Griffiths phase.}
\label{fig:figcross}
\end{figure}

\begin{figure}
\centering
\includegraphics[width=4in]{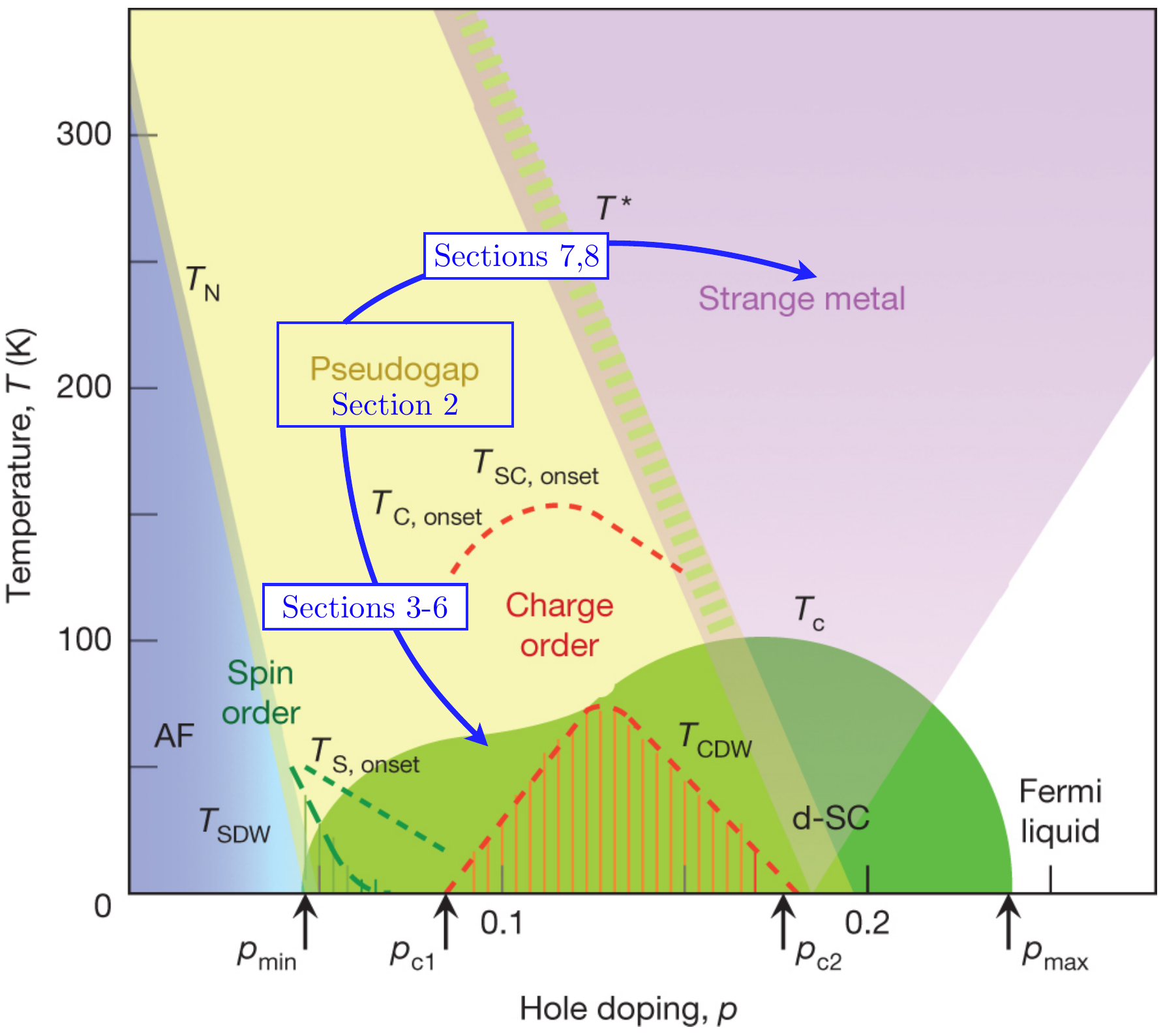}
\caption{Cuprate phase diagram from Ref.~\cite{phase_diag}. Annotations in blue have been added.
We use a theory of a fractionalized Fermi liquid (FL*) of the pseudogap to connect to the other phases in the sections noted.}
\label{fig:cupratepd1}
\end{figure}

This article presents a theoretical framework for the complex and rich phase diagram of the cuprate materials, see Figs.~\ref{fig:figcross} and \ref{fig:cupratepd1}.
Already in 1987, Anderson \cite{Anderson87} suggested that `resonating valence bond states' hold the key to the physics of the cuprates. Today, there is a well-developed theory of such gapped quantum spin liquid states with complex many-particle entanglement, which host a variety of well-defined particle-like anyonic excitations \cite{QPMbook}. 
This theory entangles the electron spins, but does not involve any motion of the electron charge, and is therefore a theory of {\it many-boson entanglement.} But needed for the cuprates are quantum states with mobile fermions, and {\it many-fermion entanglement.} The greater complexity of fermionic entanglement has been explored in some recent papers \cite{Lake19,King25}.

There was much early effort on the structure of spin liquids doped away from the insulating half-filled limit by a density of $p$ holes \cite{LeeWenRMP}. Although there were significant advances, a number of difficulties have since become apparent. Two prominent difficulties are:\\
({\it i\/}) In the underdoped `pseudogap' regime (Fig.~\ref{fig:cupratepd1}), angle-dependent magnetoresistance (ADMR) experiments \cite{Ramshaw22,Yamaji24} provide evidence for small hole pockets which can tunnel coherently between the square lattice layers, and such coherent tunneling is not possible by fractionalized charge carriers in a doped spin liquid.\\
({\it ii\/}) A $d$-wave superconductor with 4 nodal Bogoliubov quasiparticles is indeed obtained by doping a gapless spin liquid. But the observed velocities \cite{Chiao00} of these quasiparticles are highly anisotropic, with the velocity perpendicular to the Fermi surface ($v_F$) being much larger than that in the orthogonal direction ($v_\Delta$). 

In this review, we describe theoretical work on two classes of states with many-fermion entanglement: the fractionalized Fermi liquid (FL*), and quantum liquids obtained by generalizing the states of the Sachdev-Ye-Kitaev (SYK) model \cite{SY92,kitaev_talk}.

We employ the fractionalized Fermi liquid (FL*) state \cite{TSSSMV03,TSSSMV04} as a model of the pseudogap metal, and as a starting point for a description of the rest of the cuprate phase diagram. We will describe how this approach addresses the difficulties noted above. As shown in Fig.~\ref{fig:figcross}, lowering temperature from FL* leads to a theory of the $d$-wave superconductor, and other observed states with charge order. 
Upon increasing doping from the pseudogap, we described the `strange metal' by the quantum criticality of a quantum phase transition from FL* to a conventional Fermi liquid (FL) appropriate for the large doping regime. The FL*-FL transition \cite{Qi10,SSMetlitskiPunk12,YaHui-ancilla1} is an example of a metal-to-metal quantum phase transition without any associated symmetry breaking, unlike the well-known Hertz transitions \cite{hertz} which require symmetry-breaking order parameters. Metal-to-metal transitions with increasing $p$ without any symmetry breaking have also appeared in numerical cluster dynamical mean-field studies by Tremblay and collaborators \cite{AMT10,AMT11,TremblayPNAS22},
and as an `orbital-selective Mott transition in momentum space' \cite{KotliarZeros,Haule08,Ferrero09,Gull09,Ferrero10,Laad12,Laad25}. 
The FL* state realizes a finite-dimensional description of the low doping metal similar to these works, properly accounting for the non-Luttinger Fermi surface. 

The charged excitations of the FL* state are ordinary electron-like quasiparticles near the non-Luttinger Fermi surface, similar to those in a Landau Fermi liquid. However, the FL* state also has neutral spin excitations  of 
an underlying quantum spin liquid, as we summarize in Section~\ref{sec:intro_pseudogap}. For this spin liquid, we argue in Section~\ref{sec:spinliquids} for a particular gapless, critical, quantum spin liquid. Unlike the gapped spin liquid states, this critical spin liquid has no particle-like excitations.

Another distinct critical state, without quasiparticle excitations, is associated with the FL*-FL transition and the collective dynamics of electron charge, as we summarize in Section~\ref{sec:intro_strange}. This is a critical quantum `charge' liquid, in our second class of states with many-fermion entanglement, characteristic of the strange metal regime of Figs.~\ref{fig:figcross} and \ref{fig:cupratepd1}. Its theoretical description builds on concepts from the Sachdev-Ye-Kitaev (SYK) model, introduced in Section~\ref{sec:SYK}. 

\subsection{Pseudogap}
\label{sec:intro_pseudogap}

At large doping, $p$, the cuprates exhibit a `large' hole-like Fermi surface whose enclosed area has the Luttinger value of $(1+p)/2$  \cite{Dama05,Hussey08} (we measure areas as fractions of the square lattice Brillouin zone, and the factor of $1/2$ is from spin degeneracy). So it is reasonable to apply the theory of spin-fluctuation mediated pairing of Fermi liquids (FL), leading to Cooper pairs around the Fermi surface, and the formation of a BCS-type superconductor. It is known that the spin fluctuations are antiferromagnetic, and antiferromagnetic spin fluctuations indeed lead to the observed $d$-wave pairing \cite{ScalapinoRMP}.

But this powerful paradigm runs into difficulty at lower $p$, and especially in the regime of the pseudogap, where there is no complete Fermi surface above $T_c$ upon which the Cooper pairs can form, as shown in photoemission \cite{Norman98,ShenShen05,Johnson08,Johnson10,YRZEPL,Johnson11,Shen11,Kondo20,Kondo23,Damascelli25} (although indications of pocket Fermi surfaces have been discussed \cite{Johnson08,Johnson10,YRZEPL,Johnson11,PDJ18}) and scanning tunneling microscopy (STM) \cite{Hoffman14,Davis14} experiments (see Fig.~\ref{fig:arc}a). Nevertheless, below $T_c$, there is no dramatic change in the nature of the superconductivity with varying $p$, and so it is clear that the superconductor itself is always adiabatically connected to a $d$-wave BCS state.

Recent magnetotransport measurements in the pseudogap regime \cite{Ramshaw22,Yamaji24} reveal results that stand in striking contrast to earlier photoemission observations. Photoemission reveals a `Fermi arc' in the nodal regime of the Brillouin zone, along with a dispersing gapped spectrum in the anti-nodal region (see Fig.~\ref{fig:antinode}a). In contrast, magnetotransport is best modeled by complete hole pockets which can tunnel coherently between layers, and 
the observed Yamaji angle \cite{Yamaji24} is close to that obtained from hole pockets of area $p/8$. We shall review a theory in this paper which aims to resolve this tension between photoemission and magnetotransport.

Among early approaches, the `superconducting phase fluctuation' theory of the pseudogap
\cite{EmeryKivelson,Franz98,Scalapino02,Dagotto05,Berg07,Li_2010,Li11JPhys,Li11PRB,Li_2011,Sumilan17,Majumdar22,YQi23,Xiang24}  can produce the Fermi arc spectra from thermal fluctuations of the nodal quasiparticles, but has difficulty with the recent magnetotransport observations. The fluctuating spin density wave approach \cite{SchmalianPines1,SchmalianPines2,Chubukov23,Chubukov25} does yields hole pockets, but they have area $p/4$, as we will discuss further in Section~\ref{sec:singleband}.

We take a point of view for the pseudogap quite distinct from the above and other \cite{Fradkin10,Hayward:2013jna,Lee14,Nie_15,Fradkin15,Castro17,Pepin23,Fradkin25} `fluctuating order' theories. Rather than incorporating thermal fluctuations above a conventionally ordered ground state, we start from a novel, intrinsincally quantum ground state which builds in many-electron quantum entanglement.
We describe the low $p$ phases by a `fractionalized Fermi liquid' \cite{Burdin2002,TSSSMV03,Si2004,TSSSMV04,APAV04,GS10,Qi10,SSMetlitskiPunk12,Bonderson16,Assaad18,Assaad20a,Coleman22,Tsvelik24,Tsvelik25} (FL*) theory of the pseudogap metal. The FL* is a state with Fermi surfaces which do not enclose the Luttinger area, and this is possible if there is a `background' quantum spin liquid, as is reviewed in Section~\ref{sec:FL*}.  

We first present, in Section~\ref{sec:kl}, the simple mean field theory for the FL* phase in a Kondo lattice model, and its transition to the commonly observed heavy Fermi liquid FL phase by the condensation of a Higgs (`slave') boson \cite{ReadNewns83,Coleman84,AuerbachLevin86,MillisLee87,AndreiColeman,AndreiColeman2} $\Phi$. 

While the existence of a FL* phase in a two-band Kondo lattice model is well-established, the existence of FL* in single-band Hubbard-type model was initially based on general theoretical arguments \cite{RKK07,RKK08,Qi10,Moon11,Mei11,Punk12}, and a mapping to a quantum dimer model \cite{Punk15,Punk18}. But a systematic approach, based upon a mean-field theory plus fluctuations, appeared with the introduction of the Ancilla Layer Model \cite{YaHui-ancilla1}. As described in Section~\ref{sec:singleband} and \hyperref[app:sw]{Appendix A}, the ALM has a pair of ancilla qubits on each site, to yield a three-layer model: see Fig.~\ref{fig:ancilla}.
The complete ancilla gauge theory for a single-band model is described in \hyperref[app:ancilla]{Appendix B}. Furthermore, the ALM can also describe the conventional FL phase, a feature which we shall exploit in Section~\ref{sec:strange}.

The ALM allows exploration of a wider class of mean-field and gauge theories, and variational wavefunctions for the FL* state, in rough analogy to the utility of layered neural networks \cite{Hinton86}.
The mean-field theory of the single-band FL* is obtained by assuming that the top two layers form the familiar, and well established, Kondo lattice heavy Fermi liquid FL phase---this 
has a spinon Fermi surface in the {\it middle\/} layer which is unstable to hybridization with the conduction electrons with a condensed Higgs (`slave') boson $\Phi$, just as in the Kondo lattice model. But, although the resulting Fermi surface has the Luttinger area for the Kondo lattice, it does {\it not\/} have the Luttinger area for a single-band Hubbard model. Reassuringly, in this mean-field, the {\it bottom\/} ancilla layer is a decoupled spin liquid which is required to exist by the non-Luttinger area. This leads to the slogan in Fig.~\ref{fig:oneband} (the nature of the spin liquid will be discussed in Section~\ref{sec:spinliquids})
\begin{tcolorbox}
\begin{itemize}
\item
 Single-Band FL* = Kondo Lattice Heavy Fermi Liquid ~$\oplus$~ Spin Liquid.
\end{itemize}
\end{tcolorbox}

Despite its simplicity, this mean-field theory can accurately model \cite{Mascot22} the gapped electronic spectrum revealed by photoemission in the antinode in Fig.~\ref{fig:antinode}, as is discussed in Section~\ref{sec:photo_pocket}. It also predicts pockets of area $p/8$ \cite{TSSSMV03,TSSSMV04,RKK07,Qi10,Zhao_Yamaji_25,FuChun25} consistent with magnetotransport, and distinct from the fluctuating spin density theory pocket area of $p/4$. 
Moreover, the FL* state has coherent interlayer transport of electronic quasiparticles, while the spin density wave state requires inter-layer spin correlations, which are absent in HgBa$_2$CuO$_{4+\delta}$ \cite{Greven14,Greven25}.
The Yamaji effect requires interlayer quasiparticle transport, and the Yamaji angle observations of Chan {\it et al.\/} \cite{Yamaji24} in HgBa$_2$CuO$_{4+\delta}$
are therefore direct evidence for the FL* state and the presence of fractionalization in the cuprates.

Furthermore, the mean-field theory leads to a variational wavefunction for FL* \cite{YaHui-ancilla1}: numerical studies \cite{Iqbal24,HenryShiwei24,YaHui24} on such a wavefunction 
compare well with cold atom experiments on the  square lattice fermionic Hubbard model \cite{Koepsell21,Bloch24,Kendrick:2025ujy}, as we will discuss in Section~\ref{sec:wavefunction}.

But a shortcoming of the mean-field theory of Section~\ref{sec:FL*} is that the photoemission spectrum of the intermediate temperature pseudogap phase is not a Fermi arc in the nodal region.

To proceed further, we need to include fluctuations which couple the top two Kondo lattice layers to the spin liquid in the bottom layer of the ALM. We can no longer be agnostic about the nature of spin liquid, but need to make a specific choice, as that is important in determining the nature of the thermal fluctuations in the pseudogap, and its evolution upon lowering temperature. The power and flexibility of the ALM method
(over earlier approaches \cite{LeeWenRMP}) is that we are not constrained to choose the spinon Fermi surface of the middle layer, and can make an independent choice for the spin liquid of the bottom layer.
We follow the proposal by Christos {\it et al.\/} \cite{Christos:2023oru}, and choose the critical square lattice spin liquid, without quasiparticle excitations, which has dual \cite{Wang17} formulations in terms of a U(1) gauge theory of bosonic spinons \cite{NRSS89prl,NRSS90,senthil1,senthil2} (see Eq.~(\ref{cp3})), or as a SU(2) gauge theory of fermionic spinons in the $\pi$-flux phase \cite{AM88,Affleck-SU2,Fradkin88} (see Eq.~(\ref{eq:fermionhop2})).
A number of numerical works support the existence of this spin liquid over substantial intermediate energy scales on the square lattice  \cite{Yasir19,Nahum:2015vka,Becca20,Imada21,Fuzzy24,Meng24,Gu24,Chester24,Sandvik24}, and there is also some experimental evidence \cite{Hayden10,Ronnow15}.
We review the properties of this critical spin liquid in Section~\ref{sec:spinliquids}. 

Sections~\ref{sec:halffilling} and \ref{sec:dwave} show that this critical spin liquid satisfies a key constraint: as the temperature is lowered, the emergent gauge fields are confined by a transition to a $d$-wave superconductor which is adiabatically connected to the BCS state \cite{Christos:2023oru}. The onset of superconductivity at low $p$ from FL* is not a BCS-type Cooper-pairing transition, but a confinement/Higgs transition of a gauge theory \cite{Christos:2023oru,Coleman22,Sayantan25}, as indicated in Fig.~\ref{fig:figcross}. This is a specific realization of an early suggestion by Anderson \cite{Anderson87}, that cuprate superconductivity appears by exploiting the pre-existing pairing of electrons in a resonating valence bond state. On the other hand, the larger $p$ transition to superconductivity from the FL is of the BCS-type. Note that the critical singularity of the thermal superconducting transition is always described by the Kosterlitz-Thouless theory of $h/(2e)$ vortices \cite{Sayantan25}, and there is no singularity as a function of $p$ along the critical line. The distinction between small and large $p$ lies in the shorter-distance electronic structure, {\it e.g.\/} in the photoemission spectrum and the structure of the vortex core.

There were several earlier proposals \cite{Zhang88,KotliarLiu88,LeeWen96,IvanovSenthil,LeeWenRMP} for a small $p$ transition from a spin liquid to a $d$-wave superconductor. In these theories,  gapless fermionic spinons of the spin liquid directly transmute into the Bogoliubov quasiparticles of the $d$-wave superconductor with a massless Dirac dispersion. However, as noted earlier, these proposals had a significant problem as they predict a nearly isotropic velocity dispersion, with the velocities along the Brillouin zone diagonals ($v_F$, see Fig.~\ref{fig:flsdsc}B) and the orthogonal direction ($v_\Delta$) being nearly equal to each other. Experimentally, we have $v_F /v_\Delta \sim 14$ to $19$ \cite{Chiao00}. Section~\ref{sec:aniso} shows how this problem is resolved \cite{Chatterjee16,CS23} by instead considering the transition from a FL* state to a $d$-wave superconductor. In such a theory, the spinons do not transmute into Bogoliubov quasiparticles; instead, they mutually annihilate with extraneous Bogoliubov quasiparticles from the `backside' of the hole pockets.

Sections~\ref{sec:pseudogap} and \ref{sec:dwave} also review how the critical spin liquid FL* theory is connected to a number of other experiments on the underdoped cuprates:
\begin{itemize}
\item Section~\ref{sec:arc}: The `Fermi arcs' observed in photoemission \cite{Norman98,ShenShen05,Johnson08,Johnson10,YRZEPL,Johnson11,Kondo20,Kondo23,Damascelli25} and STM \cite{Hoffman14,Davis14} are realized by thermal fluctuations of the SU(2) lattice gauge theory describing the FL* state \cite{Sayantan25}. The same computations also display quantum oscillations \cite{Sayantan25}, which points to a reconciliation between photoemission and magnetotransport. We note that the conversion of the hole pockets into arcs is primarily a thermal effect. In the FL* state at $T=0$, a non-zero quasiparticle residue
is present around the hole pocket even in the presence of quantum fluctuations of the SU(2) gauge theory \cite{Mascot22}.
\item Section~\ref{sec:vortex}: The thermal SU(2) gauge theory \cite{Sayantan25} also describes the onset of superconductivity as a Kosterlitz-Thouless transition of vortices with flux $h/(2e)$ \cite{ZhangSS24}. This onset is associated with a SU(2) fundamental Higgs field $B$, which provides a fractionalized theory of intertwined superconductivity and charge orders.
As a consequence of the intertwined charge order instability of the critical spin liquid \cite{Christos:2023oru}, each vortex carries a charge order halo, similar to observations \cite{Hoffman02}.
\item Section~\ref{sec:qo}: Quantum oscillations at low $p$ and low $T$ show small electron pockets in the presence of charge density wave order \cite{Sebastian_review}. This can be explained as arising from the charge density wave acting on hole pockets only upon including the influence of the spinons of the FL* state \cite{BCS24}. 
\end{itemize}

\subsection{Strange metal}
\label{sec:intro_strange}

As indicated in Fig.~\ref{fig:figcross} and \ref{fig:cupratepd1}, Sections~\ref{sec:SYK} and \ref{sec:strange} address the strange metal phase as a crossover between the pseudogap metal and the large $p$ Fermi liquid, induced by an underlying metal-to-metal quantum phase transition {\it without\/} any symmetry breaking. Such a metal-metal transition appeared in the numerical cluster dynamic mean-field analysis of Sordi {\it et al.} \cite{AMT10,AMT11,TremblayPNAS22} and is connected to the `orbital-selective Mott transition in momentum space' \cite{KotliarZeros,Haule08,Ferrero09,Gull09,Ferrero10,Laad12,Laad25}.
We identify \cite{Qi10,SSMetlitskiPunk12} the transition with the FL*-FL transition. 

Sections~\ref{sec:SYK}-\ref{sec:griffiths} will connect the physics of the strange metal to an underlying FL*-FL transition. Our discussion of strange metals will be brief, and the reader is refereed to other reviews \cite{CGPS22,SachdevORE,SSZaanen}: our purpose here is to connect the strange metal regime of the cuprate phase diagram to that of the pseudogap in earlier sections.

A different critical quantum liquid plays an important role in the physics of the strange metal, one in which the electrons are mobile. Two solvable zero-dimensional models, the SYK model and the Yukawa-SYK model, yield much insight and their properties are reviewed in Section~\ref{sec:SYK}. 

Section~\ref{sec:strange} extends the zero-dimensional models of Section~\ref{sec:SYK} to the two-dimensional case of interest. 
We begin by the general description of a quantum phase transition in a two-dimensional metal. We focus on a possible quantum phase transition from FL* to FL with no symmetry breaking order parameters, as that is the case relevant for the hole-doped cuprates. But our methods and results apply more generally also to symmetry breaking transitions of metals \cite{Patel2}.
There is also an instability to superconductivity in the vicinity of the FL*-FL quantum phase transition \cite{YaHui-ancilla1,YaHui-ancilla2}, and this leads to the phase diagram in Fig.~\ref{fig:figcross}. 
The superconducting state is the same on both sides of the transition, and so there is no actual quantum critical point under the superconducting dome at the incipient FL*-FL transition at $p_c$ in Fig.~\ref{fig:figcross}. Nevertheless, the quantum criticality of the metal-to-metal FL*-FL transition can serve as the basis for the theory of the strange metal region above the superconducting $T_c$. It is worthwhile to note here that if the underlying phase transition in the metal involved symmetry breaking, the quantum critical point would survive the onset of superconductivity, in contrast to the FL*-FL transition we argue is appropriate for the cuprates.

For quantum phase transitions without spatial disorder, we do find the breakdown of well-defined quasiparticles {\it i.e.\/} a non-Fermi liquid. However, such clean non-Fermi liquids
described by the class of field theories considered in Section~\ref{sec:strange} have an emergent continuous translational symmetry which precludes the observed singular behavior in transport properties (see Section~\ref{sec:norandom}). 

Indeed, this absence of singular transport above is part of a more general phenomenon in the transport in clean lattice systems. It is not possible in any clean lattice model to have a non-zero resistance without the contribution of umklapp processes. The basic argument is quite simple \cite{hkms,Hartnoll:2014gba,Patel:2014jfa,Berg19}, although it has been overlooked in numerous prominent papers in the literature. Without umklapp, there is a conserved total momentum ${\bm P}$ in the system. This momentum includes the contributions of {\it all\/} excitations, including electrons, spin fluctuations, spinons, phonons, excitons, etc. It is important to note that momentum relaxation of the spins/spinons/phonons/excitons also requires their umklapp scattering. Now imagine setting up a net electrical current ${\bm J}$ in this system. The current ${\bm J}$ is not conserved even without umklapp, and ${\bm J}$ will decay from various scattering processes, although the value of ${\bm P}$ will not. (we note that, quite misleadingly, 
the statement of `charge conservation' is sometimes stated as `current conservation' in the literature.) In the long-time limit, the value of ${\bm J}$ will be the one that maximizes the entropy subject to the constraint that the value of ${\bm P}$ remains unchanged. In general, as long the cross-susceptibility $\chi_{\bm {JP}}$ is non-zero, the long-time limit of ${\bm J}$ is also non-zero (only is special systems with exact particle-hole symmetry is $\chi_{\bm {JP}}$ non-zero). So a non-zero $\chi_{\bm{JP}}$ implies that the conductivity is infinite. In situations with a nearly conserved momentum, the memory function approach yields a powerful method to compute such transport \cite{Hartnoll:2016apf}.

There can be situations without impurities, reviewed elsewhere \cite{CGPS22} and briefly in Section~\ref{sec:univ}, where the continuous translational symmetry emerges only at very low temperatures, because of the dominance of umklapp processes.
This can be the case in zero-dimensional SYK-type dynamic mean-field theories and finite dimensional quantum-critical theories of `Kondo breakdown-type' quantum phase transitions \cite{SY92,CGPS22,Sengupta97,Si1,Si2,ZhuSi,Norman07,Norman08,Norman13,Si14,SenguptaGeorges95,Burdin_2000,Burdin2002,Kotliar08,GleisKotliar24,GleisKotliar25,ChChung24,ChChung25,Lal25}, or related cluster dynamic mean-field theories of single-band models \cite{KotliarZeros,Haule08,Ferrero09,Gull09,AMT10,AMT11,Ferrero10,Laad12,TremblayPNAS22,Laad25}. 
Such models can apply at higher temperature in the `bad metal' regime \cite{CGPS22}, or in situations at low temperatures where there is an effective momentum bath {\it e.g.\/} with a heavy spinon band \cite{Norman07,Norman08,Norman13}. Moreover, it is possible that the crossover to a momentum conserving theory \cite{ChowdhuryBerg} is strongly suppressed in Kondo lattice models where the couplings between the Kondo spins are weak. Similar comments apply to holographic models \cite{schalm2016,Hartnoll:2016apf,Zaanen23,Huang:2023ihu,Schalm25} with a periodic potential.
Similarly, Shastry's extremely correlated Fermi liquid (ECFL) \cite{Shastry11,Shastry25} displays Fermi liquid transport at the lowest temperatures. But many of the dynamic mean-field models mentioned in this 
paragraph have a non-vanishing entropy density in the zero temperature limit, and this has not been observed.
Our focus, instead, is on behavior at low temperatures in situations without significant umklapp processes: the arguments reviewed in Section~\ref{sec:norandom} preclude strange metal behavior 
in such situations in the absence of impurities.

The main focus of Section~\ref{sec:strange} is on the role of impurities, which are ubiquitous in experiments, and which will prevent the emergent continuous translational symmetry.  Section~\ref{sec:random} examines impurities which do not break any symmetry of the Hamiltonian. The dominant effect of such impurities is their introduction of strongly relevant `Harris disorder' associated with spatial randomness in the local position of the quantum critical point.
We described the influence of such Harris disorder using a two-dimensional Yukawa-Sadhdev-Ye-Kitaev (2D-YSYK) model. We describe a self-averaging treatment of the spatial disorder, inspired by the structure of the solution of the SYK model. This leads to a set of universal predictions at low temperatures ($T$) in the quantum-critical `fan', which are largely independent of the particular quantum phase transition under consideration \cite{Patel2,Li:2024kxr}. These properties include a linear-in-$T$ resistivity, a $ T \ln (1/T)$ specific heat, a $\sim 1/\omega$ tail in the optical conductivity at frequency $\omega$, and marginal Fermi liquid behavior in the electronic spectrum. These are in good agreement with observations across a wide range of correlated electron materials. We also note a recent study of the Hall co-efficient of the 2D-YSYK model \cite{Davide25}, which yields a superlinear Hall effect.

However, there is a particular feature that is special to the underlying FL* to FL transition illustrated in Fig.~\ref{fig:figcross}, and not shared by symmetry breaking transitions associated with spin or charge density wave order. FL*-FL transitions have a singular particle-hole asymmetry, and this leads to singular behavior in the thermopower \cite{LPS24}. This is also consistent with observations on the cuprates \cite{Collignon21,Taillefer_Seebeck_PRX_2022,Georges_Skewed} and heavy-fermion compounds \cite{Park24}. 

Finally, we note that at sufficiently low $T$, the self-averaging treatment of disorder described above breaks down, and the contributions of regions where collective bosonic modes localize become important. 
Moreoever, we expect a crossover to confinement of gauge-charged excitations at low $T$ \cite{SSZaanen}, and so it is appropriate to deal with transitions associated with conventional spin density wave order parameter, as in Fig.~\ref{fig:figcross}. The localization of spin density wave modes is discussed in Section~\ref{sec:griffiths} (see also Ref.~\cite{SSZaanen}). Rare region effects lead to Griffiths phases \cite{TV07,PPS24,PLA24}, and the `foot' in the strange metal region, observed in transport \cite{Hussey_foot,Greene_rev} and neutron scattering \cite{Hayden25}.  There are also connections to photoemission observations \cite{DessauPLL}.

\section{Fractionalized Fermi liquids}
\label{sec:FL*}

Almost all metals are well described at low temperatures by the principles of Fermi liquid theory. This is a theory of nearly free fermionic quasiparticle excitations with the same spin and charge as an electron. The energy of these quasiparticles vanishes on a $d-1$ dimensional surface in momentum space ($d$ is the spatial dimension) known as the Fermi surface. A crucial feature for our purposes is the Luttinger constraint on the volume enclosed by the Fermi surface in momentum space. Luttinger established by a perturbative diagrammatic analysis that the enclosed volume is independent of the strength of the interactions and is determined only by the electron density $\rho$ \cite{Luttinger60}. A more precise statement is that the volume enclosed by the Fermi surface is the same as that of a free electron system with the same symmetry and the same density.

An important step forward was Oshikawa's proof of the Luttinger constraint using a 't Hooft anomaly matching argument \cite{MO00}. Oshikawa identified a mixed anomaly between translations and global U(1) symmetry of charge conservation. The value of the anomaly can be computed exactly from a knowledge of the density of electrons in the lattice Hamiltonian. Matching this anomaly with that of the low energy Fermi liquid theory, Oshikawa established the Luttinger constraint in a non-perturbative manner; see Ref.~\cite{QPMbook} for a review.

The idea of a `fractionalized Fermi liquid' (FL*) was introduced in Refs.~\cite{TSSSMV03,TSSSMV04}, as a metallic state in which the Fermi surface did {\it not\/} obey the Luttinger constraint. In the simplest case, the volume enclosed by the Fermi surface in FL* was the same as free electrons with density $\rho - 1$ (we are assuming a spin $S=1/2$ degeneracy).
The central point \cite{TSSSMV04,APAV04,Powell05,Coleman05,Qi10,SSMetlitskiPunk12,Bonderson16,SenthilElse21,QPMbook,Meng26} was that it was possible to satisfy Oshikawa's anomaly by combining the anomaly of a Fermi surface (which contributes an amount equivalent to a density $\rho-1$) with that of a fractionalized spin liquid of the type we will study in Section~\ref{sec:spinliquids} (in which the anomaly is {\it quantized\/} to an amount equivalent to a density 1). It is trivially possible to shift $\rho$ by an even integer by adding and removing filled bands, and the novelty is the shift in FL* by an odd integer.

Note that the spin liquid is {\it not\/} required to have a Fermi surface of spinons, enclosing the missing Luttinger volume. Any other spin liquid with the same Oshikawa anomaly is allowed, and we will employ a spin liquid with Dirac fermion spinons for the cuprates.

\begin{figure}[h]
\centering
\includegraphics[width=4in]{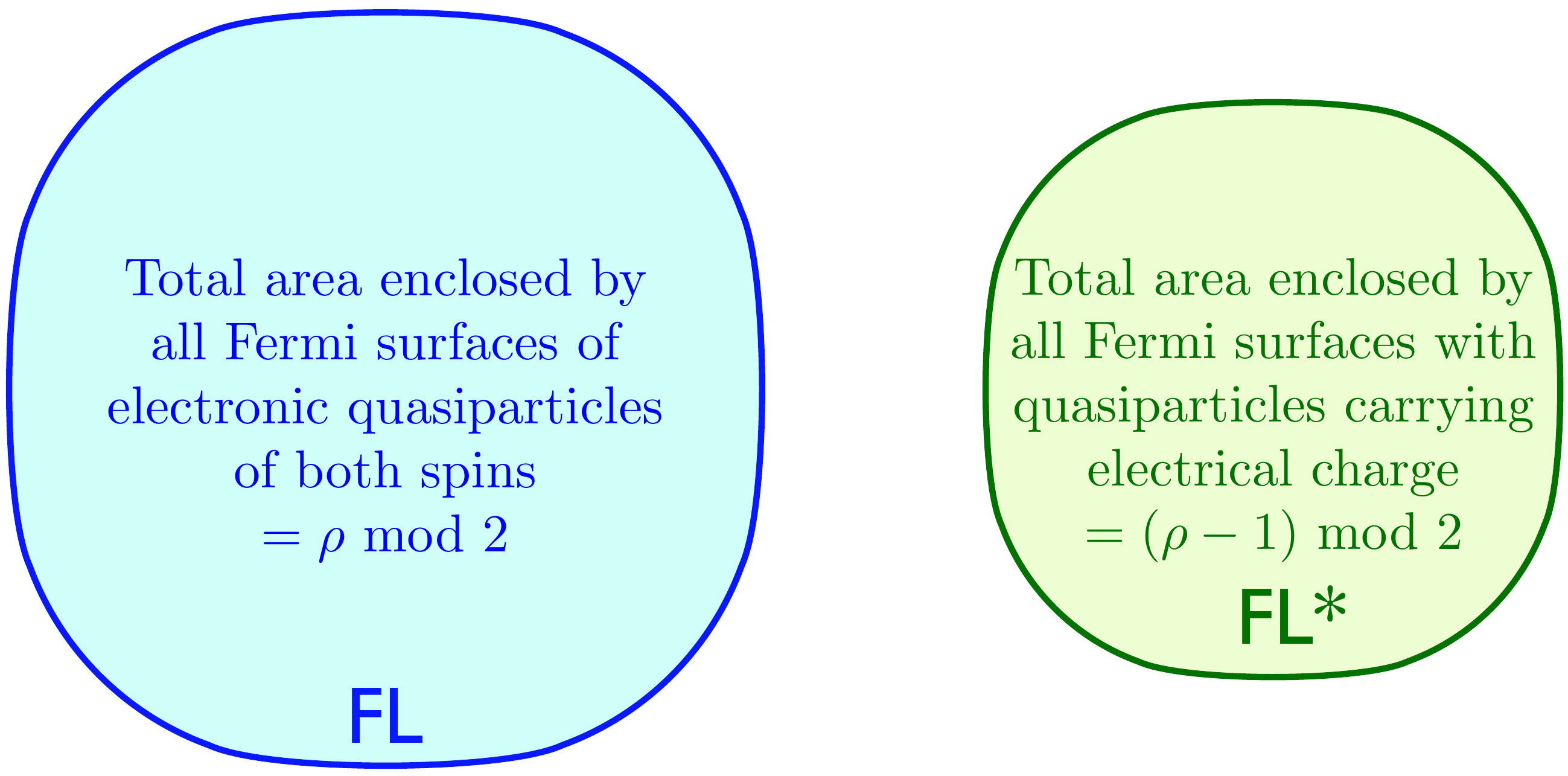}
\caption{Definitions of Fermi liquid (FL) and fractionalized Fermi liquid (FL*) for a general electronic lattice model. Here $\rho$ is the total density of electrons of both spins, and areas are measured as a fraction of Brillouin zone area. The mod 2 accounts for fully filled bands.}
\label{fig:flsdef}
\end{figure}
These ideas apply rather generally to electronic lattice models, and are summarized schematically in Fig.~\ref{fig:flsdef}. An explicit construction of FL* is simplest and well established for the Kondo lattice model, and this will be described in Section~\ref{sec:kl}. Our main interest here is the FL* phase in a single band Hubbard-like model, and we will turn to this in Section~\ref{sec:singleband}.

\subsection{Kondo lattice}
\label{sec:kl}

\begin{figure}[h]
\centering
\includegraphics[width=3.5in]{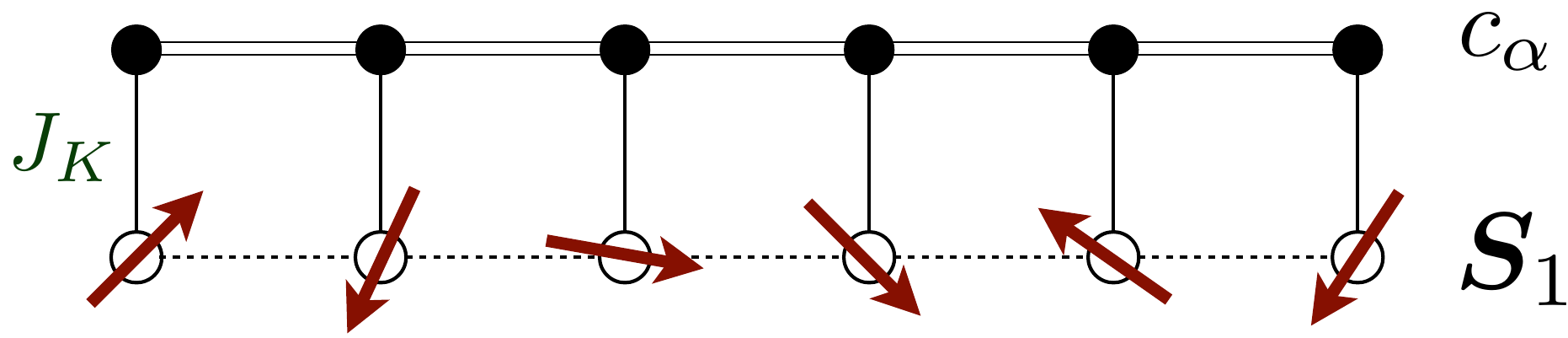}
\caption{A Kondo lattice of conduction electrons $c$ of density $p$ coupled to $S=1/2$ spins ${\bm S}_1$. All lattices are two-dimensional, although only one-dimensional projections are shown.}
\label{fig:kl}
\end{figure}

The Kondo lattice is illustrated in Fig.~\ref{fig:kl}.
A Kondo exchange interaction couples the spin model of Section~\ref{sec:spinliquids} in Eq.~(\ref{hamil}) with a Fermi surface of free electrons $c_{\alpha}$ of density $p$ in a second band:
\begin{eqnarray}
\mathcal{H}_{\rm KL} =&& \sum_{\vi < \vj} J_{1,\vi\vj} \,{\bm S}_{1\vi} \cdot {\bm S}_{1\vj} -\sum_{\vi,\vj} t_{\vi\vj} c^\dagger_{\vi\alpha} c_{\vj\alpha} 
+ \sum_{\vi} \frac{J_K}{2} {\bm S}_{1i} \cdot c_{\vi \alpha}^\dagger {\bm \sigma}_{\alpha\beta} c_{\vi \beta} \label{eq:HKL}
\end{eqnarray}
We have now written the spins as ${\bm S}_{1i}$, rather than ${\bm S}_\vi$, in anticipation of a second set of spins, ${\bm S}_{2i}$, to be introduced in the next subsection.
While $\mathcal{H}_{\rm KL}$ can have a wide variety of phases, we focus on two phases which have no broken symmetries. We proceed with the fermionic parton method of Section~\ref{sec:fermions}, replacing Eq.~(\ref{Schwingerfermion}) by 
\begin{eqnarray}
{\bm S}_{1\vi} = \frac{1}{2} f_{1\vi \alpha}^\dagger {\bm \sigma}_{\alpha\beta}^{\vphantom\dagger} f_{1\vi \beta}^{\vphantom\dagger}\,,
\label{eq:S1f}
\end{eqnarray}
where
\begin{equation}
\sum_{\alpha} f_{1\vi \alpha}^\dagger f_{1\vi \alpha} = 1 \quad, \quad \mbox{for all $\vi$}\,. \label{constraintf1}
\end{equation}
Then we decouple both sets of exchange interactions to obtain a mean-field fermion Hamiltonian \cite{ReadNewns83,Coleman84,AuerbachLevin86,MillisLee87,AndreiColeman,AndreiColeman2} (we explore such decouplings further in our discussion of insulating antiferromagnets Section~\ref{sec:fermions}) 
\begin{equation}
\mathcal{H}_{\rm KLmf} = \sum_{\vi,\vj} \left[- t_{\vi\vj} c^\dagger_{\vi\alpha} c_{\vj\alpha} -   t_{1,\vi\vj} f^\dagger_{1\vi\alpha} f_{1\vj\alpha} \right] -  \sum_{\vi}(\Phi \,  c^\dagger_{\vi\alpha} f_{1\vi\alpha}+ \Phi^\ast \, f^\dagger_{1\vi\alpha} c_{\vi\alpha}) \,. \label{HKLmf}
\end{equation}
The $\Phi$ is the decoupling field for the $J_K$ interaction, while the $t_{1,\vi\vj}$ are the decoupling fields for the $J_{1,\vi\vj}$ interactions. Unlike Section~\ref{sec:fermions}, we do not assume any flux in the $t_{1,\vi\vj}$, and make the simplest possible choice in which the $t_{1,\vi\vj}$ have the same symmetry as the lattice; consequently, the $f_{1}$ spinons initially form a spinon Fermi surface. With these choices, our Kondo lattice theory only has an emergent U(1) gauge symmetry after adding an emergent gauge field $a_{\vi\vj}$ via the Peierls factor $t_{1,\vi\vj} \rightarrow t_{1,\vi\vj} e^{i a_{\vi\vj}}$. The gauge transformations are
\begin{eqnarray}
f_{1\vi \alpha} \rightarrow e^{i \phi_\vi} f_{1\vi \alpha} \quad &,& \quad a_{\vi\vj} \rightarrow a_{ \vi\vj} + \phi_\vi -  \phi_\vj \nonumber \\
c_{\vi \alpha} \rightarrow c_{\vi \alpha} \quad &,& \quad \Phi_\vi \rightarrow \Phi_\vi e^{-i \phi_\vi}\,. \label{gaugeu1}
\end{eqnarray}
This U(1) is distinct from the global charge conservation symmetry in Eq.~(\ref{eq:charge}), under which only maps $c_{\vi \alpha} \rightarrow e^{i \theta} c_{\vi \alpha}$, while all other fields remain invariant. The condensate $\Phi$ in Eq.~(\ref{HKLmf}) higgses this gauge U(1), and so it will be safely ignored in the remaining discussion.

\begin{figure}
\centering
\includegraphics[width=5in]{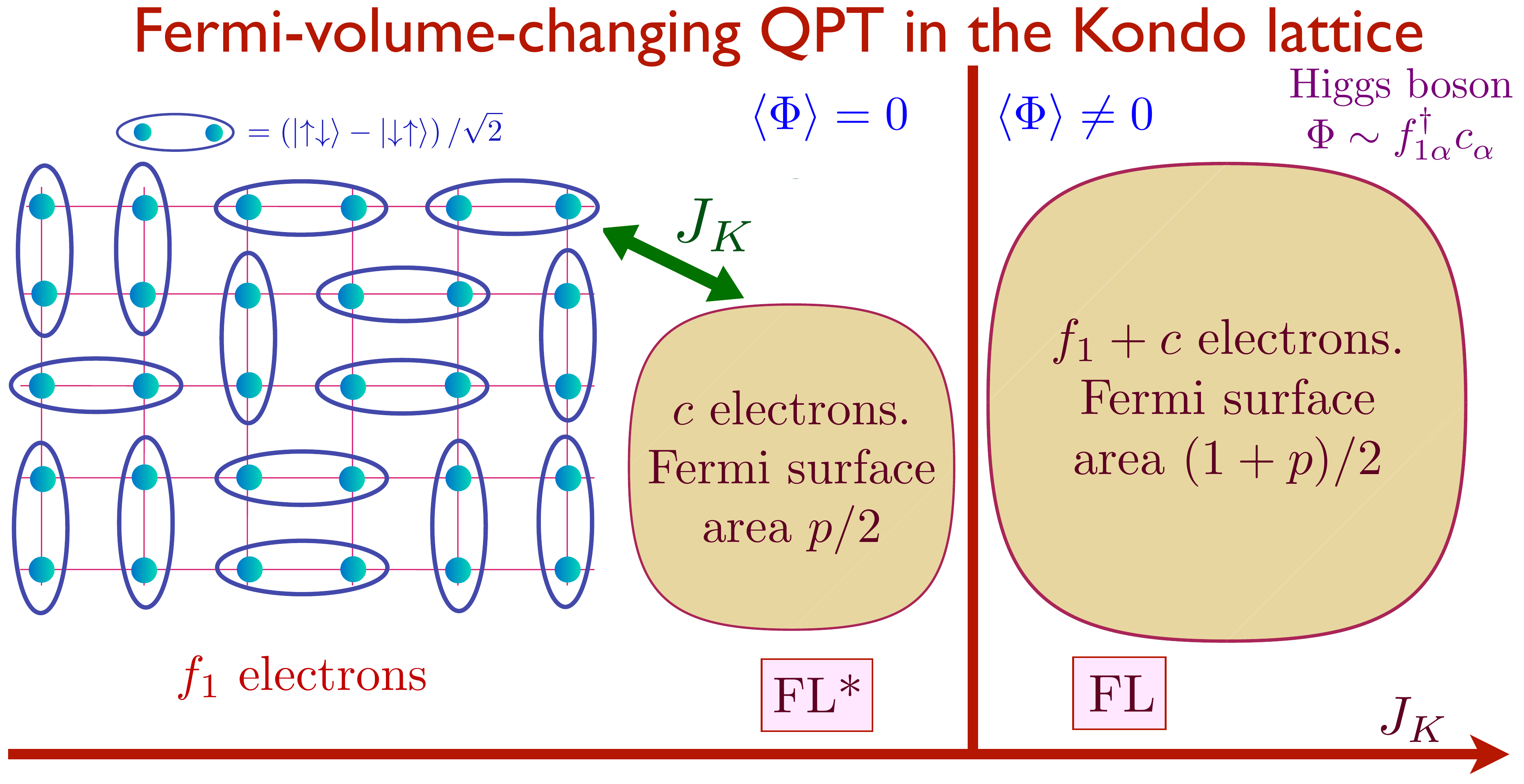}
\caption{Phase diagram of the Kondo lattice. The quoted areas are per spin. Neither phase has any symmetry breaking, but there is nevertheless a quantum phase transition (a Higgs transition in an emergent gauge theory) associated with Fermi volume change \cite{TSSSMV03,TSSSMV04}.}
\label{fig:klpd}
\end{figure}

We can now identify two distinct phases of the Kondo lattice model, neither with any symmetry breaking \cite{TSSSMV03,TSSSMV04}, as illustrated in Fig.~\ref{fig:klpd}.
\begin{itemize}
\item {\it $\langle \Phi \rangle \neq  0$, FL.} This is the conventional `heavy Fermi liquid' phase, observed in numerous heavy fermion compounds.  The Fermi surface obeys the Luttinger constraint. The condensation of the Higgs boson $\Phi$ quenches the gauge fluctuations associated with Eq.~(\ref{gaugeu1}). The Fermi surface is described by the simple two-band model of Eq.~(\ref{HKLmf}), in which $\Phi$ hybridizes the two bands. The total density of electrons is $1+p$, and if all the fermions are in the lower energy band, we obtain a Fermi surface of area $(1+p)/2$ (mod 1) per spin, as illustrated in Fig.~\ref{fig:klpd}. The heavy quasiparticle mass arises in cases where there is little direct interactions between the ${\bm S}_1$ spins: then the $J_{\vi\vj}$ and hence the $t_{1, \vi\vj}$ are very small, and we have a nearly flat band $f_1$ band hybridizing with the conduction band of $c$ near the Fermi level. 
\item {\it $\langle \Phi \rangle = 0$, FL*.} This is the novel phase, in which the $f_1$ and $c$ fermions are decoupled to leading order in $\mathcal{H}_{\rm KLmf}$. The $f_{1\alpha}$ form a spin liquid, in this case one with a spinon Fermi surface. But any other spin liquid with the same anomaly is allowed {\it i.e.\/} any other spin liquid with unit density of spinons in the ground state.  The conduction electrons now form a `small' Fermi surface of area $(\rho - 1)/2 = p/2$ (mod 1) per spin (where $\rho$ is the total density of electrons of both spins in the Kondo lattice), and this is {\it not\/} the Luttinger value. At higher order, the $f_1$ and $c$ fermions will couple with each other, but by extending the general arguments of Luttinger, we can conclude that such effects will not charge the Fermi surface area.
Note that a solution with $\langle \Phi \rangle = 0$ is not permitted in the single-spin Kondo model because $J_K$ flows to infinity under renormalization. In the lattice model, the flow of $J_K$ is cutoff at a scale determined by $J_{1\vi\vj}$ \cite{Burdin2002,qptbook}.
\end{itemize}
The above construction of FL* might appear rather too simple, as it merely decouples the spins from the conduction electrons. To appreciate the full subtlety, it is useful consider FL* in a model without spins.
\begin{itemize}
\item {\it FL* in the Anderson lattice.}
It also interesting to consider the formation of FL* in an Anderson lattice model, where the ${\bm S}_1$ spins are replaced by electronic orbitals with a large local repulsion $U$, and the total density of electrons in both bands remains at $1+p$. 
In general, the Anderson lattice model is not particle-hole symmetric, and so  the density of electrons in the $f_1$ band will deviate from unity, and consequently the actual density of electrons in the $c$ band will deviate from $p$. Nevertheless, a crucial point is that the area of the Fermi surface per spin will remain pinned at $p/2$ (mod 1) because the anomaly of the spin liquid is quantized at unity.
\end{itemize}

The existence of a FL* state in a two-band model can be extended into an essentially rigorous argument \cite{TSSSMV04,QPMbook}. We start from the decoupled limit, where the $f_1$ band forms a stable quantum spin liquid, and the conduction electrons form their own Fermi surface. This decoupled state is stable to the $J_K$ coupling because such a coupling is free of infrared divergencies and cannot destabilize the quantized anomaly of the spin liquid. Moreover, by the usual Luttinger-type analysis, the volume enclosed by the Fermi surface of the conduction electrons cannot charge to all orders in $J_K$. Related arguments appear in analyses based upon solvable Kitaev spin liquids \cite{Coleman22,Tsvelik24,Tsvelik25}.

A number of numerical studies \cite{Assaad18,Assaad20,Assaad20a,Gazit20,Assaad22a,Assaad22b,Assaad23a} support the existence of phases similar to FL and FL* in various Kondo lattice models.

\subsection{Single band model}
\label{sec:singleband}

We now turn to a construction of the FL* phase in a single band model, such as the square lattice Hubbard model (possibly with additional short-range interactions) of interest for the cuprates.
Now the situation is more complicated than in the Kondo lattice model, as there is no natural distinction between electrons that form a spin liquid, and electrons that form a Fermi surface. Nevertheless, strong physical arguments were made for the existence of FL* even in a single band model \cite{RKK07,RKK08,Qi10,Moon11,Mei11,Punk12,Punk15,Punk18}. Numerical studies of the Hubbard model have not provided complete evidence for the FL* phase, but given the small size of the pockets in the Yamaji experiment, and the limited resolution of all numerical studies, this is perhaps not surprising. We will briefly review the theoretical arguments, and then turn to the ALM as the best current method of obtaining the FL* state in a single band model.

\begin{figure}
\centering
\includegraphics[width=6in]{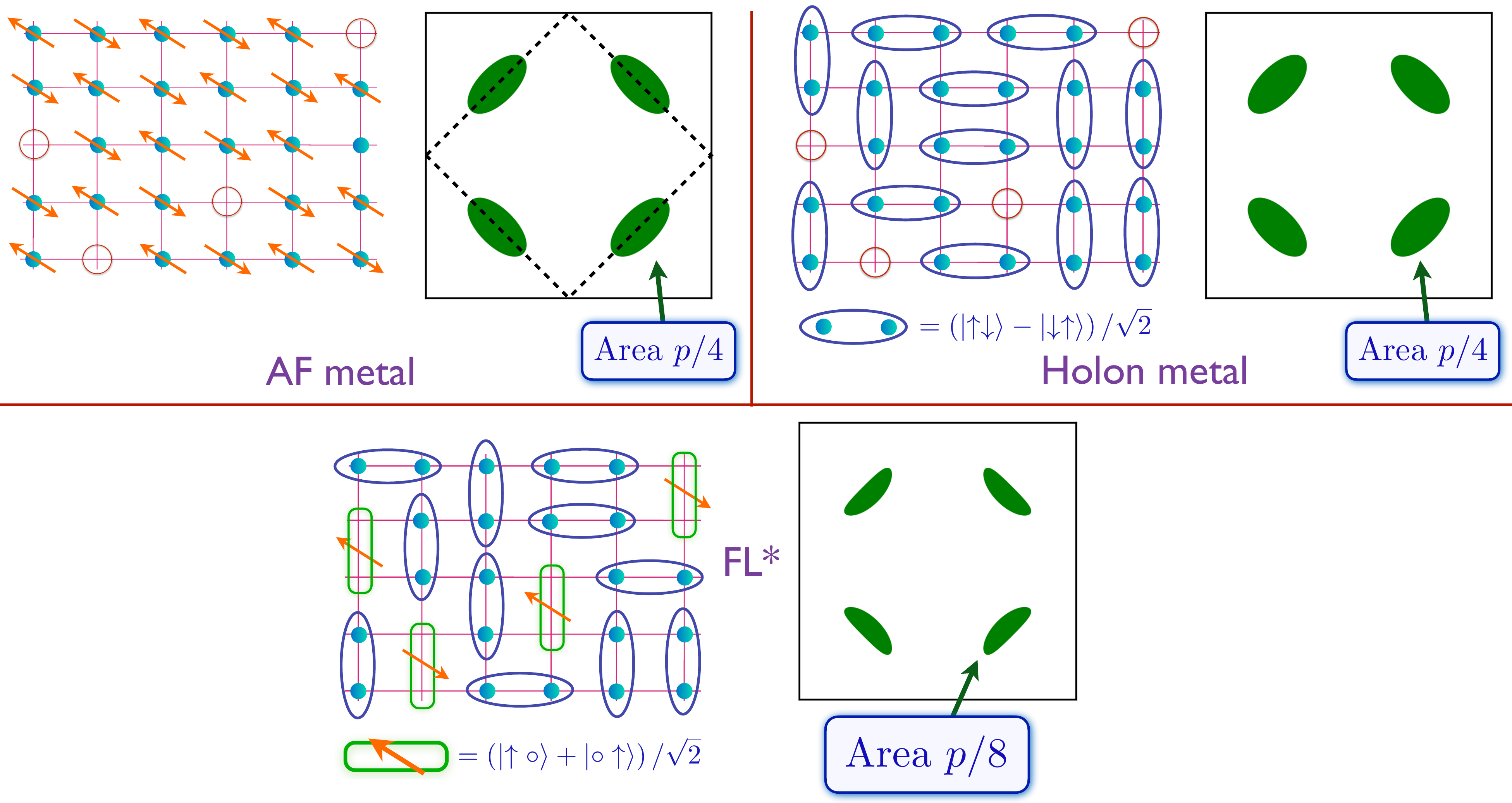}
\caption{Cartoon pictures of different states of doped antiferromagnets; adapted from Ref.~\cite{Punk15}. The areas are those that would be measured by a probe such as quantum oscillation. The AF metal has long-range antiferromagnetic order, and the reduced Brillouin zone is shown with the dashed line. The other phases do not break any symmetries. The open circles are holons, and these are assumed fermionic in the holon metal. The green dimers represent bound states of holons and spinons. The holon metal pockets do not have any symmetry associated with the reduced Brillouin zone of the AF metal. All 3 states have a Hall co-efficient of density $p$ positively charged carries.}
\label{fig:metals}
\end{figure}

At the level of cartoon pictures, we can illustrate the structure of a FL* state in Fig.~\ref{fig:metals} \cite{Punk15}.
This figure describes three distinct metallic phases in a Hubbard time model with electron density $1-p$. Recall that in the absence of a broken symmetry, the Luttinger constraint on hole Fermi surfaces is that they have a fractional area per spin of $(1+p)/2$, relative to the area of the full square lattice Brillouin zone.

\begin{itemize}
\item {\it AF Metal.} This is a state with antiferromagnetic long-range order. We can understand the Fermi surface by considering free electrons moving in a background with the same symmetry {\it i.e.\/} in a background spin-dependent potential which has a modulation at the wavevector $(\pi, \pi)$. This leads to the magnetic Brillouin zone boundary shown by the dashed line, and 4 hole pocket Fermi surfaces. Only two of these pockets are independent within the magnetic Brillouin zone. After accounting for a factor of 2 from spin, we conclude that the fractional area of each pocket is $p/4$. This Fermi surface area obeys the Luttinger constraint. Thermal fluctuations do not move Fermi surfaces, only broaden them, and so we expect that a fluctuating spin density wave state will also have pockets of area $p/4$ \cite{SchmalianPines1,SchmalianPines2,Chubukov23,Chubukov25}.
\item {\it Holon metal.} This is a state with no broken symmetry, in which the electrons have paired up in singlet bonds which resonate with each other. The dopants are realized by spinless mobile vacancies of charge $+e$, known as holons. The density of holons is $p$, and if the holons are fermions, they will form Fermi surfaces corresponding to spinless free fermions of density $p$. If there are four distinct Fermi surfaces in the Brillouin zone (as is the case in many computations), then the fractional area of each pocket will be $p/4$. Although this area is the same as that for the AF metal, the reason is very different. Now there is no broken symmetry, and the fermionic quasiparticles are spinless holons. This Fermi surface area does {\it not\/} obey the Luttinger constraint, and this is permitted because of the presence of the spin liquid. 
The holon metal pockets do not have reflection symmetry about the reduced Brillouin zone boundary shown for the AF metal.
A systematic, gauge theoretic treatment of the holon metal state is available in the literature \cite{sdw09,DCSS15b,DCSS15,CSS17,WuScheurer1,Scheurer:2017jcp,SSST19,WuScheurer2,Bonetti22,Bonetti23}, including a previous review by one of us \cite{Sachdev:2018ddg}, and applications to the Lieb lattice Hubbard model are discussed in Refs.~\cite{LiebScience,LiebPRB}. We note in passing that with the generalized definition in Fig.~\ref{fig:flsdef}, the holon metal is also a type of FL*.
\item {\it FL*.} Finally, we turn to the metallic state of interest. The holon metal state also has spinon excitations, which can be created out of the ground state. Now imagine a situation in which each holons gains energy by binding with a spinon, so that the system can pay the price for creating the spinons. 
In a $t$-$J$ model with electron hopping $t$, and exchange energy $J$, the cost for creating a spinon is order $J$, and the energy gain from the formation of the holon-spinon bound state is of order $t$, and so this bound state formation is very likely in the natural situation with $t \gg J$.
Then the ground state will change into one in which the mobile charge carriers are holon-spinon bound states \cite{Spinon-dopon05,RKK07,RKK08,Qi10,Sawatzky11,Moon11,Sawatzky11b,Mei11,Punk12,Punk15,Punk18,Grusdt18,Grusdt19,Grusdt23,Grusdt24,Balents25}. These bound states are always fermions with charge $+e$, spin $S=1/2$, just like a hole. Treating these holes as free fermions, we conclude that the total area of the Fermi surface should be $p/2$. If there are 4 distinct pockets (as there in the computation below), then each pocket will have the distinctive area of $p/8$. This Fermi surface area also does {\it not\/} obey the Luttinger constraint.
\end{itemize}
As we noted in Section~\ref{sec:intro_pseudogap}, and  will discuss further in Section~\ref{sec:pseudogap}, recent observations by Chan {\it et al.\/} \cite{Yamaji24} of the Yamaji effect in the cuprate HgBa$_2$CuO$_{4+\delta}$ show a fractional area of `approximately $1.3\%$' at a doping $p=0.1$, close to the value $p/8 = 1.25\%$ predicted for FL* \cite{TSSSMV03,TSSSMV04,RKK07,Qi10,Joshi23,Zhao_Yamaji_25}.
Moreover, the Yamaji effect requires coherent interlayer transport. This is not possible for the holon metals as the holons carry charges of distinct emergent gauge fields within each layer, but the FL* quasiparticles are gauge neutral and can indeed tunnel coherently between layers. In the AF metal coherent tunneling requires significant interlayer spin correlations, which are not observed \cite{Greven14,Greven25}.
All of these facts provide strong evidence in favor a FL* description of the pseudogap phase of the cuprates \cite{Zhao_Yamaji_25}.

We conclude this subsection by highlighting key features of the single-band FL* state of Ref.~\cite{Punk15} in Fig.~\ref{fig:metals}. We will see that these features will also play a key role in the Ancilla Layer Model of Section~\ref{sec:ancilla}, as we illustrate later in Fig.~\ref{fig:ancilla_dimer}.
\begin{tcolorbox}
  \begin{itemize}
    \item The electron quasiparticle of FL* is a (green) ``dimer'' in Fig.~\ref{fig:metals}, a bound state of a spin and vacancy.
    \item FL* has a background of spinon excitations obtained by breaking singlet bonds (the blue dimers) in Fig.~\ref{fig:metals}.
\end{itemize}  
\end{tcolorbox}

\subsubsection{Layer construction with ancilla qubits}
\label{sec:ancilla}

While much insight can be gained from the methods above, they fall short of providing a mean-field theory for the FL* phase, which could be used to study quantum phase transitions out of it. A suitable mean field theory can also lead to a trial wavefunction for the FL* state, which can be used for variational numerical computations (we will present such a wavefunction in Section~\ref{sec:wavefunction}).
To these ends, we now describe the Ancilla Layer Model (ALM) \cite{YaHui-ancilla1,YaHui-ancilla2}, which also easily ensures consistency with anomaly arguments. 

We wish to have a mean-field theory which changes a large hole-like Fermi surface of area $(1+p)/2$ to small hole-like Fermi surfaces of total area $p/2$. The simplest way to achieve this in mean-field theory is to hybridize the large Fermi surface with another band at half-filling (as in the Kondo lattice, and was done by Yang-Rice-Zhang (YRZ) \cite{YRZ,FuChun25})---this leads to a Fermi surface of the needed area $(1+p+1)/2$ (mod 1) $=p/2$. But we are {\it not allowed\/} to add a single band at half-filling because it is not `trivial' {\it i.e.} it is not anomaly-free, and its excitations cannot be integrated out because the extra band can only acquire gap with a broken symmetry. On the other hand, we {\it are allowed\/} to add two bands at half filling each (or any even number of bands), because they can form a trivial insulator with a gap. After the first added band hybridizes with the physical electrons to yield the small Fermi surfaces, the second added band is left decoupled, and it must form a spin liquid to satisfy the Oshikawa anomaly. This is the essence of the ALM.

\begin{figure}
\centering
\includegraphics[width=4.5in]{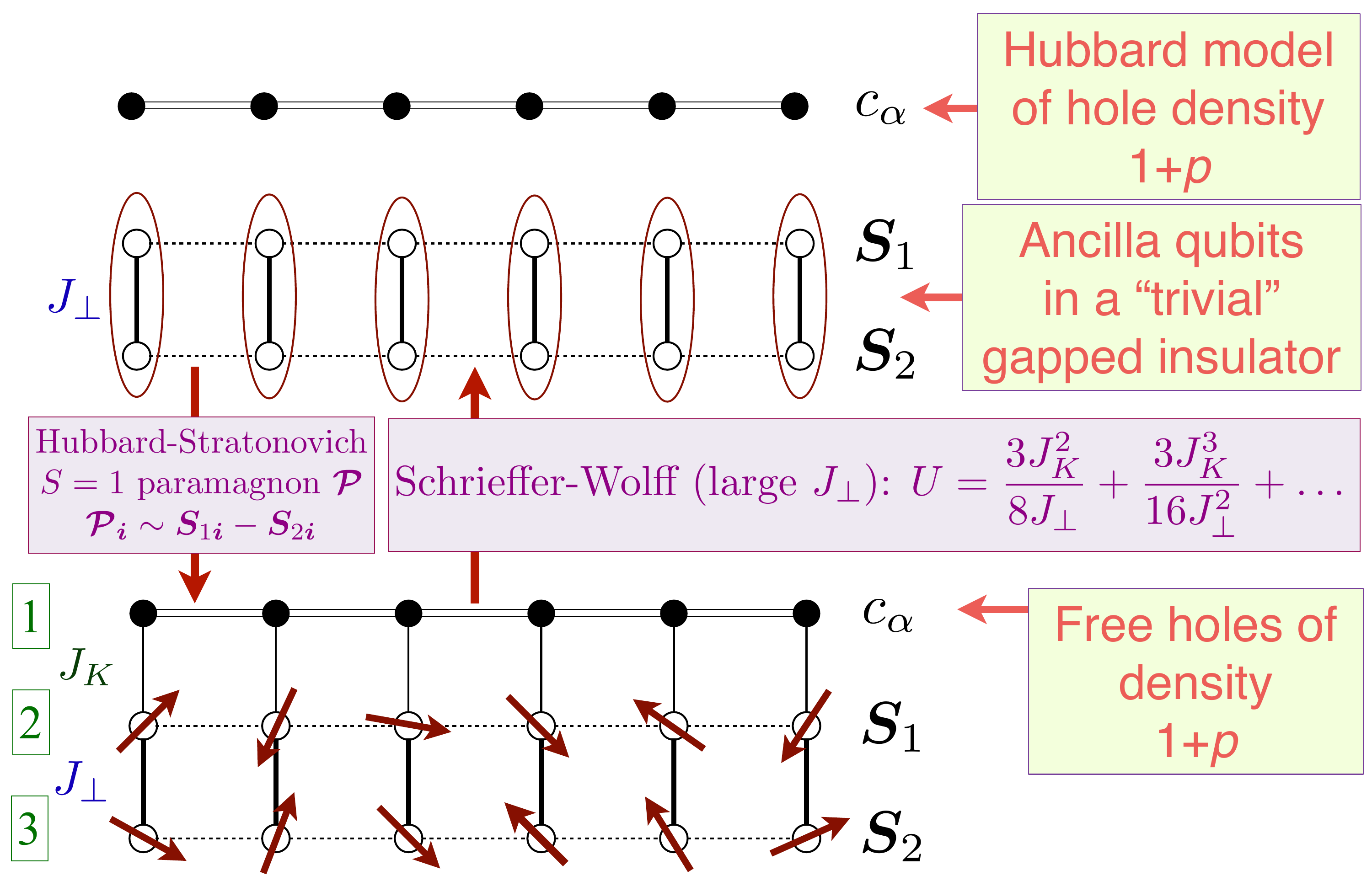}
\caption{Illustration of the mapping from a single-band Hubbard model with decoupled ancilla qubits, to a single band with free fermions coupled to a bilayer antiferromagnet. The Schrieffer-Wolff transformation is derived in Ref.~\cite{Nikolaenko:2021vlw} and reviewed in \hyperref[app:sw]{Appendix A}, and the Hubbard-Stratonovich transformation is derived in Ref.~\cite{Mascot22}. The layer numbers of the layer construction with ancilla qubits are indicated in the bottom picture; these appear in Tables~\ref{tab2} and \ref{tab3}.}
\label{fig:ancilla}
\end{figure}

We begin with a constructive derivation of the ALM starting from the familiar Hubbard model
\begin{eqnarray}
\mathcal{H}_{\rm Hubbard} &=&  -\sum_{\vi,\vj} t_{\vi\vj} c^\dagger_{\vi\alpha} c_{\vj\alpha} + U \sum_\vi 
(c_{\vi \uparrow}^\dagger c_{\vi \uparrow}) (c_{\vi \downarrow}^\dagger c_{\vi \downarrow})\,.
\label{eq:Hubbard}
\end{eqnarray}
Initially, we employ the familiar mapping of the Hubbard model to a theory of paramagnons coupled to the Fermi surface. We perform a Hubbard-Stratonovich transformation on the Hubbard interaction by a vector `paramagnon' field $\mathbfcal{P}_\vi$  \cite{Mascot22}. This is an exact transformation, which replaces the Hubbard interaction in Eq.~(\ref{eq:Hubbard}) by
    \begin{equation}
    \sum_{\vi} \left[ \frac{3U}{8} \mathbfcal{P}_\vi^2 + U\mathbfcal{P}_\vi \cdot c_{\vi \alpha}^\dagger \frac{{\bm \sigma}_{\alpha\beta}}{2} c_{\vi \beta} \right]\,. \label{PU}
    \end{equation}
    After some renormalization of the high energy states we give $\mathbfcal{P}$ some dynamics, and obtain an effective Lagrangian
    \begin{equation}
    \mathcal{L}[ \mathbfcal{P}] = \sum_\vi \left[ \frac{m_{\mathbfcal{P}}}{2} 
    \left( \partial_\tau \mathbfcal{P}_\vi \right)^2 + \frac{3U}{8}\mathbfcal{P}_\vi^2 \right] \,.
    \end{equation}
    This is a paramagnon theory with 3 local harmonic oscillators on each site. Now we take steps different from conventional paramagnon approaches.
    The ground state of the paramagnon theory is obtained when all three oscillators are in the $n=0$ state: $\left| 0,0,0 \right\rangle$. There is a triplet of first excited states:
    \begin{displaymath}
    \left| 1, 0, 0 \right\rangle  \sim \mathbfcal{P}_{\vi x} \left| 0,0,0 \right\rangle ; \left| 0, 1, 0 \right\rangle  \sim \mathbfcal{P}_{\vi y} \left| 0,0,0 \right\rangle ; \left| 0, 0, 1 \right\rangle  \sim \mathbfcal{P}_{\vi z} \left| 0,0,0 \right\rangle.
    \end{displaymath}
    We can map this low energy spectrum to that of a pair of $S=1/2$ ancilla spins with a mutual interaction $J_\perp {\bm S}_{1\vi} \cdot {\bm S}_{2\vi}$. By comparing the above matrix elements to those that couple the singlet and triplet spin states states, we obtain the operator identification
    \begin{equation}
    \mathbfcal{P}_\vi \sim {\bm S}_{1\vi} - {\bm S}_{2\vi}\,. \label{PS1S2}
    \end{equation}
    Inserting Eq.~(\ref{PS1S2}) into Eq.~(\ref{PU}), we obtain the Hamiltonian of the ALM, $\mathcal{H}_{\rm ALM}$, which is a simple augmentation of the Kondo lattice Hamiltonian $\mathcal{H}_{\rm KL}$ in Eq.~(\ref{eq:HKL}) by the layer of ${\bm S}_{2\vi}$ spins 
\begin{equation}
\mathcal{H}_{\rm ALM} 
= \mathcal{H}_{\rm KL}   + J_\perp \sum_{\vi} {\bm S}_{1\vi} \cdot {\bm S}_{2\vi}  + \sum_{\vi < \vj} J_{\vi\vj} \,{\bm S}_{2\vi} \cdot {\bm S}_{2\vj} \,.
\label{eq:Hancilla}
\end{equation}
This Hamiltonian is illustrated in the bottom half of Fig.~\ref{fig:ancilla}.
So we see that the ancilla spins are simply the fractionalization of the familiar $S=1$ paramagnon into a pair of $S=1/2$ spins \cite{Mascot22}.
The above derivation of the mapping to the ALM yields
 an additional ferromagnetic Kondo interaction between the electrons $c_{\vi\alpha}$ and  spins ${\bm S}_{2 \vi}$. Ferromagnetic Kondo couplings are expected to be irrelevant, and we will note this coupling again in Eq.~(\ref{Yukawa}).  
We have also explicitly included a direct exchange interaction between the ${\bm S}_{2\vi}$ spins, as it will be important for our purposes below. Note that there is no Hubbard interaction $U$ in $\mathcal{H}_{\rm ALM}$, and so we have replaced the $U$ in $\mathcal{H}_{\rm ancilla-decoupled}$ by the Kondo interaction $J_K$ in $\mathcal{H}_{\rm ALM}$. A complementary mapping from the ALM to the Hubbard model via the Schrieffer-Wolff transformation, also indicated in Fig.~\ref{fig:ancilla}, is described in \hyperref[app:sw]{Appendix A}.

While the derivations of $\mathcal{H}_{\rm ALM}$ are valid for large $J_\perp$, for general $J_\perp$ the mapping from the ALM to the Hubbard model requires multiple gauge fields, as is reviewed in \hyperref[app:ancilla]{Appendix B} and elsewhere \cite{QPMbook}. The additional gauge fields are higgsed in the underdoped region, and so we need not refer to them for now; we will mention them in Section~\ref{sec:strange} below Eq.~(\ref{e22}).

\begin{figure}
\centering
\includegraphics[width=5.75in]{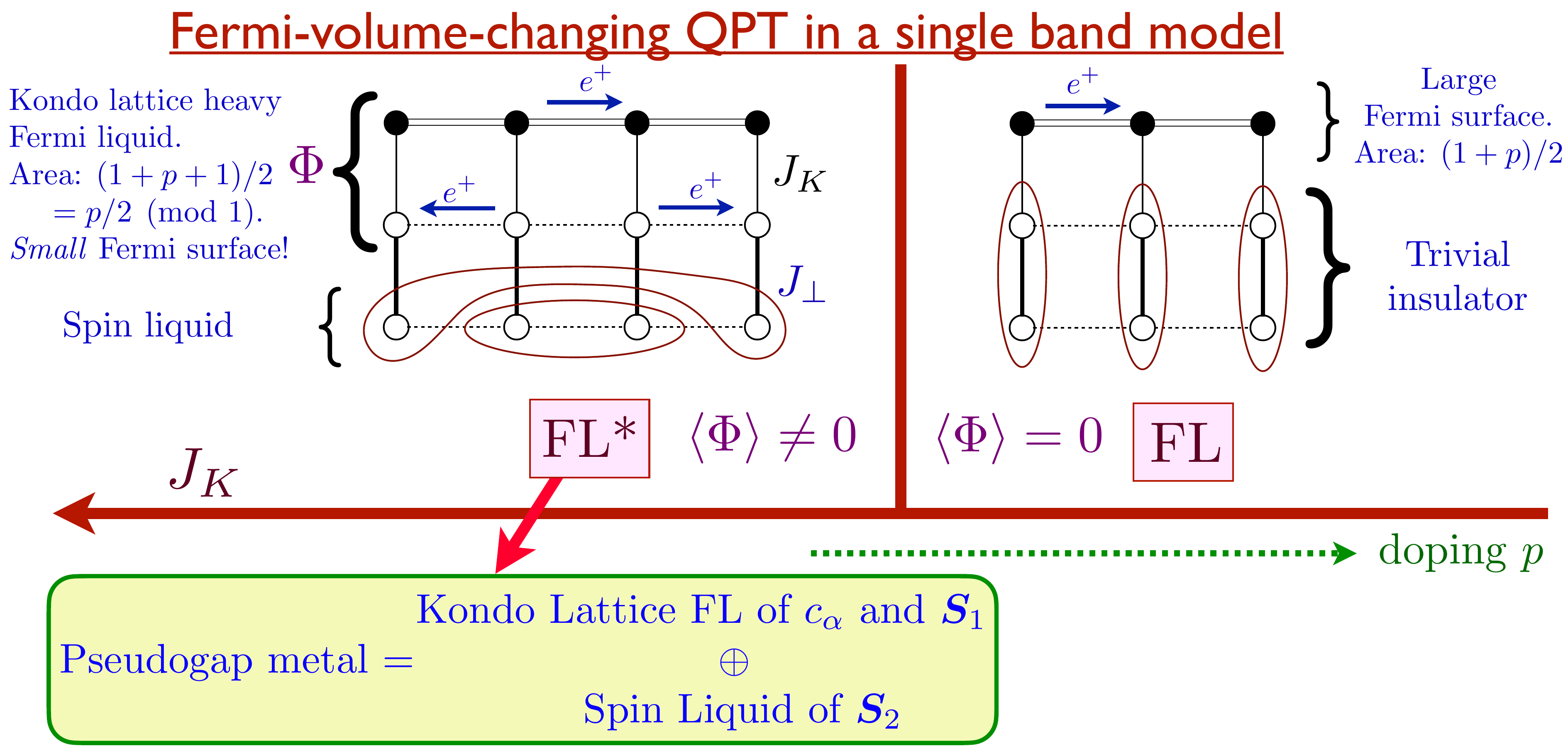}
\caption{Phases of the Ancilla Layer Model. The noted areas are per spin.
The phases are distinguished by the condensation of the boson $\Phi$, which hybridizes the conduction electrons in the top layer with the fermionic spinons in the middle layer. The ${\bm S}_2$ spin liquid will be described in Section~\ref{sec:spinliquids}.}
\label{fig:oneband}
\end{figure}

We are now ready to discuss the phase diagram of $\mathcal{H}_{\rm ALM}$ shown in Fig.~\ref{fig:oneband}.

At small $J_K$, the ancilla spins decouple into rung singlets, and we are back to a $c_{\alpha}$ state adiabatically connected to free electrons, which is the conventional Fermi liquid with a large hole pocket of area $(1+p)/2$.

At large $J_K$, we assume that the $c_\alpha$ electrons and ${\bm S}_1$ spins realize the commonly observed FL phase of $\mathcal{H}_{\rm KL}$ in Fig.~\ref{fig:klpd}. With the density of $c_\alpha$ equal to $1+p$, the total density of the `large' Fermi surface is $2+p$. As a trivial filled band with density 2 can always be removed, we obtain `small' Fermi surface associated with density $p$ in the FL phase of $\mathcal{H}_{\rm KL}$. So the total area of hole pockets is $p/2$.
While this is the Luttinger value for $\mathcal{H}_{\rm KL}$, it is {\it not\/} for $\mathcal{H}_{\rm ALM}$. The ${\bm S}_{2\vi}$ have an effective ferromagnetic Kondo coupling to the conduction electrons $c_\alpha$ (mediated by the antiferromagnetic $J_\perp$ and the antiferromagnetic $J_K$), and so we can expect them to decouple from the top two layers.
Indeed, we obtain a FL* phase for $\mathcal{H}_{\rm ALM}$ if the ${\bm S}_{2\vi}$ layer forms a spin liquid, and the effects of $J_\perp$ in Eq.~(\ref{eq:Hancilla}) can be treated perturbatively. This leads the identification in Fig.~\ref{fig:oneband} of the FL* pseudogap metal with the combination of a Kondo lattice FL state of $c_{\alpha}$ and ${\bm S}_1$, and a spin liquid of ${\bm S}_2$. 

We can now summarize the FL* construction of the Hubbard model in the following simple terms. We express the Hubbard model as a theory of antiferromagnetic paramagnons, and then fractionalize the $S=1$ paramagnon into two $S=1/2$ spins, ${\bm S}_1$ and ${\bm S}_2$. The ${\bm S}_1$ spins form a well-established Kondo lattice heavy Fermi liquid with the electrons $c_\alpha$ to yield the hole pockets. 
(We emphasize that it is the FL phase of the Kondo lattice which maps to the FL* phase of the ALM, as also noted in Fig.~\ref{fig:inverted}.)
The ${\bm S}_2$ spins form the spin liquid needed to satisfy the Oshikawa anomaly. Note that the ${\bm S}_2$ spins can form any spin liquid appropriate for $S=1/2$ spins, and we will not choose the spinon Fermi surface state for the bottom layer; instead, the suitable spin liquid will be introduced in Sections~\ref{sec:spinliquids}.

\subsubsection{Mean field theory of the cuprate pseudogap.}
\label{sec:sFL*}

We can now obtain a mean-field theory of the pseudogap by extending the Kondo lattice mean-field theory of Section~\ref{sec:kl} to the Hamiltonian $\mathcal{H}_{\rm ALM}$ in Eq.~(\ref{eq:Hancilla}).
We will discuss fluctuations about this mean-field later in Sections~\ref{sec:pseudogap} and~\ref{sec:dwave}.

We begin with the top two layers of electrons $c_\alpha$ and spins ${\bm S}_1$ in  the ALM in Fig.~\ref{fig:ancilla}.
We proceed with parton decomposition of ${\bm S}_1$ in terms of the fermionic spinons $f_{1\alpha}$ as in Eq.~(\ref{eq:S1f}), and obtain the mean-field fermion Kondo lattice Hamiltonian for the $c_\alpha$ and $f_{1\alpha}$ as $\mathcal{H}_{\rm KLmf}$ in  Eq.~(\ref{HKLmf}). At the mean-field level, we can assume the ${\bm S}_2$ spin liquid remains decoupled, and important effects of this spin liquid will be discussed in Sections~\ref{sec:pseudogap} and~\ref{sec:dwave}.

\begin{figure}
\centering
\includegraphics[width=4in]{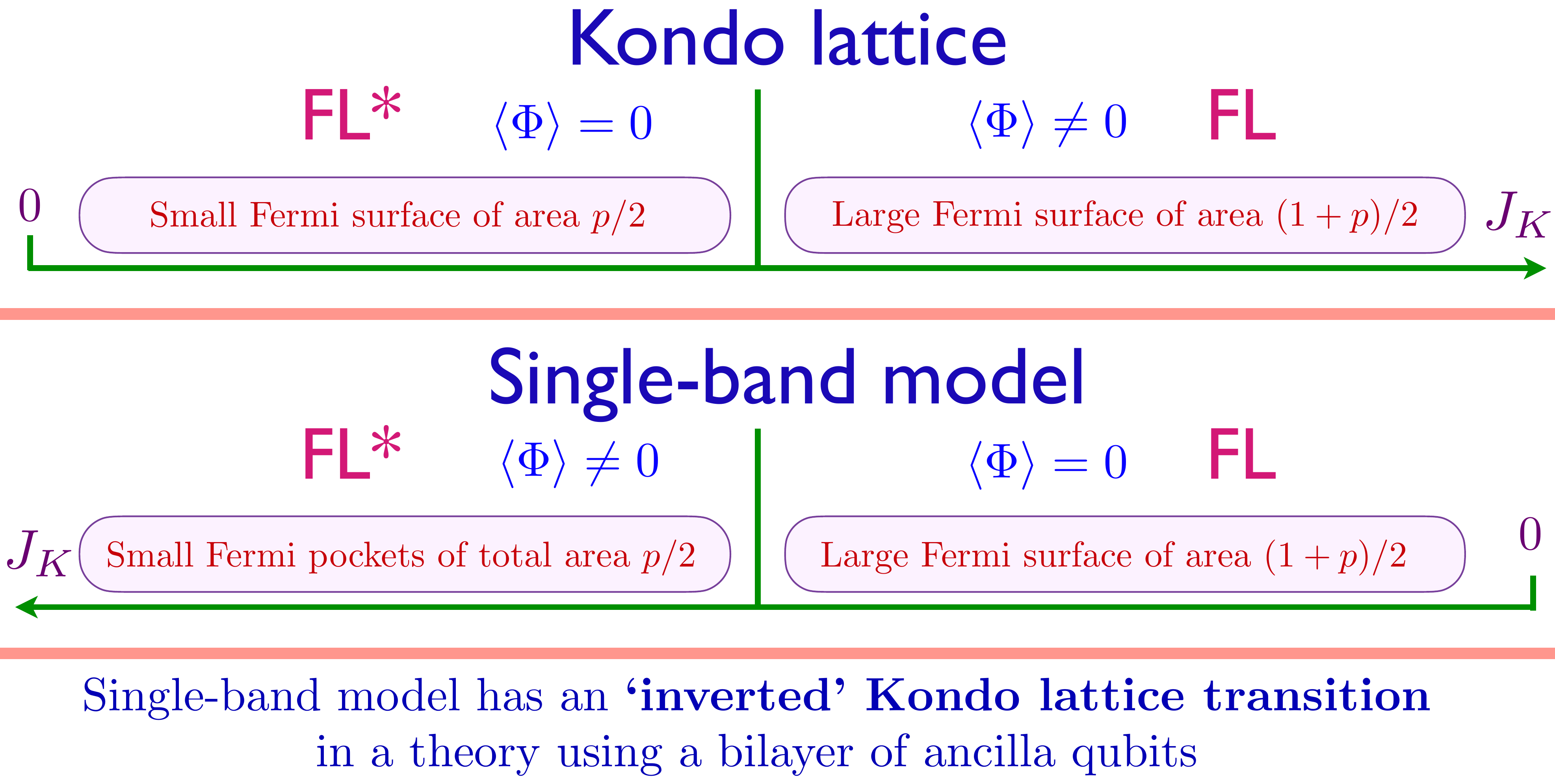}
\caption{Comparison of the phases of the Kondo lattice model in Fig.~\ref{fig:klpd}, and the ALM of the single band Hubbard model in Fig.~\ref{fig:oneband}. There is an inversion in the phase in which $\Phi$ is condensed: the well-established, $\Phi$-condensed, `large' Fermi surface phase of the Kondo lattice maps to the FL* pseudogap phase of the ALM.}
\label{fig:inverted}
\end{figure}

When considered as a theory of the Kondo lattice model $\mathcal{H}_{\rm KL}$, the FL state corresponds to the condensation of the decoupling field $\Phi$. On the other hand, for $\mathcal{H}_{\rm ALM}$, the $\Phi$ condensed phase is the FL* state. This interesting inversion is highlighted in Fig.~\ref{fig:inverted}: the single band model has an `inverted' Kondo lattice transition.
We assume in the remainder of this section, and in Sections~\ref{sec:pseudogap} and \ref{sec:dwave}, that $\Phi$ is a non-zero $c$-number. We will see in Section~\ref{sec:photo_pocket} that the magnitude of $\Phi$ determines the value of the electronic gap in the anti-nodal region of the Brillouin zone (near momenta $(\pi, 0)$, $(0, \pi)$). 

\begin{figure}
\centering
\includegraphics[width=6in]{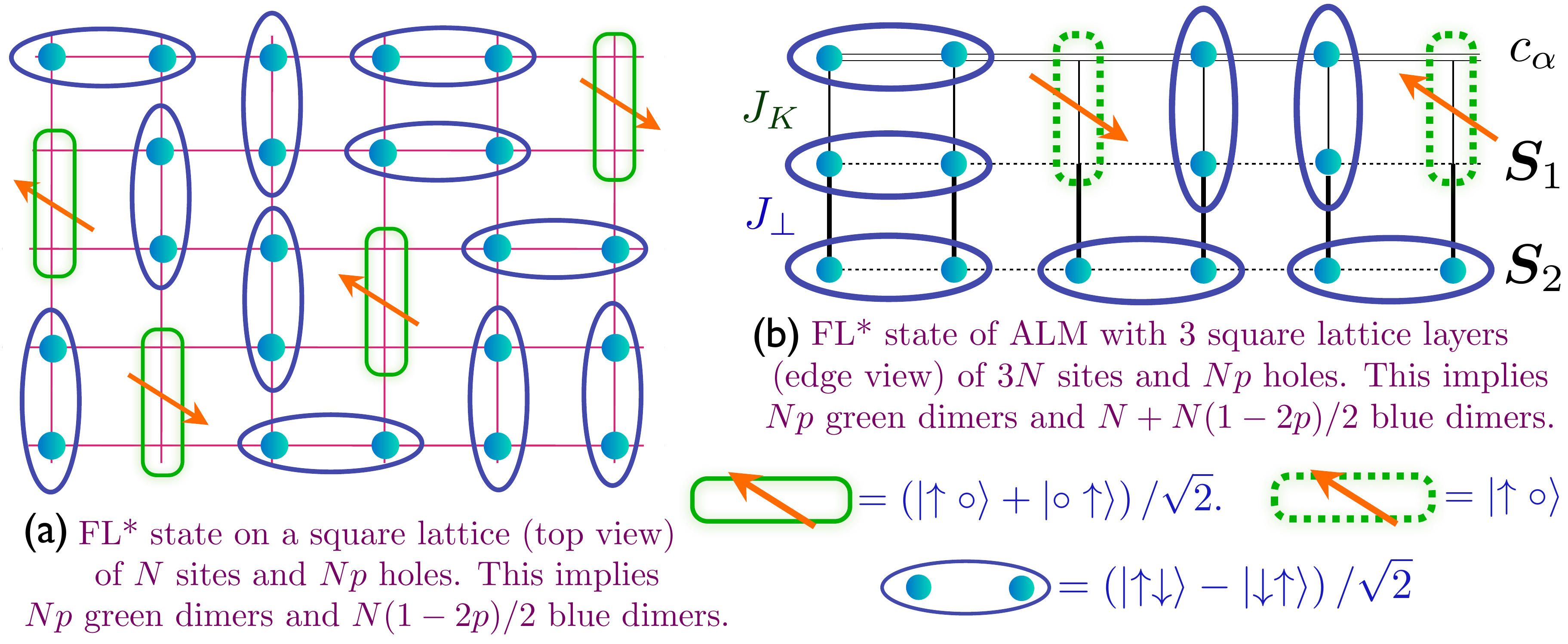}
\caption{Schematic comparison between (a) the quantum dimer theory of the square lattice FL* state \cite{Punk15} and the (b) ALM. The dashed green dimer is the state $c_{\vi\alpha}^{\vphantom\dagger} f_{1 \vi \alpha}^\dagger$ created by the hybridization $\Phi$ in the ALM. Note that the ALM has exactly $N$ blue dimers more than the dimer model, and these extra dimers are the number in a trivial rung-singlet bilayer antiferromagnet.}
\label{fig:ancilla_dimer}
\end{figure}

We can now present in Fig.~\ref{fig:ancilla_dimer} the close analogy between the FL* state obtained in the ALM and the state presented in Fig.~\ref{fig:metals} as highlighted in the box before the beginning of Section~\ref{sec:ancilla}
\begin{tcolorbox}
    \begin{itemize}
        \item 
        The electron quasiparticle of FL* is the hybridization of $c_\alpha$ and $f_{1\alpha}$ induced by $\Phi$, which creates the bound state $\sim c_{\vi\alpha}^{\vphantom\dagger} f_{1 \vi \alpha}^\dagger$ between a vacancy and a spin, analogous to the green dimers in Fig.~\ref{fig:metals}. 
%So we see that the interpretion of $f_{1\alpha}$ as a `correlation hole' in Ref.~\cite{YaHui-ancilla2} is closely connected to the quantum dimer model of Ref.~\cite{Punk15}.
        \item
        The background of spinons in FL* are obtained from the spin liquid of ${\bm S}_2$ spinons in the bottom layer, analogous the spinons obtained by breaking the blue dimers in Fig.~\ref{fig:metals}. We will later represent these spinons by $f_\alpha$ via 
        Eq.~(\ref{eq:S2f}).
    \end{itemize}
\end{tcolorbox}
A key observation is that the number of spin-singlet blue dimers in the ALM is exactly $N$ more than in the dimer model of Fig.~\ref{fig:metals}, while the number of green dimers is the same. This is acceptable because $N$ spin singlets is precisely the number that can be accommodated in a trivial rung-singlet state of a bilayer antiferromagnet.

The connection in Fig.~\ref{fig:ancilla_dimer} shows that the ALM is, in a sense, `minimal'. For a FL* mean field theory with both a small Fermi surface and spinons, we need the middle and bottom ancilla layers to provide the vacancy-spin and singlet valence bonds respectively. And a bonus, not available in the quantum dimer analysis of Ref.~\cite{Punk15,Punk18} or other approaches \cite{Spinon-dopon05,Mei11,Punk12}, is that it also provides a mean-field theory which captures the FL large Fermi when the bottom two layers are in a rung-singlet state.

\subsubsection{Wavefunction for FL* and cold atom observations.}
\label{sec:wavefunction}

A separate mean-field approach is to work with variational wavefunctions. The ALM was the first to provide a variational wavefunction for the FL* phase of the single-band Hubbard model. 
This approach couples all three layers together in
the proposed wavefunction \cite{YaHui-ancilla1}
\begin{eqnarray}
\left|{\rm FL*} \right\rangle &=&  \left[\mbox{Projection onto rung singlets of ${\bm S}_1, {\bm S}_2$} \right] \nonumber \\
&~&~~~~~~\bowtie \left| \mbox{Slater determinant of $(c,f_1)$} \right\rangle \nonumber \\
&~&~~~~~~~~~~~~~~\otimes \left| \mbox{Spin liquid of ${\bm S}_2$} \right\rangle \,. \label{wavefunction}
\end{eqnarray}
Here $f_1$ is obtained from the parton decomposition of ${\bm S}_1$ in Eq.~(\ref{eq:S1f}).
Note that this wavefunction depends only upon the co-ordinates of the $c$ electrons alone, as the co-ordinates of the $f,f_1$ fermions have been projected out. This wavefunction replaces the `vanilla' Gutzwiller projected wavefunctions of the $c$ electrons alone \cite{vanilla} in the underdoped region. We see rather explicitly in Eq.~(\ref{wavefunction}) the expressivity of the layered construction \cite{Hinton86}, allowing us to include both a small Fermi surface and a spin liquid in the same state.

The wavefunction Eq.~(\ref{wavefunction}) has been studied numerically in Refs.~\cite{Iqbal24,HenryShiwei24,YaHui24}, with the couplings in $\mathcal{H}_{\rm KLmf}$ treated as variational parameters.
The key variational parameter is the value of the Higgs field $\Phi$, which is non-zero in FL*, and vanishes in FL. 
The results of Refs.~\cite{Iqbal24,HenryShiwei24} successfully capture the evolution of 
local, multi-point spin and charge correlations with doping as measured in cold atom experiments on the square lattice fermionic Hubbard model \cite{Koepsell21,Bloch24,Kendrick:2025ujy}. The results of Ref.~\cite{YaHui24} compare well with the exact diagonalization of Hubbard models in one and two dimensions.

\subsubsection{Photoemission spectrum.}
\label{sec:photo_pocket}

We return to the mean-field theory of Section~\ref{sec:sFL*}, and describe the results of Mascot {\it et al.} \cite{Mascot22} who used $\mathcal{H}_{\rm KLmf}$ in Eq.~(\ref{HKLmf}) to model the photoemission spectra in the cuprates. Note that in such a Hamiltonian, the middle layer of the ALM is assumed to be a spinon Fermi surface (with hopping $t_{1,\vi\vj}$), which hybridizes with the top layer of electrons with the condensed Higgs boson $\Phi$. The condensation of $\Phi$ quenches the U(1) gauge field of the spinon Fermi surface.

The value $\Phi$ was fixed by the magnitude of the fermion pseudogap near wavevectors $(0,\pi)$, $(\pi, 0)$, the values of the $t_{\vi\vj}$ are known from observations at large doping, and only the values of the $t_{1,\vi\vj}$ were fitting parameters. The results provide a good fit to observations over a wide range of frequency across the Brillouin zone. 

\begin{figure}
\centering
\includegraphics[width=3.5in]{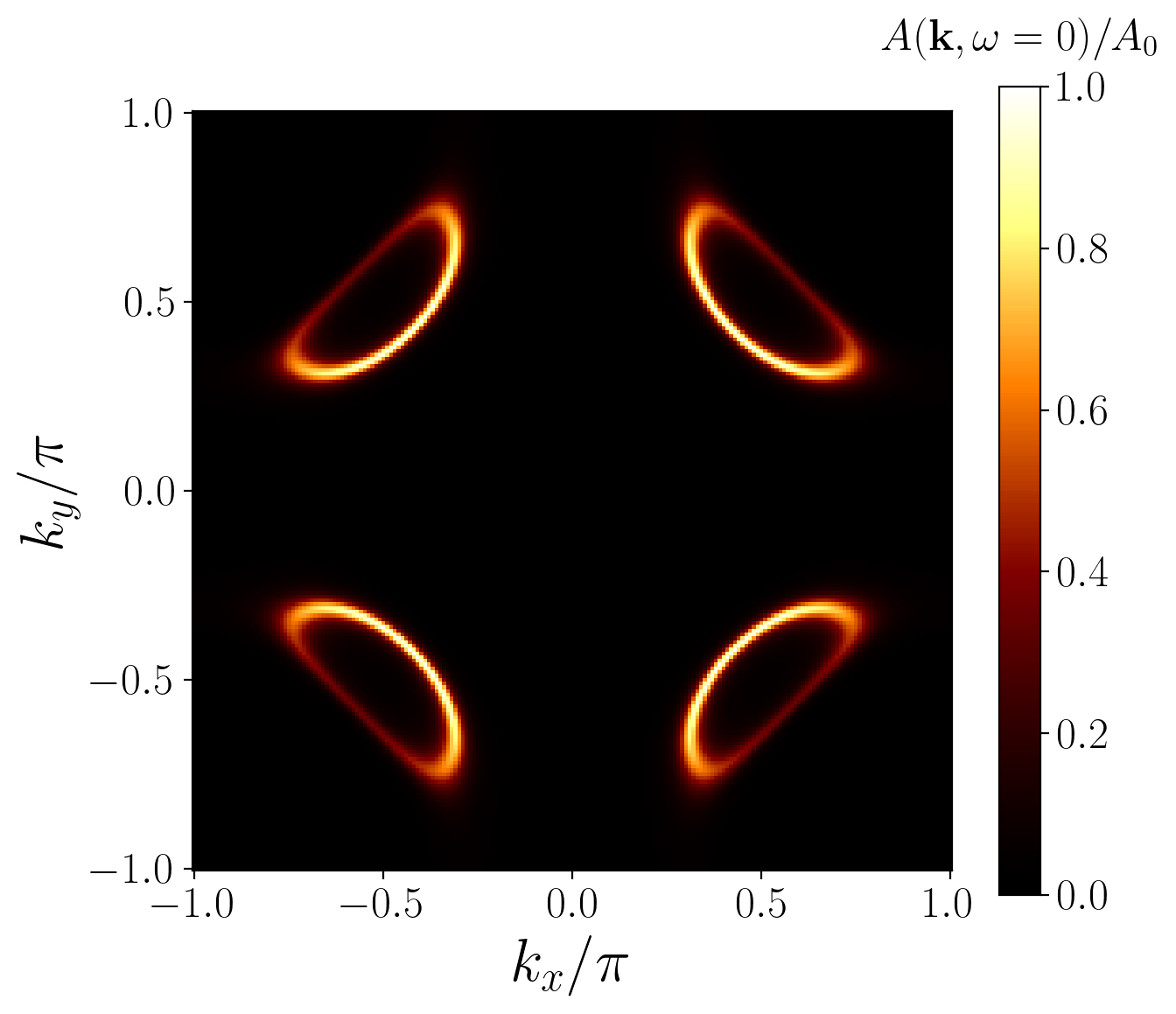}
\caption{Zero energy electronic spectral weight (as would be measured in photoemission) in the ancilla mean field theory of the FL* phase, from Ref.~\cite{Mascot22}. 
Fluctuation corrections appear later in Figs.~\ref{fig:arc} and \ref{fig:Osc}, where they lead to Fermi arcs. }
\label{fig:photo}
\end{figure}

We show the theory for the zero energy photoemission spectral weight in Fig.~\ref{fig:photo}.
There are 4 hole pockets, each of fractional area $p/8$, which are obtained from the hybridization of the $c_\alpha$ band of density $1+p$ and the $f_{1 \alpha}$ band of density 1 in Eq.~(\ref{HKLmf}). (In contrast, in spin density wave theory, we would hybridize the $c_\alpha$ band of density $1+p$, with a $(\pi, \pi)$ momentum-shifted $c_\alpha$ band also of density $1+p$ to obtain pockets of area $p/4$.)
The pockets of Fig.~\ref{fig:photo} do have faint, but visible, `backsides'. In contrast, the observations show only `Fermi arcs' corresponding to the front sides of the pockets (see Fig.~\ref{fig:arc}a later), with no clearly visible backsides (although there are some indications of backsides \cite{Johnson08,Johnson10,YRZEPL,Johnson11,PDJ18}). We will consider the influence of thermal fluctuations on this photoemission spectrum in Section~\ref{sec:arc}, where we will see that they can indeed produce the observed Fermi arcs in the intermediate temperature pseudogap.

\begin{figure}
\centering
\includegraphics[width=6in]{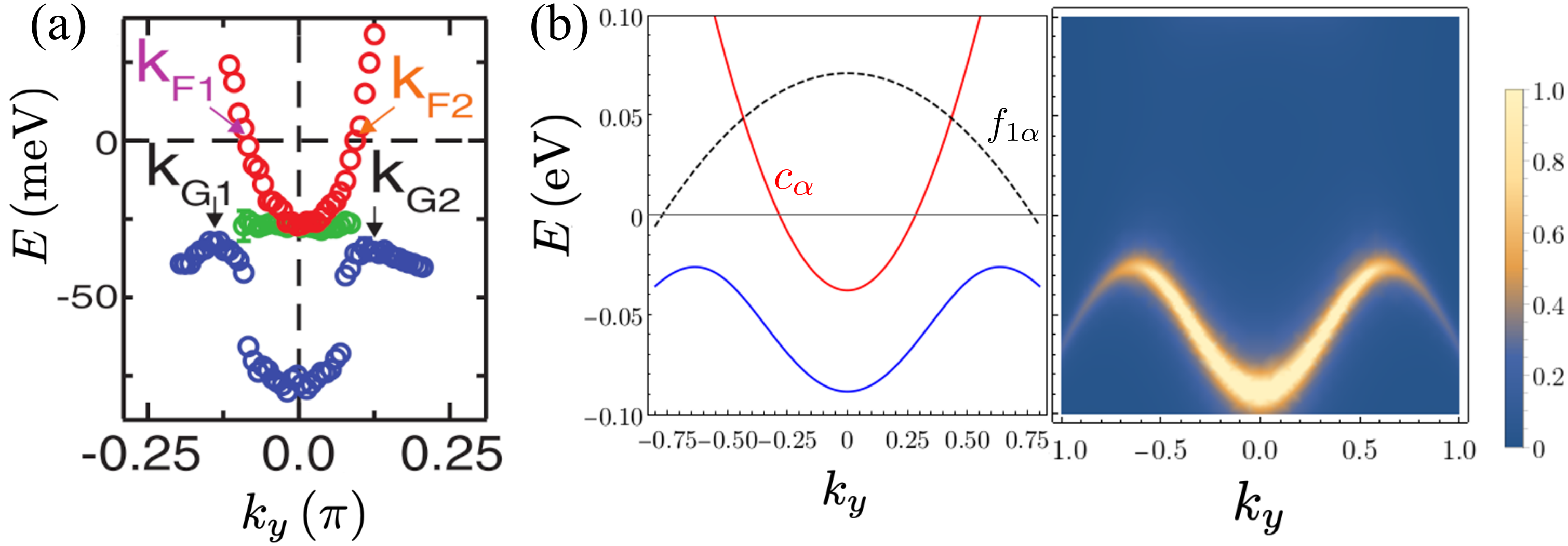}
\caption{(a) Photoemission data from Ref.~\cite{Shen11} along momenta $(\pi, k_y)$. The red points are in a region where there is no pseudogap, and so the electronic dispersion crosses the Fermi energy. The blue points show the gapped electronic excitation in the pseudogap.  The green points identify the position of a shoulder in the energy distribution curve, presumed to be the superconducting gap. (b) From Ref.~\cite{Mascot22}. The red line is the dispersion of $c_\alpha$ when $\Phi=0$ in Eq.~(\ref{HKLmf}), to be matched with the red points in the left panel. The blue line is the dispersion of $f_{1 \alpha}$ also when $\Phi = 0$ in Eq.~(\ref{HKLmf}). The blue line is the dispersion of the gapped electronic excitation when $\Phi \neq 0$ in Eq.~(\ref{HKLmf}), to be matched with the blue points in the left panel.}
\label{fig:antinode}
\end{figure}

In the antinodal region, there is a good match with the dispersion of the gapped electronic excitations, as show in Fig.~\ref{fig:antinode}.
The observations \cite{Shen11} have the notable feature that the Fermi wavevectors ($k_{F1,2}$ in Fig.~\ref{fig:antinode}a) in the Fermi surface without the pseudogap are distinct from the positions of the minimum pseudogap ($k_{G1,2}$ in Fig.~\ref{fig:antinode}a); these wavevectors would be the same if Cooper-pairing was the origin of the pseudogap. In our theory, the distinction is enabled by the distinct sizes of the $c_\alpha$ and $f_{1\alpha}$ Fermi wavevectors shown in Fig.~\ref{fig:antinode}b. A related model for the anti-nodal spectrum has been discussed in Ref.~\cite{Tsvelik_review}.

The hybridization of the $c_\alpha$ and $f_{1\alpha}$ bands in the ALM also produces a band above the Fermi level, and its momentum integrated consequences can be detected by STM. Mascot {\it et al.\/} \cite{Mascot22} presented theoretical results for the momentum integrated spectrum, and the band gap above the Fermi energy resulted in a soft, asymmetric pseudogap at positive bias in the local density of states. There is evidence for such an asymmetry in the pseudogap in STM measurements of the underdoped cuprates by Jhinhwan Lee {\it et al.\/} \cite{Davis09}: they observe a $T$-dependent evolution from a symmetric superconducting gap at low $T$, to an asymmetric pseudogap with a minimum at positive bias for $T>T_c$. Mascot {\it et al.}~\cite{Mascot22} also found a peak in the local density of states above the pseudogap from the band above the Fermi level.

\subsubsection{Magnetotransport.}
\label{sec:mtmf}

More recently, compelling evidence for hole pockets has emerged at low doping in the higher temperature pseudogap metal phase \cite{Ramshaw22,Yamaji24}.
Chan {\it et al\/} \cite{Yamaji24} have observed the remarkable Yamaji effect. 

The Yamaji effect requires coherent interlayer transport of the charge carriers, and this already places stringent constraints on the model of the pseudogap, and favoring FL*. The holon metal does not have coherent interlayer transport, while the fluctuating spin density wave state \cite{SchmalianPines1,SchmalianPines2,Chubukov23,Chubukov25} does not have significant interlayer spin correlations \cite{Greven14,Greven25}.
needed for coherent transport between SDW pockets.

Moreover, the Yamaji effect 
allows deduction of pocket area only from a knowledge of the layer spacing and the value of the Yamaji angle at which there is peak in the magnetoresistance.
At these higher temperatures there is no ordering, and the Hall effect is positive, indicating the absence of electron pockets. Their results are consistent with the FL* prediction of hole pockets of area $p/8$ \cite{TSSSMV03,TSSSMV04,RKK07,Qi10,Joshi23,Zhao_Yamaji_25,FuChun25} which can tunnel coherently between layers, as we noted earlier near Fig.~\ref{fig:metals}. 
This is to be contrasted with pockets of area $p/4$ in the SDW state.

The Yamaji angle observation of Chan {\it et al.\/} \cite{Yamaji24} is therefore the long-sought direct signature of fractionalization in the cuprates.

\section{Critical spin liquid on the square lattice}
\label{sec:spinliquids}

Our FL* theory of the pseudogap has led us to a mean-field theory described by the mean-field Kondo lattice Hamiltonian in Eq.~(\ref{HKLmf}), and a decoupled spin liquid of ${\bm S}_2$ spins in the {\it bottom\/} layer. The next step requires us to couple these sectors to each other in the ALM. In the present section, we consider the physics of the ${\bm S}_2$ layer alone by discussing the theory of spin liquids on the square lattice. 
We chose a spinon Fermi surface for the spin liquid of the ${\bm S}_1$ spins in the {\it middle\/} layer in Section~\ref{sec:singleband} to obtain the hole pockets by hybridizing it to the large Fermi surface of the top layer.
The power of the ALM method is that it gives us the flexibility to make a different choice for the spin liquid of the ${\bm S}_2$ spins in the bottom layer---we will choose it from studies of insulating square lattice antiferromagnets.

We drop the subscript 2 on the spin operators in the present section, and consider the general case of ${\bm S}_\vi$ being spin $S$ 
quantum spin operators on the sites, $\vi$, of
a square lattice with Hamiltonian
\begin{equation}
{\cal H}_J = \sum_{\vi < \vj} J_{\vi\vj} \,{\bm S}_\vi \cdot {\bm S}_\vj\,.
\label{hamil}
\end{equation}
The $J_{\vi\vj}$ are short-ranged antiferromagnetic
exchange interactions. We will mainly consider here the square lattice with nearest neighbor
interactions, but the methods generalize to a wide class of lattices and interaction ranges. 

We will begin in Section~\ref{sec:sbparton} by employing a method which fractionalizes the spin operator into bosonic partons. This leads to the low energy U(1) gauge theory with complex scalars in Eq.~(\ref{cp3}), and  to the phase diagram in Fig.~\ref{fig:Neelvbs}.  This phase diagram is in good agreement with numerical studies \cite{Becca20,Imada21,Gu24} of the square lattice antiferromagnet with first and second neighbor exchange (the $J_1$-$J_2$ model), and also of Sandvik's $J$-$Q$ model \cite{Sandvik24}.

Section~\ref{sec:fermions} fractionalizes the spin operator into fermionic partons. This leads ultimately to a seemingly different low energy theory: a SU(2) gauge theory with massless Dirac fermions in Eq.~(\ref{eq:fermionhop2}), which initially did not agree with numerical studies of square lattice antiferromagnets.
But we will argue, following Wang {\it et al.\/} \cite{Wang17}, that the bosonic and fermionic theories are equivalent, and that the confining phases of the SU(2) gauge theory lead to the same phase diagram as the bosonic partons. This equivalence is powerful, as it yields a toolbox of different approaches to study spin liquids. In particular, upon including charge fluctuations, it is the fermionic parton theory that allows study of the confining instability of the square lattice spin liquid to $d$-wave superconductivity that we study in Sections~\ref{sec:halffilling} and \ref{sec:aniso}.

\subsection{Bosonic partons}
\label{sec:sbparton}

A careful examination of the non-magnetic `spin-liquid' phases requires an
approach which is designed explicitly to be valid in a region well
separated from N\'eel long range order, and preserves SU(2) symmetry
at all stages. It should also be designed to naturally allow for neutral $S=1/2$
excitations. To this end, we introduce the Schwinger
boson description \cite{AA88}, in terms of elementary $S=1/2$ bosons.
For the group SU(2) the complete set of $(2S+1)$
states on site $\vi$ are represented as follows
\begin{equation}
|S , m \rangle \equiv \frac{1}{\sqrt{(S+m)! (S-m)!}}
(b_{\vi\uparrow}^{\dagger})^{S+m}
(b_{\vi\downarrow}^{\dagger})^{S-m} | 0 \rangle,
\end{equation}
where $m = -S, \ldots S$ is the $z$ component of the spin ($2m$ is an integer).
We have introduced two flavors of Schwinger bosons on each site, \index{Schwinger boson}
created by the canonical operator
$b_{\vi\alpha}^{\dagger}$, with $\alpha = \uparrow, \downarrow$, and
$|0\rangle$ is the vacuum with no Schwinger bosons. The total number of Schwinger bosons, $n_b$,
is the same for all the states; therefore
\begin{equation}
b_{\vi\alpha}^{\dagger}b_{\vi \alpha} = n_b
\label{boseconst}
\end{equation}
with 
\begin{equation}
n_b = 2S\,.
\end{equation}
The above
representation of the states is completely equivalent to the 
operator identity between the spin and Schwinger boson operators
\begin{equation}
{\bm S}_{\vi} =  \frac{1}{2}
b_{\vi\alpha}^{\dagger}\, {\bm \sigma}_{\alpha\beta} \, b_{\vi\beta} \label{sparton}
\end{equation}
where $\ell=x,y,z$ and the ${\bm \sigma}$ are the usual $2\times 2$ Pauli
matrices. 

The spin-states on two sites $\vi,\vj$ can combine to form a singlet in a unique
manner - the wavefunction of the (unnormalized) singlet state is particularly simple
in the boson formulation:
\begin{equation}
\left( \varepsilon_{\alpha\beta} b_{\vi\alpha}^{\dagger}
b_{\vj\beta}^{\dagger} \right)^{2S} |0\rangle
\end{equation}
Also, using the constraint in Eq.~(\ref{boseconst}), the
following Fierz-type identity can be established
\begin{equation}
\left( \varepsilon_{\alpha \beta}
b_{\vi \alpha}^{\dagger} b_{\vj \beta}^{\dagger} \right)
\left( \varepsilon_{\gamma \delta}
b_{\vi\gamma} b_{\vj\delta} \right) =
 - 2 {\bm S}_\vi \cdot {\bm S}_\vj
+ n_b^2 /2 + \delta_{\vi\vj} n_b
\label{su2}
\end{equation}
where $\varepsilon$ is the totally antisymmetric $2\times2$ tensor
\begin{equation}
\varepsilon = \left( \begin{array}{cc}
0 & 1 \\
-1 & 0 \end{array} \right)\,.
\end{equation}
This implies that ${\cal H}_J$ can be rewritten in the form (apart from
an additive constant)
\begin{equation}
{\cal H}_J = - \frac{1}{2} \sum_{\vi < \vj} J_{\vi\vj} \Bigl( \varepsilon_{\alpha \beta}
b_{\vi \alpha}^{\dagger} b_{\vj \beta}^{\dagger} \Bigr)
\Bigl( \varepsilon_{\gamma \delta}
b_{\vi\gamma} b_{\vj\delta} \Bigr)
\label{hafkag}
\end{equation}
This form makes it clear that ${\cal H}_J$ counts the number of singlet
bonds.

\subsubsection{Mean-field theory}
\label{sec:6A}

We begin by the coherent state path integral of ${\cal H}_J$ in imaginary time $\tau$ at a temperature $\beta = 1/T$
\begin{equation}
\mathcal{Z}_J = \int {\cal D} \mathcal{Q} {\cal D} b  {\cal D} \lambda
\exp \left( - \int_0^{\beta} {\cal L}_J \, d\tau \right) ,
\label{zfunct0}
\end{equation}
where
\begin{displaymath}
{\cal L}_J = \sum_{\vi} \left [
b_{\vi\alpha }^{\dagger}  \left( \frac{d}{d\tau} + i\lambda_\vi \right)
b_{\vi\alpha} - i\lambda_\vi n_b \right ]~~~~~~~~~~~~~~~~~~~~~~
\end{displaymath}
\begin{equation}
~~~~~~~~~~~+  \sum_{<\vi,\vj>} \left [
\frac{J_{\vi\vj} |\mathcal{Q}_{\vi,\vj} |^2}{2}
- \frac{J_{\vi\vj} \mathcal{Q}_{\vi,\vj}^{\ast}}{2} \varepsilon_{\alpha\beta} b_{\vi\alpha}
b_{\vj\beta}
+ H.c. \right] .
\label{zfunct}
\end{equation}
Here the $\lambda_\vi$
fix the boson number of $n_b$ at each site;
$\tau$-dependence of all fields is implicit; $\mathcal{Q}$ was introduced by
a Hubbard-Stratonovich decoupling of ${\cal H}_J$. 

This procedure is similar to that employed in deriving the Landau-Ginzburg theory of superconductivity from electron pairing, with the crucial difference that now the Lagrangian
${\cal L}_J$ has a $U(1)$ gauge invariance under which
\begin{eqnarray}
b_{\vi\alpha}^{\dagger} & \rightarrow & b_{\vi\alpha}^{\dagger} 
\exp \left( i\rho_\vi (\tau ) \right ) \nonumber \\
\mathcal{Q}_{\vi,\vj} &\rightarrow & \mathcal{Q}_{\vi,\vj} \exp \left( - i \rho_\vi (\tau ) - i
\rho_\vj (\tau ) \right) \nonumber \\
\lambda_\vi & \rightarrow & \lambda_\vi + \frac{\partial \rho_\vi}{\partial
\tau} (\tau) \,.
\label{gaugetrans}
\end{eqnarray}
The functional integral over ${\cal L}_J$
faithfully represents the partition function, but does require gauge fixing. This gauge invariance leads to emergent gauge field degrees of freedom, as we will see below.

We begin with mean-field saddle point of $\mathcal{Z}_J$ over the path integrals of $\mathcal{Q}$ and $\lambda$. The saddle-point approximation is valid in the limit of a large number of spin flavors, but we do not explore this here.
With the saddle point values $\mathcal{Q}_{\vi \vj} = \bar{\mathcal{Q}}_{\vi\vj}$, $i \lambda_\vi = \bar{\lambda}_\vi$ we obtain a 
mean-field Hamiltonian for the $b_{\vi \alpha}$
\begin{eqnarray}
{\cal H}_{J,MF} = &  \sum_{<\vi,\vj>} \left(
\frac{J_{\vi\vj} |\bar{\mathcal{Q}}_{\vi\vj} |^2}{2}
- \frac{J_{\vi\vj} \bar{\mathcal{Q}}_{\vi\vj}^{\ast}}{2} \varepsilon_{\alpha\beta} b_\vi^{\alpha}
b_\vj^{\beta}
+ H.c. \right) \nonumber \\
&
+ \sum_{\vi} \bar{\lambda}_\vi (
b_{\vi\alpha }^{\dagger} b_{\vi\alpha} -  n_b )\,.
\label{HMF}
\end{eqnarray}
This Hamiltonian is quadratic in the boson operators and all its
eigenvalues can be determined by a Bogoluibov transformation. This
leads in general to an expression of the form
\begin{equation}
{\cal H}_{J,MF} = E_{J,MF}[ \bar{\mathcal{Q}} , \bar{\lambda}] + \sum_{\mu} \omega_{\mu} [\bar{\mathcal{Q}} , \bar{\lambda}]
\gamma_{\mu\alpha}^{\dagger} \gamma_{\mu\alpha}
\end{equation}
The index $\mu$ extends over $1\ldots$number of sites in the system,
$E_{J,MF}$ is the ground state energy and is a functional of $\bar{\mathcal{Q}}$,
$\bar{\lambda}$, $\omega_{\mu}$ is the eigenspectrum of excitation energies
which is also a function of $\bar{\mathcal{Q}}$, $\bar{\lambda}$, and the
$\gamma_{\mu}^{\alpha}$ represent the bosonic eigenoperators. The
excitation spectrum thus consists of non-interacting spinor bosons.
The ground state is determined by minimizing $E_{J,MF}$ with respect to
the $\bar{\mathcal{Q}}_{\vi\vj}$ subject to the constraints
\begin{equation}
\frac{\partial E_{MF}}{\partial \bar{\lambda}_\vi } = 0
\label{sp1}
\end{equation}
The saddle-point value of the $\bar{\mathcal{Q}}$ satisfies
\begin{equation}
\bar{\mathcal{Q}}_{\vi\vj} = \langle \varepsilon_{\alpha\beta} b_{\vi\alpha} b_{\vj\beta}
\rangle
\label{sp2}
\end{equation}
Note that $\bar{\mathcal{Q}}_{\vi\vj} = - \bar{\mathcal{Q}}_{ji}$ indicating that $\bar{\mathcal{Q}}_{\vi\vj}$ is a
directed field - an orientation has to be chosen on every link.

These saddle-point equations have been solved for the square lattice with nearest neighbor exchange $J$, 
and they lead to stable 
and translationally invariant solutions for $\bar{\lambda}_\vi$ and $\bar{\mathcal{Q}}_{\vi\vj}$. The only saddle-point quantity which does not
have the full symmetry of the lattice is the orientation of the $\bar{\mathcal{Q}}_{\vi\vj}$.
Note that although it appears that such a choice of orientation appears to break inversion or reflection symmetries, such symmetries
are actually preserved: the $\bar{\mathcal{Q}}_{\vi\vj}$ are not gauge-invariant, and all gauge-invariant observables do preserve all symmetries of the 
underlying Hamiltonian. For the square lattice, we have $\bar{\lambda}_\vi = \bar{\lambda}$, $\bar{\mathcal{Q}}_{\vi,i+\hat{x}} = \bar{\mathcal{Q}}_{\vi,i+\hat{y}} = \bar{\mathcal{Q}}$.

We can also compute the dispersion $\omega_{\bm k}$ of the $\gamma_{\bm k}$ excitations. These are bosonic particles which carry
spin $S=1/2$ (`spinons'). Their dispersion is
\begin{equation}
\omega_{\bm k} = \left( \bar{\lambda}^2 - J^2 \bar{\mathcal{Q}}^2 (\sin k_x + \sin k_y)^2 \right)^{1/2}\,. \label{eq:bosondisp}
\end{equation}
\begin{figure}
\centering
\includegraphics[width=4in]{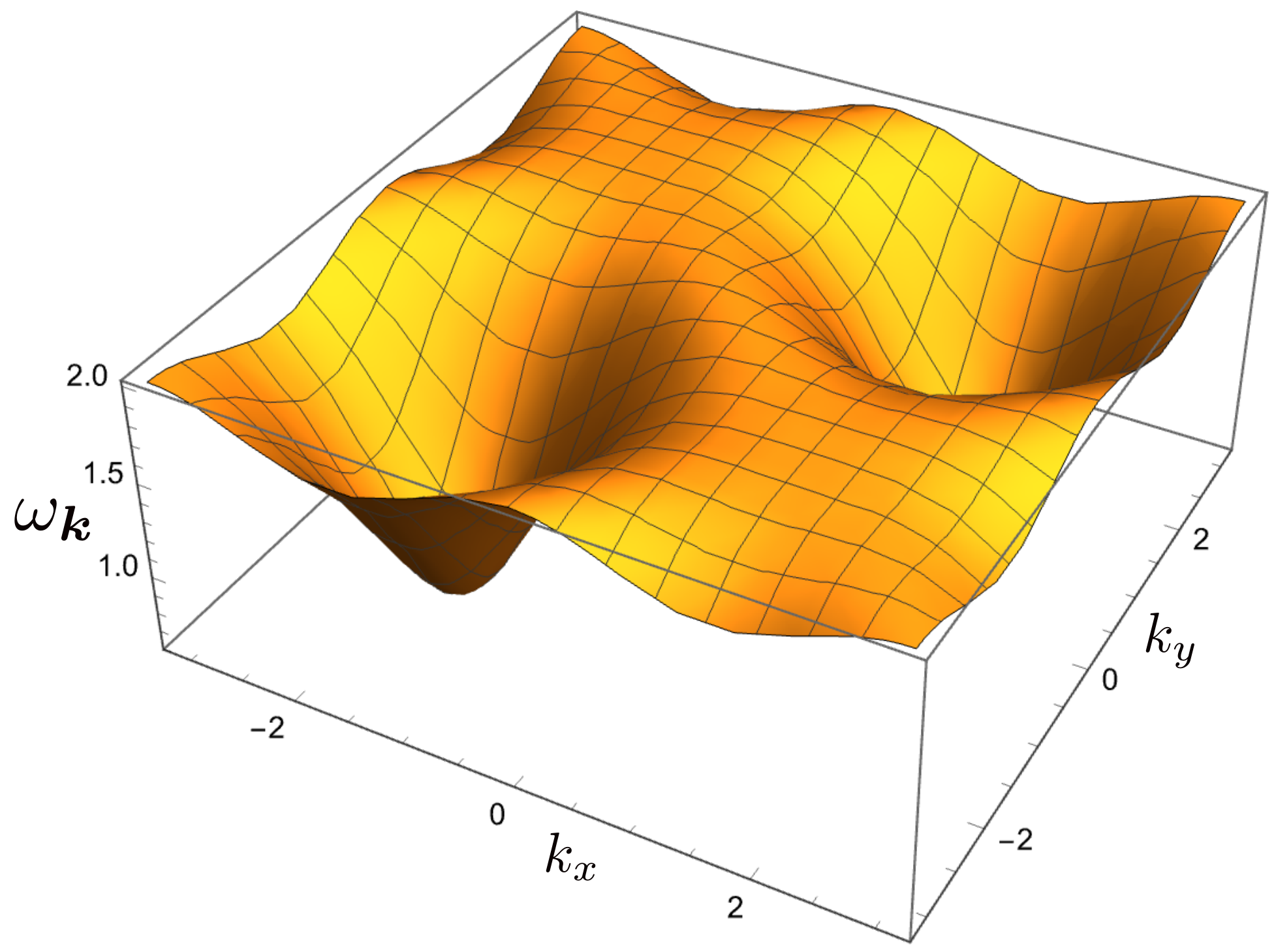}
\caption{Dispersion of bosonic spinons in a square lattice spin liquid, from Eq.~(\ref{eq:bosondisp}).}
\label{fig:bosondisp}
\end{figure}
We plot the dispersion in Fig.~\ref{fig:bosondisp}. Note the minima
at ${\bm k} = \pm (\pi/2, \pi/2)$ with an energy gap of $\left( \bar{\lambda}^2 - 4 J^2 \bar{\mathcal{Q}}^2 \right)^{1/2}$.

\subsubsection{Low energy U(1) gauge theory}
\label{sec:U1}

We now examine the low energy theory in the regime where the energy gap of the spinon excitations is small. Here, we can take a continuum limit for the spinons, and also account for the fluctuations of $\mathcal{Q}$ and $\lambda$. For the spinons, we introduce the wavevector at the minimum spinon gap
${\bm k}_0 = (\pi/2 , \pi/2 )$ and parameterize on the checkerboard $A$ and $B$ sublattices (with $\vi_x + \vi_y$ even and odd)
\begin{eqnarray}
b_{A\vi\alpha} &=& \psi_{1\alpha} ({\bm r}_\vi ) e^{i {\bm k}_0 \cdot {\bm r}_\vi }
\nonumber \\
b_{B\vi\alpha} &=& -i \varepsilon_{\alpha\beta} \psi_{2\beta} ({\bm r}_\vi )
e^{i {\bm k}_0 \cdot {\bm r}_\vi } \, .
\label{bosepar}
\end{eqnarray}
For $\mathcal{Q}$ and $\lambda$, we anticipate that the fluctuations will be un-important unless associated with the gauge symmetry in Eq.~(\ref{gaugetrans}). So we focus only on the phases of the $\mathcal{Q}_{\vi\vj}$ and parameterize
\begin{eqnarray}
\mathcal{Q}_{\vi, i + \hat{x}} &=& \bar{\mathcal{Q}} \exp \left(i \Theta_{\vi x} \right) \nonumber \\
\mathcal{Q}_{\vi, i + \hat{y}} &=& \bar{\mathcal{Q}} \exp \left(i \Theta_{\vi y} \right)\,,
\label{gaugetrans1a_sl}
\end{eqnarray}
and express the phases in terms of continuum field $(a_\tau, a_x, a_y)$ via
\begin{eqnarray}
\Theta_{\vi x} (\tau) &&= \eta_\vi a_x ({\bm r}, \tau) \nonumber \\
\Theta_{\vi y} (\tau) &&= \eta_\vi a_y ({\bm r}, \tau) \nonumber \\
\lambda_\vi &&= - i \bar{\lambda} - \eta_\vi a_\tau ({\bm r}, \tau)
\label{gaugetrans3_sl}
\end{eqnarray}
where 
\begin{equation}
\eta_\vi = (-1)^{\vi_x + \vi_y}
\end{equation}
identifies the checkerboard sublattices. 
Next, we insert these parameterizations into the spinon action,
perform a gradient expansion, and transform the Lagrangian ${\cal L}_J$ into ($a$ is the lattice spacing)
\begin{eqnarray}
{\cal L}_z &=& \int \frac{d^2 r}{2a^2} \left [
\psi_{1\alpha}^{\ast} \left( \frac{d}{d\tau} + i a_{\tau}
\right)
\psi_{1\alpha} +
\psi_{2\alpha}^{\ast} \left( \frac{d}{d\tau} - i a_{\tau}
\right)
\psi_{2\alpha} \right.
\nonumber \\
&& \qquad\qquad
 + \bar{\lambda} \left( |\psi_{1\alpha} |^2
+ |\psi_{2\alpha} |^2 \right) %\right.
%\end{displaymath}
%\begin{displaymath}
%~~~~~~~~~~~~~~~~~~~~~~
-2 J \bar{\mathcal{Q}} \left ( \psi_{1\alpha}\psi_{2\alpha} +
\psi_{1\alpha}^{\ast}\psi_{2\alpha}^{\ast}
\right )
\nonumber \\
&&\qquad\qquad
+  (J/2) \bar{\mathcal{Q}} a^2 \left [
\left ( {\bm \nabla}  + i  {\bm a} \right ) \psi_{1\alpha}
\left ( {\bm \nabla}  - i {\bm a} \right ) \psi_{2\alpha} \right.
%\end{displaymath}
%\begin{equation}
%~~~~~~~~~~~~~~~~~~~~~~~~~~~~~~~~~~~~~~~~~~~~~~~~~~~~~~~~~~~~~
\nonumber \\
&& \qquad\qquad\qquad
+ \left.
\left ( {\bm \nabla} - i {\bm a} \right )
\psi_{1\alpha}^{\ast}
\left ( {\bm \nabla} + i {\bm a} \right )
\psi_{2\alpha}^{\ast} \right ]  \Biggr] \, .
\label{charge}
\end{eqnarray}
We now introduce the fields
\begin{eqnarray*}
z_{\alpha} & = & (\psi_{1\alpha} +
\psi_{2\alpha}^{\ast})/\sqrt{2} \\
\pi_{\alpha} & = & (\psi_{1\alpha} -
\psi_{2\alpha}^{\ast})/\sqrt{2} \, ,
\end{eqnarray*}
to map Eq.~(\ref{charge}) to
\begin{eqnarray}
{\cal L}_z &=& \int \frac{d^2 r}{2a^2} \left [
\pi_{\alpha}^{\ast} \left( \frac{d}{d\tau} + i a_{\tau}
\right)
z_{\alpha} -
\pi_{\alpha} \left( \frac{d}{d\tau} - i a_{\tau}
\right)
z_{\alpha}^\ast \right.
\nonumber \\
&& \qquad\qquad
 + \bar{\lambda} \left( |z_{\alpha} |^2
+ |\pi_{\alpha} |^2 \right) %\right.
-2 J \bar{\mathcal{Q}} \left ( |z_\alpha|^2 - |\pi_\alpha|^2 
\right )
\nonumber \\
&&\qquad\qquad
+  (J/2) \bar{\mathcal{Q}} a^2 \left [
\left|\left ( {\bm \nabla}  + i  {\bm a} \right ) z_{\alpha} \right|^2 - 
\left|\left ( {\bm \nabla}  + i  {\bm a} \right ) \pi_{\alpha} \right|^2   \right ]  \Biggr] \, .
\label{charge2}
\end{eqnarray}
From Eq.~(\ref{charge2}), it is clear that the
the $\pi$ fields have `mass' $\bar{\lambda} + 2J \bar{\mathcal{Q}}$,
while the $z$ fields
have a mass $\bar{\lambda} - 2 J \bar{\mathcal{Q}}$ which vanishes at a quantum phase transition where the $z_\alpha$ condense, leading to N\'eel order. The $\pi$ fields can therefore
be safely integrated out,
and ${\cal L}_z$ yields
the following effective action, valid at distances much
larger than the lattice
spacing~\cite{NRSS89prl,NRSS90}:
\begin{equation}
S_{\rm eff} =
\int \frac{d^2 r}{4\sqrt{2}a} \int
d \tau \left\{
|(\partial_{\mu} - ia_{\mu})z_{\alpha}|^2
+ \frac{\Delta^2}{c^2}
|z^{\alpha} |^2\right\} \,.
\label{sefp}
\end{equation}
Here $\mu$ extends over $x,y,\tau$,
$c = \sqrt{2}J \bar{\mathcal{Q}} a$ is the spin-wave velocity, we have rescaled $\tau \rightarrow \tau/c$, and 
$\Delta = (\bar{\lambda}^2 - 4J^2 \bar{\mathcal{Q}}_1^2 )^{1/2}$ is the gap
towards spinon excitations.
Thus the long-wavelength theory describes a spin liquid with 
of a massive, spin-1/2, relativistic, boson $z_{\alpha}$ (spinon) excitation
coupled to a U(1) gauge field $a_\mu$. 

The continuum theory also makes it easy to determine the fate of the antiferromagnet when the spin energy gap vanishes. We expect that $z_\alpha$ will bose condense, and this will break the spin rotation symmetry; a term quartic in $z_\alpha$ will be needed to stabilize the condensate. But $z_\alpha$ carries a U(1) gauge charge, and so is not directly observable. 
Following the definitions of the underlying spin operators, it is not difficult to show
that the gauge-invariant composite
\begin{equation}
\mathbfcal{N} = z_\alpha^{\ast} {\bm \sigma}_{\alpha\beta} z_\beta \sim \eta_\vi {\bm S}_\vi \label{neelz}
\end{equation}
is just the N\'eel order parameter.

However, there is an important ingredient that our low energy theory has not yet considered. These are non-perturbative fluctuations of $a_\mu$ which are Dirac monopoles in 2+1 dimensional spacetime. We will not carry out a full analysis here, and merely summarize some important consequences. An important result is that the spin liquid noted above is  ultimately not a spin liquid. It is unstable to proliferation of monopoles, and ultimately confines a valence bond solid. But monopoles do not have a significant effect on the N\'eel state.

\subsubsection{Quantum criticality}
\label{sec:dqcp}

On general symmetry grounds, we extend Eq.~(\ref{sefp}) to a theory for the vicinity of the quantum critical point at which
the spinon gap vanishes \cite{SJ90}:
\begin{eqnarray}
\mathcal{S}_{U(1)} &=& \int d^3 x \, \left( \mathcal{L}_z + \mathcal{L}_{\rm monopole} \right) + \mathcal{S}_B  \nonumber \\
\mathcal{L}_z &=&
|(\partial_\mu -  i a_\mu) z_\alpha|^2 + g |z_\alpha|^2 + u \left(|z_\alpha|^2\right)^2 + K (\epsilon_{\mu\nu\lambda} \partial_\nu a_\lambda)^2 
\nonumber \\
 \mathcal{L}_{\rm monopole} &=& - y \left( \mathcal{M}_a + \mathcal{M}_a^\dagger \right) \nonumber \\
\mathcal{S}_B &=&  i 2S \sum_\vi \eta_\vi \int d \tau \, a_{\vi \tau} \,. \label{cp11}
\end{eqnarray}
The theory $\mathcal{L}_z$ is also known as the $\mathbb{CP}^1$ model. We have
included monopoles $\mathcal{M}_a$ in the gauge field $a_\mu$, and also the Berry phase of the spinons in the ground state. \begin{figure}
\centering
\includegraphics[width=4in]{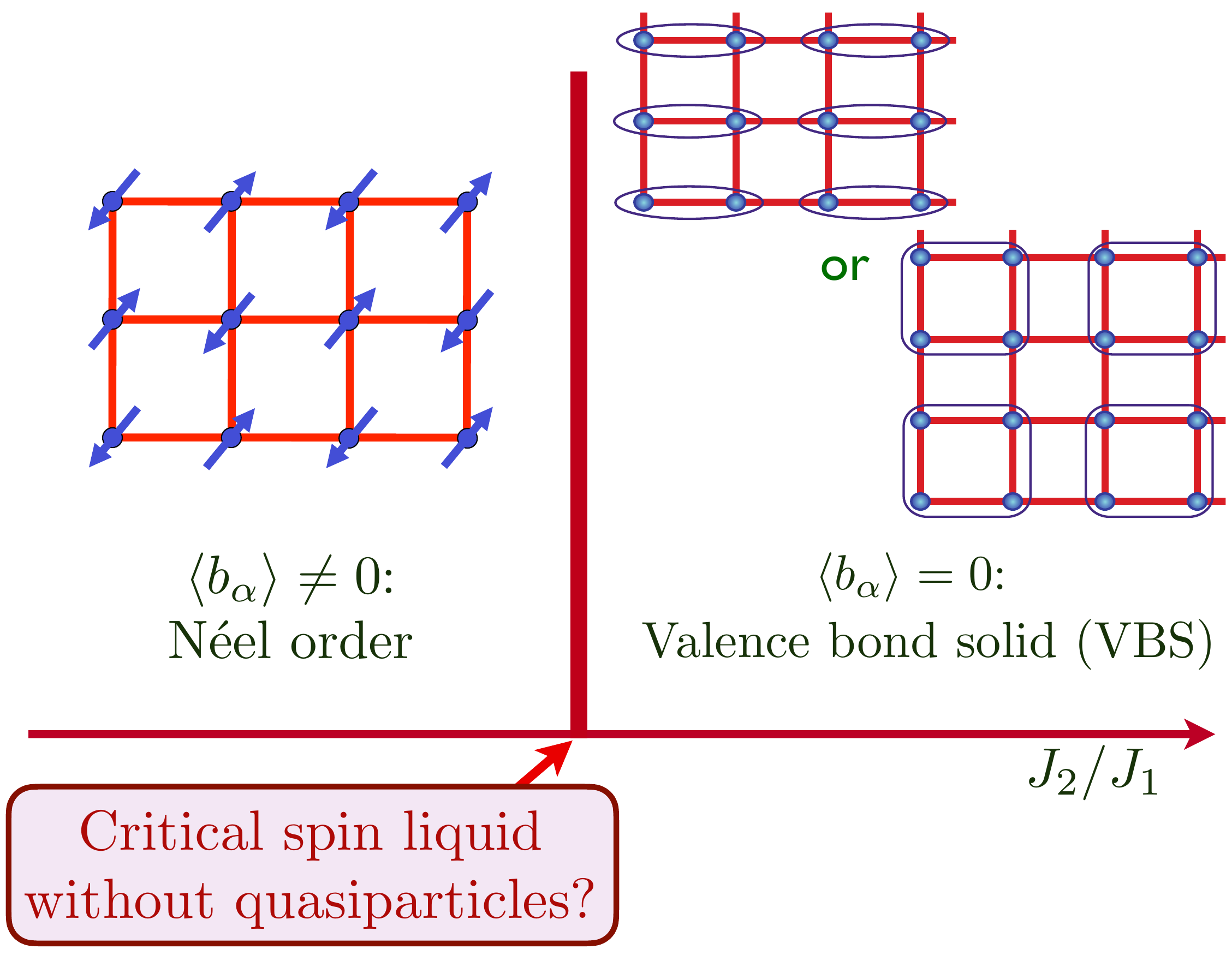}
\caption{Phase diagram of the U(1) gauge theory with bosonic spinons, Eq.~(\ref{cp3}. The N\'eel order appears in a Higgs phase where the bosonic spinons are condensed. The VBS order appears in the confining phase, and is induced by the Berry phases of the confining monopoles. The same phase diagram applies to the fermionic spinon theory in Eq.~(\ref{e105}), and the SO(5) $\sigma$-model with the WZW term in Eq.~(\ref{wzw}).}
\label{fig:Neelvbs}
\end{figure}
As we tune the coupling $g$ in Eq.~(\ref{cp11}), we can expect the 2 phases shown in Fig.~\ref{fig:Neelvbs}:\\
({\it i\/}) N\'eel phase, $g<g_c$: the spinon $z_\alpha$ condenses in a Higgs phase with $\langle z_\alpha \rangle \neq 0$. The $a_\mu$ gauge field is Higgsed, and spin rotation symmetry is broken by opposite polarization of the spins on the two sublattices.\\
({\it ii\/}) Valence bond solid (VBS), $g> g_c$: the spinons are gapped. For half-integer spin $S$, there is broken translational symmetry by the crystallization of valence bonds in the pattern shown in Fig.~\ref{fig:Neelvbs}.

We now obtain a potential gapless spin liquid if there is a continuous quantum phase transition at $g=g_c$. For half-integer spin $S$, the single monopole terms in Eq.~(\ref{cp11}) average to zero at long wavelengths from the Berry phases, and only quadrupoled monopole terms survive. So we can simplify the continuum theory for the vicinity of the quantum critical point to \cite{senthil1,senthil2} 
\begin{equation}
\!\!\!\!\!\!\! \!\!\!\!\!\!\! \!\!\!\! 
\mathcal{L}_z =
|(\partial_\mu -  i a_\mu) z_\alpha|^2 + g |z_\alpha|^2 + u \left(|z_\alpha|^2\right)^2 + K (\epsilon_{\mu\nu\lambda} \partial_\nu a_\lambda)^2  - y_4 \left( \mathcal{M}_a^4 + \mathcal{M}_a^{\dagger 4} \right)\,, \label{cp3}
\end{equation}
where $y_4$ is the quadrupoled monopole fugacity. \index{monopoles} There is ample numerical evidence that $y_4$ is irrelevant near a possible critical point, and so the question reduces to whether the theory $\mathcal{L}_z$ at $y_4=0$ exhibits a critical point which realizes a conformal field theory in 2+1 dimensions. This is a question that has been studied extensively in numerics, and it is clear that a `deconfined critical' description is suitable over a substantial length scale: with fractionalized spinons interacting with a U(1) gauge field in the absence of monopoles \cite{Nahum:2015vka,Meng24,Gu24,Chester24,Sandvik24}. 

\subsection{Fermionic partons}
\label{sec:fermions}

We now present an alternative analysis of the square lattice antiferromagnet in Eq.~(\ref{hamil}), replacing the bosonic partons in Eq.~(\ref{sparton}) by fermionic partons. This will ultimately lead to the same phase diagram as in Fig.~\ref{fig:Neelvbs}, but with a dual description of the phases and the criticality. This dual fermionic description turns out to be the most efficient way to describe the connection between the critical spin liquid and $d$-wave superconductivity in the doped antiferromagnet, as we will see in Section~\ref{sec:halffilling}.

The following Schwinger {\it fermion} representation applies only for $S=1/2$
\begin{equation}
{\bm S}_{\vi} =  \frac{1}{2}
f_{\vi\alpha}^{\dagger}\, {\bm \sigma}_{\alpha\beta} \, f_{\vi \beta}
\label{Schwingerfermion}
\end{equation}
where $f_{\vi \alpha}$ are canonical fermions obeying the constraint
\begin{equation}
\sum_{\alpha} f_{\vi \alpha}^\dagger f_{\vi \alpha} = 1 \quad, \quad \mbox{for all $\vi$}. \label{constraintf}
\end{equation}

While the bosonic parton representation led to the U(1) gauge symmetry in Eq.~(\ref{gaugetrans}), it turns out the Eqs.~(\ref{Schwingerfermion}) and (\ref{constraintf}) have a larger SU(2) gauge invariance, and this will be crucial to our results. The analysis is clearest upon introducing a matrix notation for the fermions 
\begin{equation}
\mathcal{F}_\vi \equiv \left(
\begin{array}{cc}
f_{\vi \uparrow} & - f_{\vi \downarrow} \\
f_{\vi \downarrow}^\dagger & f_{\vi \uparrow}^\dagger
\end{array}
\right) \label{Fmatrix}
\end{equation}
This matrix obeys the `reality' condition
\begin{equation}
\mathcal{F}_{\vi}^\dagger = \sigma^y \mathcal{F}_{\vi}^T \sigma^y. \label{fdft}
\end{equation}
Now we can write Eq.~(\ref{Schwingerfermion}) as
\begin{equation}
{\bm S}_{\vi} = -\frac{1}{4} \mbox{Tr} ( \mathcal{F}_{\vi}^{\phantom\dagger} \sigma^z {\bm \sigma}^T \sigma^z \mathcal{F}_{\vi}^{\dagger} ) \,. \label{NambuST}
\end{equation}
The SU(2) gauge symmetry is now associated with a SU(2) matrix $V_\vi$ under which \cite{Affleck-SU2,Fradkin88,AndreiColeman2}
\begin{equation}
\mathcal{F}_\vi \rightarrow V_\vi \, \mathcal{F}_\vi \,, \label{SU2gauge}
\end{equation}
which is easily seen to leave the spin operator in Eq.~(\ref{NambuST}) invariant. The global spin rotation symmetry is however
\begin{equation}
\mathcal{F}_\vi \rightarrow  \mathcal{F}_\vi \, \sigma^z R^T \sigma^z  \,. \label{su212b}
\end{equation}
where $R$ is the $S=1/2$ spin rotation matrix defined by
\begin{equation}
\left( \begin{array}{c}
 f_{\vi \uparrow} \\ f_{\vi \downarrow} \end{array} \right) \rightarrow R  \left( \begin{array}{c}
 f_{\vi \uparrow} \\ f_{\vi \downarrow} \end{array} \right)\,.
 \end{equation}

\begin{figure}
\centering
\includegraphics[width=4.5in]{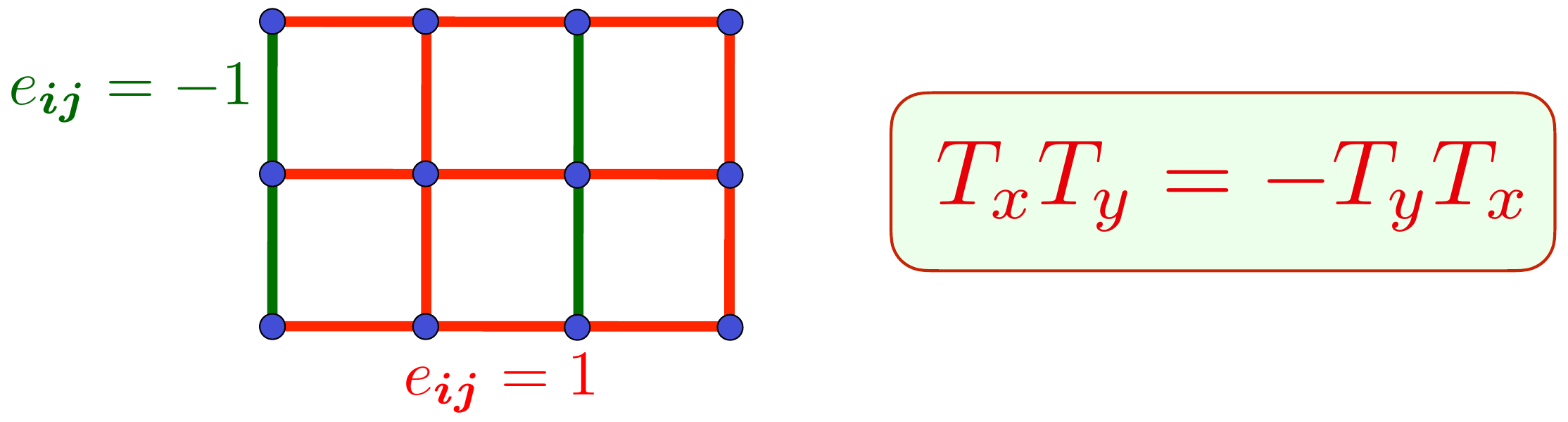}
\caption{Background $\pi$ flux acting on the spinons $f$, and also on the chargons $B$.}
\label{fig:eij}
\end{figure}

Next we insert Eq.~(\ref{NambuST}) into Eq.~(\ref{hamil}), and perform Hubbard-Stratonovich transformation to obtain an effective Hamiltonian for the spinons, following the same procedure as for bosonic spinons. We skip the intermediate steps, and focus directly on the fermion bilinear Hamiltonian on symmetry grounds; we obtain the mean-field nearest-neighbor spin liquid Hamiltonian for the spinons of the $\pi$-flux phase \cite{AM88}:
\begin{eqnarray}
    \mathcal{H}_{SLf} & = & \frac{iJ}{2} \sum_{\langle \vi \vj \rangle} e_{\vi \vj} \left[  \mbox{Tr} \left( \mathcal{F}_{\vi}^{\dagger} \mathcal{F}_{\vj }^{\phantom\dagger} \right) -  \mbox{Tr} \left( \mathcal{F}_{\vj }^{\dagger} \mathcal{F}_{\vi }^{\phantom\dagger} \right) \right] \nonumber \\
&=& i J \sum_{\langle \vi\vj\rangle} e_{\vi\vj} \left( \Psi_\vi^\dagger  \Psi_\vj - \Psi_\vj^\dagger  \Psi_\vi \right); \quad
\Psi_\vi \equiv \left( \begin{array}{c} f_{\vi \uparrow} \\ f_{\vi \downarrow}^{\dagger} \end{array} 
\right),
    \label{eq:fermionhop}
\end{eqnarray}
where $e_{\vi \vj} = \pm 1$ represents $\pi$-flux on the fermions as shown in Fig.~\ref{fig:eij}. We choose
$e_{\vi\vj} =- e_{\vj \vi}$ and
\begin{equation}
 e_{\vi,\vi+\hat{{\bm x}}}  =  1 \,,\quad
 e_{\vi,\vi+\hat{{\bm y}}}  =  (-1)^{x}      \,, \label{su2ansatz}
\end{equation}
where $\vi = (x,y)$, $\hat{\bm x} = (1,0)$, $\hat{\bm y} = (0,1)$.
In the chosen form, it is evident that $\mathcal{H}_{Slf}$ is invariant under the global spin rotation in Eq.~(\ref{su212b}), and also at least the gobal gauge transformation associate with Eq.~(\ref{SU2gauge}). We have chosen the hopping to be pure imaginary because of the identity
\begin{equation}
     \mbox{Tr} \left( \mathcal{F}_{\vi}^\dagger  \mathcal{F}_{\vj }^{\phantom\dagger} \right) = -\mbox{Tr} \left( \mathcal{F}_{\vj }^\dagger  \mathcal{F}_{\vi }^{\phantom\dagger} \right)\,.
\end{equation}

\begin{figure}
\centering
\includegraphics[width=4in]{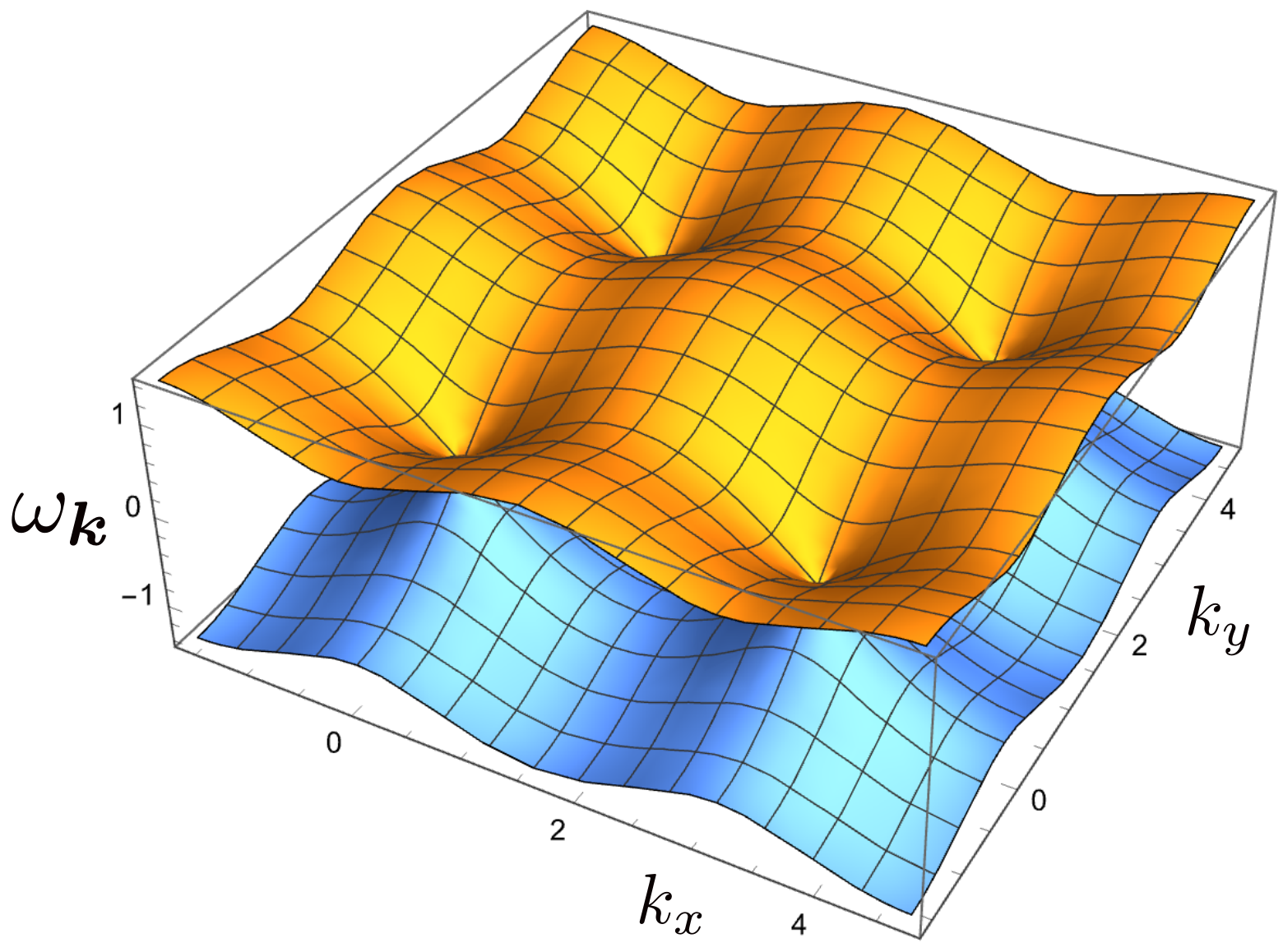}
\caption{Dispersion of fermionic spinons in Eq.~(\ref{eq:fermiondisp}).}
\label{fig:fermiondisp}
\end{figure}

We can now easily diagonalize the Hamiltonion in Eq.~(\ref{eq:fermionhop}), and obtain the fermionic dispersion spectrum analogous to Eq.~(\ref{eq:bosondisp})
\begin{equation}
\omega_{\bm k} = \pm 2J \left( \sin^2 k_x + \sin^2 k_y \right)^{1/2}\,. \label{eq:fermiondisp}
\end{equation}
We show a plot of this analogous to Fig.~\ref{fig:bosondisp} in Fig.~\ref{fig:fermiondisp}.
Unlike the bosonic spinons, the energy of the fermionic spinons is allowed to be negative, and the negative energy fermion states are occupied in the ground state. The constraint in Eq.~(\ref{constraintf}) is then automatically satisfied. Notice the two independent nodal Dirac points at ${\bm k}_v$, $v=1,2$ with
\begin{equation}
{\bm k}_1 = (0,0) \quad, \quad {\bm k}_2 = (0, \pi)\,.
\end{equation}
The index $v$ is the `valley'.

\subsubsection{Low energy SU(2) gauge theory.}
\label{sec:su2theory}

Going beyond mean field theory, while still remaining on the lattice, we extend the mean field Hamiltonian in Eq.~(\ref{eq:fermionhop}) to be invariant under space-dependent SU(2) gauge symmetries. To achieve this, we need to introduce a SU(2) gauge field $U_{\vi\vj} = U_{\vj \vi}^\dagger$ on each lattice link upon which the SU(2) gauge transformation acts as
\begin{equation}
U_{\vi\vj} \rightarrow V_\vi^{\phantom\dagger} U_{\vi \vj} V_\vj^{\dagger}\,.
\end{equation}
Then, by gauge invariance, we extend Eq.~(\ref{eq:fermionhop}) to
\begin{eqnarray}
    \mathcal{H}_{SLf} & = & \frac{iJ}{2} \sum_{\langle \vi \vj \rangle} e_{\vi \vj} \left[  \mbox{Tr} \left( \mathcal{F}_{\vi}^{\dagger} U_{\vi \vj}^{\phantom\dagger} \mathcal{F}_{\vj }^{\phantom\dagger} \right) -  \mbox{Tr} \left( \mathcal{F}_{\vj }^{\dagger} U_{\vj \vi}^{\phantom\dagger} \mathcal{F}_{\vi }^{\phantom\dagger} \right) \right] \nonumber \\
&=& i J \sum_{\langle \vi\vj\rangle} e_{\vi\vj} \left( \Psi_\vi^\dagger U_{\vi \vj}^{\phantom\dagger} \Psi_\vj - \Psi_\vj^\dagger U_{\vj \vi}^{\phantom\dagger} \Psi_\vi \right); \quad
\Psi_\vi \equiv \left( \begin{array}{c} f_{\vi \uparrow} \\ f_{\vi \downarrow}^{\dagger} \end{array} 
\right),
    \label{eq:fermionhop2}
\end{eqnarray}
The first form in terms of $\mathcal{F}$ makes both the SU(2) gauge invariance and the SU(2) spin rotation invariance explicit, while in the second form in terms of $\Psi$ only the gauge invariance is explicit.

If we had not used the pure imaginary hopping in Eq.~(\ref{eq:fermionhop}), then the mean-field Hamiltonian would break (`Higgs') the SU(2) gauge symmetry to a smaller symmetry. A `staggered flux' ansatz which breaks the SU(2) down to U(1) has commonly been used in the literature \cite{LeeWenRMP}. However, it is now known that this U(1) spin liquid allows single monopole perturbations \cite{Alicea08,Song1} (unlike the quadrupole monopole perturbations in Eq.~(\ref{cp3})), and such single monopole terms are expected to drive a strong instability to confinement. So we don't consider the staggered-flux U(1) spin liquid.

The continuum formulation of this theory can be obtained by following the same procedure as in Section~\ref{sec:U1}, but we have to carefully account for the SU(2) gauge symmetry. 
First, neglecting gauge fluctuations of $U_{\vi\vj}$ for now, let us write Eq.~(\ref{eq:fermionhop}) in momentum space, in terms of the fermions $\mathcal{F}_s ({\bm k})$, where $s=A,B$ is a sublattice 
index. Now the sublattices refer to sites with $\vi_x$ even and odd, which are the two sites in the unit cell. We obtain
\begin{eqnarray}
\mathcal{H}_{SLf} = -J \sum_{\bm k} \mbox{Tr} \left( \mathcal{F}^\dagger ({\bm k}) \left[ \rho^x \sin (k_x) +\rho^z \sin(k_y) \right]  \mathcal{F}^\dagger ({\bm k}) \right)\,, \label{e102}
\end{eqnarray}
where $\rho^\ell$, with $\ell=x,y,z$, are Pauli matrices in sublattice space. Next, analogous to Eq.~(\ref{bosepar}), 
we take the continuum limit near the valley momenta in terms of $\mathcal{X}_{sv} ({\bm r}, \tau)$ 
\begin{eqnarray}
\mathcal{F}_{A\vi} &=& \sum_v \mathcal{X}_{Av} ({\bm r}, \tau) e^{i {\bm k}_v \cdot {\bm r}_i} \nonumber \\
\mathcal{F}_{B\vi} &=& \sum_v \mathcal{X}_{Bv} ({\bm r}, \tau) e^{i {\bm k}_v \cdot {\bm r}_i}\label{e101}
\end{eqnarray}
for $\vi$ on the A and B sublattices respectively. This yields the imaginary time Lagrangian density
\begin{equation}
\mathcal{L}_{\mathcal{X}} = \frac{1}{2} \mbox{Tr} \left(\mathcal{X}^\dagger \left[ \partial_\tau + 2 J i \rho^x \partial_x + 2 J i \rho^z \mu^z \partial_y \right] 
\mathcal{X} \right)  \,, \label{e103}
\end{equation}
where $\mu^\ell$ are the Pauli matrices in valley space.
We recall that the fermion $\mathcal{X}_{sv}$ has four-components, and each component is a $2 \times 2$ matrix which obeys the reality condition in Eq.~(\ref{fdft}). We can write this in a relativistic Dirac form 
\begin{equation}
\mathcal{L}_{\mathcal{X}} = \frac{i}{2} \mbox{Tr} \left(\bar{\mathcal{X}} \gamma^\mu \partial_\mu \mathcal{X} \right) \,, \label{e104}
\end{equation}
with the definitions
\begin{equation}
\bar{\mathcal{X}} = -i \mathcal{X}^\dagger \gamma^0 \quad, \quad \gamma^0 = \rho^y \mu^z \quad, \quad \gamma^x =  \rho^z \mu^z \quad, \quad \gamma^y = -\rho^x  \,,
\end{equation}
where we have absorbed factor of $c=2J$ for the velocity of light.
Finally, it is a simple matter to include the SU(2) gauge field by taking the continuum limit by writing
\begin{equation}
U_{\vi \vj} = \exp \left(-i A_{ij}^\ell \sigma^\ell \right)
\label{eq:UA}
\end{equation}
(where $\sigma^\ell$ are the Pauli matrices in SU(2) gauge space) and expanding the exponential. We then obtain 
\begin{equation}
\mathcal{L}_{\mathcal{X}} = \frac{i}{2} \mbox{Tr} \left(\bar{\mathcal{X}} \gamma^\mu \left[\partial_\mu -i A_\mu^\ell \sigma^\ell \right]  \mathcal{X} \right) \,.\label{e105}
\end{equation}

The theory in Eq.~(\ref{e105}) is the analog of the $\mathcal{L}_z$ in Eq.~(\ref{cp11}) for bosonic spinons. The latter theory was a U(1) gauge theory with two relativistic complex scalars $z_\alpha$. In the present case, we have a SU(2) gauge theory with $N_f =2$ massless Dirac fermions, associated with valley index $v$. The global symmetry of $z_\alpha$ was just spin rotations $z \rightarrow R z$. In contrast, here we have emergent global symmetry which combines spin and valley rotations. A first guess is a SU(4) symmetry generalizing Eq.~(\ref{su212b})
\begin{equation}
\mathcal{X} \rightarrow \mathcal{X} L\,,
\end{equation}
where $L$ acts on spin and valley space with $L^\dagger L = 1$. However, imposition of the reality condition Eq.~(\ref{fdft}) shows that we also need
\begin{equation}
L^T = \sigma^y L^\dagger \sigma^y\,,
\end{equation}
and so the symmetry is only Sp(4)=SO(5)$/\mathbb{Z}_2$ \cite{RanWen06,Wang17}.
In terms of the Hermitian Lie algebra elements $M$, with $L=e^{i M}$, the reality condition is
\begin{equation}
M^T = - \sigma^y M \sigma^y\,.
\end{equation}
Requiring that $M$ commute with the $\gamma^\mu$, we can now write down the 10 elements of the Lie algebra of Sp(4)=SO(5)$/\mathbb{Z}_2$
\begin{equation}
M = \{\sigma^x, \sigma^y, \sigma^z, \mu^z \sigma^x, \mu^z \sigma^y, \mu^z \sigma^z, \rho^x \mu^y, \rho^x \mu^x \sigma^x, \rho^x \mu^x \sigma^y,\rho^x \mu^x  \sigma^z\}\,. \label{e200}
\end{equation}

The remaining 5 SU(4) generators which commute with the $\gamma^\mu$ are ($t = 1\ldots 5$)
\begin{equation}
\Gamma^t = \{ \mu^z, \rho^x \mu^x, \rho^x \mu^y \sigma^x, \rho^x \mu^y \sigma^y, \rho^x \mu^y \sigma^z \}\,.
\end{equation}
The $\Gamma^t$ all anti-commute with each other, and transform as a SO(5) vector under the generators in Eq.~(\ref{e200}).
It is now straightforward to check by working back to the lattice operators from the information above that the vector $i \mbox{Tr} \left( \bar{\mathcal{X}} \Gamma^t \mathcal{X} \right)$ corresponds precisely to the 5 components of the orders parameters shown in Fig.~\ref{fig:Neelvbs}: the first two components are the VBS order, and the last 3 components are the N\'eel order $\mathbfcal{N}$ in Eq.~(\ref{neelz}) \cite{RanWen06,Wang17}.

Wang {\it et al.} \cite{Wang17} have argued that the likely fate of the SU(2) gauge theory upon confinement is a state which the SO(5) symmetry is spontaneously broken with  $\langle i \mbox{Tr} \left( \bar{\mathcal{X}} \Gamma^t \mathcal{X} \right) \rangle \neq 0$. The lattice model does not have exact SO(5) symmetry, and the choice between the N\'eel and VBS components of $\Gamma^t$ is made by additional 4-fermi terms that can be added to Eq.~(\ref{e105}). So the ultimate fate of the theory is essentially identical to the fate of the bosonic spinon theory in Section~\ref{sec:sbparton}, as illustrated in Fig.~\ref{fig:Neelvbs}. This is essentially the reason for the duality between the theories in Section~\ref{sec:dqcp} and \ref{sec:su2theory}, and Wang {\it et al.\/} have provided additional topological arguments.

\subsection{SO(5) non-linear $\sigma$-model}
\label{sec:SO5}

\begin{figure}
\centering
\includegraphics[width=5.5in]{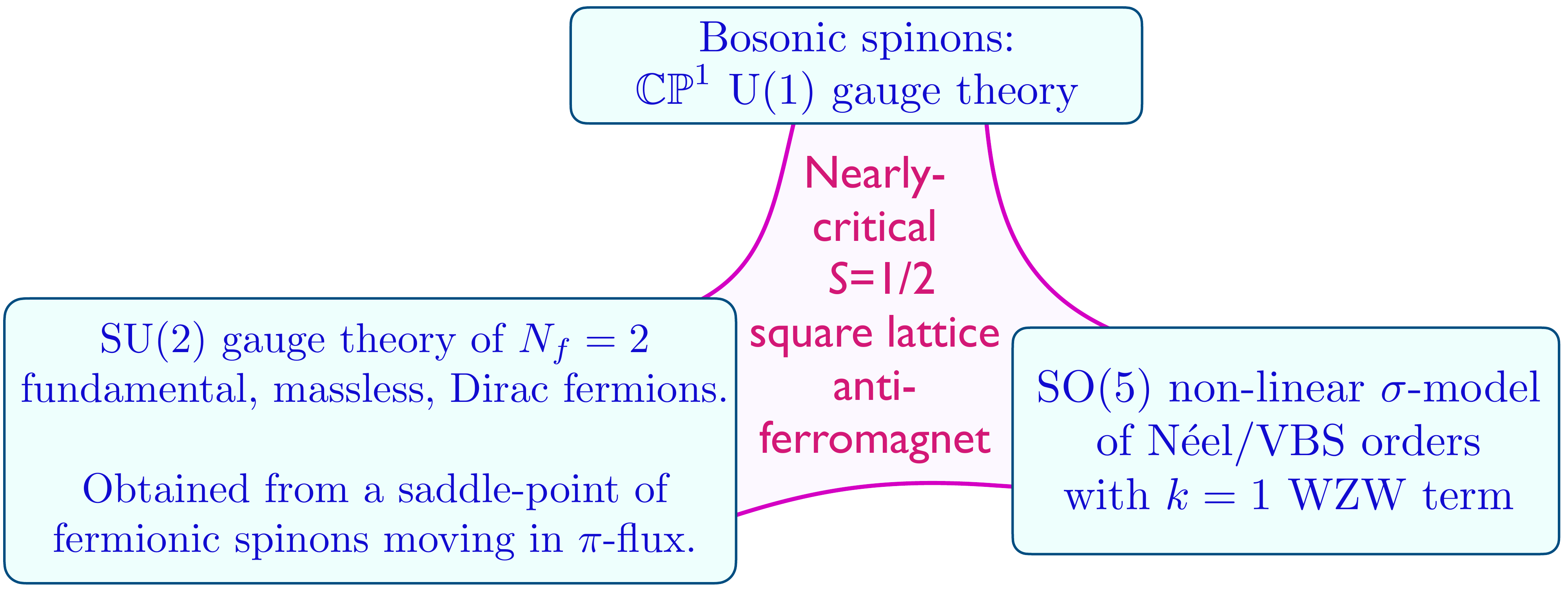}
\caption{Three field-theoretical formulations of the $S=1/2$ square lattice antiferromagnet near the N\'eel-VBS transition in Fig.~\ref{fig:Neelvbs}. All three are valid descriptions \cite{Wang17}, but the SU(2) gauge theory of Dirac fermions is the most convenient starting point to describe the connection to $d$-wave superconductivity.}
\label{fig:triality}
\end{figure}

There is a third formulation of the theories in Section~\ref{sec:dqcp} and \ref{sec:su2theory} which is useful for some purposes, as we illustrate in Fig.~\ref{fig:triality}.
This is obtained most simply by coupling Eq.~(\ref{e105}) to the SO(5) vector order parameter, and integrating out the fermions. Introducing the SO(5) fundamental unit length field $n_t$, $n_t n_t=1$ to Eq.~(\ref{e105})
\begin{equation}
\mathcal{L}_{\mathcal{X}n} = \frac{i}{2} \mbox{Tr} \left(\bar{\mathcal{X}} \gamma^\mu \left[\partial_\mu -i A_\mu^\ell \sigma^\ell \right]  \mathcal{X} \right)  - i n_t \mbox{Tr} \left( \bar{\mathcal{X}} \Gamma^t \mathcal{X} \right) \,, \label{e106}
\end{equation}
we integrate out the Dirac fermions following the analysis of Ref.~\cite{Abanov:1999qz} and obtain 
\begin{equation}
\mathcal{L}_n = \frac{1}{2g} \left( \partial_\mu n_t \right)^2 + 2 \pi i \Gamma[n_t] \label{wzw}
\end{equation}
The last term is the Wess-Zumino-Witten (WZW) term at level 1: it is a Berry phase associated with spacetime textures of $n_t$, a higher dimensional analog of the Berry phase of a single spin which is proportional to area enclosed by a spherical path \cite{LS15,Wang17}: an explicit expression of $\Gamma[n_t]$ requires 4+1 dimensions with an emergent spatial direction. 
Upon reduction to a O(3) non-linear sigma model for the N\'eel order parameter $\mathbfcal{N}$ in Eq.~(\ref{neelz}), the WZW term reduces \cite{TanakaHu} to the Berry phases of the monopoles noted near Eq.~(\ref{cp3}).

Also, note that while the SO(5) symmetry is explicit in the fermionic spinon theory in Eq.~(\ref{e105}), it is not explicit in the bosonic spinon theory in Eq.~(\ref{cp3}), but expected to be emergent \cite{SenthilFisher06}.

The form in Eq.~(\ref{wzw}) has been exploited in recent numerical work on the fuzzy sphere \cite{Fuzzy24}. 
Their results, and those of a number of other numerical works \cite{Yasir19,Nahum:2015vka,Becca20,Imada21,Meng24,Gu24,Chester24,Sandvik24} show that the critical spin liquid defined by 
Eq.~(\ref{cp3}), Eq.~(\ref{e105}), or (\ref{wzw}) is stable over a substantial intermediate energy and length scales, before ultimately confining into a N\'eel or VBS state. This intermediate range stability is not a bug, but a feature ideal for our purposes of defining a FL* state at intermediate temperatures, which ultimately confines to variety of other states at low temperatures.

\subsection{Direct observation of spinons}

We now discuss possible experimental evidence for the spinons of the deconfined critical state of Fig.~\ref{fig:triality} in the cuprate compounds. 

In the insulating antiferromagnet, higher energy neutron scattering has been interpreted using spinons by Headings {\em et al.\/} \cite{Hayden10} and Dalla Piazza {\it et al.\/} \cite{Ronnow15}.

For the doped system, it has recently been argued \cite{BCS24,SSZaanen} that RIXS measurements \cite{Keimer11,Hayden19} do indeed provide support for the presence of spinons. At half-filling, there is sharp spin wave excitation extending to high energy $\sim 400$meV. This excitation survives upon doping even at high energies, but becomes much broader. This broad spectrum has similarities to that expected from the $\pi$-flux spin liquid, as computed near the destruction of the N\'eel order in the insulating $J_1$-$J_2$ model by Ref.~\cite{Becca18}. See Figs.~8 and 9 in Ref.~\cite{SSZaanen} for a comparison.  Ref.~\cite{Alex_hourglass} has considered the influence of charge order on the fermionic spinons, and shown that it can reproduce the `hourglass' structure of the incommensurate spin corrections.

Another issue in the comparison with the cuprates is that the free fermion theory of the $\pi$-flux phase predicts a gapless spin spectrum near momenta $(\pi, 0)$ and $(0,\pi)$, and low energy spin excitations have not been observed at these momenta. However, Ref.~\cite{BCS24} has argued that this spectrum is suppressed by SU(2) gauge fluctuations. Their argument is based upon the scaling dimensions of the associated operators as measured by fuzzy-sphere analysis \cite{Fuzzy24} of the SO(5) theory reviewed in Section~\ref{sec:SO5}. 

In the following discussion of the doped system, we will see that 
the spinons $f_\alpha$ in $\mathcal{H}_{SLf}$ play a crucial role in the theory of the observed electronic spectrum in Sections~\ref{sec:arc}, \ref{sec:aniso},  and \ref{sec:qo}. We can interpret this successful description of the electronic spectrum as an indirect observation of spinons.

\section{Confinement to $d$-wave superconductivity and charge order at half-filling}
\label{sec:halffilling}

We have so far considered the square lattice antiferromagnet as an insulator with an essentially infinite gap to electrically charged excitations. In the following discussion, we will build on the low energy theory of such an antiferromagnet as a SU(2) gauge theory coupled to fermionic spinons in Eq.~(\ref{eq:fermionhop2}). We have argued that the phase diagram of such a theory is as in Fig.~\ref{fig:Neelvbs} {\it i.e.\/} except possibly in a critical region, the SU(2) gauge theory confines at low energies, and we obtain either the N\'eel or VBS states.

We now wish to consider a more general situation in which the gap to charged excitations can vanish \cite{ChristosLuo24}. In the cuprates, gapless charged excitations appear when we dope the antiferromagnet. We will consider this important situation in the following sections. But for now we consider the simpler case where the charge gap vanishes while the electronic density remains the same as in an insulator. We emphasize that the present section describes such a charge gap vanishing transition without using the ALM, and using only symmetry arguments. We will resume explicit use of the ALM in Section~\ref{sec:pseudogap} when we consider the doped case, although we can also proceed there with only symmetry arguments.

We work at half-filling and
also assume particle-hole symmetry as it simplifies the analysis. We can decrease the charge gap by describing the insulator by an underlying Hubbard model with on-site repulsion $U$, and reducing the value of $U$, or by adding additional off-site interactions. Such models have been considered in numerical studies \cite{AssaadImada,Assaad22,Assaad24,Assaad25,Xu:2020qbj,Scaletter21,Scaletter22,HongYao21,HongYao22,HongYao25,Zhaoyu22,Zhaoyu24}. We will now show that the SU(2) gauge theory of Eq.~(\ref{eq:fermionhop2}) has other possible fates once charged excitations are included, the most interesting of which is a $d$-wave superconductor with gapless nodal quasiparticles. 

In terms of adiabatic continuity, this $d$-wave superconductor is precisely the superconductor observed in the cuprates. However, the $d$-wave superconductor obtained in this section has one significant quantitative difference from the observations, noted in Section~\ref{sec:intro}: it has a Lorentz-invariant form of its dispersion, with the two velocities only the square lattice diagonals, $v_F$ and $v_\Delta$, being equal to each other (see Fig.~\ref{fig:flsdsc}B). The cuprates instead have $v_F \gg v_\Delta$. We will resolve this problem in an interesting manner in Section~\ref{sec:aniso} when we consider the transition from FL* to a $d$-wave superconductor in the doped case.

The only matter field in Section~\ref{sec:fermions} is the fermion $\mathcal{F}$, which has electrical charge 0, spin $1/2$, and is a gauge SU(2) fundamental. As we are allowing for charged fluctuations, we need to define an electron operator, which has charge $-e$, spin $1/2$, and is a gauge SU(2) singlet. This directly leads us to introducing a boson $B$ which has charge $+e$, spin 0, 
and is a gauge SU(2) fundamental, so that a composite of $\mathcal{F}$ and $B$ will have the same quantum numbers as the electron. We now show that this information is basically sufficient to deduce an effective action for $B$, and to reach our main conclusions. We will give a more microscopic definition of the field $B$ in the doped case later near Eq.~(\ref{Yukawa}).

A boson with the same quantum numbers as our $B$ was introduced in earlier work \cite{Fradkin88,AndreiColeman2,LeeWen96,HermeleHoneycomb}, but with an important difference. In the earlier work, the expectation value of $B^\dagger B$ was set to be equal to the doping density. That is not the case in our work, as the doping density also includes the density of fermionic holes (see Section~\ref{sec:FL*}). In this section, we are considering the particle-hole symmetric case at half-filling, and $\langle B^\dagger B \rangle$ is non-zero even though there is no doping (this is similar to the situation in the superfluid-insulator transition of bosons \cite{QPT}).

Similar to Eq.~(\ref{Fmatrix}), we introduce a matrix notation for the electron $\mathcal{C}$ and the boson $B$:
\begin{eqnarray}
& \mathcal{C}_\vi \equiv \left(
\begin{array}{cc}
c_{\vi \uparrow} & - c_{\vi \downarrow} \\
c_{\vi \downarrow}^\dagger & c_{\vi \uparrow}^\dagger
\end{array}
\right) \,, \quad B_\vi \equiv \left( \begin{array}{c} B_{1\vi}  \\ B_{2\vi} \end{array} \right) \,, \quad  \mathcal{B}_\vi \equiv \left( \begin{array}{cc} B_{1\vi} & - B_{2\vi}^\ast \\ B_{2\vi} & B_{1\vi}^\ast \end{array} \right) \,, \label{defB}
\end{eqnarray}
all of which obey the reality condition analagous to Eq.~(\ref{fdft}).
Then the generalization of the SU(2) gauge transformation in Eq.~(\ref{SU2gauge}) is 
\begin{eqnarray}
\mathcal{C}_\vi \rightarrow \mathcal{C}_\vi \quad & , \quad 
\mathcal{F}_\vi \rightarrow V_\vi \, \mathcal{F}_\vi   \nonumber \\
\mathcal{B}_\vi \rightarrow V_\vi \, \mathcal{B}_\vi \quad & , \quad 
U_{\vi\vj} \rightarrow  V_\vi \, U_{\vi\vj} \, 
V_\vj^{\dagger} \,, \label{eq:gauge}
\end{eqnarray}
while the generalization of the global SU(2) spin rotation in Eq.~(\ref{su212b}) is
\begin{eqnarray}
\mathcal{C}_\vi \rightarrow \mathcal{C}_\vi \, \sigma^z R^T \sigma^z \quad & , \quad 
\mathcal{F}_\vi \rightarrow \mathcal{F}_\vi \, \sigma^z R^T  \sigma^z \nonumber \\
\mathcal{B}_\vi \rightarrow  \mathcal{B}_\vi \quad & , \quad 
U_{\vi\vj} \rightarrow  U_{\vi\vj}  \,.
\label{eq:spin}
\end{eqnarray}
Finally, the U(1) charge conservation symmetry acts as
\begin{eqnarray}
\mathcal{C}_\vi \rightarrow  \Theta \,\mathcal{C}_\vi \, \quad & , \quad 
\mathcal{F}_\vi \rightarrow \mathcal{F}_\vi   \nonumber \\
\mathcal{B}_\vi \rightarrow  \mathcal{B}_\vi \, \Theta^\dagger \quad & , \quad 
U_{\vi\vj} \rightarrow  U_{\vi\vj}  \,,
\label{eq:charge}
\end{eqnarray}
where 
\begin{eqnarray}
    \Theta = \left( \begin{array}{cc}
        e^{i\theta} & 0 \\
         0 & e^{-i \theta}
    \end{array}
    \right)\,.
\end{eqnarray}
See also Table~\ref{tab2} later for a summary of these gauge and symmetry transformations.

By matching these gauge, spin rotation, and charge conservation symmetries we deduce that the operator correspondence between the electrons, the Higgs boson $B$, and the fermionic spinons must be
\begin{equation}
    \mathcal{C}_{\vi}^{\vphantom\dagger} \sim \mathcal{B}_\vi^\dagger  \,\mathcal{F}_\vi^{\vphantom\dagger} \,. \label{eq:CBF}
\end{equation}
In terms of the matrix components, we can write Eq.~(\ref{eq:CBF}) as 
\begin{equation}
c_{\vi \alpha}^{\dagger} \sim
B_{1\vi}^{\vphantom\dagger} f_{\vi \alpha}^\dagger + B_{2\vi}^{\vphantom\dagger} \varepsilon_{\alpha\beta}^{\vphantom\dagger} f_{\vi \beta}^{\vphantom\dagger}  \,, \label{cBf}
\end{equation}
where $\varepsilon_{\alpha\beta}$ is the unit antisymmetric tensor for spin SU(2).

We now obtain an energy functional for $B$ in a Landau-type expansion \cite{Christos:2023oru}. Such a functional must also involve the gauge field $U_{\vi \vj}$ to maintain gauge invariance. The fermion $f$ experiences a $\pi$ flux with pure imaginary hopping, while the electron $c$ has purely real hopping with zero flux (in the absence of an applied physical magnetic field). From these facts and Eq.~(\ref{eq:CBF}) we reach the important conclusion that the boson $B$ must also have purely imaginary hopping with $\pi$-flux (the $i w$ term in Eq.~(\ref{Bfunctional}) below). So the relation
\begin{equation}
T_x T_y = - T_y T_x\,, \label{txty}
\end{equation}
realizing the $\pi$-flux applies both to the spinons and to $B$.
We can also reach these conclusions, and obtain other constraints, by examining the action of all symmetry operators of $f$, and use Eq.~(\ref{eq:CBF}) to deduce the action of symmetry operations on $B$: the results are summarized in Table~\ref{tab1}.
\begin{table}
    \centering
    \begin{tabular}{|c|c|c|}
\hline
Symmetry & $f_{\alpha}$ & $B_a $  \\
\hline 
\hline
$T_x$ & $(-1)^{y} f_{ \alpha}$ & $(-1)^{y} B_a $ \\
$T_y$ & $ f_{ \alpha}$ & $ B_a $ \\
$P_x$ & $(-1)^{x} f_{ \alpha}$  & $(-1)^{x} B_a $ \\
$P_y$ & $(-1)^{y} f_{ \alpha}$  & $(-1)^{y} B_a $ \\
$P_{xy}$ & $(-1)^{xy} f_{ \alpha}$ & $(-1)^{x y} B_a $ \\
$\mathcal{T}$ & $(-1)^{x+y} \varepsilon_{\alpha\beta} f_{ \beta}$  & $(-1)^{x + y} B_a $\\
\hline
\end{tabular}
    \caption{Projective transformations of the $f$ spinons and $B$ chargons on lattice sites $\vi = (x,y)$ 
    under the symmetries $T_x: (x,y) \rightarrow (x + 1, y)$; $T_y: (x,y) \rightarrow (x,y + 1)$; 
    $P_x: (x,y) \rightarrow (-x, y)$; $P_y: (x,y) \rightarrow (x, -y)$; $P_{xy}: (x,y) \rightarrow (y, x)$; and time-reversal $\mathcal{T}$.
    The indices $\alpha,\beta$ refer to global SU(2) spin, while the index $a=1,2$ refers to gauge SU(2). 
    }
    \label{tab1}
\end{table}
These considerations lead to the energy functional $\mathcal{E}_2 [B, U] +  \mathcal{E}_4 [B, U]$ with terms quadratic and quartic in $B$ respectively:
\begin{eqnarray}
\mathcal{E}_2 [B, U] &=&  (r + 2 \sqrt{2} w) \sum_\vi B^\dagger_\vi B_\vi +  i w \sum_{\langle \vi \vj\rangle} e_{\vi \vj} \left( B_\vi^\dagger U_{\vi \vj} B_\vj - B_\vj^\dagger U_{\vj \vi} B_\vi \right) 
\nonumber \\ &~& + \kappa \sum_{\square} 
\left\{ 1- \frac{1}{2} \mbox{Re} \mbox{Tr} \prod_{\vi\vj \in \square} U_{\vi \vj} \right\} \nonumber \\
\mathcal{E}_4 [B, U] &=& \frac{u}{2} \sum_{\vi} \rho_{\vi}^2   + V_1 \sum_{\vi} \rho_{\vi} \left( \rho_{\vi + \hat{\bm x}} + \rho_{\vi + \hat{\bm y}} \right) 
 + g \sum_{\langle \vi \vj \rangle} \left| \Delta_{\vi\vj} \right|^2 
 + J_1 \sum_{\langle \vi \vj \rangle}  Q_{\vi\vj}^2 \nonumber \\
 &~&+ K_1 \sum_{\langle \vi \vj \rangle}  J_{\vi\vj}^2 + V_{11} \sum_{\vi} \rho_{\vi} \left( \rho_{\vi + \hat{\bm x}+ \hat{\bm y}} + \rho_{\vi +\hat{\bm x}- \hat{\bm y}} \right) \nonumber \\
&~&+ V_{22} \sum_{\vi} \rho_{\vi} \left( \rho_{\vi + 2\hat{\bm x}+ 2\hat{\bm y}} + \rho_{\vi + 2\hat{\bm x}- 2\hat{\bm y}} \right)\,.
\label{Bfunctional}
\end{eqnarray}
The quartic terms are expressed as products of bilinears of $B$ which are associated with various gauge-invariant observables as identified below
\begin{eqnarray}
&&\mbox{site charge density:~}\left\langle c_{\vi \alpha}^\dagger c_{\vi \alpha}^{\vphantom\dagger} \right\rangle \sim \rho_{\vi} = B^\dagger_\vi B_\vi^{\vphantom\dagger} \nonumber \\
&&\mbox{bond density:~} \left\langle c_{\vi \alpha}^\dagger c_{\vj \alpha}^{\vphantom\dagger} + c_{\vj \alpha}^\dagger c_{\vi \alpha}^{\vphantom\dagger} \right\rangle~\sim Q_{\vi \vj} = Q_{\vj\vi} = \mbox{Im} \left(B^\dagger_\vi e_{\vi \vj\vphantom\dagger} U_{\vi\vj\vphantom\dagger} B_\vj \right) \nonumber \\
&&\mbox{bond current:~} i\left\langle c_{\vi \alpha}^\dagger c_{\vj \alpha}^{\vphantom\dagger} - c_{\vj \alpha}^\dagger c_{\vi \alpha}^{\vphantom\dagger} \right\rangle \sim J_{\vi \vj} = - J_{\vj\vi} =  \mbox{Re} \left( B^\dagger_\vi e_{\vi \vj\vphantom\dagger} U_{\vi \vj\vphantom\dagger} B_\vj^{\vphantom\dagger} \right) \nonumber \\
&&\mbox{Pairing:~} \left\langle \varepsilon_{\alpha\beta} c_{\vi \alpha} c_{j \beta} \right\rangle \sim \Delta_{\vi \vj} = \Delta_{\vj \vi} = \varepsilon_{ab} B_{a\vi} e_{\vi \vj} [U_{\vi \vj}]_{bc} B_{c\vj}\,. 
\label{latticeorders}
\end{eqnarray}
These relations are the key to $B$ providing a fractionalized description of intertwined orders. We have retained terms involving nearest neighbor sites in Eq.~(\ref{Bfunctional}), and a few terms with longer-range density-density interactions. 
\begin{figure}
\centering
\includegraphics[width=6in]{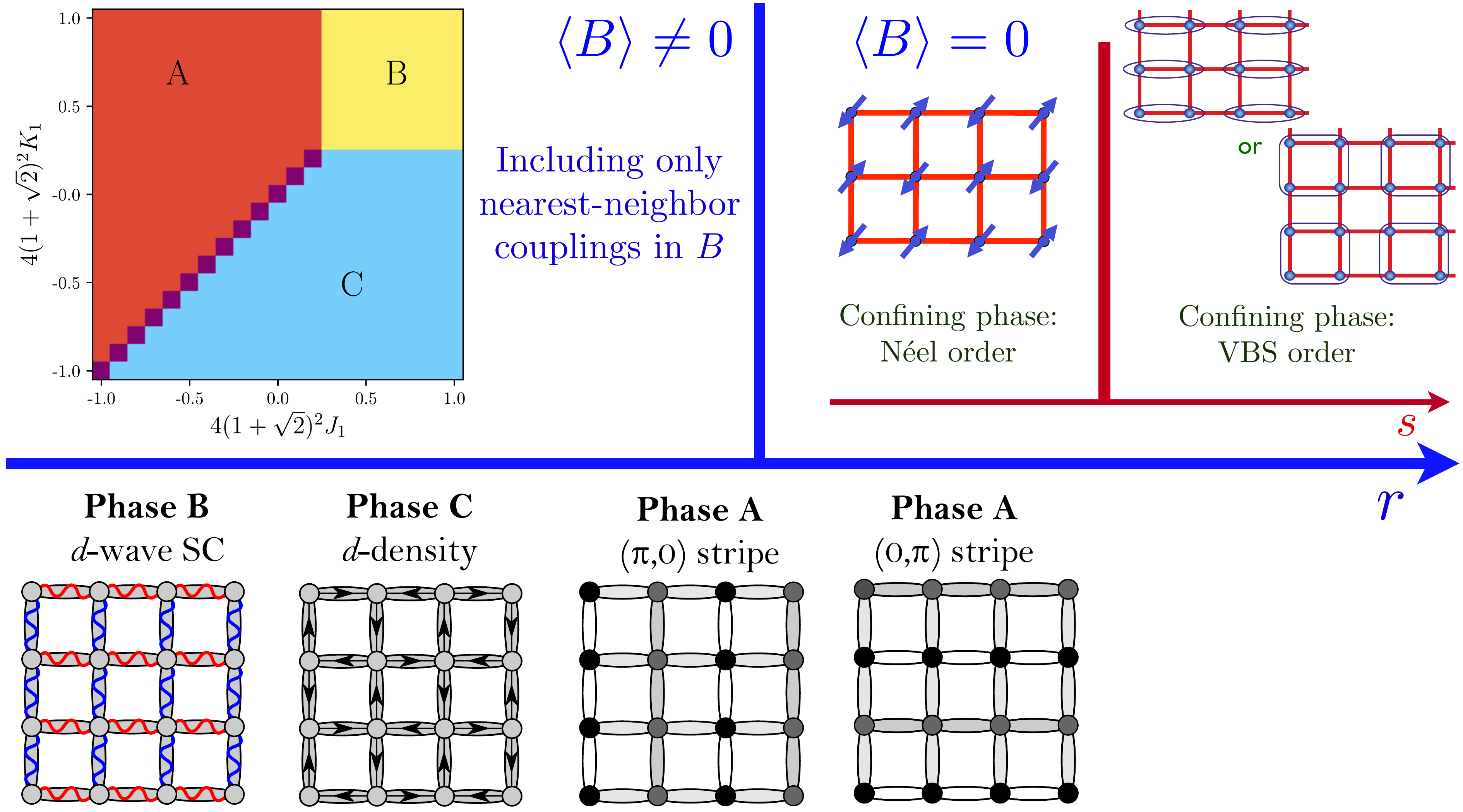}
\caption{Mean field phase diagram obtained by minimization of the Higgs potential of $B$, $\mathcal{E}_2 + \mathcal{E}_4$ (from Ref.~\cite{Christos:2023oru}). For $r>0$, there is no Higgs condensate $\langle B \rangle =0$, and we obtain the same phases as in the insulator from the confinement of the fermionic spinons described by Eq.~(\ref{eq:fermionhop2}). For $r<0$, $\langle B \rangle\neq 0$, and we minimized the Higgs potential with only nearest neighbor interactions by setting $V_{11}=V_{22}=0$. Modifications with further neighbor interactions appear in Fig.~\ref{fig:period4}.}
\label{fig:pd1}
\end{figure}

Fig.~\ref{fig:pd1} shows a phase diagram obtained by minimizing the energy functional with nearest-neighbor interactions only ($V_{11} = V_{22}=0$). Three phases are found, also illustrated in Fig.~\ref{fig:pd1}:
\begin{enumerate}[label=\Alph*.]
\item This state A charge stripe order with period 2, centered on the sites.
\item A $d$-wave superconductor, with $\Delta_{\vi,\vi+\hat{\bm x}} = - \Delta_{\vi,\vi+\hat{\bm y}}$.
\item A ``$d$-density wave'' state which has a staggered pattern of spontaneous current.
\end{enumerate}
An interesting feature is that these orders are degenerate in the quadratic energy functional $\mathcal{E}_2$, and the degeneracy is broken only at quartic order in $\mathcal{E}_4$. The fact that the leading term is degenerate provides a rationale for nearly-degenerate multiple competing or `intertwined' orders \cite{Christos:2023oru}; more conventional Landau theory approaches \cite{Fradkin10,Hayward:2013jna,Lee14,Nie_15,Castro17,Fradkin15,Pepin23,Fradkin25} do not have any term in which the degeneracy is exact without fine-tuning.

Our primary interest for now is phase B. Remarkably, the structure of the $\pi$-flux spin liquid, and consequently  the $\pi$-flux on $B$ leads to $d$-wave pairing, and not $s$-wave pairing. Also, once $B$ is condensed, we can identify $c \sim f$ via Eq.~(\ref{eq:CBF}), and so the electron spectral function will inherit nodal Bogoliubov quasiparticles from the massless Dirac spinons. 
The main phenomenological difficulty, as noted earlier, is that the Bogoliubov quasiparticles will have isotropic dispersion, as in Eq.~(\ref{eq:fermiondisp}) and Fig.~\ref{fig:fermiondisp}. However, other features of the $d$-wave state obtained from the energy functional in 
Eq.~(\ref{Bfunctional}) do match observations, including vortices with flux $h/(2e)$ (despite the boson $B$ having charge $e$), and intertwined charge order halos of vortex cores. As these features apply also to the doped case, we defer their discussion to Section~\ref{sec:dwave}, where we will also fix the difficulty with the anisotropic velocities in Section~\ref{sec:aniso}. 

\begin{figure}
\centering
\includegraphics[width=3.5in]{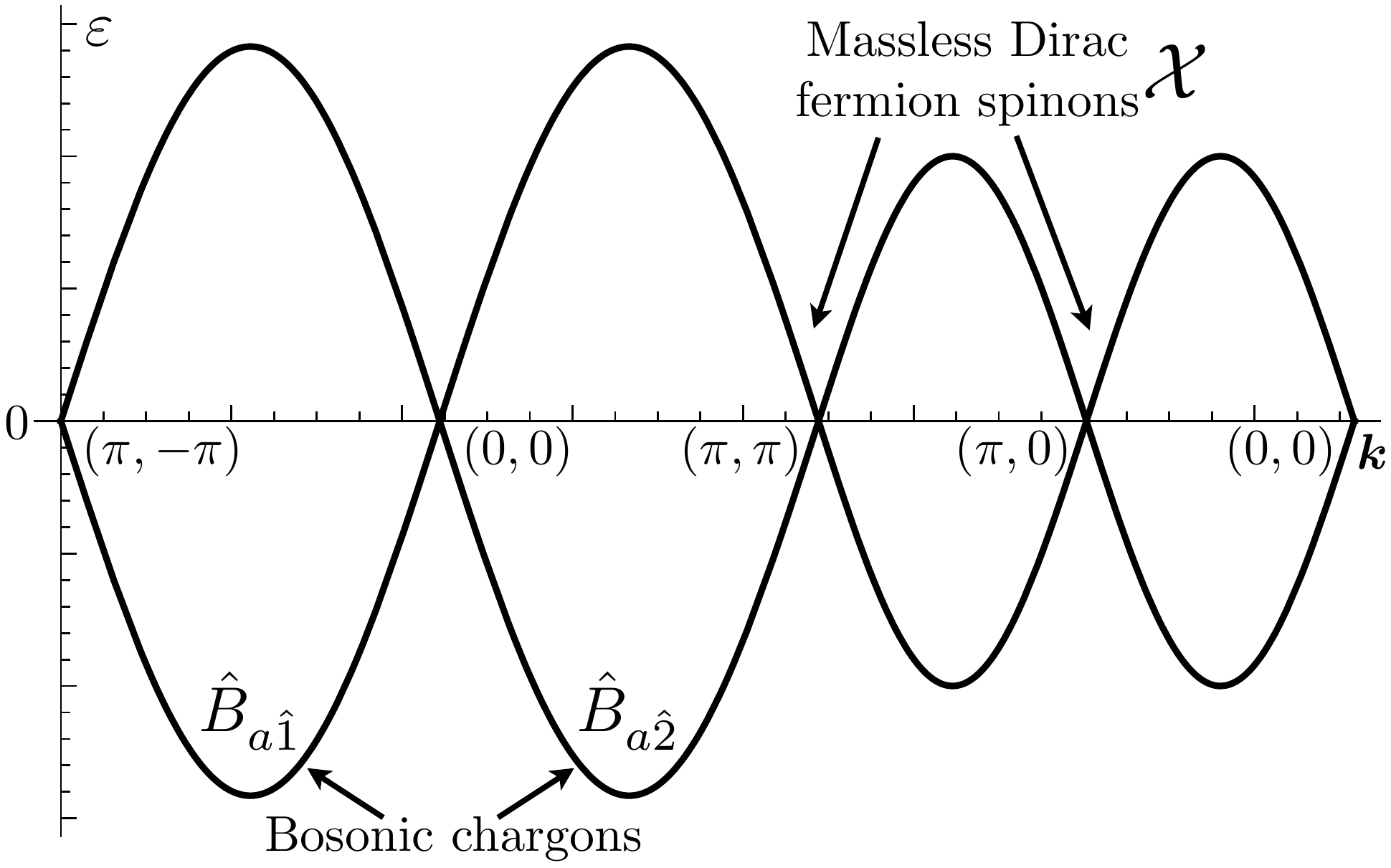}
\caption{Common dispersion of the fermionic spinons $f$, and the bosonic chargons $B$. The continuum fermionic fields $\mathcal{X}$ are defined at zero energy, while the continuum bosonic fields $\hat{B}$ are defined at the minimum energy.}
\label{fig:chargondisp}
\end{figure}

We now discuss more global aspects of the phase diagram shown in Fig.~\ref{fig:pd1}, as a function of the tuning parameter $r$, which is the `mass' of $B$. 
\begin{itemize}
\item
When $r$ is large and positive, then we can ignore the $B$ sector, and revert to the spinon only theory of Section~\ref{sec:fermions}. The low energy theory is Eq.~(\ref{e105}), and we expect a confining insulator with either N\'eel or VBS order as the ground state. 
\item 
When $r$ is negative, $B$ condenses, and this has the salutary effect of making the gauge field $A$ massive, as in the Higgs phenomenon. In this case, a mean-field treatment of interactions in the bosonic sector only is qualitatively valid, and we obtain one of three states A,B,C listed above.
\end{itemize}
It is also interesting to consider the nature of the low energy theory when we approach the Higgs condensation transition \cite{ChristosLuo24}. The dispersion of the $B$ bosons is the same as that of the $f$ fermions, apart from an overall constant - see Fig.~\ref{fig:chargondisp}. 
While the low energy fermions are at the Fermi level in the middle of the band, the low energy bosons will condense near the bottom of the band. Because there are two such minima in the $\pi$-flux band structure, we can identify two valleys $v=\hat{1}, \hat{2}$, and introduce a continuum field $\hat{B}_{av}$ for the low energy theory, where $a=1,2$ is a SU(2) gauge space index (this is similar to Section~\ref{sec:U1}). From the lattice transformations in Table~\ref{tab1}, we can also relate the order parameters A,B,C to gauge-invariant bilinears of $\hat{B}$
\begin{eqnarray}
\mbox{$d$-wave superconductor} &&:~~ \varepsilon_{ab} \hat{B}_{a \hat{1}} \hat{B}_{b \hat{2}} \equiv \Delta \nonumber \\
\mbox{$x$-CDW} &&:~~  \hat{B}_{a\hat{1}}^\ast \hat{B}_{a\hat{1}}^\pdagger-  \hat{B}_{a\hat{2}}^\ast \hat{B}_{a\hat{2}}^\pdagger \equiv \hat{B}^\dagger \mu^z \hat{B}^\pdagger \nonumber  \\
\mbox{$y$-CDW}&&:~~ \hat{B}_{a\hat{1}}^\ast \hat{B}_{a\hat{2}}^\pdagger +  \hat{B}_{a\hat{2}}^\ast \hat{B}_{a\hat{1}}^\pdagger \equiv \hat{B}^\dagger \mu^x \hat{B}^\pdagger  \nonumber \\
\mbox{$d$-density wave}&&:~~ i \left( \hat{B}_{a\hat{1}}^\ast \hat{B}_{a\hat{2}}^\pdagger -  \hat{B}_{a\hat{2}}^\ast \hat{B}_{a\hat{1}}^\pdagger\right) \equiv -\hat{B}^\dagger \mu^y \hat{B}^\pdagger  \,.
\label{contorders}
\end{eqnarray}
Here $\mu^\ell$ are the Pauli matrices acting in the boson valley space of Fig.~\ref{fig:chargondisp}.
\begin{figure}
\centering
\includegraphics[width=3.5in]{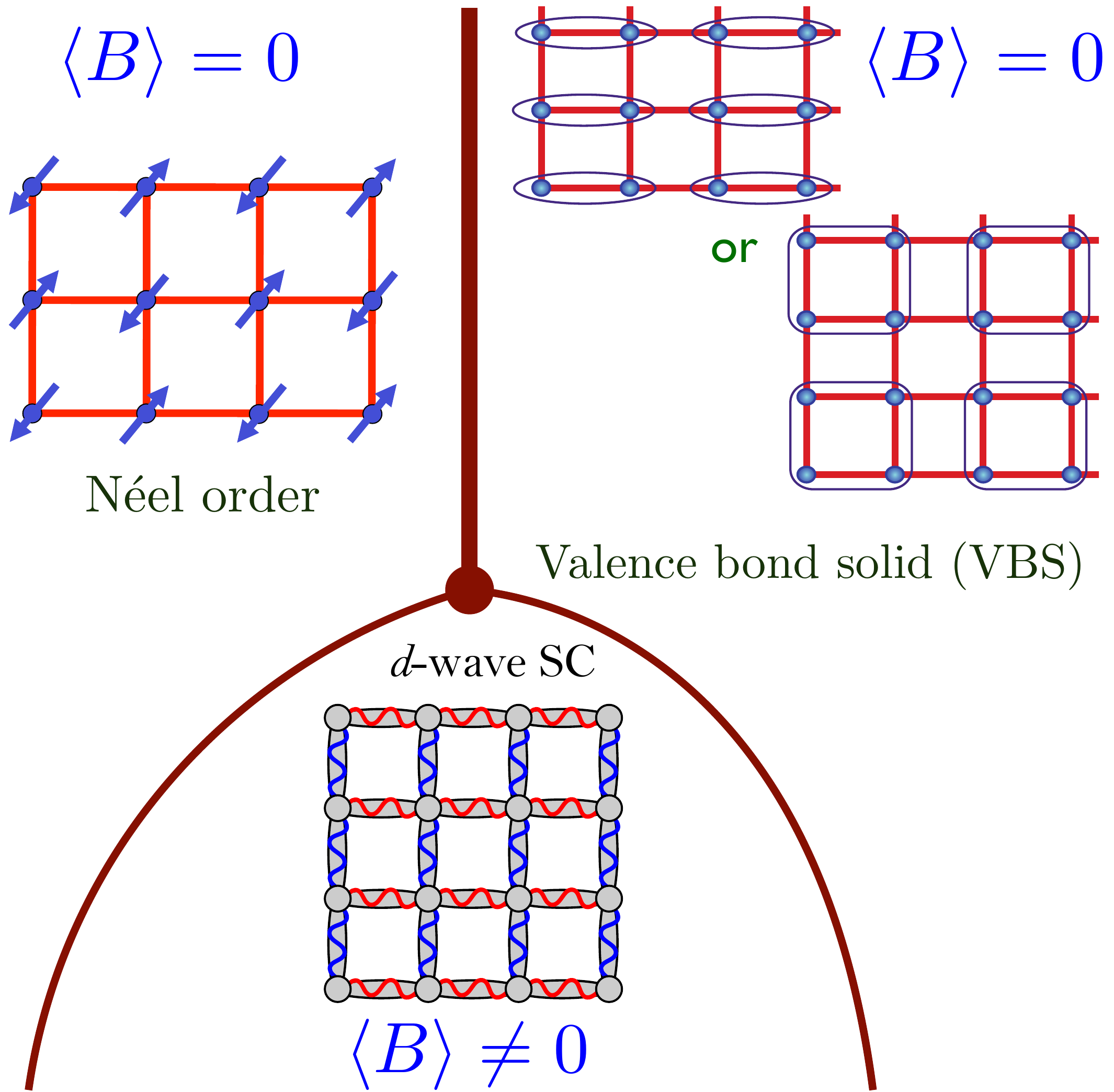}
\caption{Proposed phase diagram of $\mathcal{L}_{\mathcal{X}\hat{B}}$.}
\label{fig:pd2}
\end{figure}

We now assume that the quartic couplings of $B$ are such that the ground state in the Higgs phase is a $d$-wave superconductor. Then, we can sketch the phase diagram in Fig.~\ref{fig:pd2} with the 3 important phases - N\'eel, VBS, and $d$-wave superconductivity. 
Near the onset of the $B$ condensate, we extend the spinon continuum theory of Eq.~(\ref{e105}) to also include phases A,B,C by adding a continuum Lagrangian for $\hat{B}$: 
\begin{eqnarray}
&& \mathcal{L}_{\mathcal{X}\hat{B}} =   \mathcal{L}_\mathcal{X} + \left| \left( \partial_\mu - i A_\mu^\ell \sigma^\ell \right) \hat{B} \right|^2 + r |\hat{B}|^2 + \bar{u} |\hat{B}|^4  \nonumber \\
    && + v_1 \left( \hat{B}^\dagger \mu^z \hat{B} \right)^2 + v_1 \left(   \hat{B}^\dagger \mu^x \hat{B} \right)^2  + v_2 \left(  \hat{B}^\dagger \mu^y \hat{B} \right)^2 + v_3 \left|  \varepsilon_{ab} \hat{B}_{a 1} \hat{B}_{b 2} \right|^2 \,. \label{LXB}
\end{eqnarray}
We have added only a relativistic time derivative term for $\hat{B}$, which is the allowed term at half-filling with particle-hole symmetry.

\section{SU(2) gauge theory of the cuprate pseudogap}
\label{sec:pseudogap}

Now we return to the analysis of the pseudogap using the ALM in Eq.~(\ref{eq:Hancilla}). In Section~\ref{sec:sFL*}, we presented a mean field analysis in terms of decoupled Kondo lattice and spin liquid models. This section will couple them using the methods developed in Sections~\ref{sec:spinliquids} and \ref{sec:halffilling}.

On the spin liquid layer of ${\bm S}_2$ spins we write a parton decomposition which parallels that in Eq.~(\ref{Schwingerfermion})
\begin{eqnarray}
{\bm S}_{2\vi} = \frac{1}{2} f_{\vi \alpha}^\dagger {\bm \sigma}_{\alpha\beta}^{\vphantom\dagger} f_{\vi \beta}^{\vphantom\dagger}\,.
\label{eq:S2f}
\end{eqnarray}
Then the analysis of the exchange interactions within the ${\bm S}_2$ layer is precisely that in Section~\ref{sec:fermions}.
W decouple the $J_{ \vi\vj}$ term in Eq.~(\ref{eq:Hancilla}) to $\mathcal{H}_{SLf}$ in Eq.~(\ref{eq:fermionhop2}) realizing a $\pi$-flux state of the ${\bm S}_2$ spins with 
a SU(2) gauge field. 

To couple this spin liquid to the Kondo lattice, 
we have to decouple the 
$J_\perp$ term coupling the $f_{1}$ and $f$ spinons. Given the SU(2) gauge structure of the ${\bm S}_2$ layer, it pays to decouple the $J_\perp$ term in a manner which keeps the SU(2) gauge invariance explicit. In fact, the needed decoupling field is precisely the boson $\mathcal{B}_\vi$ introduced in 
Eq.~(\ref{defB}). We also introduce a matrix fermion operator $\mathcal{F}_{1\vi}$
\begin{equation}
\mathcal{F}_{1\vi} \equiv \left(
\begin{array}{cc}
f_{1\vi \uparrow} & - f_{1\vi \downarrow} \\
f_{1\vi \downarrow}^\dagger & f_{1\vi \uparrow}^\dagger
\end{array}
\right), \label{F1matrix}
\end{equation}
whose transformations under the symmetries in Eqs.~(\ref{eq:gauge},\ref{eq:spin},\ref{eq:charge}) are the same as those of $\mathcal{C}_\vi$.
We summarize the gauge and symmetry properties in Table~\ref{tab2}.
\begin{table}
    \centering
    \begin{tabular}{|c|c||c||c|c|}
\hline
\multirow{2}{*}{Field} & \multirow{2}{*}{Layer} & Gauge & \multicolumn{2}{c|}{Global} \\
\cline{3-5}
 & & SU(2) & SU(2) & U(1) \\ \hline
 $c$ or $\mathcal{C}$ & 1 & ${\bm 1}$ & ${\bm 2}_R$ & -1 \\
 \hline
 $f_1$ or $\mathcal{F}_1$ & 2 & ${\bm 1}$ & ${\bm 2}_R$ & -1 \\ \hline
 $f$ or $\mathcal{F}$ & 3 & ${\bm 2}_L$ & ${\bm 2}_R$ & 0 \\ \hline
 $B$ or $\mathcal{B}$ & $2 \leftrightarrow 3$ & ${\bm 2}_L$ & ${\bm 1}$ & 1 \\ \hline
\end{tabular}
    \caption{Summary of gauge and global symmetry transformations for the fields of the ALM in the FL* phase. The representations of the SU(2) are indicated by their dimension; the subscripts $L$/$R$ indicate whether the SU(2) acts by left/right multiplication in the matrix form of the field. The representations of the global U(1) is the electrical charge in units of $e$. For the fermions, the layer column indicates the layer number is Fig.~\ref{fig:ancilla}. For the bosons,  the layer column indicates the layers between which there is a Yukawa coupling to the fermions.}
    \label{tab2}
\end{table}

Then, from the $J_\perp$ term, symmetry considerations are sufficient to constrain the structure of the Yukawa term between ${B}$ and the fermions \cite{Christos:2023oru}, which follows from Eq.~(\ref{eq:CBF}) and is illustrated in Fig.~\ref{fig:PhiB}:
\begin{figure}
\centering
\includegraphics[width=3in]{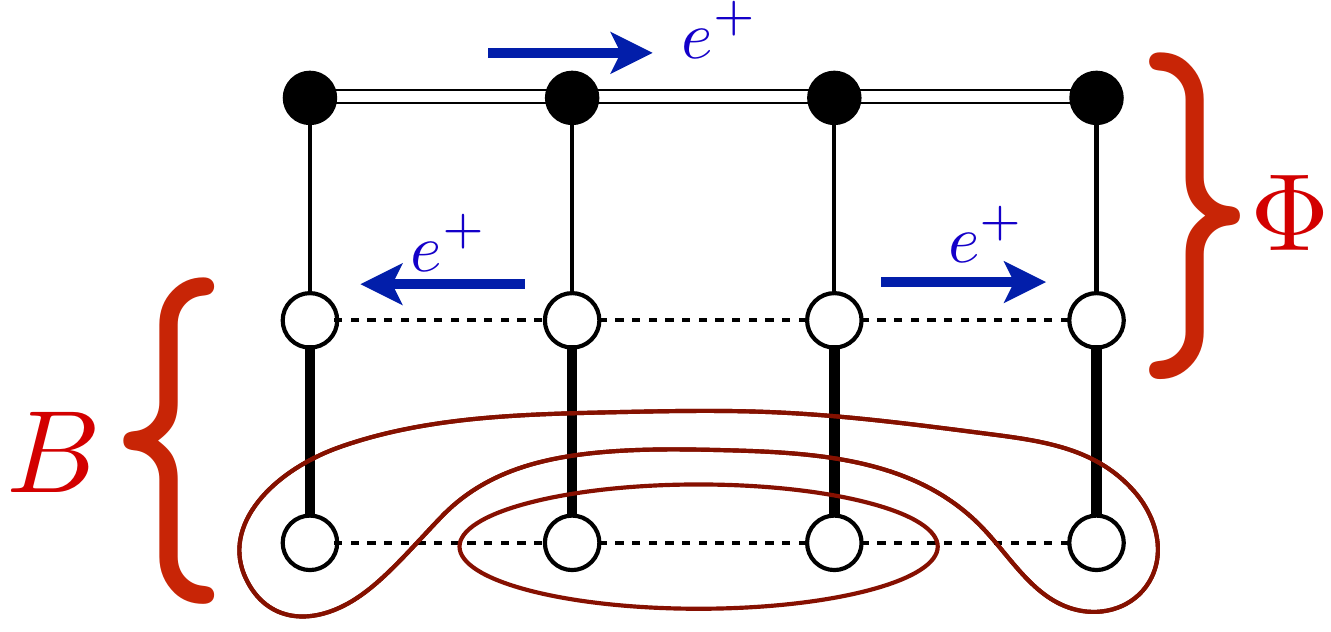}
\caption{The two distinct Higgs fields in the ancilla layer theory of the single band Hubbard model. $\Phi$ hybridizes conduction electrons in the top layer with spinons in the middle layer. $B$ couples the spinons of the bottom layer to the upper layers.}
\label{fig:PhiB}
\end{figure}
\begin{eqnarray}
   \mathcal{H}_Y  && = -\frac{1}{2} \sum_\vi \left[ i \, \mbox{Tr} \left(\mathcal{F}_{1\vi}^\dagger \mathcal{B}_{\vi}^\dagger \mathcal{F}_{i}^{\vphantom\dagger} \right) +  i g \, \mbox{Tr} \left(\mathcal{C}_{\vi}^\dagger \mathcal{B}_{\vi}^\dagger \mathcal{F}_{i}^{\vphantom\dagger}\right) +\mbox{H.c.} \right] \nonumber \\
   && =  \sum_\vi \left[  i \left( B_{1\vi}^{\vphantom\dagger} f_{\vi\alpha}^\dagger f_{1\vi \alpha}^{\vphantom\dagger} - B_{2 \vi}^{\vphantom\dagger} \varepsilon_{\alpha\beta}^{\vphantom\dagger} f_{\vi \alpha}^{\vphantom\dagger} f_{1\vi \beta}^{\vphantom\dagger} \right)
   + \mbox{H.c.}\right. \nonumber \\
   && ~~~~~~~~~~~~~~~ \left.   + i g \left( B_{1\vi}^{\vphantom\dagger} f_{\vi\alpha}^\dagger c_{\vi \alpha}^{\vphantom\dagger} - B_{2 \vi}^{\vphantom\dagger} \varepsilon_{\alpha\beta}^{\vphantom\dagger} f_{\vi \alpha}^{\vphantom\dagger} c_{\vi \beta}^{\vphantom\dagger} \right)
   + \mbox{H.c.}\right] \,,
   \label{Yukawa}
\end{eqnarray}
We have also included a Yukawa coupling to $c_\alpha$ from an allowed term $\sim {\bm S}_{2\vi} \cdot c_{\vi \alpha}^\dagger {\bm \sigma}_{\alpha\beta} c_{\vi \beta}$, which descends from the Kondo coupling noted below Eq.~(\ref{PS1S2}).

We can now collect all terms to write down the complete Hamiltonian needed for our analysis of the pseudogap metal, and its low temperature instabilities.
\begin{equation}
\mathcal{H}_{\rm pseudogap} = \mathcal{H}_{\rm KLmf} + \mathcal{H}_{SLf} + \mathcal{H}_Y + \mathcal{E}_2 [B,U] + \mathcal{E}_4 [B,U] \label{eq:WS}
\end{equation}
specified in Eqs.~(\ref{HKLmf}), (\ref{eq:fermionhop2}), (\ref{Yukawa}), (\ref{Bfunctional}). This Hamiltonian has 3 fermions $c_\alpha$, $f_{1 \alpha}$, $f_\alpha$ whose transformations under SU(2) gauge, spin rotation, and electromagnetic charge symmetries are in Eqs.~(\ref{eq:gauge}), (\ref{eq:spin}), (\ref{eq:charge}), with $f_{1\alpha}$ transforming just like $c_\alpha$. As we noted earlier, the boson $\Phi$ will be treated as a $c$-number constant for now, although we will consider its quantum and thermal fluctuations in Section~\ref{sec:strange} when we consider the transition to the FL phase at large doping. The completion of $\mathcal{H}_{\rm pseudogap}$ for the case of dynamic $\Phi$ is presented in \hyperref[app:ancilla]{Appendix B}.

The boson $B$ couples the Kondo Lattice to the spin liquid on the bottom layer, with SU(2) gauge field $U_{\vi \vj}$. If we ignore this couping, the Kondo lattice yields the `small' heavy Fermi liquid Fermi surface of the FL* phase described in Section~\ref{sec:photo_pocket}. Section~\ref{sec:arc} will consider fluctuations of $B$ to obtain a complete description of the electronic spectrum of the FL* phase. When $B$ condenses, we quench the SU(2) gauge field $U_{\vi \vj}$, and the resulting confining phases will be discussed in Section~\ref{sec:dwave}.

There is a remarkable similarity between Eq.~(\ref{eq:WS}), and the Weinberg-Salam theory of weak interactions \cite{Christos:2023oru}. Although the dispersions of the fermions and bosons have a lattice structure, the SU(2)$\times$ U(1) gauge structure (we treat the electromagnetic U(1) as global), and the Yukawa couplings between the Higgs and the fermions are similar, with the spinons mapping to neutrinos, and the electrons mapping to  electrons.

\subsection{Photoemission spectrum}
\label{sec:arc}

\begin{figure}
\centering
\includegraphics[width=3.5in]{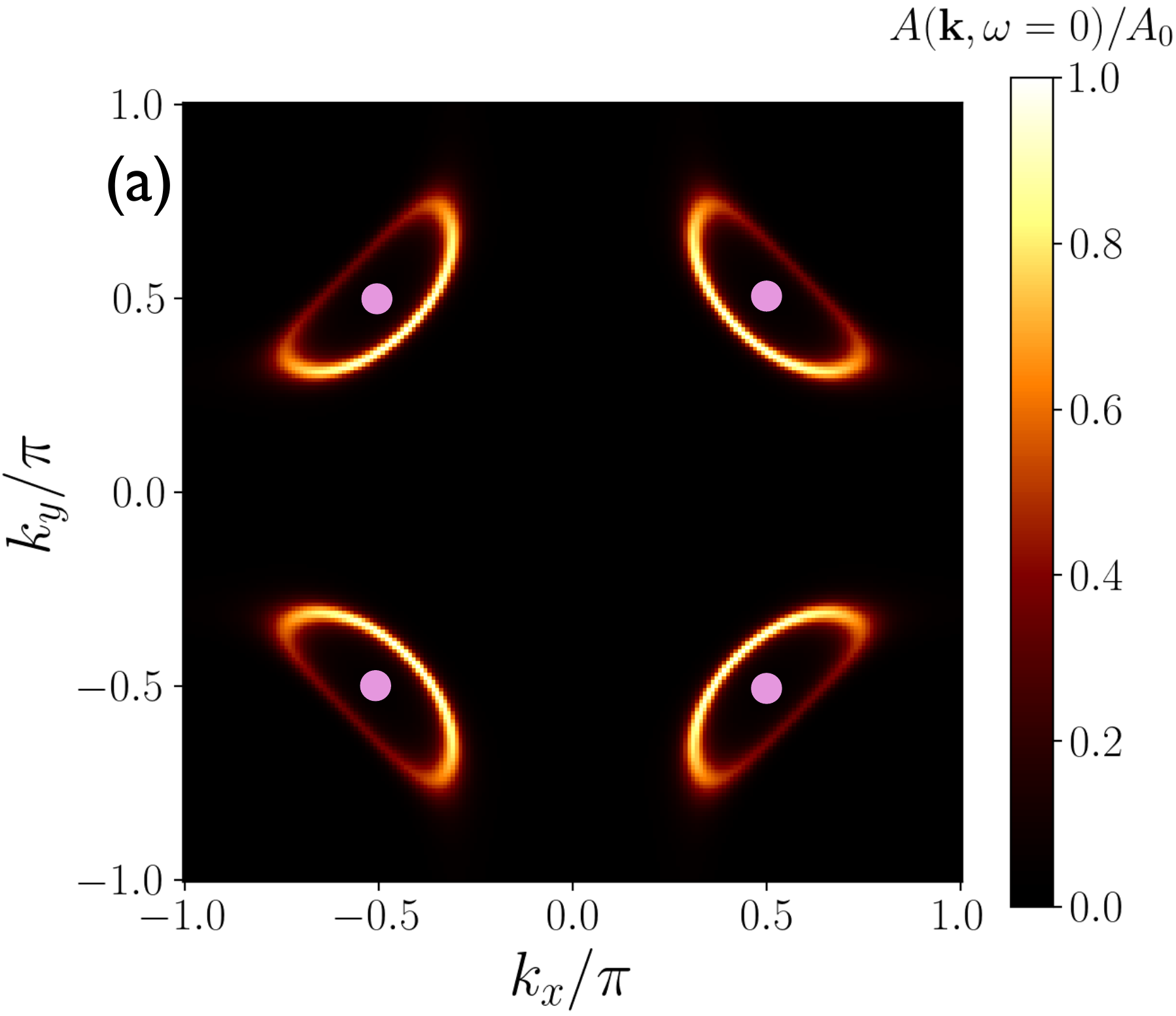}
\caption{From Ref.~\cite{Sayantan25}. Low energy spectrum of FL*, showing the photoemission spectrum of hole pockets of Fig.~\ref{fig:photo}, and the positions of the nodal points of the $f_\alpha$ Dirac spinons (pink dots). The spinon spectrum is as in Fig.~\ref{fig:chargondisp}, but in a gauge where the $B$ spectrum has a minimum at zero momentum. The nodal points are at $(\pm \pi/2, \pm \pi/2)$. The spinons couple to the hole pockets via the fluctuations of $B$ in the Yukawa coupling in Eq.~(\ref{Yukawa}).}
\label{fig:fls_spinon}
\end{figure}

In the mean-field analysis of Section~\ref{sec:photo_pocket}, we obtained hole pocket description of the Fermi surface in Fig.~\ref{fig:photo}, along with the dispersion antinodal pseudogap in Fig.~\ref{fig:antinode}. Here, we will consider modification of the Fermi surface from the coupling to the bottom layers of the ALM, where the spinons reside. Should the FL* state survive at $T=0$, we expect the coupling of the hole pockets to the spinons to be relatively innocuous, as their interaction is mediated by the gapped $B$ bosons, and their zero energy modes lie at distinct momenta (see Fig.~\ref{fig:fls_spinon})---there remains a non-zero quasiparticle residue around the hole pocket at $T=0$ \cite{Mascot22}.

\begin{figure}
    \centering
    \includegraphics[width=3.5in]{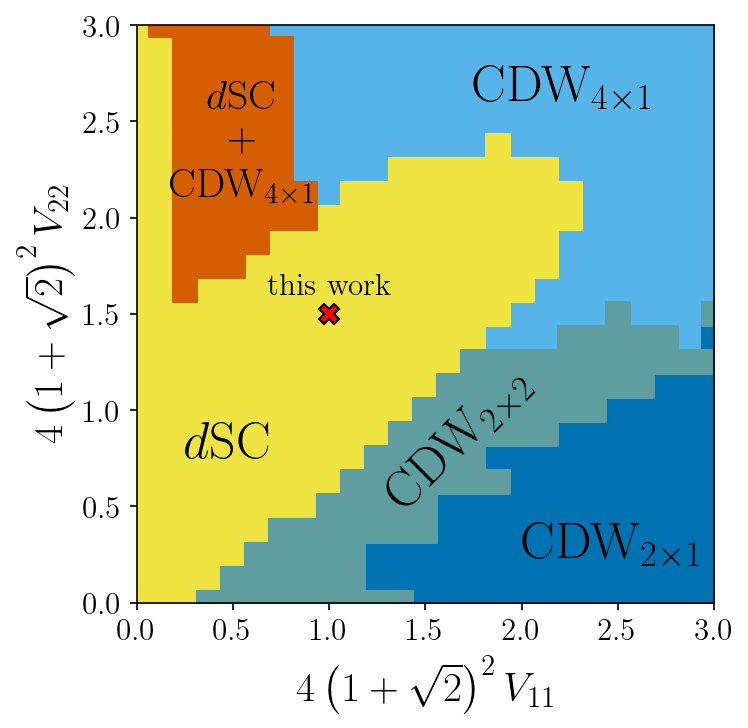}
    \caption{From Ref.~\cite{Sayantan25}. Mean-Field phase diagram of Eq.~(\ref{Bfunctional}) a function of the further neighbor interactions $V_{11}$ and $V_{22}$, extending that in Fig.~\ref{fig:pd1}. The other parameters are
    $r=-0.732$, $w=0.40$, $u=0$, $V_1=0$, $g=0.021446$, $J_1=K_1={2}/({4(1+\sqrt{2})^2})$.
    $d$SC is $d$-wave superconductivity, CDW$_{n\times m}$ is a charge density wave with a supercell with $n\times m$ lattice sites. The red cross marks the parameter values chosen for the Monte Carlo simulations of Eq.~(\ref{Z20}) in Ref.~\cite{Sayantan25}.}
    \label{fig:period4}
\end{figure}

However, there is a stronger effect at non-zero $T$, above low $T$ phases (such as those in Fig.~\ref{fig:period4}) where the spinons are confined.
The computations of Pandey {\it et al} \cite{Sayantan25} of thermal fluctuations included the coupling to the spinons in $\mathcal{H}_{\rm pseudogap}$ in Eq.~(\ref{eq:WS}). This can be expected to have a stronger effect along the Brillouin zone diagonals proximate to the spinon nodes, as shown in Fig.~\ref{fig:fls_spinon}.
Indeed they found that upon including thermal fluctuations the backside of the hole pockets are suppressed, and spectral weight turned into Fermi arcs, similar to observations.

They carried out Monte Carlo simulations of the classical, thermal, lattice gauge theory for $B$ and $U_{\vi \vj}$ defined by the partition function
\begin{equation}
 \mathcal{Z}_{2+0} = \int \prod_i \mathcal{D} B_i \int \prod_{\langle ij \rangle} \mathcal{D} U_{ij} \exp \left( - 
(\mathcal{E}_2 [B, U] + \mathcal{E}_4 [B, U])/T \right)\,, \label{Z20}
\end{equation}
where the energy functionals are defined in Eq.~(\ref{Bfunctional}). They observe a Kosterlitz-Thouless transition to a $d$-wave superconductor at low energies. The choice of the coupling constants in $\mathcal{E}_4 [B, U]$ was motivated by the phase diagram shown in Fig.~\ref{fig:period4}: at the chosen point the ground state is a $d$-wave superconductor, while the next best state has period 4 charge order. 

\begin{figure}
\centering
\includegraphics[width=5.2in]{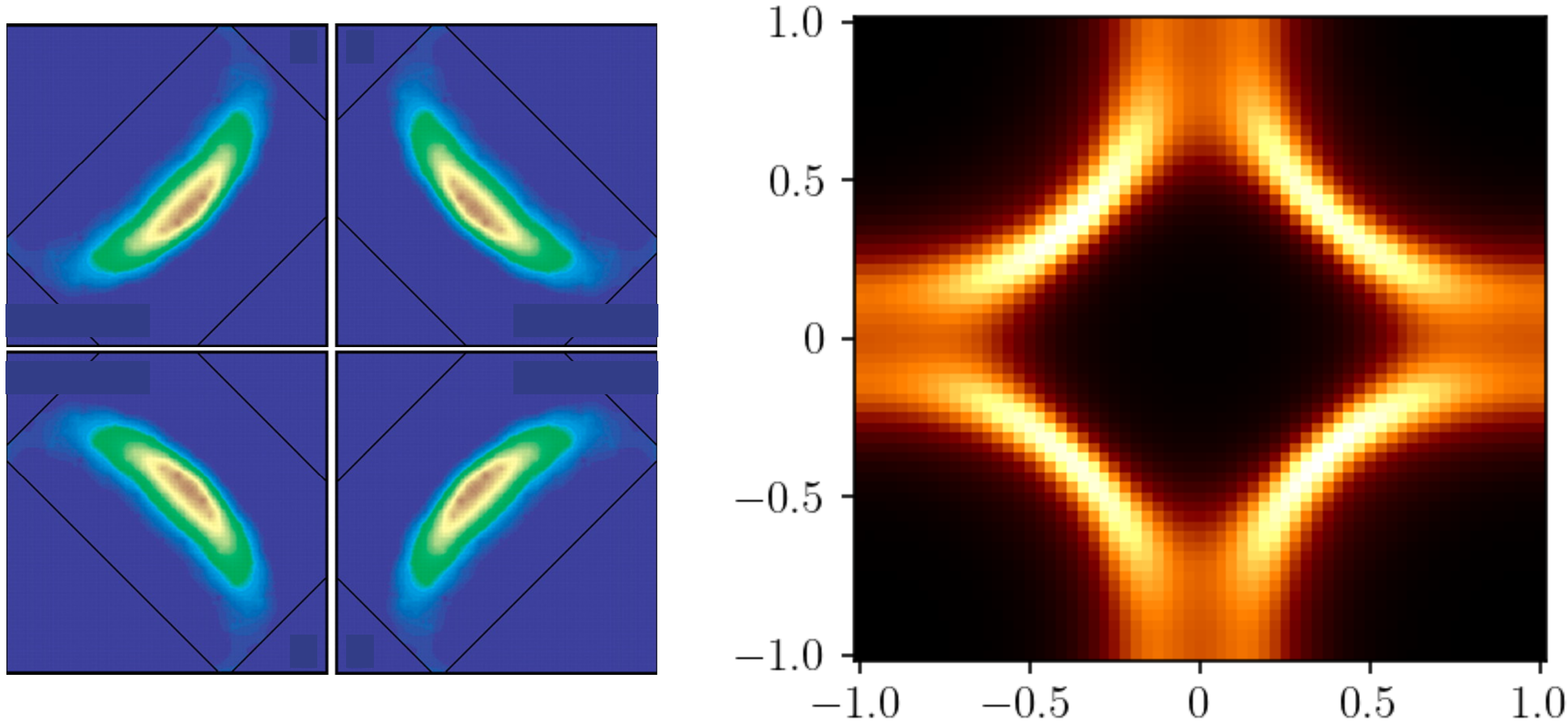}
\caption{The right panel shows the electron spectral function measured in photoemission computed in Ref.~\cite{Sayantan25} from $\mathcal{H}_{\rm KLmf} + \mathcal{H}_{SLf} + \mathcal{H}_Y$ after averaging over the Monte Carlo simulation of the classical, thermal fluctuations of $B$ and $U_{\vi \vj}$ in Eq.~(\ref{Z20}). The spectrum with $B=U=0$ is in Fig.~\ref{fig:fls_spinon}, showing both the hole pockets and the spinons. The left panel is experimental photoemission data from Ref.~\cite{ShenShen05}.}
\label{fig:arc}
\end{figure}

Fig.~\ref{fig:photo} showed the prediction for the photoemission spectrum in the pseudogap phase without the coupling to the ${\bm S}_2$ spin liquid.
Pandey {\it et al.\/} \cite{Sayantan25} included the coupling to thermal fluctuations of $B$ via the Yukawa coupling in Eq.~(\ref{Yukawa}). The thermal fluctuations of $B$ and $U_{\vi \vj}$ were included via the ensemble in Eq.~(\ref{Z20}). Here the Born-Oppenheimer procedure is to choose random samples of the $B_\vi$ and $U_{\vi\vj}$, diagonalize $\mathcal{H}_{\rm KLmf} + \mathcal{H}_{SLf} + \mathcal{H}_Y$ for each sample, and average over the spectral functions. The results are shown in Fig.~\ref{fig:arc}.
Note that the back side of the pocket in Fig.~\ref{fig:photo} is now invisible, and the resulting spectrum is similar to the `Fermi arc' observed in experiments.

\subsection{Magnetotransport}
\label{sec:mt}

With the success of the computation in Section~\ref{sec:arc}, we need to return to the tension between these observations and the picture emerging from recent magnetotransport observations \cite{Ramshaw22,Yamaji24} which favor hole pocket Fermi surfaces. In this section we discuss how this tension is resolved by a theory based upon 
$\mathcal{H}_{\rm pseudogap}$ in Eq.~(\ref{eq:WS}).

\begin{figure}
    \centering
    \includegraphics[width=5.5in]{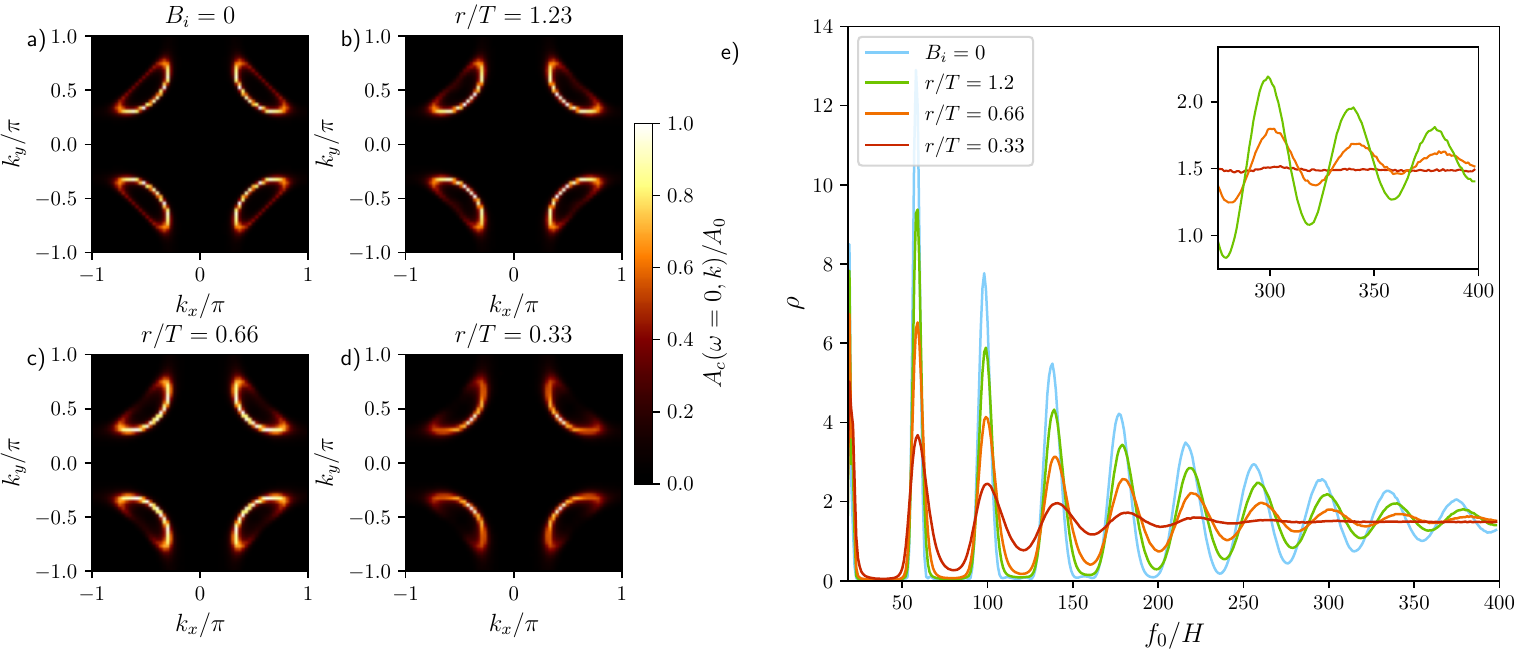}
    \caption{From Ref.~\cite{Sayantan25}. (a) The photoemission electron spectral function of Fig.~\ref{fig:fls_spinon} averaged over gaussian fluctuations of $B$ with energy $\mathcal{E}_2$, as described by $\mathcal{H}_{\rm KLmf} + \mathcal{H}_{SLf} + \mathcal{H}_Y$ in Eqs.~(\ref{HKLmf}), (\ref{eq:fermionhop2}), (\ref{Yukawa})
    for different values of the boson `mass' $r$. (b) Density of states $\rho$ in the presence of a magnetic field $H$ which couples minimally to $c,f_1$ for the same Hamiltonian and gaussian average. 
    The inset is an expanded view of the data at smaller values of $H$.The frequency $f_{0}={h}/({ea_0^2})$ corresponds to the area of the Brillouin zone; for the cuprates $f_0 \approx 28600$ T.}
    \label{fig:Osc}
\end{figure}
Given the role of thermal $B$ fluctuations in removing the pocket backsides in the photoemission spectrum in Fig.~\ref{fig:arc}, it is legitimate to ask if the thermal $B$ fluctuations will also wash out quantum oscillations. The latter probe does not eject electrons form the sample, unlike photoemission, and so its behavior can be distinct.
This question was also investigated by Pandey {\it et al.} \cite{Sayantan25}: they averaged the quantum oscillations associated with $\mathcal{H}_{\rm KLmf} + \mathcal{H}_{SLf} + \mathcal{H}_Y$ over gaussian fluctations of $B$. They found that the quantum oscillations are damped by thermal fluctuations but survive in regimes where the
`back sides' of the pockets are faint, as shown in Fig.~\ref{fig:Osc}.

\section{From the pseudogap to $d$-wave superconductivity and intertwined orders}
\label{sec:dwave}

This section addresses the fate of the FL* pseudogap as the temperature is lowered, upon including the coupling to the ${\bm S}_2$ spin liquid in $\mathcal{H}_{\rm pseudogap}$ in Eq.~(\ref{eq:WS}).
This Hamiltonian specifically choses the $\pi$-flux spin liquid of Section~\ref{sec:fermions} for the ${\bm S}_2$ layer.
We will show that for this spin liquid there is a transition to a conventional BCS-type $d$-wave superconductor, with anisotropic nodal velocities for the Boboliubov quasiparticles, and $h/(2e)$ vortices. Nevertheless the transition itself is not of the BCS type with a Cooper-pairing instability of a Fermi surface. Instead, the transition is driven by the confinement of the fractionalized excitations of the ${\bm S}_2$ spin liquid. We also find nearby instabilities to charge ordering, consistent with observations. These instabilities are driven by the condensation of the Higgs field $B$, which serves as a fractionalized order parameter whose composites describe the intertwined orders.

We note here recent numerics which support the idea of the $d$-wave superconductor emerging from the doping-induced confinement of the $\pi$-flux spin liquid. As we have discussed in Section~\ref{sec:fermions}, the $\pi$-flux spin liquid is one description of the quantum-criticality between the N\'eel and VBS states. The numerical studies of Refs.~\cite{Jiang21,Jiang23} examined the $J_1$-$J_2$ square lattice antiferromagnet near the N\'eel-VBS transition, and indeed found $d$-wave superconductivity upon doping.

\subsection{Anisotropic velocities in the $d$-wave superconductor}
\label{sec:aniso}

We now show how the problem of isotropic quasiparticle velocities, noted in Sections~\ref{sec:intro} and \ref{sec:halffilling}, is resolved by the presence of the pocket Fermi surfaces described by $\mathcal{H}_{\rm KLmf}$ in Eq.~(\ref{HKLmf}). The discussion below is based on the detailed computations presented in Refs.~\cite{Chatterjee16,CS23}.

\begin{figure}
\centering
\includegraphics[width=4.5in]{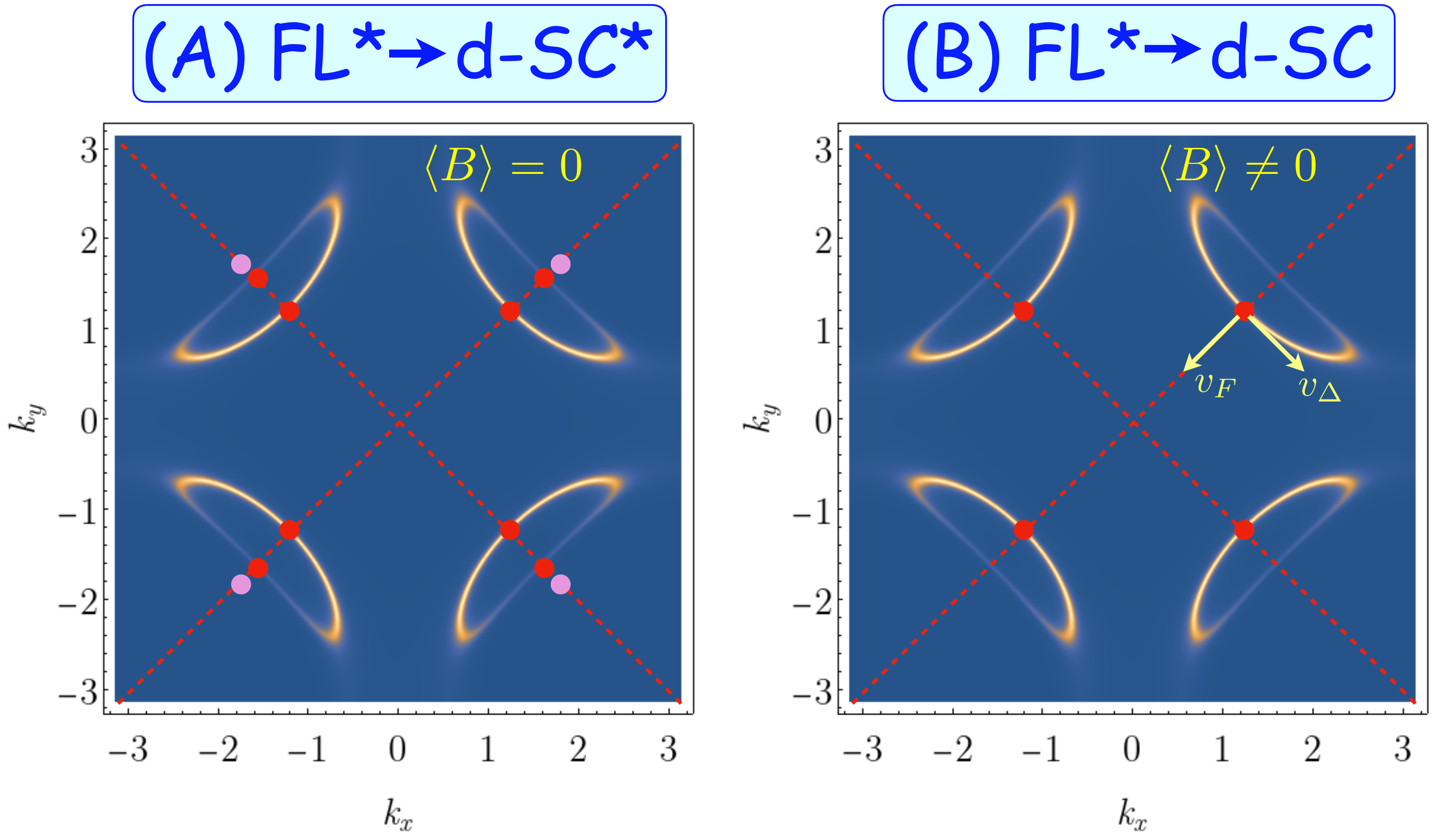}
\caption{(A) Cooper pairing the Fermi surface quasiparticles in FL* leads to d-SC* state, with 8 nodal Bogoliubov quasiparticles (red), and 4 nodal spinons (pink) shown earlier in Fig.~\ref{fig:fls_spinon}.
(B) Upon condensing $B$, the spinons mutually annihilate 4 of the Bogoliubov quasiparticles, leaving 4 Bogoliubov quasiparticles with $v_F \gg v_\Delta$. }
\label{fig:flsdsc}
\end{figure}

Given the pocket Fermi surfaces and spinons in Fig.~\ref{fig:fls_spinon}, we imagine imposing a BCS type pairing on the Fermi surface excitations. 
If the pairing is $d$-wave, it would lead to 8 nodal Bogoliubov points as shown in Fig.~\ref{fig:flsdsc}A. 
However this state also has the 4 nodal quasiparticles of the ${\bm S}_2$ spin liquid, associated with the dispersion in Fig.~\ref{fig:fermiondisp}. So strictly speaking, this state remains fractionalized, and is {\it not\/} a conventional $d$-wave superconductor. It would be appropriate to call it d-SC*.

However, if we induce the pairing by the $B$ condensate in Eq.~(\ref{Yukawa}), the SU(2) gauge field is higgsed. Morever, the Yukawa coupling allows the nodal quasiparticles of the ${\bm S}_2$ spin liquid to hybridize with the Bogoliubov quasiparticles of the pocket Fermi surfaces.  The net result, sketched in Fig.~\ref{fig:flsdsc}B, is that the $B$ condensate can enable the nodal points on the `backsides' of the pocket Fermi surfaces to mutually annihilate with the spinons of the ${\bm S}_2$ spin liquid. We are then left with the 4 nodal quasiparticles on front sides of the pocket Fermi surfaces. The number of nodal points are the same as those obtained in conventional BCS theory from $d$-wave pairing of a Fermi liquid. These remaining nodal points are associated with pairing on the pocket Fermi surfaces, and these is no reason for their velocity to be isotropic (unlike the spinons).

\begin{figure}
    \centering
    \includegraphics[width=\linewidth]{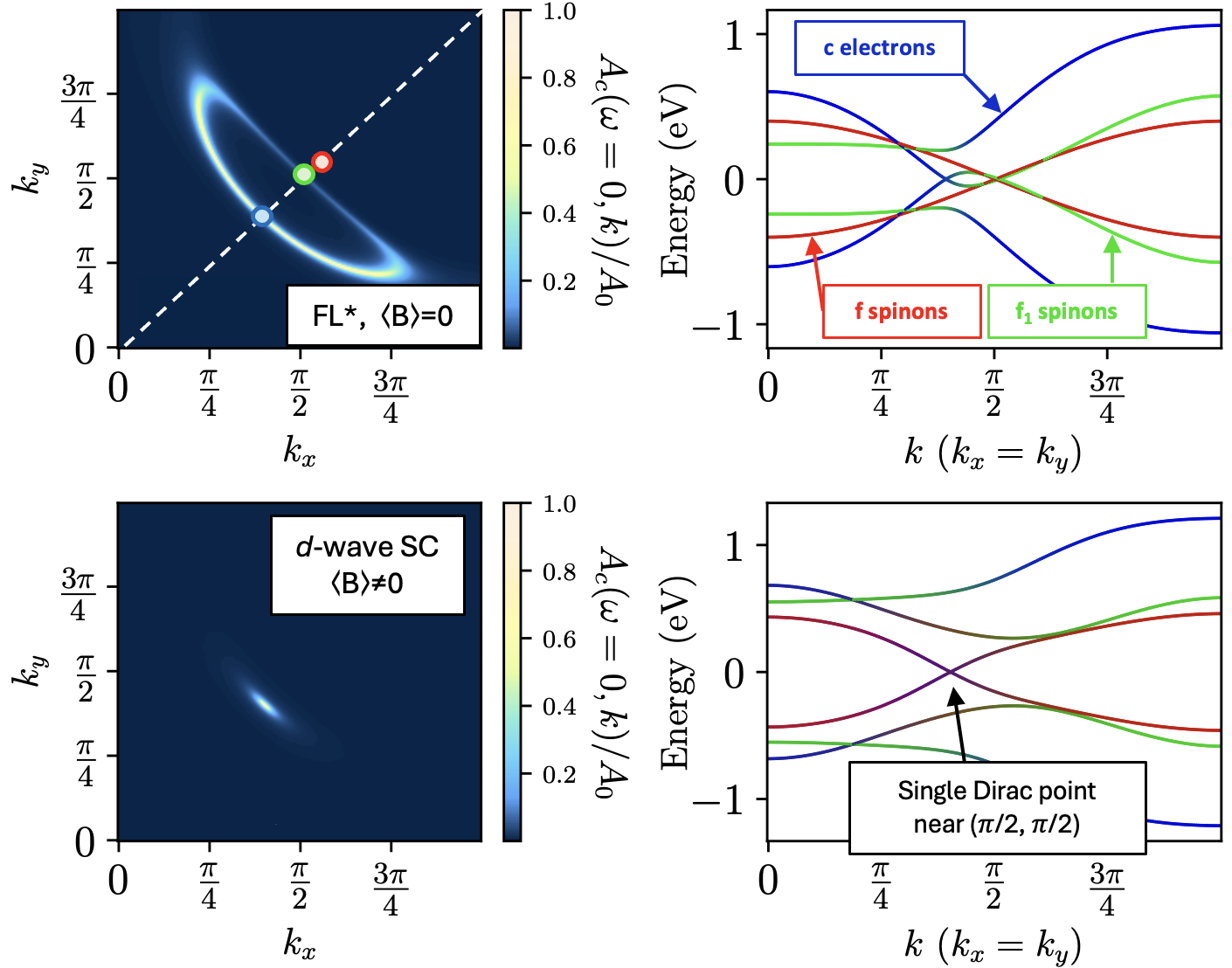}
    \caption{From Ref.~\cite{CS23}. We show how Higgs condensing $B$ leads to a $d$-wave superconducting state with four nodes for the case of the hole doped cuprates. We show the normal state spectral density at the Fermi level (top left) where we indicate the gapless points along the Brillouin zone diagonal associated with the $c$ electron (blue circle), $f_1$ spinons (green circle) and $f$ spinons (red circle). The spin liquid Dirac point colored red will annhilate with the gapless state on the backside pocket originating from the $f$ spinons when $B$ condenses. We also show the mean-field band structure for this normal state along the Brillouin zone diagonal. The bottom left shows the spectral density at the Fermi level after $B$ is condensed with a single node near $\left(\pi/2,\pi/2\right)$. The bottom right shows the corresponding mean-field band structure when $B$ is condensed in the superconducting state.}
    \label{fig:node_annhilate}
\end{figure}

We illustrate in Fig.~\ref{fig:node_annhilate} how this annihilation occurs via hybridization between the electrons and spinons within a mean-field band structure of $\mathcal{H}_{\rm KLmf} + \mathcal{H}_{SLf} + \mathcal{H}_Y$ in Eqs.~(\ref{HKLmf}), (\ref{eq:fermionhop}), and (\ref{Yukawa}) .
 Indeed, computations which diagonalize this Hamiltonian with $B$ fixed and $U=1$ do indeed yield anisotropic velocities similar to those observed. Unlike the situation in Section~\ref{sec:halffilling}, the spinons do not become the Bogoliubov quasiparticles in the doped case, although the spinons are needed to annihilate the extraneous Bogoliubov quasiparticles.

\begin{figure}
    \centering
    \includegraphics[width=\linewidth]{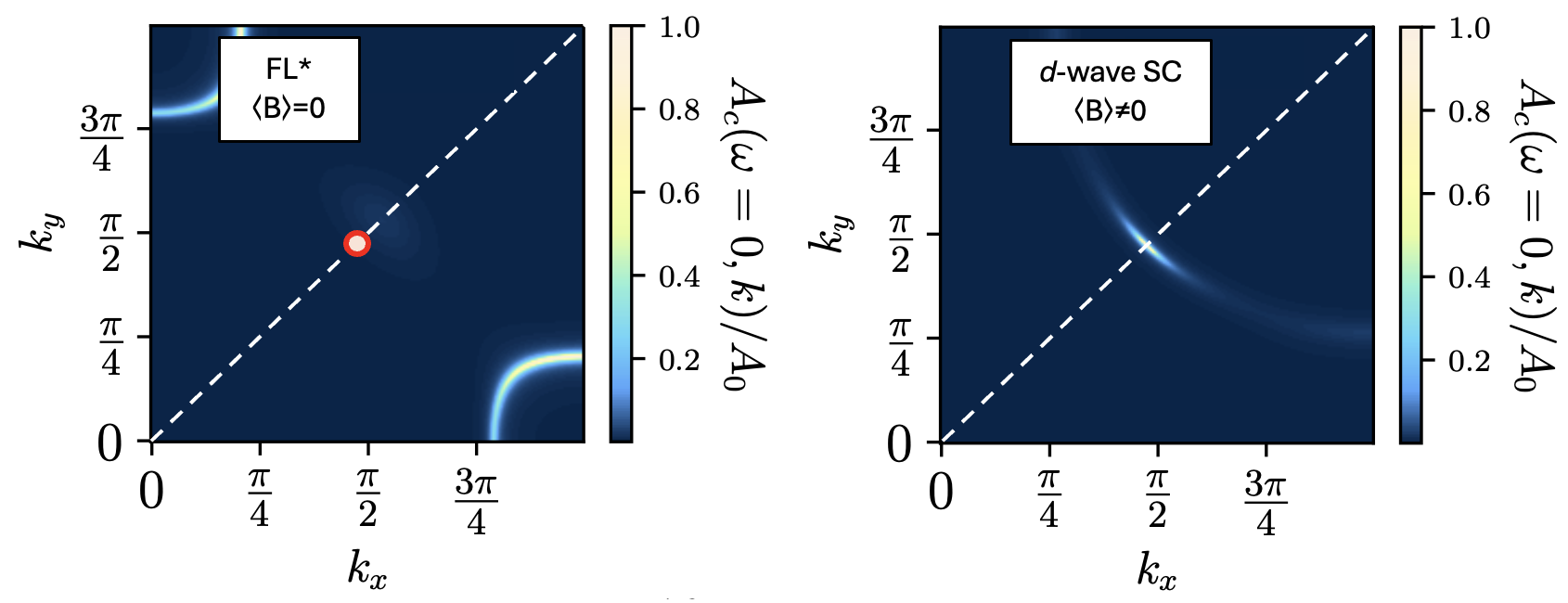}
    \caption{From Ref.~\cite{CS23}. We show an example of the momentum resolved spectral function at the Fermi level of an example normal state at small doping in the electron doped cuprates (left) and the corresponding spectral function after $B$ condenses and drives a phase transition to a $d$-wave superconductor (right). In both cases, we denote the Brillouin zone diagonal with a dashed white line and the Dirac point associated with the spin liquid  (for a particular choice of gauge) is indicated with a purple circle.}
    \label{fig:electrondoped}
\end{figure}

An interesting prediction can be made in the particular case of the electron-doped cuprates. In these materials, photoemission experiments have observed a normal state to superconductivity which is gapped near $(\pi/2,\pi/2)$ and has spectral weight only near electron pockets at the antinode \cite{PhysRevLett.88.257001}. If $d$-wave superconductivity were to emerge as a BCS instability from such a normal state, the resulting superconductor should be gapped. However, the FL$^*$ theory yields a different prediction. Similar to how the spinon degrees of freedom emerge to annhilate with the backside pocket when $B$ Higgs condenses in the hole-doped case, the Dirac node of the spin liquid will appear with a finite spectral weight near $(\pi/2,\pi/2)$ when $B$ condenses, leading to a nodal superconductor, as shown in Fig.~\ref{fig:electrondoped}. Such a prediction can be explored in future photoemission experiments in the electron doped cuprates \cite{Xu_2023} and could serve as an experimental test of the FL$^*$ theory.

\subsection{Vortices with flux $h/(2e)$ and charge order halos}
\label{sec:vortex}

We have seen in Sections~\ref{sec:halffilling} and \ref{sec:aniso} that $d$-wave superconductivity is induced by the condensation of $B$, a boson which carries electrical charge $e$. So we might worry that the elementary flux quantum of such a superconductor is $h/e$. However, that is not correct, and the SU(2) gauge field $A^\ell_\mu$ (introduced in Eq.~(\ref{eq:UA})) plays a central in establishing the presence of vortices with flux $h/(2e)$ \cite{ZhangSS24}.

The argument only requires $A^z_\mu$ to be non-zero. In terms of the two components of $B = (B_1, B_2)$, we can write the following gradient term in the action far from the vortex core
\begin{equation}
|({\bm \nabla} - i {\bm A}^z - i {\bm a}) B_1|^2 + |({\bm \nabla} + i {\bm A}^z - i {\bm a}) B_2|^2 
\end{equation}
where ${\bm A}^z$ is the spatial component of $A^z_\mu$, and ${\bm a}$ is the electromagnetic vector potential (we are using units here with $\hbar=e=1$). Let us assume that the phases of $B_{1,2}$ wind by $2\pi n_{1,2}$ around the vortex core, where $n_{1,2}$ are integers. Then, finiteness of the energy of the vortex requires
\begin{eqnarray}
\int d^2 r \,  \left({\bm \nabla} \times {\bm A}^z + {\bm \nabla} \times {\bm a}\right)\cdot \hat{\bm z} &=& 2 \pi n_1 \nonumber \\
\int d^2 r \,  \left({\bm \nabla} \times {\bm A}^z - {\bm \nabla} \times {\bm a}\right)\cdot \hat{\bm z} &=& 2 \pi n_2
\end{eqnarray}
By choosing $n_1 = 1$, $n_2 = 0$, we obtain a vortex with $\int \left({\bm \nabla} \times {\bm a}\right)\cdot \hat{\bm z} = \pi$, which corresponds to flux $h/(2e)$.

There is an interesting feature of the vortex solution near its core. Recall from Eqs.~(\ref{latticeorders},\ref{contorders}) that different orientations of the complex vector $(B_1, B_2)$ correspond to different local orders. At the vortex core, it is preferential for the orientation of $B$ to rotate from that preferring d-SC, to one of the other orders \cite{ZhangSS24}. Explicit solutions of the continuum theory Eq.~(\ref{LXB}) were obtained in Ref.~\cite{ZhangSS24} with period 2 charge order.

\begin{figure}
\centering
\includegraphics[width=5.2in]{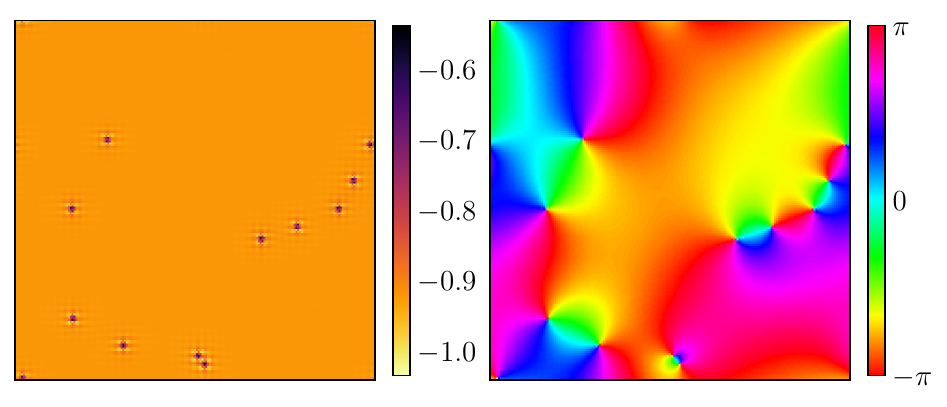}
\caption{From Ref.~\cite{Sayantan25}. The bond density (left panel) and the distribution of the phase of the superconducting order 
    parameter (right panel) on a $192\times 192$ lattice at low temperature in the thermal ensemble defined by Eq.~(\ref{Z20}).}
\label{fig:halos}
\end{figure}

Fig.~\ref{fig:halos} shows a snapshot from the simulation of Ref.~\cite{Sayantan25} at low $T$ at the parameter value in Fig.~\ref{fig:period4}.
Note the vortices with the phase winding of $2\pi$ in the superconducting order parameter (corresponding to $h/(2e)$ vortices), and the period 4 charge order halos around each vortex. These halos are remarkably similar to those observed by Hoffman {\it et al.\/} \cite{Hoffman02}. Several other features of the vortex solution were also found in Ref.~\cite{ZhangSS24} to correspond STM observations. 

\subsection{Quantum oscillations}
\label{sec:qo}

We discussed observations in high magnetic field in the intermediate temperature pseudogap phase in Section~\ref{sec:mt}. Here we turn to observations at very low $T$.

At large doping, Vignolle {\it et al.\/} \cite{Hussey08} observed quantum oscillations are compatible with a `large' Fermi surface of area $(1+p)/2$. 

At low doping and low temperatures, quantum oscillations show evidence for small electron pockets \cite{Sebastian_review}. The formation of electron pockets is believed to be associated with the appearance of charge density wave order at high magnetic fields. But computations on charge density wave models had difficulty reproducing the observed spectrum. 

\begin{figure}
\centering
\includegraphics[width=3in]{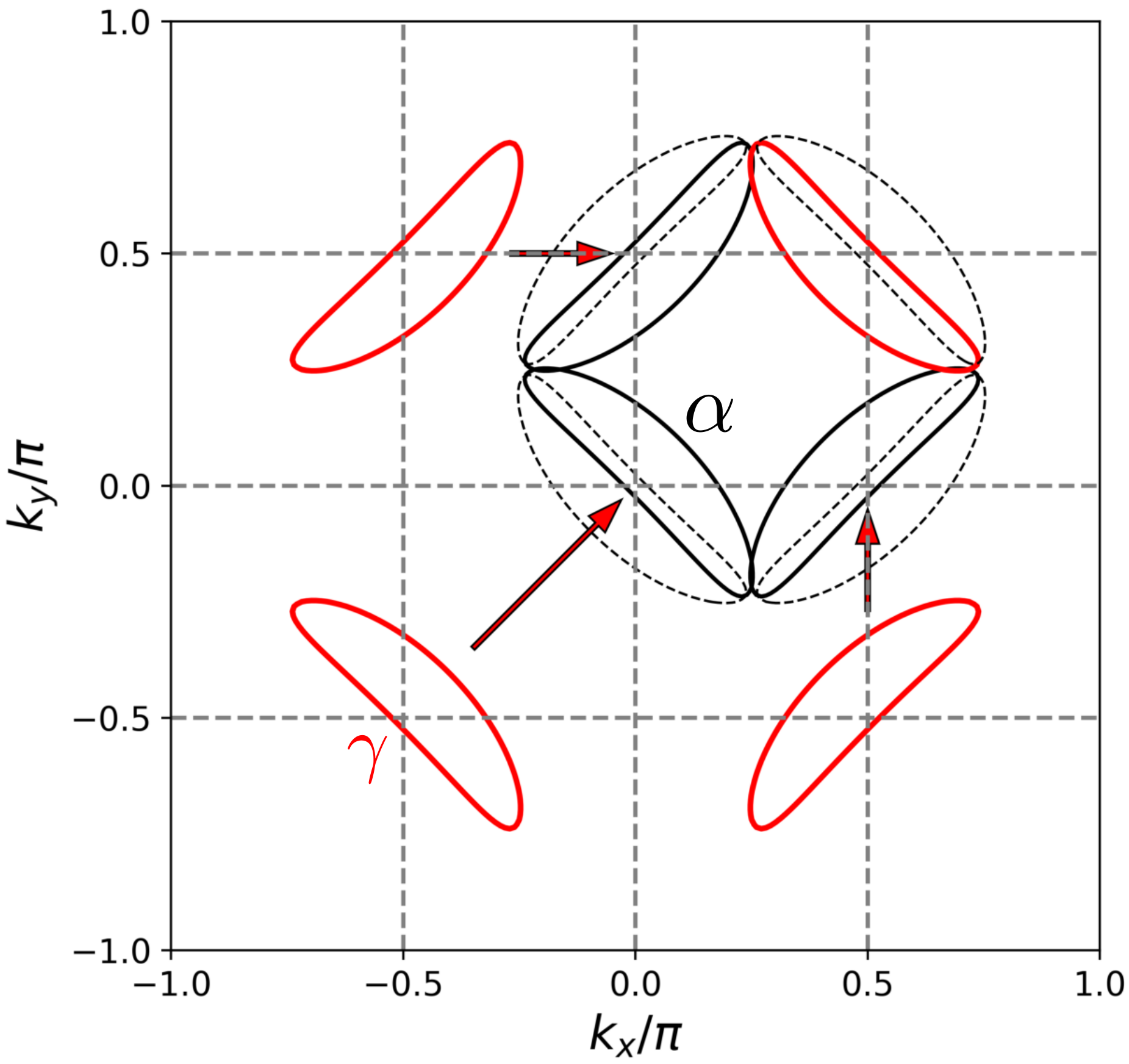}
\caption{From Ref.~\cite{BCS24}. 
Illustration of the $\alpha$ electron pocket of Harrison and Sebastian \cite{HS11}, from the combination of the 4 Fermi arcs of
the $\gamma$ hole pockets of Fig.~\ref{fig:photo},
after shifts by the CDW wavevectors.}
\label{fig:hs_pocket}
\end{figure}

Harrison and Sebastian \cite{HS11} proposed a phenomenological model of the electron pocket, in which the Fermi arcs combined to form an $\alpha$ electron pocket after shifts by the wavevectors of the charge density wave (CDW). This is illustrated in Fig.~\ref{fig:hs_pocket}

\begin{figure}
\centering
\includegraphics[width=5.5in]{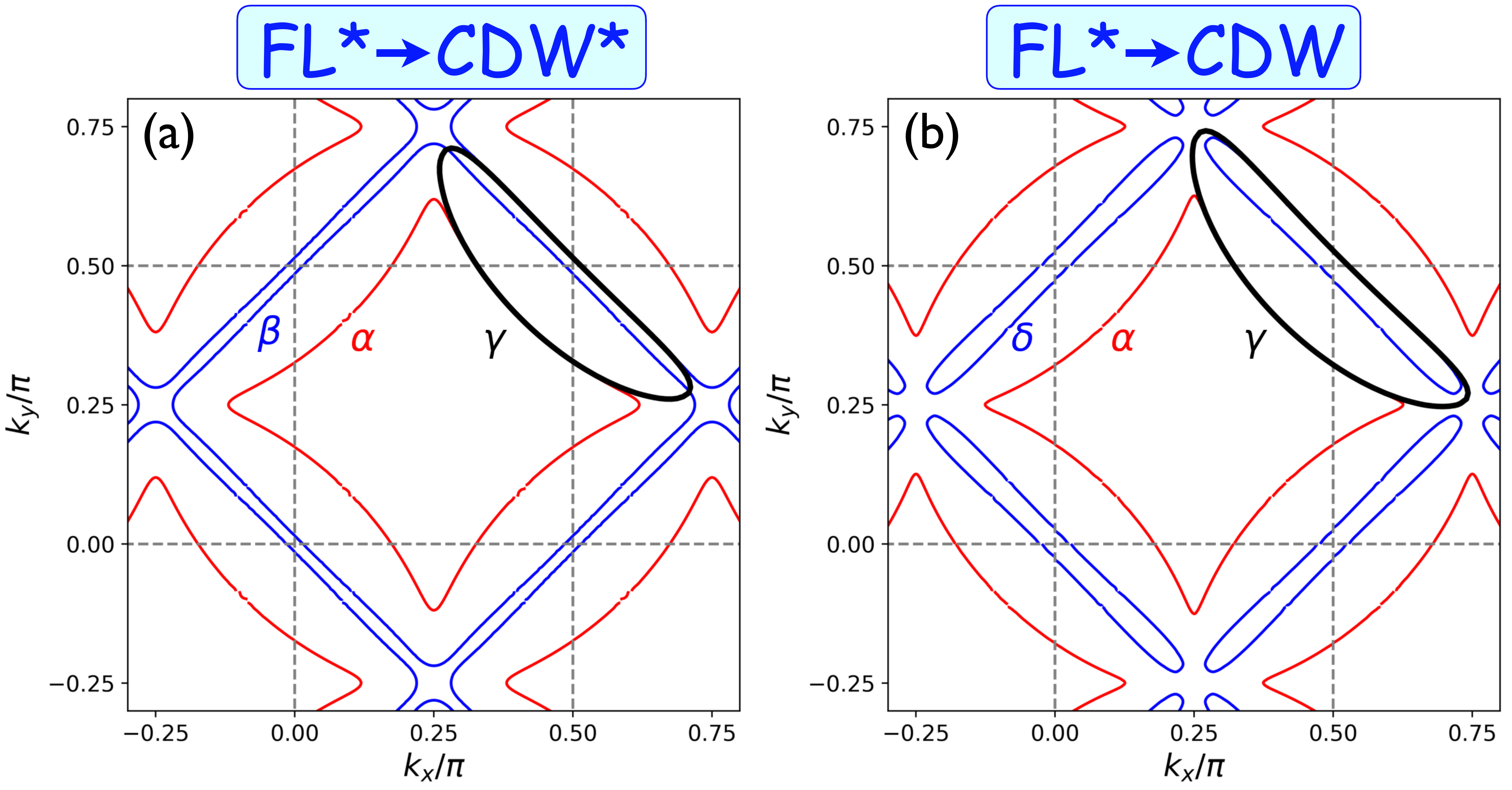}
\caption{From Ref.~\cite{BCS24}. 
(a) A computation similar to that in Ref.~\cite{Zhang_2016}, by applying a charge density wave to $\mathcal{H}_{\rm KLmf}$ in Eq.~(\ref{HKLmf}); this computation does not include spinons. In addition to the observed $\alpha$ pocket, there is a large $\beta$ hole pocket which is not observed. (b) A computation including spinons. A modulated $B$ condensate is applied to $\mathcal{H}_{\rm KLmf} + \mathcal{H}_{SLf} + \mathcal{H}_Y$ in Eqs.~(\ref{HKLmf}), (\ref{eq:fermionhop2}), (\ref{Yukawa}). The mixing with the spinons replaces the $\beta$ pocket by the $\delta$ pocket, with a size too small to be seen in quantum oscillations.
}
\label{fig:BCS_pocket}
\end{figure}

Zhang and Mei \cite{Zhang_2016} worked with a model of pocket Fermi surfaces (similar to $\mathcal{H}_{\rm KLmf}$) in the presence of CDW order: in addition to the observed $\alpha$ electron pocket, their results show an unobserved additional oscillation frequency from a $\beta$ Fermi surface arising from the backsides of the hole pockets completing the Fermi arcs. This is illustrated in Fig.~\ref{fig:BCS_pocket}a. Note that this computation leads to a CDW* state, because the spinons remain deconfined.

Bonetti {\it et al.\/} \cite{BCS24} extended the computations of Zhang and Mei to include spinons by working with  $\mathcal{H}_{\rm KLmf} + \mathcal{H}_{SLf} + \mathcal{H}_Y$ in Eqs.~(\ref{HKLmf}), (\ref{eq:fermionhop2}), (\ref{Yukawa}).
They included the charge density wave by a $B$ condensate which had spatial modulations but no superconducting order, as defined by the local mappings in Eq.~(\ref{latticeorders}). The presence of the $B$ condensate implies that the SU(2) gauge field is fully higgsed, and the resulting state is a conventional CDW with no fractionalized excitations.
As shown in Fig.~\ref{fig:BCS_pocket}b, the unobserved $\beta$ pocket is replaced by the $\delta$ pockets which are very small and hence compatible with current quantum oscillation data.
The role of the spinons here in removing unobserved fermionic excitaions is rather analogous to their role in Sections~\ref{sec:arc} and \ref{sec:aniso}.

\section{The SYK model}
\label{sec:SYK}

We now turn to a different critical quantum liquid, associated with mobile electrons, the SYK model. This is a zero-dimensional model for which the absence of quasiparticle excitations is well-established. A direct and extensive application of the zero-dimensional SYK model has been to the low energy quantum theory of charged black holes, and this is reviewed elsewhere \cite{SachdevMEF}. There are also connections to experiments on graphene flakes \cite{ShackletonGraphene,Kim_flake}. But our primary interest here will be extensions to two-dimensional models relevant to the strange metal state of the cuprates and other correlated electron materials---we will turn to this in Section~\ref{sec:strange}, where the zero-dimensional SYK models will motivate a theory of realistic FL*-FL transitions in two-dimensional metals.
We also note other reviews \cite{CGPS22,SachdevORE,SSZaanen} for a more in-depth discussion of the topics in Sections~\ref{sec:SYK}-\ref{sec:griffiths}.

\begin{figure}
\begin{center}
\includegraphics[width=3in]{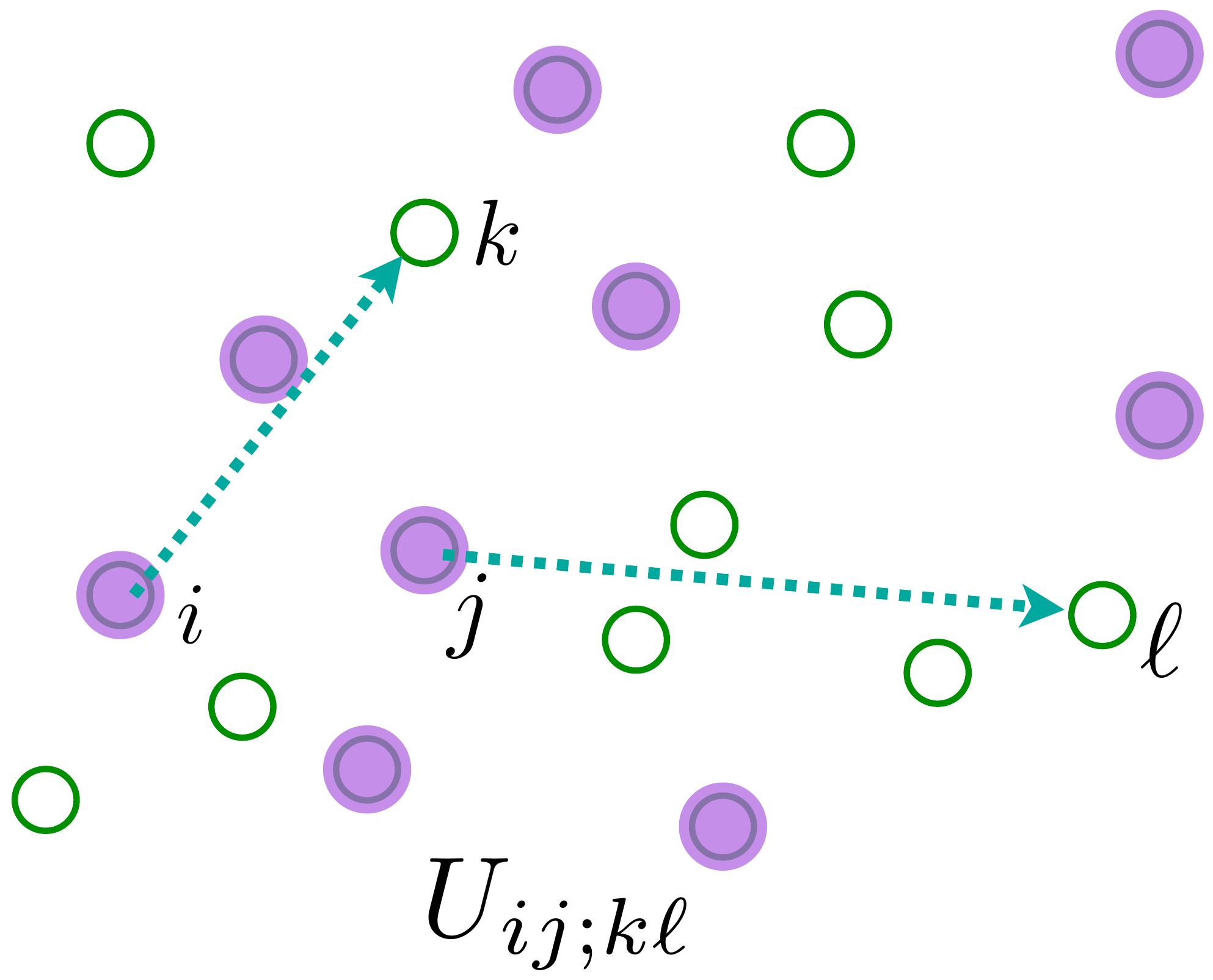}
\end{center}
\caption{The SYK model: fermions undergo the transition (`collision') shown with quantum amplitude $U_{\vi\vj;k\ell}$.}
\label{fig4}
\end{figure}

The Hamiltonian of a version of a SYK model is illustrated in Fig.~\ref{fig4}. A system with fermions $c_\vi$, $i=1\ldots N$ states is assumed. Depending upon physical realizations, the label $\vi$ could be position or an orbital, and it is best to just think of it as an abstract label of a fermionic qubit with the two states $\left|0 \right\rangle$ and $c_\vi^\dagger \left|0 \right\rangle$. $\mathcal{Q} N$ fermions are placed in these states, so that a density $\mathcal{Q} \approx 1/2$ is occupied, as shown in Fig.~\ref{fig4}. 
The quantum dynamics is restricted to {\it only\/} have a `collision' term between the fermions, analogous to the right-hand-side of the Boltzmann equation. However, in stark contrast to the Boltzmann equation, statistically independent collisions are not assumed, and quantum interference between successive collisions is accounted for: this is the key to building up a many-body state with non-trivial entanglement. So a collision in which fermions move from sites $\vi$ and $j$ to sites $k$ and $\ell$ is characterized not by a probability, but by a quantum amplitude $U_{\vi\vj;k\ell}$, which is a complex number.

The model so defined has a Hilbert space of order $2^N$ states, and a Hamiltonian determined by order $N^4$ numbers $U_{\vi\vj;k\ell}$. Determining the spectrum or dynamics of such a Hamiltonian for large $N$ seems like an impossibly formidable task. But with the assumption that the $U_{\vi\vj;k\ell}$ are statistically independent random numbers, remarkable progress is possible. Note that an ensemble of SYK models with different $U_{\vi\vj;k\ell}$ is not being considered, but a single fixed set of $U_{\vi\vj;k\ell}$. Most physical properties of this model are self-averaging at large $N$, and so as a technical tool, they can be rapidly obtained by computations on an ensemble of random $U_{\vi\vj;k\ell}$. In any case, the analytic results described below have been checked by numerical computations on a computer for a fixed set of $U_{\vi\vj;k\ell}$.
Recall that, even for the Boltzmann equation, there was an ensemble average over the initial positions and momenta of the molecules that was implicitly performed.

Specifically, the Hamiltonian in a chemical potential $\mu$ is 
\begin{eqnarray}
&\mathcal{H} = \frac{1}{(2 N)^{3/2}} \sum_{\vi,\vj,k,\ell=1}^N U_{\vi\vj;k\ell} \, c_\vi^\dagger c_\vj^\dagger c_k^{\vphantom \dagger} c_\ell^{\vphantom \dagger} 
-\mu \sum_{\vi} c_\vi^\dagger c_\vi^{\vphantom \dagger} \nonumber \\
& ~~~~~~c_\vi c_\vj + c_\vj c_\vi = 0 \quad, \quad c_\vi^{\vphantom \dagger} c_\vj^\dagger + c_\vj^\dagger c_\vi^{\vphantom \dagger} = \delta_{\vi\vj} \nonumber \\
&~~~~\mathcal{Q} = \frac{1}{N} \sum_\vi c_\vi^\dagger c_\vi^{\vphantom \dagger} \, ; \quad
[\mathcal{H}, \mathcal{Q}] = 0\, ; \quad  0 \leq \mathcal{Q} \leq 1\,,  \label{HH}
\end{eqnarray}
and its large $N$ limit is most simply taken graphically, order-by-order in $U_{\vi\vj;k\ell}$, and averaging over $U_{\vi\vj;k\ell}$ as independent random variables with $\overline{U_{\vi\vj;k\ell}} = 0$ and $\overline{|U_{\vi\vj;k\ell}|^2} = U^2$.
\begin{figure}
\begin{center}
\includegraphics[width=3in]{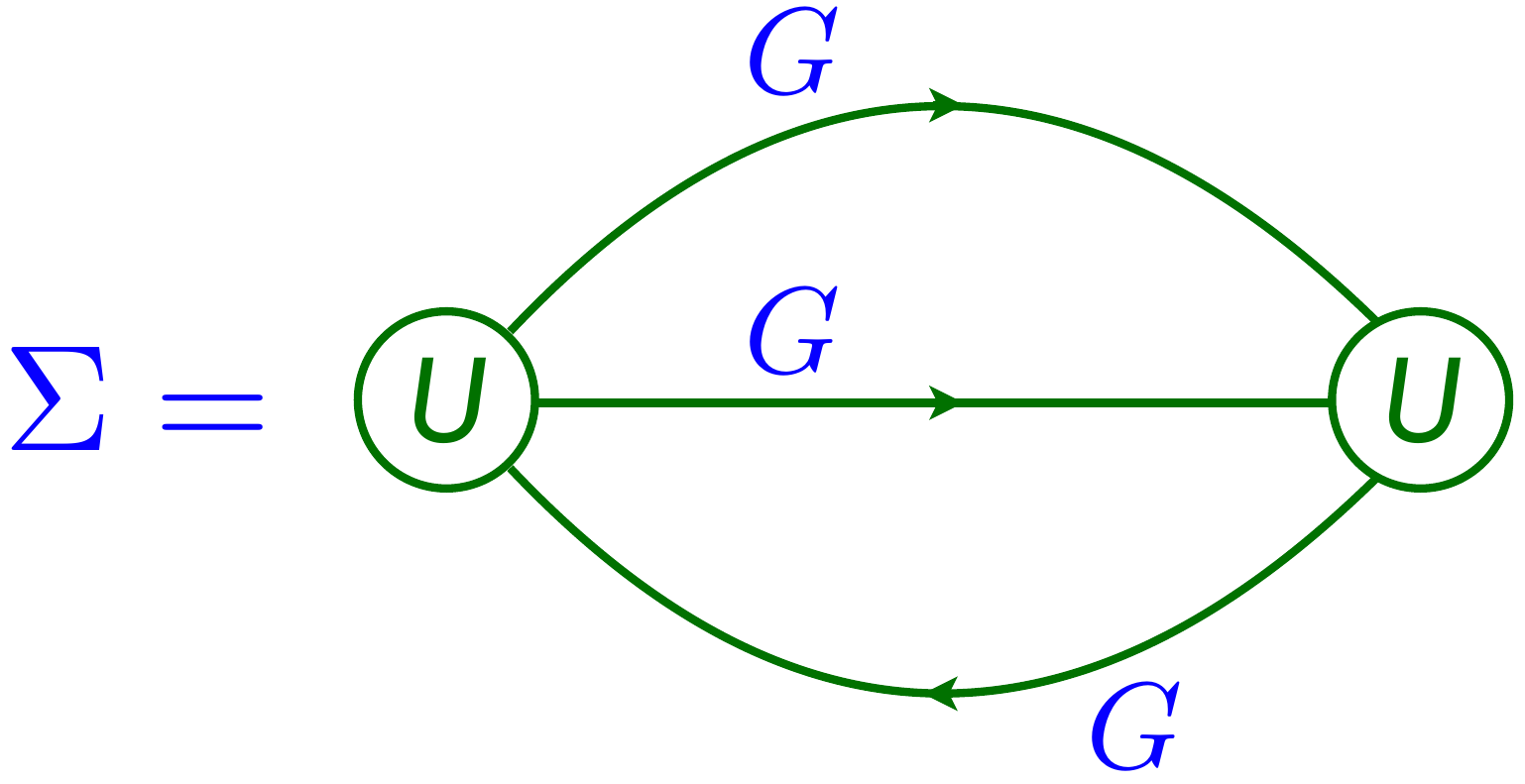}
\end{center}
\caption{Self-energy for the fermions of $\mathcal{H}$ in Eq.~(\ref{HH}) in the limit of large $N$. The intermediate Green's functions are fully renormalized.}
\label{fig:sygraph}
\end{figure}
This expansion can be used to compute graphically the Green's function in imaginary time $\tau$
\begin{eqnarray}
G(\tau) = - \frac{1}{N} \sum_\vi \overline{\left\langle \mathcal{T} \left( c_\vi (\tau) c_\vi^\dagger (0) \right) \right\rangle}\,, \label{G1}
\end{eqnarray}
where $\mathcal{T}$ is the time-ordering symbol, the angular brackets are a quantum average for any given $U_{\vi\vj;k\ell}$, and the over-line denotes an average over the ensemble of $U_{\vi\vj;k \ell}$. (It turns out that  the last average is not needed for large $N$, because the quantum observable is self-averaging.)
In the large $N$ limit, only the graph for the Dyson self energy, $\Sigma$, in Fig.~\ref{fig:sygraph} survives, and the on-site fermion Green's function is given by the solution of the following equations
\begin{eqnarray}
G(i\omega_n) &= \frac{1}{i \omega_n + \mu - \Sigma (i\omega_n)} \nonumber \\ 
\Sigma (\tau) & = -  U^2 G^2 (\tau) G(-\tau) \nonumber \\
G(\tau = 0^-) & = \mathcal{Q}\,, \label{sy1}
\end{eqnarray}
where  $\omega_n$ is a fermionic Matsubara frequency. The first equation in Eq.~(\ref{sy1}) is the usual Dyson relation between the Green's function and self energy in quantum field theory, the second equation in Eq.~(\ref{sy1}) is the Feynman graph in Fig.~\ref{fig:sygraph}, and the last determines the chemical potential $\mu$ from the charge density  $\mathcal{Q}$.  These equations can also be obtained as saddle-point equations of the following exact representation of the disordered-averaged partition function, expressed as a `$G-\Sigma$' theory \cite{GPS2,Sachdev15,kitaevsuh,Maldacena_syk}:
\begin{eqnarray}
\mathcal{Z} &= \int \mathcal{D} G(\tau_1, \tau_2) \mathcal{D} \Sigma (\tau_1, \tau_2) \exp (-N I) \nonumber \\
I &= \ln \det \left[ \delta(\tau_1 - \tau_2) (\partial_{\tau_1} + \mu) - \Sigma (\tau_1, \tau_2) \right] \nonumber \\
&~~~
+ \int d \tau_1 d \tau_2  \left[ \Sigma(\tau_1, \tau_2) G(\tau_2, \tau_1) + (U^2/2) G^2(\tau_2, \tau_1) G^2(\tau_1, \tau_2) \right] \label{GSigma1}
\end{eqnarray}
This is a path-integral over bi-local in time functions $G(\tau_1, \tau_2)$ and $\Sigma(\tau_1, \tau_2)$, whose saddle point values are the Green's function $G(\tau_1 - \tau_2)$,  and the self energy $\Sigma (\tau_1 - \tau_2)$. This bi-local $G$ can be viewed as a composite quantum operator corresponding to an on-site fermion bilinear
\begin{eqnarray}
G(\tau_1, \tau_2) = - \frac{1}{N} \sum_\vi  \mathcal{T} \left( c_\vi (\tau_1) c_\vi^\dagger (\tau_2) \right) 
\end{eqnarray}
that is averaged in Eq.~(\ref{G1}).

For general $\omega$ and $T$, the equations in Eq.~(\ref{sy1}) have to be solved numerically. But an exact analytic solution is possible in the limit $\omega, T \ll U$. 
At $T=0$, the asymptotic forms can be obtained straightforwardly \cite{SY92}
\begin{eqnarray}
G(i \omega) \sim -i \mbox{sgn} (\omega) |\omega|^{-1/2} \quad, \quad \Sigma(i \omega) - \Sigma (0) \sim -i \mbox{sgn} (\omega) |\omega|^{1/2}\,,
\label{sy10}
\end{eqnarray}
and a more complete analysis of Eq.~(\ref{sy1}) gives the exact form at non-zero $T$ ($\hbar = k_B = 1$) \cite{Parcollet1}
\begin{eqnarray}
G (\omega)  = \frac{-i C e^{-i \theta}}{(2 \pi T)^{1/2}}
\frac{\Gamma \left( \displaystyle \frac{1}{4} - \frac{i  \omega}{2 \pi T} + i \mathcal{E} \right)}
{\Gamma \left(  \displaystyle \frac{3}{4} - \frac{i \omega }{2 \pi T} + i \mathcal{E} \right)} \quad\quad  |\omega|, T \ll U \,. \label{sy2}
\end{eqnarray}
Here, $\mathcal{E}$ is a dimensionless number which characterizes the particle-hole asymmetry of the spectral function; both $\mathcal{E}$ and the pre-factor $C$ are determined by an angle $-\pi/4 < \theta < \pi/4$
\begin{eqnarray}
e^{2 \pi \mathcal{E}} = \frac{\sin(\pi/4 + \theta)}{\sin(\pi/4 - \theta)} \quad, \quad  C = \left( \frac{\pi}{U^2 \cos (2 \theta) }\right)^{1/4}\,,
\end{eqnarray}
and the value of $\theta$ is determined by a Luttinger relation to the density $\mathcal{Q}$ \cite{GPS2}
\begin{eqnarray}
\mathcal{Q} = \frac{1}{2} - \frac{\theta}{\pi} - \frac{\sin(2 \theta)}{4}\,.
\end{eqnarray}

A notable property of Eq.~(\ref{sy2}) at $\mathcal{E}=0$ is that it equals the temporal Fourier transform of the spatially local correlator of a fermionic field of dimension 1/4 in a conformal field theory in 1+1 spacetime dimensions. A theory in 0+1 dimensions is considered here, where conformal transformations map the temporal circle onto itself, as reviewed in Appendices A and B of Ref.~\cite{CGPS22}; such transformations allow a non-zero $\mathcal{E}$. An important consequence of this conformal invariance is that Eq.~(\ref{sy2}) is a scaling function of $\hbar \omega/(k_B T)$ (after restoring fundamental constants); in other words, the characteristic frequency scale of Eq.~(\ref{sy2}) is determined solely by $k_B T/\hbar$, and is independent of the value of $U/\hbar$. A careful study of the consequences of this conformal invariance have established the following properties of the SYK model (more complete references to the literature are given in other reviews \cite{CGPS22,QPMbook}):
\begin{itemize} 
\item There are no quasiparticle excitations, and the SYK model exhibits quantum dynamics with a Planckian relaxation time of order $\hbar/(k_B T)$ at $T \ll U$. In particular, the relaxation time is {\it independent\/} of $U$, a feature not present in any ordinary metal with quasiparticles. While the Planckian relaxation in Eq.~(\ref{sy2}) implies the absence of quasiparticles with the same quantum numbers as the $c$ fermion, it does not rule out the possibility that $c$ has fractionalized into some emergent quasiparticles; this possibility is ruled out by the exponentially large number of low energy states, as discussed below.
\item At large $N$, the many-body density of states at fixed $\mathcal{Q}$ is \cite{Cotler16,Bagrets17,Maldacena_syk,kitaevsuh,StanfordWitten,GKST} (see Fig.~\ref{fig5}a)
\begin{equation}
D(E) \sim \frac{1}{N} \exp (N s_0) \sinh \left( \sqrt{2 N \gamma E} \right)\,, \label{de}
\end{equation}
where the ground state energy has been set to zero.
Here $s_0$ is a universal number dependent only on $\mathcal{Q}$ ($s_0 = 0.4648476991708051 \ldots$ for $\mathcal{Q}=1/2$), $\gamma \sim 1/U$ is the only parameter dependent upon the strength of the interactions, and the $N$ dependence of the pre-factor is discussed in Ref.~\cite{GKST}.
\begin{figure}
\begin{center}
\includegraphics[width=6in]{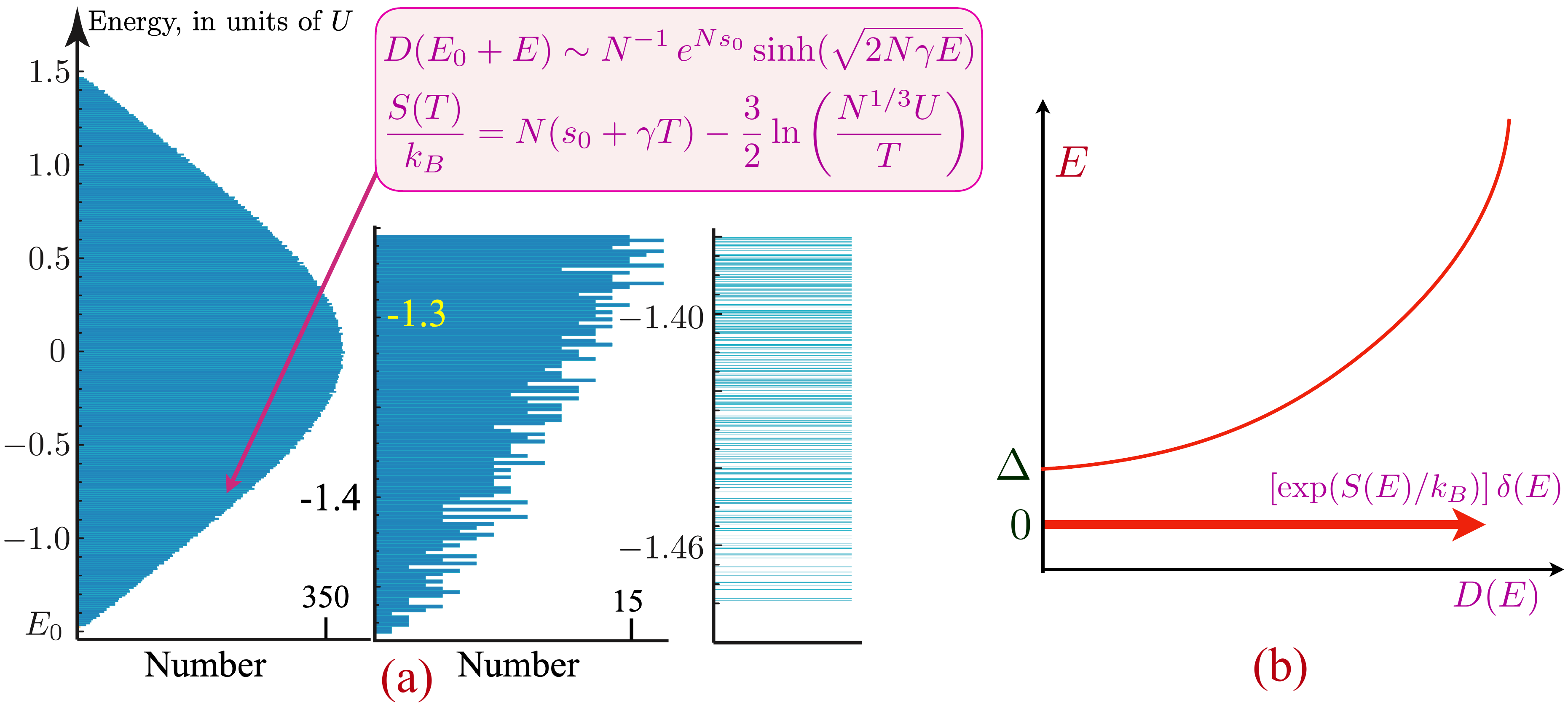}
\end{center}
\caption{(a) From Ref.~\cite{CGPS22}. Plot of the 65536 many-body eigenvalues of a $N = 32$ Majorana SYK Hamiltonian; however, the analytical results quoted here are for the SYK model with complex fermions which has a similar spectrum. The coarse-grained low-energy and 
low-temperature behavior is described by Eq.~(\ref{de}) and Eq.~(\ref{SSYK}). 
(b) Schematic of the lower energy density of states of a supersymmetric generalization of the SYK model \cite{Fu16,StanfordWitten}. There is a delta function at $E=0$, and
the energy gap $\Delta$ is proportional to the inverse of $S(E=0)$.}
\label{fig5}
\end{figure}
Given $D(E)$, the partition function can be computed from
\begin{equation}
\mathcal{Z}(T) = \int dE D(E) e^{-E/(k_B T)}\,,
\end{equation}
and hence the low-$T$ dependence of the entropy at fixed $Q$ is given by
\begin{equation}
\frac{S(T)}{k_B} = N(s_0 + \gamma \, k_B T) - \frac{3}{2}\ln \left(\frac{U}{k_B T} \right) - \frac{\ln N}{2} + \ldots \,. \label{SSYK}
\end{equation}
The thermodynamic limit $\lim_{N \rightarrow \infty} S(T)/N$ yields the microcanonical entropy 
\begin{eqnarray}
S(E)/k_B = Ns_0 + \sqrt{2N \gamma E}\,, 
\end{eqnarray}
and this connects to the extensive $E$ limit of Eq.~(\ref{de}) after using Boltzmann's formula $D(E) \sim \exp \left( S(E)/k_B \right)$.
The limit $\lim_{T \rightarrow 0} \lim_{N \rightarrow \infty} S(T)/(k_B N) = s_0$ is non-zero, implying an energy-level spacing exponentially small in $N$ near the ground state: the density of states Eq.~(\ref{de}) implies that any small energy interval near the ground state contains an exponentially large number of energy eigenstates (see Fig.~\ref{fig5}a).
This is very different from systems with quasiparticle excitations, whose energy level spacing vanishes with a positive power of $1/N$ near the ground state, as quasiparticles have order $N$ quantum numbers. The exponentially small level spacing therefore rules out the existence of quasiparticles in the SYK model. 
\item
However, it important to note that there is no exponentially large degeneracy of the ground state itself in the SYK model, unlike that in a supersymmetric generalization of the SYK model (see Fig.~\ref{fig5}b) and the ground states in Pauling's model of ice \cite{Pauling}. Indeed, the SYK model is the first system to exhibit an extensive zero temperature entropy without an exponentially large ground state degeneracy.
 Obtaining the ground-state degeneracy requires the opposite order of limits between $T$ and $N$, and numerical studies show that the entropy density does vanish in such a limit for the SYK model.  The many-particle wavefunctions of the 
low-energy eigenstates in Fock space change chaotically from one state to the next, providing a realization of maximal many-body quantum chaos \cite{Maldacena16} in a precise sense.
This structure of eigenstates is very different from systems with quasiparticles, for which the lowest energy eigenstates differ only by adding and removing a few quasiparticles. 
\item
The $E$ dependence of the density of states in Eq.~(\ref{de}) is associated with a time reparameterization mode, and Eq.~(\ref{de}) shows that its effects are important when $E \sim 1/N$. The low energy quantum fluctuations of Eq.~(\ref{GSigma1}) can be expressed in terms of a path integral which reparameterizes imaginary time $\tau \rightarrow f(\tau)$, in a manner analogous to the quantum theory of gravity being expressed in terms of the fluctuations of the spacetime metric. There are also quantum fluctuations of a phase mode $\phi (\tau)$, whose time derivative is the charge density, and the path integral in Eq.~(\ref{GSigma1}) reduces to the partition function
\begin{equation}
\mathcal{Z}_{SYK-TR} = e^{N s_0} \int \mathcal{D} f \mathcal{D} \phi \exp \left( - \frac{1}{\hbar} \int_0^{\hbar/(k_B T)}\!\!\!\!\!\!  d \tau \, \mathcal{L}_{SYK-TR} [ f,\phi] \right) \label{feynsyk}
\end{equation}
The Lagrangian $\mathcal{L}_{SYK-TR}$ is known, and involves a Schwarzian of $f(\tau)$. Remarkably, despite its non-quadratic Lagrangian, the path integral in Eq.~(\ref{feynsyk}) can be performed exactly \cite{StanfordWitten}, and leads to Eq.~(\ref{de}).
\end{itemize}

\subsection{The Yukawa-SYK model}
\label{YSYK}

The SYK model defined above is a 0+1 dimensional theory with no spatial structure, and so cannot be directly applied to transport of strange metals in non-zero spatial dimensions. A great deal of work has been undertaken on generalizing the SYK model to non-zero spatial dimensions \cite{CGPS22}, but this effort has ultimately not been successful: although `bad metal' states (see Section~\ref{sec:strange}) have been obtained, low $T$ strange metals have not. But another effort based upon a variation of the SYK model, the 0+1 dimensional `Yukawa-SYK' model \cite{Fu16,Murugan:2017eto,Patel:2018zpy,Marcus:2018tsr,Wang:2019bpd,Ilya1,Wang:2020dtj,Altman20,WangMeng21,Schmalian1,Schmalian2,Schmalian3}, has been a much better starting point for a non-zero spatial dimensional theory, as shown in Section~\ref{sec:strange}. 

\begin{figure}
\begin{center}
\includegraphics[width=2.25in]{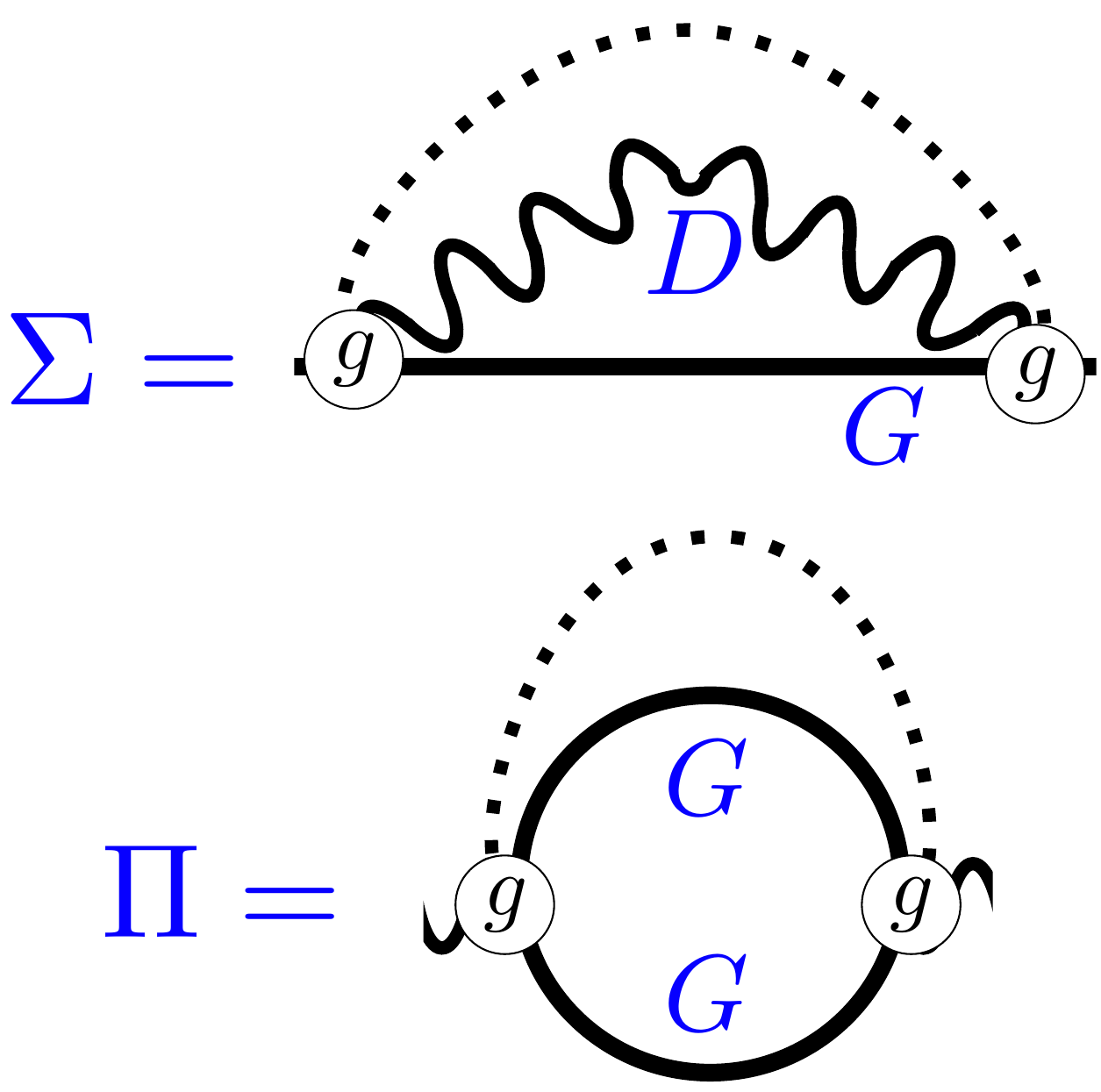}
\end{center}
\caption{Self-energies of the fermions and bosons in the Hamiltonian $\mathcal{H}_Y$ in Eq.~(\ref{HY}). The intermediate Green's functions are fully renormalized.}
\label{fig:yukawa}
\end{figure}

In the spirit of Eq.~(\ref{HH}), a model of fermions $c_\vi$ ($i=1 \ldots N$) and bosons $\Phi_\ell$ ($\ell = 1 \ldots N$) with a Yukawa coupling $g_{\vi\vj\ell}$ between them is now considered
\begin{eqnarray}
\mathcal{H}_Y = -\mu \sum_{\vi} c_\vi^\dagger c_\vi^{\vphantom\dagger} + \sum_{\ell} \frac{1}{2} \left( \pi_\ell^2 + \omega_0^2 \Phi_\ell^2 \right) + \frac{1}{N}\sum_{\vi\vj\ell} g_{\vi\vj\ell}^{\vphantom\dagger} c_\vi^\dagger c_\vj^{\vphantom\dagger} \Phi_\ell^{\vphantom\dagger} \,,\label{HY}
\end{eqnarray}
with $g_{\vi\vj\ell}$ independent random numbers with zero mean and r.m.s. value $g$. The bosons are oscillators with the same frequency $\omega_0$, while the fermions have no one-particle hopping. The large $N$ limit of Eq.~(\ref{HY}) can be taken just as for the SYK model in Eq.~(\ref{HH}). The self-energy graph in Fig.~\ref{fig:sygraph} is replaced by those in Fig.~\ref{fig:yukawa}: the phonon Green's function is $D$, while the phonon self-energy is $\Pi$.

Continuing the parallel with the SYK model, 
the disorder-averaged partition function of the Yukawa-SYK model is a bi-local $G$-$\Sigma$-$D$-$\Pi$ theory, analogous to Eq.~(\ref{GSigma1}):
\begin{eqnarray}
  \mathcal{Z} && = \int \mathcal{D} G \, \mathcal{D} \Sigma \, \mathcal{D} D \, \mathcal{D} \Pi \exp( - N S_{\rm all}) \nonumber \\
    S_{\rm all} && = -\ln\det(\partial_\tau-\mu+\Sigma)+\frac{1}{2}\ln\det(-\partial_\tau^2+\omega_0^2-\Pi) \nonumber \\
       && +\int d\tau  \int d \tau' \left[- \Sigma(\tau';\tau)G(\tau,\tau')+\frac{1}{2}\Pi(\tau',\tau)D(\tau,\tau') \right. \nonumber \\
       &&~~~~~~~~~~~~~~~~~~~~~~~\left. + \frac{g^2}{2}G(\tau,\tau')G(\tau',\tau)D(\tau,\tau')   \right]\,. \label{Sall}
\end{eqnarray}
The large $N$ saddle-point equations replacing Eq.~(\ref{sy1}) are:
\begin{eqnarray}
G(i \omega_n) = \frac{1}{i \omega_n + \mu - \Sigma (i \omega_n)} \quad &, \quad D(i \omega_n) = \frac{1}{\omega_n^2 + \omega_0^2 - \Pi (i \omega_n)} \nonumber \\
\Sigma (\tau) = g^2 G(\tau) D(\tau) \quad &, \quad \Pi (\tau) = - g^2 G(\tau) G(-\tau) \label{ysyk1}
\end{eqnarray}

The solution of Eqs.~(\ref{Sall}) and (\ref{ysyk1}) leads to a critical state with properties very similar to that of the SYK model \cite{Ilya1,Schmalian1,Schmalian2,Schmalian3}. Only the low-frequency behavior of the Green's functions at $T=0$, is quoted analogous to Eq.~(\ref{sy10}):
\begin{eqnarray}
G(i \omega) \sim -i \mbox{sgn} (\omega) |\omega|^{-(1-2 \Delta)} \quad, \quad D(i \omega) \sim |\omega|^{1-4 \Delta} \quad , \quad \frac{1}{4} < \Delta < \frac{1}{2}\,. \label{ysyk10}
\end{eqnarray}
Inserting the ansatz Eq.~(\ref{ysyk10}) into Eq.~(\ref{ysyk1}) fixes the value of the critical exponent $\Delta$. 
\begin{eqnarray}
\frac{4 \Delta - 1}{2(2 \Delta - 1) [ \sec(2 \pi \Delta) - 1 ]} = 1 \quad , \quad \Delta = 0.42037 \ldots \label{Deltaval}
\end{eqnarray}
Although the fermion Green's function has an exponent which differs from that of the SYK model, the thermodynamic properties have the same structure as that of the SYK model, including the presence of the Schwarzian mode and the form of the many-body density of states.

\section{From the SYK model to strange metals}
\label{sec:strange}

We now turn to the `strange metal' regime of Fig.~\ref{fig:figcross} and \ref{fig:cupratepd1}. A similar regime is found in numerous correlated electron materials,
and we will present here a theory \cite{Patel2,Li:2024kxr} which applies to a wide variety of quantum phase transitions, and so can explain the universality in the observations. 

Some of the key properties of strange metals, as observed in recent experiments, are first summarized \cite{Hartnoll21}:
\begin{enumerate}
\item The resistivity, $\rho (T)$, of strange metals has a linear-$T$ dependence at low temperatures:
\begin{eqnarray}
\rho(T) = \rho_0 + A T + \ldots \quad, \quad T \rightarrow 0. \label{p1}
\end{eqnarray}
Importantly, this resistivity is below the Mott-Ioffe-Regel bound \cite{Hartnoll21}, so $\rho (T) < h/e^2$ in $d=2$ spatial dimensions.
Metals with $\rho (T) > h/e^2$ are {\it bad\/} metals, and are not discussed here. 
Bad metals can be described by lattice models of coupled SYK `islands' \cite{SongBalents,PatelArovas,ChowdhuryBerg}, as reviewed elsewhere \cite{CGPS22}.
\item Ordinary metals have low $T$ specific heat which vanishes linearly with $T$, but in a strange metal the 
specific heat is enhanced to $\sim T \ln (1/T)$ as $T \rightarrow 0$.
\item Careful analyses of optical data in the cuprates over wide ranges of frequencies and temperatures \cite{NormanChubukov,Michon22} has shown that the optical conductivity can be accurately described by the following form
\begin{eqnarray}
\sigma (\omega ) = \frac{K}{\displaystyle \frac{1}{\tau_{\rm trans} (\omega)} - i \omega \frac{m_{\rm trans}^\ast (\omega)}{m}} ~~; ~~ \frac{1}{\tau_{\rm trans} (\omega)} \sim |\omega| \Phi_\sigma \left( \frac{\hbar \omega}{k_B T} \right)\,, \label{p2}
\end{eqnarray}
where $K$ is a constant, and the transport scattering rate $1/\tau_{\rm trans}$ scales linearly with the larger of $|\omega|$ and $k_B T/\hbar$. The frequency dependence of the effective transport mass $m_{\rm trans}^\ast$ is then determined by a Kramers-Kronig connection to that of $1/\tau_{\rm trans}$, which leads to a logarithmic frequency dependence in $m_{\rm trans}^\ast (\omega) $. 
\item 
Photoemission experiments on the cuprates have measured the electron self-energy near the nodal point in the Brillouin zone. This was found to obey the scaling form \cite{DessauPLL}
\begin{eqnarray}
\frac{1}{\tau_{\rm in} (\omega)} = 2 \, \mbox{Im} \Sigma (\omega) \sim |\omega|^{2\alpha} \Phi_{\Sigma} \left( \frac{\hbar \omega}{k_B T} \right) \label{p3}
\end{eqnarray}
with an exponent $\alpha \approx 1/2$ near optimal doping. The value $\alpha=1/2$ corresponds to a `marginal Fermi liquid' \cite{Varma89}, at least as far as the self energy is concerned. But an important point is that there is no direct theoretical connection between the single-particle scattering rate $1/\tau_{\rm in} (\omega)$ in Eq.~(\ref{p3}), and the value of transport scattering rate  $1/\tau_{\rm trans}(\omega)$ in Eq.~(\ref{p2}), although they are observed to have the same exponent. As seen below, the transport and single-particle scattering rates can be very different in some common models.
\item In experimental observations \cite{Bruin13,Legros19,Gael21}, the value of the overall constant $K$ in Eq.~(\ref{p2}) is often fixed by writing the d.c. conductivity in the Drude form
\begin{eqnarray}
\sigma = \frac{ n e^2 \tau_{\rm trans}}{m^\ast}\,,
\label{eq:Drude}
\end{eqnarray}
where $n$ is the known conduction electron density, and $m^\ast$ is an electronic effective mass. In some experiments, the transport mass $m_{\rm trans}^\ast$ of Eq.~(\ref{p2}) is used in Eq.~(\ref{eq:Drude}), while other experiments use the $m^\ast$ determined from thermodynamic measurements. In the form Eq.~(\ref{eq:Drude}), the absolute value of $\tau_{\rm trans}$ can be deduced from experimental observations. In the strange metal, such a value is found to obey `Planckian' behavior with \cite{Bruin13,Legros19,Gael21,Shekhter24,Hussey24,Alloul24,Proust25}
\begin{eqnarray}
\frac{1}{\tau_{\rm trans}} = \widetilde{\alpha} \, \frac{k_B T}{\hbar}\,, \label{Planckian}
\end{eqnarray}
\end{enumerate}
with $\widetilde{\alpha}$ a numerical constant of order unity. Measurements of $1/\tau_{\rm trans}$ in La$_{1.6-x}$Nd$_{0.4}$Sr$_x$CuO$_4$ in angle-dependent magnetotransport show $\widetilde{\alpha} = 1.2 \pm 0.4$ \cite{Gael21} upon using the thermodynamic $m^\ast$.

\subsection{Universal model}
\label{sec:univ}

This subsection will present a simple and universal generalization of the Yukawa-SYK model of Section~\ref{YSYK} to spatial dimension $d=2$ which reproduces all five of the above observed properties \cite{Patel2,Li:2024kxr}.

We will discuss the model appropriate for the cuprates here, although the results can be extended to other metallic quantum phase transitions with or without disorder. As we have noted in Section~\ref{sec:intro_strange}, in the clean case when umklapp scattering is suppressed, although there can be no quasiparticles around the Fermi surface, the transport is nevertheless Fermi liquid-like: this will be described in Section~\ref{sec:norandom}. So we will focus on the disordered case, with the most important disorder being of the Harris type, which produces local variations in the value of the coupling tuning the system across the quantum critical point. 

Our model for the hole-doped cuprates is a FL*-FL quantum phase transition (see Fig.~\ref{fig:flsfan}) without symmetry breaking, and in the clean limit this is distinct from metallic quantum phase transitions with symmetry breaking. The latter are further distinguished in the clean limit by cases in which the order parameters reside at zero or non-zero momenta \cite{hertz} which have dynamic critical exponents $z=3$ and $z=2$ respectively. The remarkable fact is that spatial disorder unifies these 3 classes into essentially the same universal theory \cite{Patel2} which we summarize below for the case of the FL*-FL transition, and in Section~\ref{sec:griffiths} for the spin density wave transitions. However, we note that there is an enhancement of thermopower which is special to the FL*-FL transition, as discussed elsewhere \cite{LPS24}, and this compares well with experiments on the cuprates \cite{Collignon21,Taillefer_Seebeck_PRX_2022,Georges_Skewed}.

\begin{figure}
\centering
\includegraphics[width=6in]{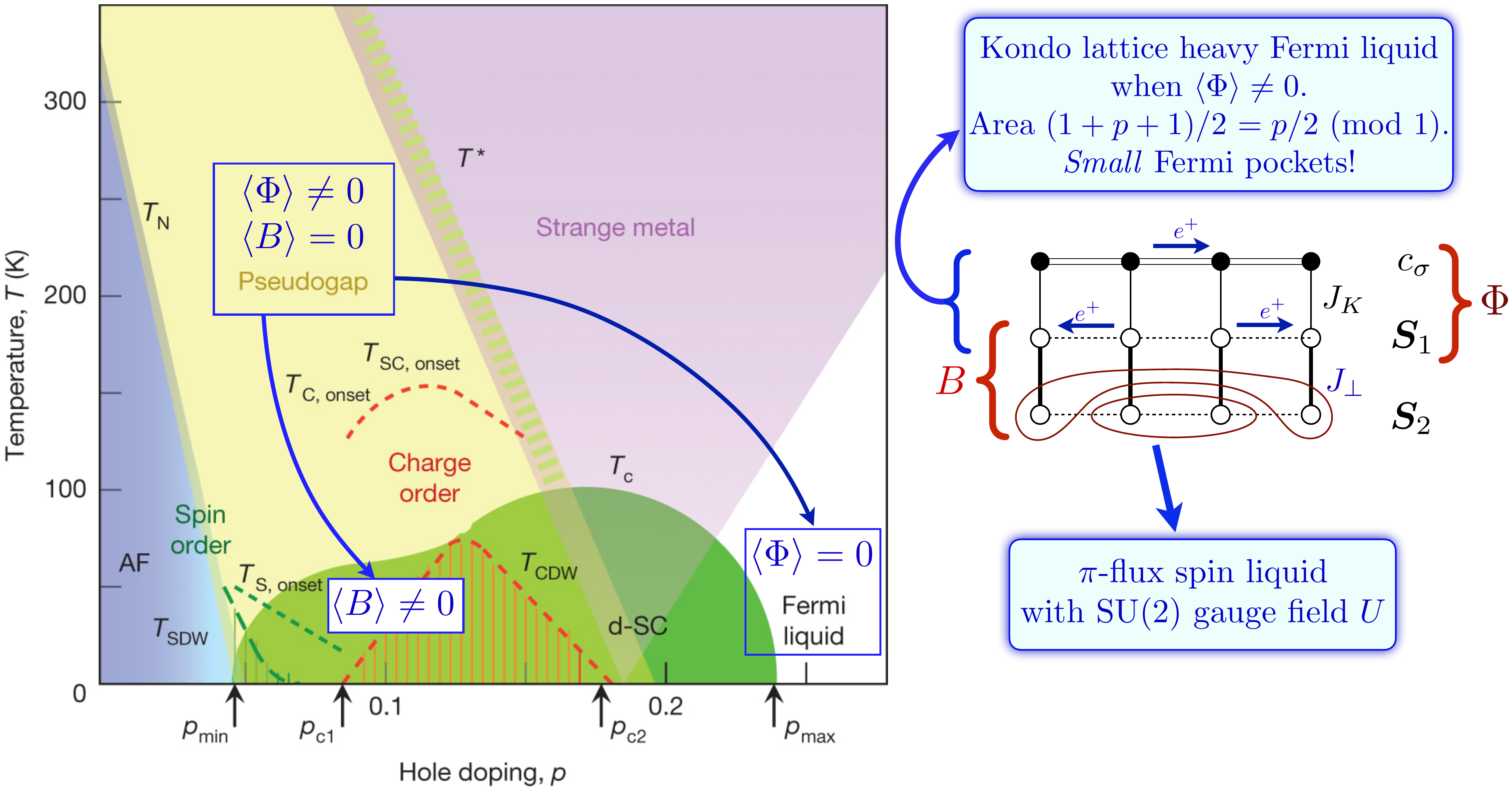}
\caption{On the left is the cuprate phase diagram from Ref.~\cite{phase_diag}; annotations in blue have been added. 
On the right is the ALM, with the Higgs fields $\Phi$ and $B$.
In Section~\ref{sec:strange} we consider the transition from the FL* pseudogap to the Fermi liquid focusing
on the Higgs field $\Phi$, while setting $B=0$. The transition from the pseudogap to the d-SC was discussed in Section~\ref{sec:dwave} as a theory of the dynamics of $B$, while setting $\Phi$ to a non-zero constant which determined the magnitude of the pseudogap. See \hyperref[app:ancilla]{Appendix B} for the complete gauge theory of $B$ and $\Phi$ for both transitions.}
\label{fig:cupratepd2}
\end{figure}

\begin{figure}
\begin{center}
\includegraphics[width=4in]{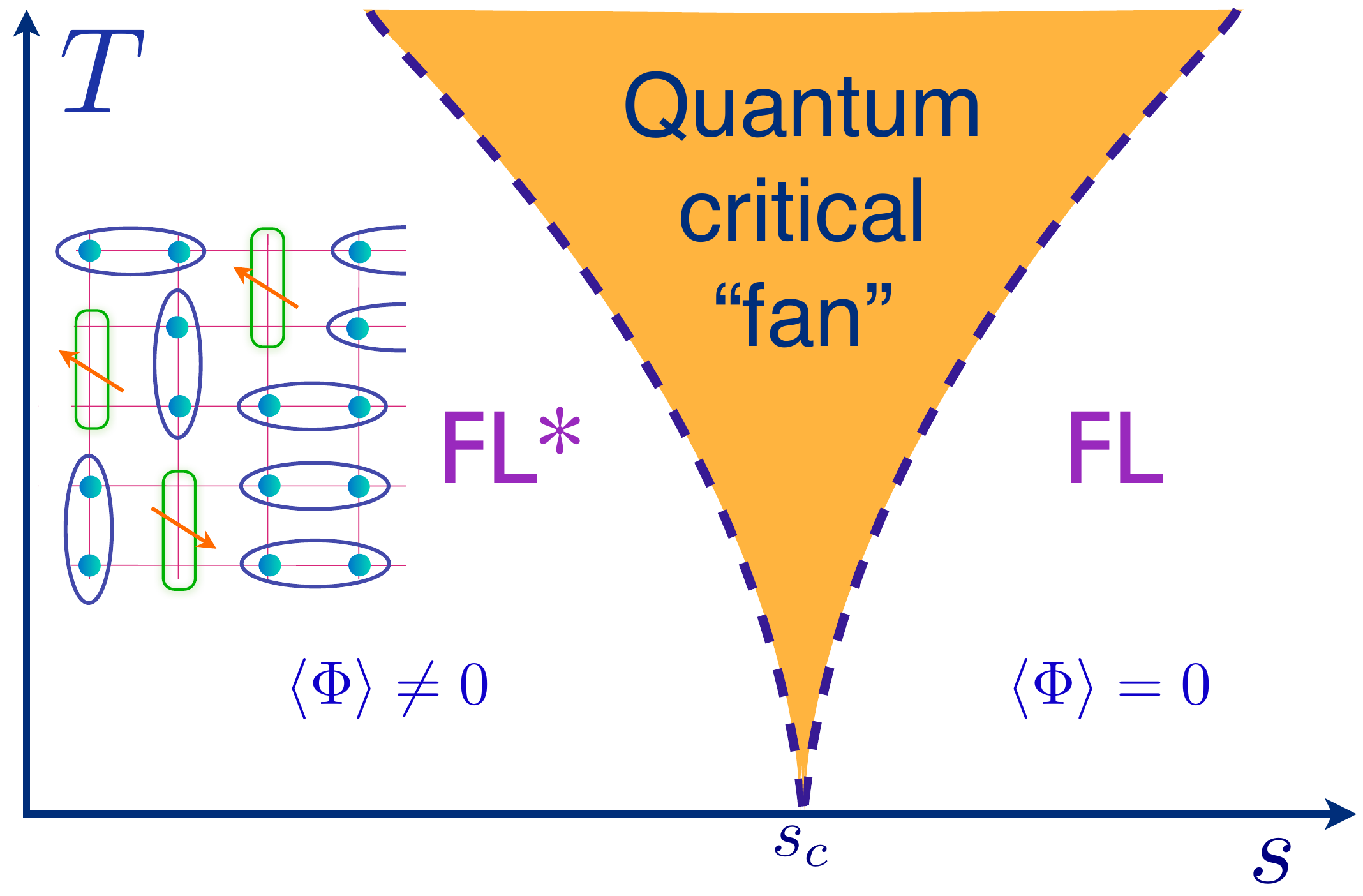}
\end{center}
\caption{Schematic phase diagram of a single-band metal with a Fermi surface volume changing quantum transition without a broken symmetry on either side of the transition. For the Kondo lattice, the transition is `inverted' as shown in Fig.~\ref{fig:inverted}, with the Higgs field $\Phi$ condensed on the FL side. $\Phi$ was the key variational parameter in the wavefunction description of the FL*-FL transition in Section~\ref{sec:wavefunction}. The `naked' quantum critical point at $s_c$
is the same as that at $p_c$ in Fig.~\ref{fig:figcross}. This quantum critical point is expected to the unstable to a superconducting dome \cite{YaHui-ancilla1,YaHui-ancilla2}, with the superconducting state smoothly connected across $s_c$, as shown in Fig.~\ref{fig:figcross} at $p_c$. This is an important difference from quantum phase transitions in metals with symmetry breaking, for which the quantum phase transition is also present inside the superconductor.}
\label{fig:flsfan}
\end{figure}

For the cuprates, we present the phase diagram of Fig.~\ref{fig:cupratepd1} again in Fig.~\ref{fig:cupratepd2}, but now with annotations updated with reference to the ALM in Fig.~\ref{fig:PhiB}.
In Section~\ref{sec:dwave} we have considered the low temperature fate of the FL* pseudogap associated with condensation of the Higgs field $B$. In this analysis, we treated $\Phi$ simply as a c-number constant. In the present section, $\Phi$ and $B$ will exchange roles. In considering the transition from FL* to FL in Fig.~\ref{fig:oneband} and Fig.~\ref{fig:cupratepd2}, we consider the dynamics of the $\Phi$ field, and the fermions in the top two layers in Fig.~\ref{fig:PhiB}. In FL*, we can safely set $B=0$ (apart for some thermal fluctuations which were needed in Section~\ref{sec:arc}). In the FL phase of Fig.~\ref{fig:oneband}, we see that the $f_{1}$ fermions of the middle layer form a trivial rung-singlet state with the $f$ fermions in the bottom layer: this confinement of $f_1$ and $f$ is not associated with $B$, but by the confinement of a $\widetilde{\rm SU}(2)$ gauge field \cite{YaHui-ancilla1,YaHui-ancilla2,Nikolaenko:2021vlw} (see \hyperref[app:ancilla]{Appendix B}). Consequently, it is safe to ignore $B$ in the quantum criticality of the FL* to FL transition \cite{YaHui-ancilla1,YaHui-ancilla2,LZDC20}. So we will only consider the $\Phi$ boson here, along with the $c_\alpha$ and $f_1$ fermions in the top two layers of the ALM. 
%There are additional gauge fields to which $\Phi$ couples that do have to be considered for the FL* to FL transition \cite{YaHui-ancilla1,YaHui-ancilla2,LZDC20} (see the \hyperref[app:ancilla]{Appendix}), but we will ignore them for simplicity, as they are not crucial in the criticality of the disordered case. 
%This is the approximation in which the inversion mapping to the Kondo lattice in Fig.~\ref{fig:inverted} becomes exact.

We therefore consider the $T>0$ quantum criticality of the FL* to FL transition \cite{Qi10,SSMetlitskiPunk12} sketched in Fig.~\ref{fig:flsfan}, associated with the Higgs transition in $\Phi$ \cite{YaHui-ancilla1}. Note that $\Phi$ was also the key variational parameter in the trial wavefunction description of the FL* to FL transition in Section~\ref{sec:wavefunction}.

{\it Dynamic mean-field theories.} We could study this transition using the dynamic mean field theories \cite{SY92,CGPS22,Sengupta97,Si1,Si2,AMT10,AMT11,ZhuSi,Norman07,Norman08,Norman13,Si14,SenguptaGeorges95,Burdin_2000,Burdin2002,Kotliar08,TremblayPNAS22,GleisKotliar24,GleisKotliar25,ChChung24,ChChung25,Lal25}, which map the transition to self-consistent Green's functions similar to those studied in Section~\ref{sec:SYK}. Such theories are in the strong umklapp limit, and could apply in very clean Kondo lattice systems over a significant intermediate temperature range. However, we expect umklapp to be suppressed at low temperatures, and the emergence of a conserved momentum leads to absence of strange metal behavior, as we discuss in Section~\ref{sec:norandom}; similar comments apply to other numerical studies \cite{Fratini24,Navinder25}. Moreover, the dynamic mean-field theories also have the serious issue of a non-vanishing entropy density in the zero temperature limit \cite{CGPS22}.

Here, we will instead treat the spatial dimensionality seriously, and focus on the low energy, long-wavelength structure. Spatially random couplings play a crucial role in such an analysis, and lead to new Griffiths effects associated with the `foot' of strange metal behavior \cite{PPS24,PLA24,SSZaanen} which we will discuss in Section~\ref{sec:griffiths}.

As an aside, we note that the FL*-FL quantum critical point is expected to be unstable to $d$-wave superconductivity \cite{YaHui-ancilla1,YaHui-ancilla2}, with the BCS state on the FL side adiabatically connected to the superconductivity obtained from FL*, as we have seen in Sections~\ref{sec:halffilling} and \ref{sec:FL*}. So the FL*-FL quantum critical point will disappear under the superconducting dome, as shown earlier in Fig.~\ref{fig:figcross}. This is an important distinction from quantum transitions in metals with symmetry breaking: then there are two distinct superconducting states, and the symmetry breaking transition is also present within the superconducting dome.

At the mean-field level the FL* to FL transition is described by $\mathcal{H}_{\rm KLmf}$ in Eq.~(\ref{HKLmf}). Starting from $\mathcal{H}_{\rm KLmf}$, we include spatial dependence in all fields, allow for quenched spatial disorder, and obtain the (universal) Lagrangian of the two-dimensional Yukawa-Sachdev-Ye-Kitaev (2D-YSYK) model
\begin{eqnarray}
 \mathcal{L}_\Phi && = \sum_{\bm k} c_{{\bm k}\alpha}^\dagger \left( \frac{\partial}{\partial \tau} + \varepsilon ({\bm k})\right) c_{{\bm k}\alpha} + \sum_{\bm k} f_{1{\bm k}\alpha}^\dagger \left( \frac{\partial}{\partial \tau} + \varepsilon_1 ({\bm k})\right) f_{1{\bm k}\alpha} \nonumber \\
&& + \int d^2 {\bm r} \Bigl\{ s \,|\Phi({\bm r})|^2 +  [g + {\color{darkgreen} g'({\bm r})}] \, c_{\alpha}^{\dagger}({\bm r})  f_{1\alpha} ({\bm r}) \, \Phi({\bm r}) + \mbox{H.c.} \nonumber \\
&& + K \, | \nabla_{\bm r} \Phi ({\bm r})|^2 +u  \,|\Phi({\bm r})|^4  + {\color{darkgreen} v({\bm r})} \left[c_\alpha^\dagger ({\bm r}) c_\alpha  ({\bm r}) +
f_{1\alpha}^\dagger (\bm{r}) f_{1\alpha} (\bm{r}) \right]\Bigr\}\,.
\label{e22}
\end{eqnarray}
We are using the convention that quenched random variables (with no dynamics and a fixed, random dependence on ${\bm r}$) are colored in green. For the Kondo lattice case, 
there should also be a U(1) gauge field associated with Eq.~(\ref{gaugeu1}) in the Lagrangian in Eq.~(\ref{e22}), but we have dropped it as it does not significantly modify the critical behavior \cite{Guo2022,Altman20}. For the single-band case, there is a $\widetilde{\rm SU}(2)\times{\rm U}(1)$ gauge field \cite{YaHui-ancilla1,YaHui-ancilla2,LZDC20}, as we review in \hyperref[app:ancilla]{Appendix B} (the tilde is to distinguish from the distinct SU(2) gauge field of Sections~\ref{sec:spinliquids}-\ref{sec:dwave}); the U(1) gauge field is the local constraint in the ${\bm S}_1$ layer (as in Eq.~(\ref{gaugeu1})), and the $\widetilde{\rm SU}(2)$ gauge field is the singlet projection between the ${\bm S}_1$ and ${\bm S}_2$ layers, and we ignore these gauge fields for similar reasons.
The dispersions $\varepsilon ({\bm k})$ and $\varepsilon_1 ({\bm k})$ arise from the $t_{\vi\vj}$ and $t_{1\vi\vj}$ in Eq.~(\ref{HKLmf}), $s$ is the tuning parameter in Fig.~\ref{fig:flsfan}, and we have added various in an effective action for $\Phi$.

The most common form of spatial disorder in studies of metallic transport is a random potential from impurities, and this realized by {\color{darkgreen} $v({\bm r})$}. This spatial disorder is averaged over after
assuming it is uncorrelated at different points in space
\begin{eqnarray}
{\color{darkgreen} \overline{v ({\bm r})} = 0 \quad, \quad \overline{v({\bm r}) v({\bm r'})} = v^2 \delta({\bm r}-{\bm r'})\,. }\label{f5}
\end{eqnarray}
One of the main new points made in Refs.~\cite{Guo2022,Patel2}, as we will see in Section~\ref{sec:norandom}, is that random potential disorder is also not sufficient to produce a strange metal, and the effects of spatial randomness in the interactions \cite{Altman20} must also be considered.
Spatial disorder in the Kondo exchange interaction $J_K$ in Eq.~(\ref{eq:HKL}) translates into the spatially random Yukawa coupling {\color{darkgreen} $g' ({\bm r})$} which we take to obey the disorder average
\begin{eqnarray}
{\color{darkgreen} \overline{g'({\bm r})} = 0 \quad, \quad \overline{g'({\bm r}) g'({\bm r'})} = g'^2 \delta({\bm r}-{\bm r'}).} \label{f7}
\end{eqnarray}
In the underlying electronic model, the exchange $J_K \sim t^2/U$, where $t$ is some hopping--so randomness in $t$ (which is a form of random potential disorder) translates to randomness in $J_K$, and hence in the Yukawa coupling.
We can also see that upon integrating out the fermions, the coupling {\color{darkgreen} $g' ({\bm r})$} generates spatial randomness in $s$. The latter is usually identified as `Harris disorder', associated with spatial randomness in the position of the phase transition. But as long as we are in a self-averaging regime, the lesson from the Yukawa-SYK model is that it is more convenient to keep the randomness in the Yukawa coupling. Moreover, we can transfer random mass disorder to random Yukawa coupling disorder simply by rescaling $\Phi ({\bm r})$ with a ${\bm r}$-dependent factor: so keeping it in {\color{darkgreen} $g' ({\bm r})$} is equivalent to working in the boson eigenstates of the random mass term. At very low energies, the random mass does eventually lead to boson localization and the new physics \cite{PPS24,PLA24,SSZaanen} of the `foot', and this is discussed in Section~\ref{sec:griffiths}.

The properties of this strange metal theory will be determined by directly extending the methods used to solve the Yukawa-SYK model. This extension can be viewed as simply solving the equations in Fig.~\ref{fig:yukawa}, while using the propagators in $\mathcal{L}_\Phi$. Alternatively, a fictitious flavor index on all fields ranging over $N$ values can be introduced and the large $N$ limit taken, assuming couplings are random in this flavor space. For simplicity, we ignore the difference in dispersion between the $c$ and $f_1$ fermions in the remaining discussion: this does not significantly modify the results as long as there is random spatial potential acting on the fermions. Then we obtain a 
 $G$-$\Sigma$-$D$-$\Pi$ theory which is a direct generalization of Eq.~(\ref{Sall}) to Green's functions that are bilocal in {\it both\/} space and time
\begin{eqnarray}\label{Sall2}
&&  \mathcal{Z}  = \int \mathcal{D} G \, \mathcal{D} \Sigma \, \mathcal{D} D \, \mathcal{D} \Pi \exp( - N S_{\rm all}) \nonumber \\
 &&   S_{\rm all} = -\ln\det(\partial_\tau+\varepsilon({{\bm k}})-\mu+\Sigma)+\frac{1}{2}\ln\det(-\partial_\tau^2+{{\bm q}}^2+m_b^2-\Pi) \nonumber \\
       && +\int d\tau d^2 r \int d \tau' d^2 r'\left[- \Sigma(\tau',{\bm r}';\tau,{\bm r})G(\tau,{\bm r};\tau',{\bm r}')+ \right. \nonumber \\
      &&\frac{1}{2}\Pi(\tau',{\bm r}';\tau,{\bm r})D(\tau,{\bm r};\tau',{\bm r}') +
        \frac{g^2}{2}G(\tau,{\bm r};\tau',{\bm r}')G(\tau',{\bm r}';\tau,{\bm r})D(\tau,{\bm r};\tau',{\bm r}') \nonumber \\
        &&   +  \frac{{\color{darkgreen}v^2}}{2}G(\tau,{\bm r};\tau',{\bm r}')G(\tau',{\bm r}';\tau,{\bm r})\delta({\bm r}-{\bm r}') \nonumber \\
        &&  +  \left. \frac{{\color{darkgreen} g'^2}}{2}G(\tau,{\bm r};\tau',{\bm r}')G(\tau',{\bm r}';\tau,{\bm r})D(\tau,{\bm r};\tau',{\bm r}')\delta({\bm r}-{\bm r}') \right]\,.
\end{eqnarray}
For compactness of notation and analysis, we have assumed that the $c$ and $f_1$ fermions Green's functions are both equal to $G$, but it is not difficult to treat the more general case. 
Note that the spatially random couplings lead to an additional $\delta({\bm r} - {\bm r}')$ in their contributions arising from the disorder averages in Eqs.~(\ref{f5}) and (\ref{f7}). The saddle point of Eq.~(\ref{Sall2}) leads to equations for the Green's functions which can also be derived from Fig.~\ref{fig:yukawa}:
\begin{eqnarray}
&\Sigma(\tau,{\bm r}) = g^2 D(\tau,{\bm r})G(\tau,{\bm r}) +  {\color{darkgreen}v^2} G(\tau,{\bm r}) \delta^2({\bm r}) +  {\color{darkgreen}{g'}^2}G(\tau,{\bm r})D(\tau,{\bm r})\delta^2({\bm r}), \nonumber \\
& \Pi(\tau,{\bm r}) = -g^2 G(-\tau,-{\bm r})G(\tau,{\bm r}) -  {\color{darkgreen}{g'}^2}G(-\tau,{\bm r})G(\tau,{\bm r})\delta^2({\bm r}), \nonumber \\
&G(i\omega,{\bm k}) = \frac{1}{i\omega-\varepsilon({\bm k})+\mu-\Sigma(i\omega,{\bm k})}, \nonumber \\
&D(i\Omega,{\bm q}) = \frac{1}{\Omega^2+{\bm q}^2+m_b^2-\Pi(i\Omega,{\bm q})}. 
\label{eq:saddle_pt_eqs}
\end{eqnarray}
We also need an equation to determine the boson mass $m_b$: this is determined by the Hartree contribution from the $u$ term in Eq.~(\ref{e22}) \cite{Patel1}.

This is a good point to note the importance of the elastic scattering term {\color{darkgreen} $v({\bm r})$}. This broadens the Fermi surfaces, and so effects from Fermi surface nesting or `hotspots' are quenched. This is a key reason for the universality of ${\bm L}_\Phi$, which applies to symmetry breaking transitions too, along with the non-symmetry breaking FL*-FL transition we are considering. This is also the reason the difference in dispersion between $c$ and $f_1$ is mostly not important, and has been neglected above.
However, the particle-hole asymmetry of the FL*-FL transition is important for the thermopower, and we comment on this near Eq.~(\ref{g3p}).

\begin{figure}
\begin{center}
\includegraphics[width=3in]{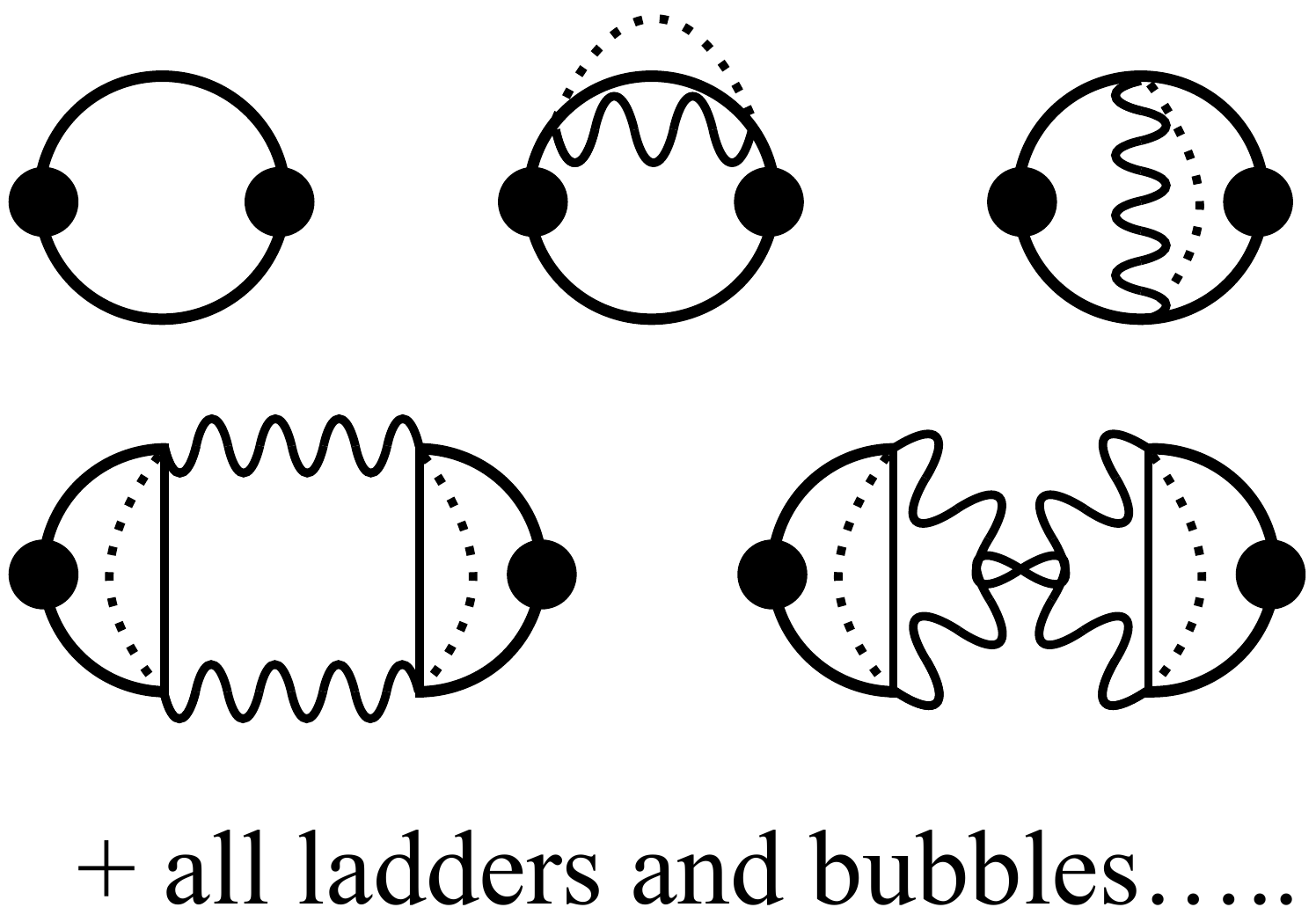}
\end{center}
\caption{Diagrams for the conductivity for the theory $ \mathcal{L}_\Phi$ in Eq.~(\ref{e22}).}
\label{fig:ladders}
\end{figure}

Before discussing the solution of Eq.~(\ref{eq:saddle_pt_eqs}), the computation of response functions of fermion bilinears, such as the conductivity, is described. These can be obtained by inserting external sources into Eq.~(\ref{Sall2}) and then taking the variational derivatives with respect to them. This leads to the graphs shown in Fig.~\ref{fig:ladders}, which have to evaluated with fully renormalized Green's functions.

The following subsections discuss the solutions of the equations in Eq.~(\ref{eq:saddle_pt_eqs}) and Fig.~\ref{fig:ladders} for the cases without and with spatial randomness.

\subsection{No spatial randomness}
\label{sec:norandom}

The solution of Eq.~(\ref{eq:saddle_pt_eqs}) with $g \neq 0$, but $g'=0$ and $v=0$, is considered. This corresponds to the quantum phase transition without disorder, and has been much studied in the literature.
At the quantum critical point, Eq.~(\ref{eq:saddle_pt_eqs}) yields a non-Fermi liquid form for the fermion Green's function, and a Landau-damped form for the boson Green's function \cite{PALee89,Patel1}
\begin{eqnarray}
\Sigma (i \omega, {\bm k}) \sim -i \mbox {sgn} (\omega) |\omega|^{2/3} \,, & \quad G( i\omega,{\bm k}) = \frac{1}{i \omega - \varepsilon({\bm k}) - \Sigma (i \omega, {\bm k})} \nonumber \\ \quad D(i \Omega,{\bm q} ) & = \frac{1}{\Omega^2 + {\bm q}^2 + \gamma |\Omega|/q}\,, \label{g1}
\end{eqnarray}
The fermion Green's function has a sharp Fermi surface in momentum space, and ${\bm k}$ in Eq.~(\ref{g1}) is assumed to be close to the Fermi surface. But $G$ is diffusive in frequency space, indicating the absence of well-defined fermionic quasiparticles.

However, an important point is that essentially none of this non-Fermi liquid structure feeds into the conductivity, which remains very similar to that of a Fermi liquid \cite{Maslov12,Maslov17a,Maslov17b,Berg19,Guo2022,SenthilShi22,GuoIII} with the form:
\begin{eqnarray}
\sigma (\omega) \sim \frac{1}{-i \omega} + |\omega|^0 + \cdots \quad \mbox{($\omega^{-2/3}$ term has vanishing co-efficient)} \label{g2}
\end{eqnarray}
There has been a claim \cite{YBK94} of a $\omega^{-2/3}$ contribution to the conductivity, but its co-efficient vanishes after evaluation of all the graphs in Fig.~\ref{fig:ladders} \cite{SenthilShi22,Guo2022}. This cancellation can be understood as a consequence of Kohn's theorem \cite{Kohn61}, which states that in a Galilean-invariant system only the first term of the right-hand-side of Eq.~(\ref{g2}) is non-zero.
A Galilean-invariant system is not considered here, but all contributions to the possible $\omega^{-2/3}$ term arise from long-wavelength processes in the vicinity of patches of the Fermi surface, and these patches can be embedded in a system which is Galilean-invariant also at higher energies.

\subsection{With spatial randomness}
\label{sec:random}

The absence of the $\omega^{-2/3}$ term in Eq.~(\ref{g2}) is a strong indication that the resolution of the strange metal problem cannot come from the clean limit of quantum-critical theories described by models of the class in Eq.~(\ref{e22}). There can be umklapp processes which dissipate momentum, but these require special features of the Fermi surface to survive at low momentum. Recent nonlinear THz 2D coherent spectroscopy observations on the cuprates \cite{Armitage25} observe a rapid momentum relaxation rate, supporting models in which disorder plays a central role.

A theory with only potential scattering disorder as in Eq.~(\ref{f5}), {\it i.e.} $g \neq 0$, {\color{darkgreen}$v \neq 0$}, but {\color{darkgreen}$g'=0$}, is also not sufficient \cite{Guo2022,Foster2022}: it leads to marginal Fermi-liquid behavior in the electron self energy, but no strange metal behavior in transport.
So for a generic and universal theory of strange metals, the influence of disorder with $g$, {\color{darkgreen}$g'$}, and {\color{darkgreen}$v$} all non-zero should be considered.
The solution of Eq.~(\ref{eq:saddle_pt_eqs}) yields a boson Green's function which has a diffusive form at the critical point \cite{HLR} 
\begin{eqnarray}
D(i\Omega, {\bm q}) \sim \frac{1}{{\bm q}^2 + \gamma |\Omega|} \,.\label{g3}
\end{eqnarray}

This is a good point to mention, in passing, a special feature of the FL*-FL transition, not shared by phase transitions of two-dimensional metals with symmetry breaking order parameters. The $\Phi$ propagator of the FL*-FL transition is sensitive to particle-hole asymmetry, as it also carries electrical charge (for $\mathcal{L}_\Phi$, the asymmetry requires $\varepsilon ({\bm k}) \neq \varepsilon_{1} ({\bm k})$). 
This allows a more general form for the boson propagator than Eq.~(\ref{g3}) \cite{Altman20,LPS24}
\begin{equation}
D(i\Omega, {\bm q}) \sim \frac{1}{{\bm q}^2 + \gamma |\Omega| - i \bar{\gamma}  \Omega}\,. \label{g3p}
\end{equation}
The $\bar{\gamma}$ term is responsible for the singular thermopower response, as discussed in Ref.~\cite{LPS24}, and this connects to observations on Fermi volume changing transitions in cuprates \cite{Collignon21,Taillefer_Seebeck_PRX_2022,Georges_Skewed} and heavy-fermion compounds \cite{Park24}.

Inserting Eq.~(\ref{g3}) or (\ref{g3p}) into the fermion Green's function gives a marginal Fermi liquid form \cite{HLR, Patel2}
\begin{eqnarray}
&& G (\omega)  \sim  \frac{1}{ \displaystyle 
\omega \, \frac{m^\ast (\omega)}{m} - \varepsilon ({\bm k}) + 
i \left(\frac{1}{\tau_e} + \frac{1}{\tau_{\rm in} (\omega)} \right) \mbox{sgn} (\omega) } \label{g4} \\
&& \frac{1}{\tau_{e}} \sim  {\color{darkgreen}v^2} \,; \quad  \frac{1}{\tau_{\rm in} (\omega)} \sim \left( \frac{g^2}{{\color{darkgreen}v^2}} + {\color{darkgreen}g^{\prime 2}} \right) |\omega | \, ; \quad  
\frac{m^\ast (\omega)}{m} \sim \frac{2}{\pi} \left( \frac{g^2}{{\color{darkgreen} v^2}} + {\color{darkgreen} g^{\prime 2}} \right) \ln (\Lambda/\omega)\,. \nonumber
\end{eqnarray}
The expressions in the second line are schematic, and show only the dependence upon $g$, {\color{darkgreen} $g'$} and {\color{darkgreen} $v$} without numerical constants.
This result matches the photoemission observations in Eq.~(\ref{p3}) for $\alpha = 1/2$. Note that there are two distinct contributions to the singular $|\omega|$ electron inelastic scattering rate $1/\tau_{\rm in}$: one from the combination of impurity scattering $v$ with the spatially uniform interaction $g$ \cite{HLR}, and the other from the spatially random interaction {\color{darkgreen} $g'$} \cite{Altman20,Patel2}.

Inserting these solutions for the Green's functions into the action in Eq.~(\ref{Sall2}), gives a $T \ln (1/T)$ specific heat \cite{Patel1}.

Turning to the evaluation of the conductivity graphs in Fig.~\ref{fig:ladders}, the key property of the strange metal, the conductivity, is given by the form in Eq.~(\ref{p2}), with \cite{Altman20,Patel2}
\begin{eqnarray}
 \frac{1}{\tau_{\rm trans} (\omega)} \sim {\color{darkgreen} v^2} & + {\color{darkgreen} g^{\prime 2}} |\omega|  \quad; \quad 
\frac{m_{\rm trans}^\ast (\omega)}{m} \sim \frac{2 {\color{darkgreen} g^{\prime 2}}}{\pi} \ln (\Lambda/\omega) \label{g5}
\end{eqnarray}
This expression shows that the residual resistivity $\rho_0$ at $T=0$ is determined by the elastic scattering rate $1/\tau_e \sim {\color{darkgreen} v^2}$,  as in a disordered Fermi liquid. The inelastic processes lead to a frequency and temperature dependence which matches precisely with the observational form in Eq.~(\ref{p2}). An important feature is that of the two processes contributing to the electron inelastic scattering rate $1/\tau_{\rm in}$ in Eq.~(\ref{g4}), only one contributes to the inelastic transport rate $1/\tau_{\rm trans}$. The processes involving the spatially uniform interaction $g$ and the impurity potential $v$ {\it cancel out\/} in the computation of the conductivity from Fig.~\ref{fig:ladders}, and {\it only\/} those involving the spatially random interaction {\color{darkgreen} $g'$} survive \cite{Patel2}. A consequence of this cancellation is that the constant $\widetilde{\alpha}$ in Eq.~(\ref{Planckian}) approaches $\widetilde{\alpha} = \pi/2$ for the quasiparticle $m^\ast$ in the limit ${\color{darkgreen} g'} \gg g$ \cite{Patel1}, and decreases from this value as $g$ is increased \cite{Patel2}.

To summarize, the conductivity of the theory $\mathcal{L}_\Phi$ yields the strange metal conductivity in Eq.~(\ref{p1}), 
with $\rho_0 \sim {\color{darkgreen} v^2}$ and $A \sim {\color{darkgreen} g^{\prime 2}}$. Note that the value of $g$ does not make a direct difference to the value of the linear-$T$ resistivity, although it does affect the marginal Fermi liquid behavior of the electron self energy, as noted in Eq.~(\ref{g4}). It is also notable that the residual resistivity and linear-$T$ resistivity slope are determined by different sources of disorder: those in Eqs.~(\ref{f5}) and (\ref{f7}) respectively. This distinction should be important in understanding trends in observations \cite{Ong91,IISC94}.
In the quantum critical region, the approach outlined above \cite{Patel2}, and the more complete quantum Monte Carlo studies \cite{PLA24} show that the slope $\widetilde{\alpha}$ of the transport scattering rate in Eq.~(\ref{Planckian}) is largely independent of the strength of disorder.
\begin{figure}
\begin{center}
\includegraphics[width=5.5in]{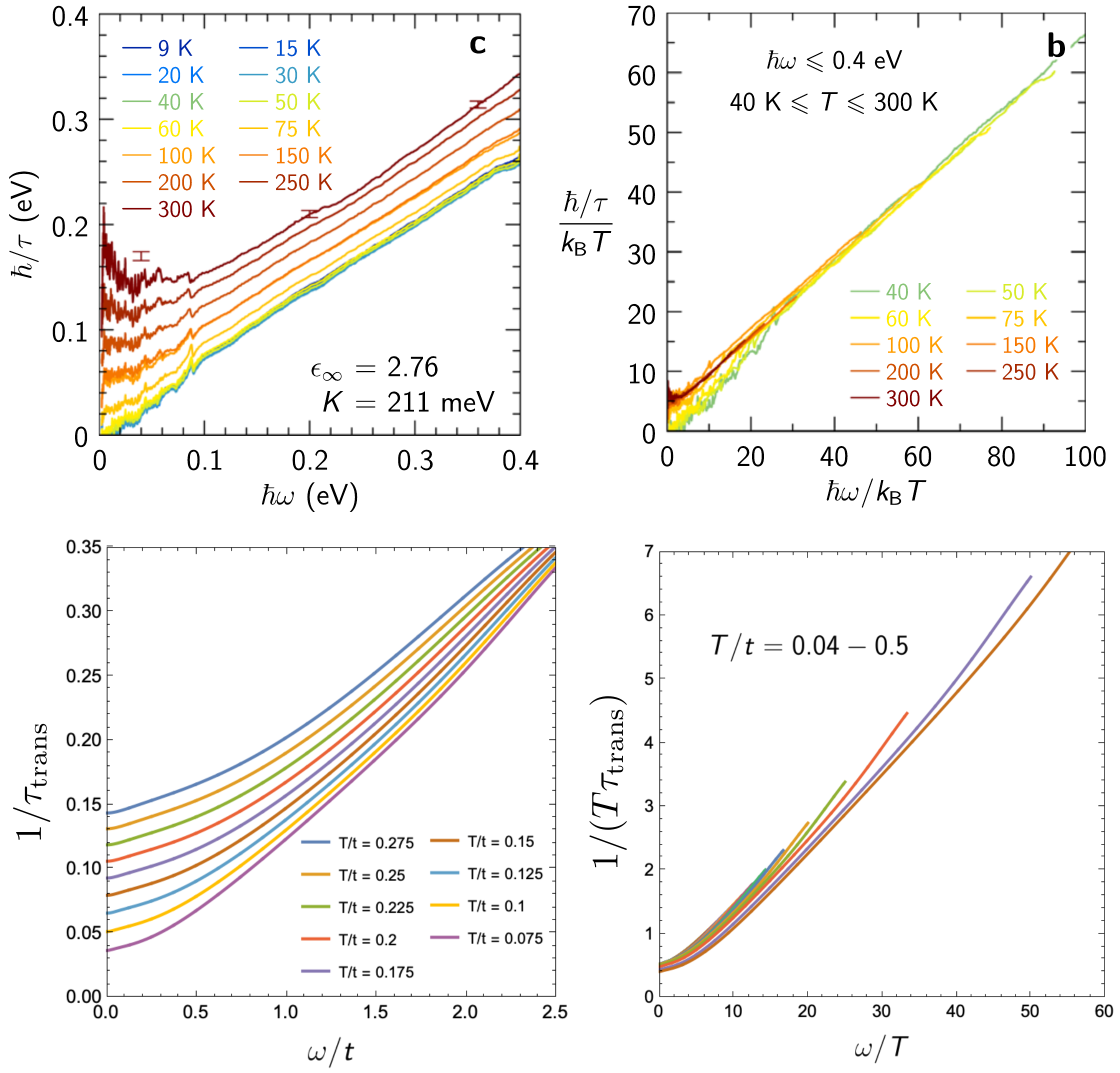}
\end{center}
\caption{The top panels display the measurements of transport relaxation time, $\tau_{\rm trans}$, obtained from the optical conductivity in Michon {\it et al.\/}~\cite{Michon22}.
The bottom panels show computation of the same quantity in the 2D-YSYK model by Li {\it et al.\/}~\cite{Li:2024kxr}.
}
\label{fig:tau}
\end{figure}

We note that a full numerical solution of Eq.~(\ref{eq:saddle_pt_eqs}) at $g=0$ has been presented by Li {\it et al.\/} \cite{Li:2024kxr}, including results for the conductivity and the onset of superconductivity.
Their results for the transport scattering time $\tau_{\rm trans}$ are shown in Fig.~\ref{fig:tau}, along with the corresponding optical conductivity observations of Michon {\it et al\/} \cite{Michon22}. Both obey the Planckian scaling form in Eq.~(\ref{p2}) reasonably well. The scaling is better in the experimental observations, and we suspect this is the due to the logarithmic violations in the theory, such as those in Eq.~(\ref{g5}). It would be interesting to extend the study of optical conductivity to the quantum Monte Carlo approach \cite{PLA24} in future work.

The key role of spatial randomness in the Yukawa coupling in this theory implies a prediction: correlated electron systems will not exhibit low $T$ strange metal behavior in sufficiently clean samples. Evidence in support of this prediction has appeared in recent experiments on graphene: while twisted bilayer graphene has a strange metal phase \cite{CaoStrange}, the much cleaner system of rhombohedral trilayer graphene does not \cite{YoungNotStrange}.

Finally, note that a recent computation \cite{PatelNoise} of shot noise in the {\color{darkgreen} $g'$-$v$} model yields results in agreement with observations \cite{NatelsonNoise}. The {\color{darkgreen} $g'$-$v$} model has also been used to study non-linear optical response \cite{Kryhin24}, and there are interesting connections to recent observations \cite{Armitage25}

\section{Boson localization and Griffiths phase}
\label{sec:griffiths}

This section turns to the behavior of the optimally and overdoped cuprates at low temperatures. When the superconductivity is suppressed by a strong magnetic field, the transport properties show an extended regime of non-Fermi liquid behavior, commonly associated with the `foot' in the phase diagram of Ref.~\cite{Hussey_foot}, and also observed in Refs.~\cite{Greene_rev,DessauPLL,Hayden25}.

We address this behavior by examining non-self-averaging features of the model of Section~\ref{sec:strange}. At lower $T$, we expect a crossover from the fractionalized degrees of freedom to confinement: so we will replace the Higgs boson $\Phi$ by a spin density wave (SDW) order parameter \cite{SSZaanen}, and consider the disordered version of the transition at $p_{\rm sdw}$ in Fig.~\ref{fig:figcross}.

We write the spin density wave order as ($a = x,y,z$)
\begin{equation}
S_a ({\bm r}) = \sum_\ell \phi_{\ell a} e^{i {\bm Q}_\ell \cdot {\bm r}}
\label{Sphi}
\end{equation}
where $\ell = 1 \dots 4$ labels the 4 ordering wavevectors ${\bm Q}_\ell$ at $(\pi (1 \pm \delta), \pi)$ and $(\pi, \pi(1 \pm \delta))$. We are interested in fluctuations of the SDW order parameters $\phi_{\ell a}$ coupled to electrons $c_{{\bm k} \sigma}$ with dispersion $\varepsilon ({\bm k})$ which has a Fermi surface. 
We describe this with a 2D-YSYK model \cite{Patel1,Patel2} with imaginary time ($\tau$) Lagrangian (replacing $\mathcal{L}_\Phi$ in Eq.~(\ref{e22}))
\begin{eqnarray}
\mathcal{L}_\phi & = \sum_{{\bm k}} c_{{\bm k}\alpha}^\dagger \left( \frac{\partial}{\partial \tau} + \varepsilon ({\bm k})\right) c_{{\bm k}\alpha}  +
\int d^2 {\bm r} \, \Bigl\{
s \, [\phi({\bm r})]^2  \nonumber \\
& +  \sum_{\ell} [g + {\color{darkgreen} g' ({\bm r})}] c^{\dagger}_\alpha ({\bm r})  \sigma^a_{\alpha \beta} c_{\beta} ({\bm r}) \, \phi_{\ell a} ({\bm r}) e^{i {\bm Q}_\ell \cdot {\bm r}} + K \, [ \nabla_{\bm r} \phi ({\bm r})]^2 +u  \,[\phi({\bm r})]^4 \nonumber \\
& ~~~~~~~~~+ {\color{darkgreen} v({\bm r})} c_\alpha^\dagger ({\bm r}) c_\alpha  ({\bm r}) \Bigr\} \,. \label{e11}
\end{eqnarray}
The self-averaging behavior of this theory is essentially the same as that of  $\mathcal{L}_\Phi$ in Eq.~(\ref{e22}), leading to equations closely related to those in Eq.~(\ref{eq:saddle_pt_eqs}). We will instead examine the low $T$ physics here, where there is a breakdown of self-averaging, and the local environment near each impurity can behave differently.

The dominant effect associated with breakdown of self-averaging in disordered metals is the localization of the electronic quasiparticle eigenstates. This will also be present in the theory in Eq.~(\ref{e11}), associated with the random potential {\color{darkgreen} $v({\bm r})$}. However, in $d=2$, such electron localization effects are only marginally relevant at weak disorder \cite{LeeRama}.
In contrast, the effects due to the random interaction {\color{darkgreen} $g' ({\bm r})$} are significantly stronger \cite{CastroNeto1,Schmalian02,Schmalian05}. The latter effects are associated with local shifts in the deviation from criticality, and their relevance is determined by the Harris criterion \cite{CCFS,Harris}: the exponent of critical point shifts is $2/\nu - d$. The disorder free theory has $\nu=1/2$, implying a strong violation of the Harris criterion, and strong relevance of the disorder associated with $g' ({\bm r})$---we will focus on this disorder in the remainder of this section. We note that a focus on bosonic disorder and localization is also a feature of works postulating two-level systems \cite{Berg24a,Berg24b,Bashan25}.

\subsection{Strong disorder renormalization group}
\label{sec:dasguptama}

Given the strong relevance of weak {\color{darkgreen} $g'$}, it is natural to examine the limit of strong disorder. This was examined in Refs.~\cite{TV07,TV09,TV13}. They proceeded by integrating out the fermions, assuming the fermions remain extended over the relevant energy scales. This leads to the following effective action for the boson $\phi$:
\begin{eqnarray}
\mathcal{S}  &= & \mathcal{S}_{K\phi} +  \mathcal{S}_{\phi d} \nonumber \\
    \mathcal{S}_{K\phi} &=& \int d \tau \Biggl[ -\sum_{ \vi < \vj } {\color{darkgreen} K_{\vi\vj}}\phi_{\vi a} \phi_{\vj a} +  \sum_{\vj} \biggl\{ \frac{{\color{darkgreen} s_{\vj}}}{2} \phi_{\vj a}^2   + \frac{u}{4M} \left( \phi_{\vj a}^2 \right)^2 \biggr\}\Biggr] \nonumber \\
    \mathcal{S}_{\phi d} &=& \frac{T}{2} \sum_{\Omega} \sum_\vj \left(\gamma |\Omega| + \Omega^2/c^2 \right)|\phi_{\vj a} (i\Omega)|^2 \,, \label{e10}
\end{eqnarray}
We have combined the indices $\ell,a$ in Eq.~(\ref{Sphi}) into a single index $a = 1 \ldots M$, with $M=12$ the case of interest. The Landau damping from the fermions is represented by $\mathcal{S}_{\phi d}$, and the disorder in {\color{darkgreen} $g' ({\bm r})$} has led to the disordered couplings {\color{darkgreen} $K_{\vi\vj}$} and {\color{darkgreen} $s_\vj$} in $\mathcal{S}_{K\phi}$. The Dasgupta-Ma \cite{DasguptaMa} strong disorder renormalization group of  proceeds by identifying the sites with the largest values of {\color{darkgreen} $K_{\vi\vj}$} or {\color{darkgreen} $s_\vi$}, and integrates out the fluctuations of corresponding $\phi_\vi$---the details of the computation are presented in Ref.~\cite{TV09}. If the largest coupling is {\color{darkgreen} $s_2$}, then we can integrate out the fluctuations of $\phi_2$, and the remaining couplings renormalize as 
\begin{equation}
{\color{darkgreen}
\widetilde{K}_{\vi\vj}} = {\color{darkgreen} K_{\vi\vj}} + {\color{darkgreen} \frac{K_{\vi 2} K_{2 \vj}}{s_2}} \quad, \quad \vi,\vj \neq 2
\,. \label{IRG1}
\end{equation}
If the largest coupling is {\color{darkgreen} $K_{23}$} then we replace $\phi_2$ and $\phi_3$ with a single boson $\widetilde{\phi}_2$ with renormalized couplings
\begin{equation}
{\color{darkgreen}
\widetilde{s}_{2}} = 2 {\color{darkgreen} \frac{s_{2} s_{3}}{K_{23}}} \quad, \quad 
{\color{darkgreen} \widetilde{K}_{2\vj}} = {\color{darkgreen} K_{2\vj}} +  {\color{darkgreen} K_{3\vj}}\quad, \quad \vj \neq 2,3
\,.
\label{IRG2}
\end{equation}

Remarkably, these are the same renormalization group equations as those for the quantum Ising model with random exchange and transverse field \cite{TV07,TV09,TV13}
\begin{equation}
H_{\rm RIM} = - \sum_{\vi < \vj} {\color{darkgreen} K_{\vi \vj}} Z_\vi Z_\vj - \sum_{\vi} {\color{darkgreen} s_\vi} X_\vi \,, \label{HRIM}
\end{equation}
where $X_\vi$, $Z_\vi$ are Pauli operators on qubits on sites $\vi$. The solutions of renormalization group equations of this model in two spatial dimensions have been studied numerically in Ref.~\cite{OM00}---they found a critical point controlled by an infinite randomness fixed point, along with Griffiths phase with continuously varying exponents on the disordered side.

This mapping between theories in Eqs.~(\ref{e10}) and (\ref{HRIM}) does appear rather surprising. The former has a multicomponent order parameter with a continuous symmetry which is Landau-damped, whereas the latter has a discrete symmetry and no Landau-damping. But the connection between them can be seen by examining the rare-region arguments in Refs.~\cite{Thill95,RSY95}. Consider a rare region of size $L$ in which is deep within the ordered phase. This will occur with a probability $\sim \exp (- c_1 L^d)$, and have an effective moment of size $L^d$. Considering the temporal fluctuations of this block spin for the Ising case as a one-dimensional classical Ising model, we obtain a correlation time of order $\xi_\tau \sim \exp (c_2 L^d) $. Correspondingly, for the continuous symmetry case, the Landau-damping implies a $1/|\tau|^2$ temporal interaction for the block spin, and this also has a correlation time of order 
$\xi_\tau \sim \exp (\bar{c}_2 L^d) $ \cite{Kosterlitz76}. Adding the contribution of these rare regions by the saddle-point method, we obtain a power-law Griffiths singularity in the local density of states \cite{Thill95,RSY95} in both cases.

\subsection{Large $M$ theory}
\label{sec:largeM}

While the mapping to the quantum Ising model yields considerable insight on the strong disorder limit, our interest is also in systems with weak disorder, and of the crossover to the self-averaging physics studied in Section~\ref{sec:strange}. To this end, we follow the large $M$ analysis of Refs.~\cite{Adrian08,PPS24}. For convenience, we retain disorder only in the random `mass' term, as that will generate disorder in other couplings; so we replace $\mathcal{S}_{K \phi}$ in Eq.~(\ref{e10}) by
\begin{equation}
 \mathcal{S}_\phi = \int d \tau \Biggl[ \frac{K}{2}\sum_{\langle \vi \vj \rangle} \left( \phi_{\vi a} -\phi_{\vj a} \right)^2 +  \sum_{\vj} \biggl\{ \frac{s+ {\color{darkgreen} \delta s_{\vj}}}{2} \phi_{\vj a}^2   + \frac{u}{4M} \left( \phi_{\vj a}^2 \right)^2 \biggr\}\Biggr] 
\,. \label{e12a}
\end{equation}
Then the large $M$ saddle point equations are
\begin{eqnarray}
    \widetilde{\mathcal{S}}_\phi =& \int d \tau \Bigl[ \frac{K}{2}\sum_{\langle \vi \vj \rangle} \left( \phi_{\vi a} -\phi_{\vj a} \right)^2 +  \sum_{j} \frac{ {\color{darkgreen} \widetilde{s}_\vj}}{2} \phi_{\vj a}^2   \Bigr] \nonumber \\
    & {\color{darkgreen} \widetilde{s}_\vj} = s + {\color{darkgreen} \delta s_\vj} + \frac{u}{M}
    \sum_a \left\langle \phi_{\vj a}^2 \right\rangle_{\widetilde{\mathcal{S}}_\phi + \mathcal{S}_{\phi d}} \nonumber \\
   &~~~ = s + {\color{darkgreen} \delta s_\vj} + u T \sum_\Omega \sum_{\alpha} \frac{\psi_{\alpha \vi} \psi_{\alpha \vj}}{\gamma |\Omega| + \Omega^2/c^2 + e_\alpha}\,, \label{e12}
\end{eqnarray}
where $e_\alpha$ and $\psi_{\alpha \vj}$ are eigenvalues and eigenfunctions of the $\phi$ quadratic form in $\widetilde{\mathcal{S}}_\phi$, labeled by the index $\alpha = 1 \ldots L^2$ for a $L \times L$ sample. For each disorder realization {\color{darkgreen} $\delta s_j$}, the values of {$\color{darkgreen} \widetilde{s}_j$} are determined by numerically solving (\ref{e12}), and this also yields results for the eigenvalues $e_\alpha$ and the eigenvectors $\psi_{\alpha j}$. 

\begin{figure}
\begin{center}
\includegraphics[width=5.5in]{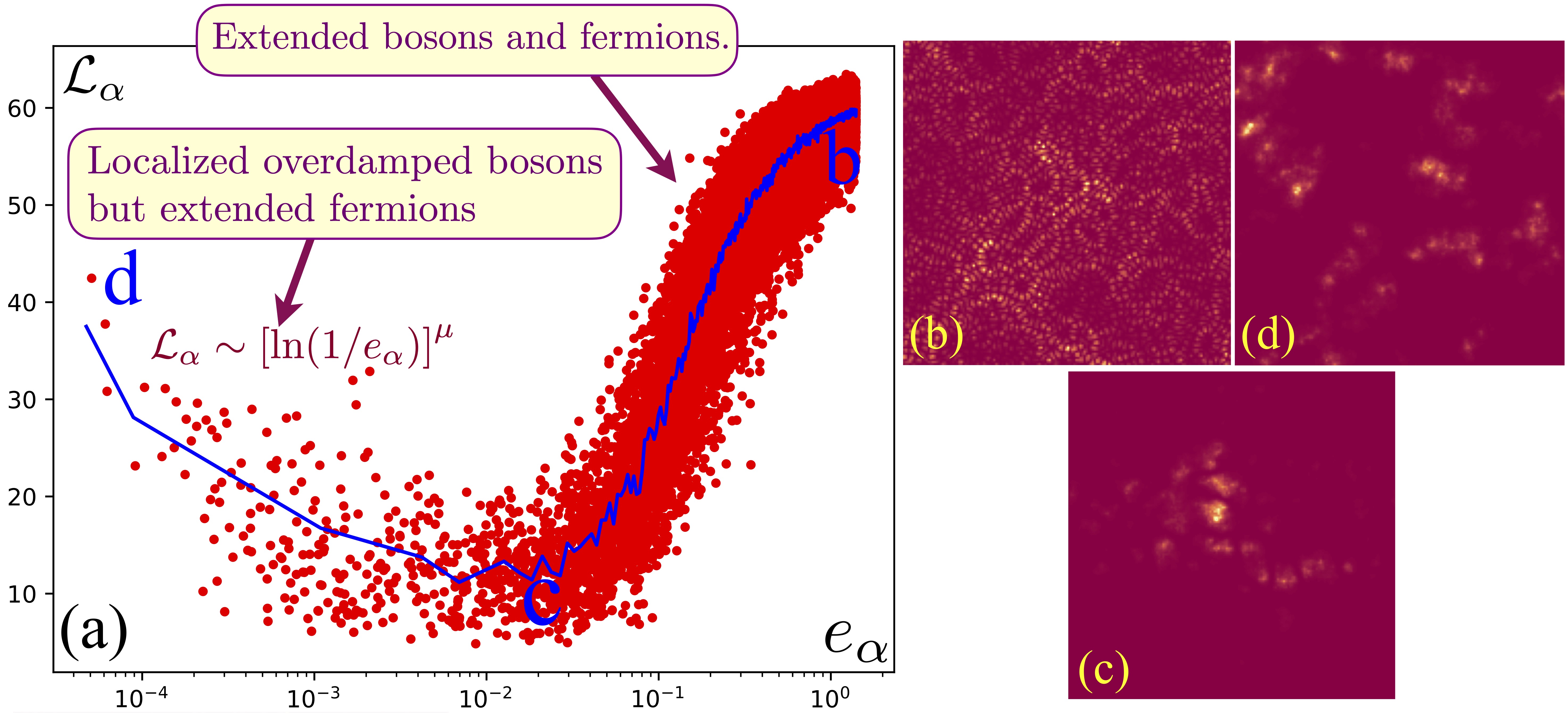}
\end{center}
\caption{From Ref.~\cite{SSZaanen} and adapted from Patel {\it et al.\/}~\cite{PPS24}. (a) Localization length $\mathcal{L}_\alpha$ of overdamped bosonic eigenmodes of $\widetilde{\mathcal{S}}_\phi$ in Eq.~(\ref{e12}) as a function of their energy $e_\alpha$. (b,c,d) Pictures of the corresponding bosonic eignfunctions.  The `foot' is described by the localized bosons.
The universal, self-averaging, 2D-YSYK theory of the `fan' by Patel {\it et al.\/}~\cite{Patel2} and Li {\it et al.\/}~\cite{Li:2024kxr} in Section~\ref{sec:strange}, applies to the regime of extended bosons.}
\label{fig:eigenmodes}
\end{figure}
Numerical results obtained from the solution of Eq.~(\ref{e12}) at the quantum critical value $s=s_c$ appear in Fig.~\ref{fig:eigenmodes}. We show the localization length $\mathcal{L}_\alpha$ of the eigenmodes at energy $e_\alpha$. At large $e_\alpha$, the value of $\mathcal{L}_\alpha$ is of order the system size, implying extended bosonic eigenmodes, and the applicability of the self-averaging 2D-YSYK theory of Section~\ref{sec:strange}. As we lower $e_\alpha$, there is a minimum in $e_\alpha$, associated with localization of the bosonic eigenmodes. Below the minimum, we find a logarithmic increase of $\mathcal{L}_\alpha$ with decreasing $e_\alpha$: this logarithmic increase is precisely that expected from the infinite-randomness fixed point of the Ising model in a transverse field \cite{MO00} in Eq.~(\ref{HRIM}).

\subsection{Quantum Monte Carlo}
\label{sec:qmc}
\begin{figure}
\begin{center}
\includegraphics[width=3.5in]{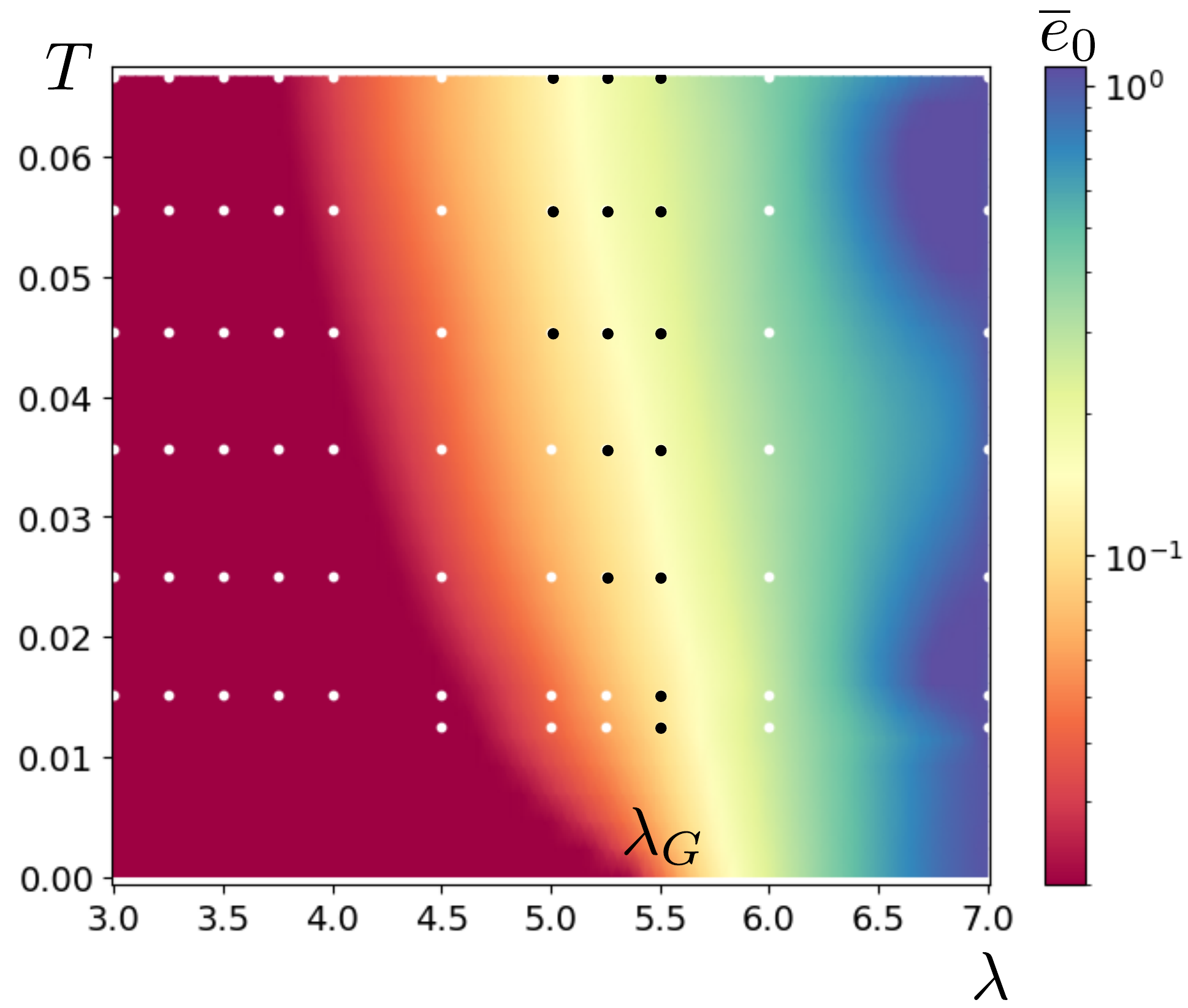}
\end{center}
\caption{From the quantum Monte Carlo study of the coupled boson-fermion model of Ref.~\cite{PLA24}. The disordered averaged energy of the lowest boson eigenmode $\overline{e}_0$. The coupling $\lambda$ is the analog of the tuning parameter $s$ in Eq.~(\ref{e11}). Note the extended Griffiths phase at low $T$.}
\label{fig:boson_gap}
\end{figure}
Ref.~\cite{PLA24} examined the complete original model of interacting bosons and fermions in Eq.~(\ref{e11}) by quantum Monte Carlo simulations. They employed a fermion dispersion to allow a sign-problem-free study \cite{metlitski4}, and set the average Yukawa coupling to $g=0$. The determined the bosonic eigenmodes by digaonalizing the zero frequency susceptibility matrix for each realization of the disorder
\begin{equation}
\chi_{\vi \vj} (i\Omega = 0) = \int_0^{1/T} d \tau
\left\langle \phi_{\vi a} (\tau) \phi_\vj (0) \right\rangle \,.
\end{equation}
Their results for the boson localization length are very similar to those of the large $M$ computation in Fig.~\ref{fig:eigenmodes}. We show instead the Monte Carlo results for the averaged energy of the lowest boson eigenmode $\overline{e}_0$ in Fig.~\ref{fig:boson_gap}. The results show a low temperature `foot' of very small gaps, which we interpret as the analog of the Griffiths phase of the quantum Ising model \cite{MO00}.

\subsection{Neutron scattering}
\label{sec:neutrons}

\begin{figure}
\begin{center}
\includegraphics[width=5.5in]{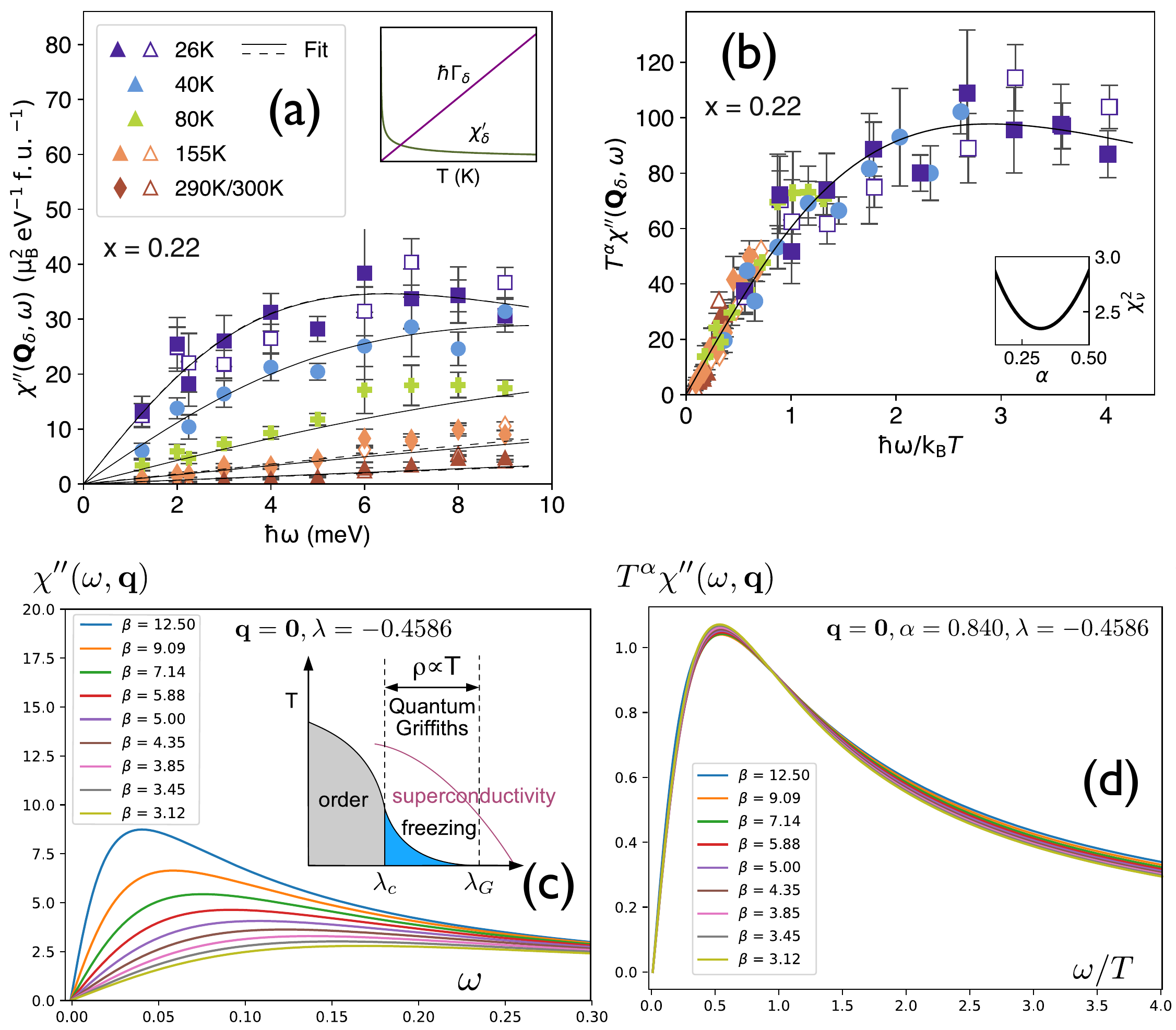}
\end{center}
\caption{From Ref.~\cite{Hayden25}. (a) Neutron scattering observations on La$_{2-x}$Sr$_x$CuO$_4$ at $x=0.22$ at the ordering wavevectors ${\bm Q}_\ell$. (b) Scaling fit of observations to Eq.~(\ref{scalechi}). (c) Dynamic susceptibility computed from the large-$M$ equations in Eq.~(\ref{e12}). (d) Scaling fits of large-$M$ results to Eq.~(\ref{scalechi}).  }
\label{fig:hayden}
\end{figure}
Neutron scattering experiments on La$_{2-x}$Sr$_x$CuO$_4$ have observed critical spin fluctuations at a wide range of dopings \cite{Aeppli98,Hayden25}. Fig.~\ref{fig:hayden}a,b show good `Planckian' scaling fits of the dynamic scaling of the suspectibility at the ordering wavevectors ${\bm Q}_{\ell}$ at doping $x=0.22$
\begin{equation}
\chi^{\prime\prime} (\omega) \sim T^{-\alpha} \Phi_\chi \left( \frac{\hbar \omega}{k_B T} \right) \,. \label{scalechi}
\end{equation}
These results have been interpreted \cite{Hayden25} in terms of the Griffiths phase of the spin density wave ordering transition. Fig.~\ref{fig:hayden}c,d show results obtained by the exact analytic continuation of the numerical solution of the large $M$ equations in Eq.~(\ref{e12}), which are also found to obey Planckian scaling to a reasonable accuracy. The data in Fig.~\ref{fig:hayden}d are for the quantum critical point, where the value of the exponent $\alpha$ is larger than the experimental value. However, the large-$M$ value of $\alpha$ decreases when we move into the Griffiths phase, as is appropriate for the doping of $x=0.22$. The full Monte Carlo simulation of Eq.~(\ref{e11}) \cite{PLA24} yields a value of $\alpha$ close to the experimental observation \cite{Hayden25}.

\subsection{Photoemission}
\label{sec:dessau}

\begin{figure}
\begin{center}
\includegraphics[width=5.5in]{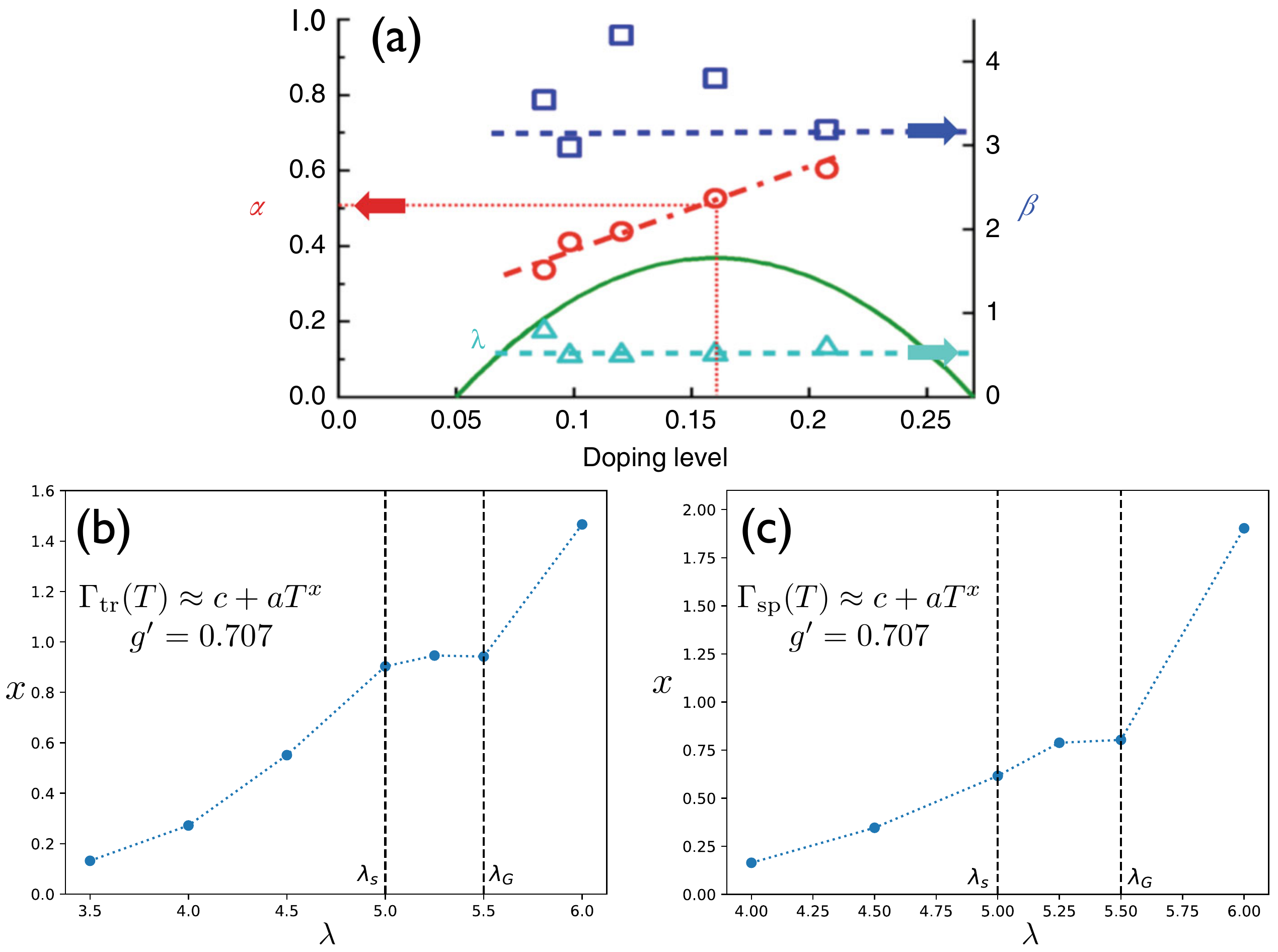}
\end{center}
\caption{(a) From Ref.~\cite{DessauPLL}. Fit parameters of the self energy of the `power-law liquid' in Eq.~(\ref{eq:pll}) of Bi$_2$Sr$_2$CaCu$_2$O$_{8+\delta}$. (b) Exponent $x$ of the transport relaxtion rate $\Gamma_{\rm tr} (T)$ computed from the results of Ref.~\cite{PLA24} as a function of their tuning parameter $\lambda$. A plateau appears in the exponent in the Griffiths phase $\lambda_s < \lambda < \lambda_G$. (c) Exponent $x$ of the single particle relaxation rate, also computed from Ref.~\cite{PLA24}.}
\label{fig:dessau}
\end{figure}
Finally, we turn to the remarkable photoemission data of Reber {\it et al.\/} \cite{DessauPLL}. From measurements of the electron spectral function along the Brillouin zone diagonals across a wide range of temperatures, frequencies, and dopings in Bi$_2$Sr$_2$CaCu$_2$O$_{8+\delta}$, they argue for a `power-law liquid' across a wide range of dopings. They
fit the imaginary part of the electron self energy to the form (corresponding to the scaling form in Eq.~(\ref{p3}))
\begin{equation}
\Sigma'' (\omega) = \Gamma_0 + \lambda \frac{[(\hbar \omega)^2 + (\beta k_B T)^2]^{\alpha}}{(\hbar \omega_N)^{2 \alpha-1}}\,,
\label{eq:pll}
\end{equation}
and the values of their fit parameters $\alpha, \lambda, \beta$ are shown in Fig.~\ref{fig:dessau}a. 

The electron self energy and optical conductivity was also computed in the Monte Carlo study of Ref.~\cite{PLA24} across the Griffiths phase. The value of their exponent $x$ of the $T$ dependence of the relaxation rates is shown in Fig.~\ref{fig:dessau}b. This is in reasonable agreement with the expected relation $x = 2 \alpha$ with the experimental data. But note that there is a plateau in the value of $x$ in the theory, that is (as yet) not visible in the experimental data.

\ack
This research was supported by NSF Grant DMR-2245246 and by the Simons Collaboration on Ultra-Quantum Matter which is a grant from the Simons Foundation (651440, S. S.). 
P.M.B. acknowledges support by the German National Academy of Sciences Leopoldina through Grant No.~LPDS 2023-06 and the Gordon and Betty Moore Foundation’s EPiQS Initiative Grant GBMF8683.
The Flatiron Institute is a division of the Simons Foundation. M.C. acknowledges funding from Amazon Web Services, AWS Quantum Program, and the National Science Foundation (PHY-2317110). The Institute for Quantum Information and Matter is an NSF Physics Frontiers Center.

\appendix
\section*{Appendix A. Schrieffer-Wolff transformation from the ALM to the Hubbard model}
\addcontentsline{toc}{section}{Appendix A. Schrieffer-Wolff transformation from the ALM to the Hubbard model}
\phantomsection
\label{app:sw}

Here we present an alternative derivation of the connection between the Hubbard model and the ALM, which is also illustrated in Fig.~\ref{fig:ancilla}. 
We begin with a Hubbard model and augment it with ancillas which are decoupled from the Hubbard model.
It is important that the new degrees of freedom have a trivial ground state with an energy gap, so that its extra excited states can be eliminated by a canonical transformation. For the ancillas, we choose a pair of qubits associated with every site of the square lattice, which we identify as $S=1/2$ spins ${\bm S}_{1\vi}$ and ${\bm S}_{2\vi}$. There is an antiferromagnetic exchange interaction $J_{\perp}$  between the two spins at each $\vi$, and so the ancillas form a bilayer square lattice antiferromagnet. We take $J_\perp$ to be large enough so that the ground state of the bilayer is smoothly connected to the `trivial' rung singlet state. The decoupled ancilla Hamiltonian is therefore
\begin{equation}
\mathcal{H}_{\rm ancilla-decoupled} =  \mathcal{H}_{\rm Hubbard} + J_\perp \sum_{\vi} {\bm S}_{1\vi} \cdot {\bm S}_{2\vi}  + \ldots \,,
\label{eq:Hancillad}
\end{equation}
where the ellipses represent exchange interactions within the ancilla layers. As indicated in Fig.~\ref{fig:ancilla}, we proceed in a direction opposite to that of the paramagnon derivation. Starting from $\mathcal{H}_{\rm ALM}$, we perform a Schrieffer-Wolff transformation to map to $\mathcal{H}_{\rm ancilla-decoupled}$ \cite{Nikolaenko:2021vlw}. 
    This Schrieffer-Wolff transformation eliminates all the excited states of the ancilla spins in powers of $1/J_\perp$, and returns to the single band Hubbard model for the $c_{\vi \alpha}$, with the value of $U$ shown below in Eq.~(\ref{Uval}) and Fig.~\ref{fig:ancilla} \cite{Nikolaenko:2021vlw}.
    At leading order in $1/J_\perp$, we can demonstrate this mapping by considering only a single value of $\vi$. Then at each $\vi$, the ancilla Hamiltonian is
    \begin{equation}
    H_1 = J_\perp {\bm S}_1 \cdot {\bm S}_2 - \mu c_\alpha^\dagger c_\alpha + J_K {\bm S}_1 \cdot c_\alpha^\dagger \frac{{\bm \sigma}_{\alpha\beta}}{2} c_{\beta}
    \end{equation}
    At large $J_\perp$, the ${\bm S}_{1,2}$ form a spin singlet, and there are 4 low-lying states of the $c_\alpha$. Upon considering perturbation theory in $J_K$, the energies $E_n$ of the state with $n$ fermions are
    \begin{eqnarray}
    E_0 &=& -\frac{3J_\perp}{4},~~E_2 = -\frac{3J_\perp}{4} -2\mu\,, \nonumber \\ 
    E_1 &=& -\frac{3J_\perp}{4} -\mu - \frac{3 J_K^2}{16 J_\perp}  + \mathcal{O}(1/J_\perp^2) \,.  
    \end{eqnarray}
    These states map onto those of a Hubbard model 
    \begin{equation}
    H_2 = - \mu' c_\alpha^\dagger c_\alpha + U (c_\uparrow^\dagger c_\uparrow )(c_\downarrow^\dagger c_\downarrow )
    \end{equation}
    with parameters
    \begin{equation}
    U = \frac{3 J_K^2}{8 J_\perp}, \quad
    \mu' = \mu +  \frac{3 J_K^2}{16 J_\perp}\,. \label{Uval}
    \end{equation} 
Extending this mapping to all sites, we obtain the canonical (Schrieffer-Wolff) transformation connecting $\mathcal{H}_{\rm ALM}$ in Eq.~(\ref{eq:Hancilla}) to $\mathcal{H}_{\rm ancilla-decoupled}$ in Eq.~(\ref{eq:Hancillad}).

\section*{Appendix B. Complete gauge theory of the ALM}
\addcontentsline{toc}{section}{Appendix B. Complete gauge theory of the Ancilla Layer Model}
\phantomsection
\label{app:ancilla}

\setcounter{table}{2}
\setcounter{equation}{0}
\renewcommand{\theequation}{A.\arabic{equation}}

The ancilla representation of the single-band Hubbard model has
a ${\rm SU}(2) \times \widetilde{\rm SU}(2) \times {\rm U}(1)$ gauge structure \cite{YaHui-ancilla1,YaHui-ancilla2,QPMbook}, as we indicated below Eq.~(\ref{e22}). 
These 3 gauge fields are connected to three distinct constraints that need to be imposed on the three-layer ALM so that it reduces to the single-layer Hubbard model.
\begin{itemize}
\item
The SU(2) gauge field $U_{\vi\vj}$ was that introduced in Section~\ref{sec:spinliquids}, and it implements the local 
constraint Eq.~(\ref{constraintf}) in the bottom layer of ${\bm S}_2$ spins defined by Eq.~(\ref{eq:S2f}).
\item
The U(1) gauge field $e^{i a_{\vi\vj}}$ is that associated with Eq.~(\ref{gaugeu1}) and the constraint Eq.~(\ref{constraintf1}). 
\item 
While the two gauge fields above have quite familiar origins from on-site constraints, the third  
$\widetilde{\rm SU}(2)$ gauge field has a different origin. It is associated with the requirement that the ${\bm S}_{1\vi}$ and ${\bm S}_{2\vi}$ have singlet correlations, and their triplet state should only be a virtual state. This requirement can be implemented 
by transforming them to a common rotating reference frame in spin space, which introduces a $\widetilde{\rm SU}(2)$ gauge field~\cite{sdw09}; performing the integral over the $\widetilde{\rm SU}(2)$ gauge field averages over rotating reference frames, and hence projects ${\bm S}_{1\vi}+{\bm S}_{2\vi}$ to the singlet. This $\widetilde{\rm SU}(2)$ gauge field is closely related to that in other approaches to the pseudogap in the single-band Hubbard model \cite{DCSS15b,DCSS15,CSS17,WuScheurer1,Scheurer:2017jcp,Sachdev:2018ddg,SSST19,WuScheurer2,Bonetti22,Bonetti23}. Sections~\ref{sec:spinliquids}-\ref{sec:dwave} considered a limiting case of the ancilla layer theory, appropriate for the present FL* theory of the pseudogap, where the $\widetilde{\rm SU}(2)$ gauge field is fully higgsed, and so can be neglected. But the $\widetilde{\rm SU}(2)$ gauge field is needed for a proper description of the FL state, so that it can confine ${\bm S}_{1\vi}$ and ${\bm S}_{2\vi}$ to rung-singlets in the FL state of Fig.~\ref{fig:oneband} at large doping.
\end{itemize}

Next, we list the matter content of the ALM.
\begin{itemize}
\item 
The fermionic electrons $c_{\vi\alpha}$ in the top layer. These are also represented by $\mathcal{C}_\vi$ in Eq.~(\ref{defB}).
\item 
The fermionic spinons $f_{1\vi \alpha}$ representing the ${\bf S}_{1\vi}$ spins of the middle layer via Eq.~(\ref{eq:S1f}). These are also represented by $\mathcal{F}_{1\vi}$ defined in Eq.~(\ref{F1matrix}).
\item 
The fermionic spinons $f_{\vi \alpha}$ representing the ${\bf S}_{2 \vi}$ spins of the bottom layer via Eq.~(\ref{eq:S2f}). These are also represented by $\mathcal{F}_{\vi}$ defined in Eq.~(\ref{Fmatrix}).
\item 
The Higgs boson $\Phi_{\vi \alpha\beta}$ obtained by decoupling the exchange interaction $J_K$ between the top two layers \cite{YaHui-ancilla1}.
In Sections~\ref{sec:FL*}-\ref{sec:dwave} we assumed this boson was condensed with $\Phi_{\vi \alpha\beta} = \Phi_\vi \delta_{\alpha\beta}$. In Section~\ref{sec:strange}, we considered fluctuations of $\Phi$, but only of its diagonal component for simplicity. This boson is the analog of the `slave boson' which appears in theories of the heavy Fermi liquid state of the Kondo lattice, where it has only a single component.
\item The boson $B_\vi$ obtained by decoupling the exchange interaction $J_\perp$ between the lower two layers of ${\bm S}_{1\vi}$ and ${\bm S}_{2\vi}$ spins \cite{Christos:2023oru}. This is also represented by $\mathcal{B}_\vi$, as defined in Eq.~(\ref{defB}).
\end{itemize}

We are now ready to specify gauge and global symmetry transformations of all the fields, which are summarized in Table~\ref{tab3}.
\begin{table}
    \centering
    \begin{tabular}{|c|c||c|c|c||c|c|}
\hline
\multirow{2}{*}{Field} & \multirow{2}{*}{Layer} & \multicolumn{3}{c||}{Gauge} & \multicolumn{2}{c|}{Global} \\
\cline{3-7}
 & & SU(2) & U(1) &   \rule{0pt}{2.8ex} $\widetilde{\rm SU}(2)$ & SU(2) & U(1) \\ \hline
 $c$ or $\mathcal{C}$ & 1 & ${\bm 1}$ & 0 & ${\bm 1}$ & ${\bm 2}_R$ & -1 \\
 \hline
 $f_1$ or $\mathcal{F}_1$ & 2 & ${\bm 1}$ & -1 & ${\bm 2}_R$ &  ${\bm 1}$ & 0 \\ \hline
 $f$ or $\mathcal{F}$ & 3 & ${\bm 2}_L$ & 0 & ${\bm 2}_R$ & ${\bm 1}$ & 0 \\ \hline
 \rule{0pt}{2.6ex} $\Phi$ or $\tilde{\Phi}$ & $1 \leftrightarrow 2$ & ${\bm 1}$ & 1 & $\bar{\bm 2}_R$ & ${\bm 2}_L$ & -1 \\ \hline
 $B$ or $\mathcal{B}$ & $2 \leftrightarrow 3$ & ${\bm 2}_L$ & 1 & ${\bm 1}$ & ${\bm 1}$ & 0 \\ \hline
\end{tabular}
    \caption{As in Table~\ref{tab2}, but for all phases of the ALM. These transformations reduce to those in Table~\ref{tab2} when $\Phi$ is condensed. The representations of the SU(2) are indicated by their dimension; the subscripts $L$/$R$ indicate whether the SU(2) acts by left/right multiplication in the matrix form of the field. The representations of the global U(1) is the electrical charge in units of $e$. For the fermions, the layer column indicates the layer number in Fig.~\ref{fig:ancilla}. For the bosons,  the layer column indicates the layers between which there is a Yukawa coupling to the fermions---see also Figs.~\ref{fig:PhiB} and \ref{fig:cupratepd2}. In Sections~\ref{sec:FL*}-\ref{sec:dwave} we assume $\Phi$ is condensed and fixed, while $B$ is dynamical. In Section~\ref{sec:strange}, we only include the dynamics of $\Phi$ while assuming $B=0$.}
    \label{tab3}
\end{table}

The action of the SU(2) gauge transformation $V_\vi$ remains the same as that in Eq.~(\ref{eq:gauge})
\begin{eqnarray}
U_{\vi\vj} & \rightarrow &  V_\vi \, U_{\vi\vj} \, 
V_\vj^{\dagger} \nonumber \\
\mathcal{C}_\vi & \rightarrow & \mathcal{C}_\vi \nonumber \\
\Phi_{\vi\alpha\beta} & \rightarrow & \Phi_{\vi\alpha\beta} \nonumber \\
\mathcal{F}_{1\vi} & \rightarrow & \mathcal{F}_{1\vi}   \nonumber \\
\mathcal{B}_\vi &\rightarrow& V_\vi \, \mathcal{B}_\vi \nonumber \\
\mathcal{F}_\vi &\rightarrow& V_\vi \, \mathcal{F}_\vi   \,. \label{eq:gaugeapp}
\end{eqnarray}
The action of the U(1) gauge transformation $\phi_\vi$ in Eq.~(\ref{gaugeu1}) also remains the same:
\begin{eqnarray}
a_{\vi\vj} & \rightarrow &  a_{\vi\vj} + \phi_\vi - \phi_\vj \nonumber \\
\mathcal{C}_\vi & \rightarrow & \mathcal{C}_\vi \nonumber \\
\Phi_{\vi\alpha\beta} & \rightarrow & e^{- i\phi_\vi} \, \Phi_{\vi\alpha\beta} \nonumber \\
f_{1\vi\alpha} & \rightarrow & e^{i \phi_\vi} \, f_{1\vi\alpha}   \nonumber \\
B_\vi &\rightarrow& e^{-i \phi_i} B_\vi \nonumber \\
\mathcal{F}_\vi &\rightarrow&  \mathcal{F}_\vi   \,. \label{gaugeu1app}
\end{eqnarray}
Finally, we describe the action of the $\widetilde{\rm SU}(2)$ gauge transformation which transforms ${\bm S}_{1,2}$ into a common rotating reference frame, and is generated by the SU(2) matrix $\widetilde{V}_i$ \cite{YaHui-ancilla1,YaHui-ancilla2,QPMbook}.
\begin{eqnarray}
\widetilde{U}_{\vi\vj} & \rightarrow &  \widetilde{V}_\vi  \, \widetilde{U}_{\vi\vj} \, 
 \widetilde{V}_\vj^\dagger  \nonumber \\
\mathcal{C}_\vi & \rightarrow & \mathcal{C}_\vi \nonumber \\
\Phi_{\vi\alpha\beta} & \rightarrow & \Phi_{\vi\alpha\gamma} \widetilde{V}^\dagger_{\gamma\beta} \nonumber \\
\mathcal{F}_{1\vi} & \rightarrow & \mathcal{F}_{1\vi} \, \sigma^z \widetilde{V}_\vi^{T}  \sigma^z \mbox{~or equivalently~} f_{1\vi \alpha} \rightarrow \widetilde{V}_{\vi\alpha\beta} f_{1 \vi \beta} \nonumber \\
\mathcal{B}_\vi &\rightarrow&  \mathcal{B}_\vi \nonumber \\
\mathcal{F}_\vi &\rightarrow&  \mathcal{F}_\vi \, \sigma^z \widetilde{V}_\vi^T \sigma^z \,. \label{eq:gaugeapp2}
\end{eqnarray}
Note that $\mathcal{F}_\vi$ transforms under both SU(2) and $\widetilde{\rm SU}(2)$ gauge transformations, under left multiplcation for the former, and under right multiplication for the latter. The associativity of matrix multiplication ensures that the two gauge transformations commute with each other.

Turning to the global symmetries, there is now a significant change from the action of global spin rotation $R$ in Eq.~(\ref{eq:spin}). The transformation to the rotating reference frame transfers the action of $R$ from the spinons $f_1$ and $f$ to the Higgs field $\Phi$:
\begin{eqnarray}
c_{\vi\alpha} & \rightarrow & R_{\alpha\beta} \,c_{\vi\beta} \mbox{~or equivalently~} \mathcal{C}_\vi \rightarrow \mathcal{C}_i \, \sigma^z R^T \sigma^z \nonumber \\
\Phi_{\vi\alpha\beta} & \rightarrow & R_{\alpha\gamma}\, \Phi_{\vi\gamma\beta}  \nonumber \\
\mathcal{F}_{1\vi} & \rightarrow & \mathcal{F}_{1\vi}  \nonumber \\
\mathcal{B}_\vi &\rightarrow&  \mathcal{B}_\vi \nonumber \\
\mathcal{F}_\vi &\rightarrow&  \mathcal{F}_\vi \,. \label{eq:spin2}
\end{eqnarray}
Similarly, the global electromagnetic charge transformation of Eq.~(\ref{eq:charge}) now acts on $\Phi$:
\begin{eqnarray}
c_{\vi\alpha} & \rightarrow & e^{i \theta} \,c_{\vi\alpha} \nonumber \\
\Phi_{\vi\alpha\beta} & \rightarrow & e^{i \theta} \Phi_{\vi\alpha\beta}  \nonumber \\
\mathcal{F}_{1\vi} & \rightarrow & \mathcal{F}_{1\vi}  \nonumber \\
\mathcal{B}_\vi &\rightarrow&  \mathcal{B}_\vi \nonumber \\
\mathcal{F}_\vi &\rightarrow&  \mathcal{F}_\vi \,. \label{eq:charge2}
\end{eqnarray}
When we condense $\langle \Phi_{\vi\alpha\beta} \rangle = \Phi \, \delta_{\alpha\beta}$, then the U(1) and $\widetilde{\rm SU}(2)$ gauge symmetries are higgsed and tied to global symmetries, and the global symmetries revert back to those in Eqs.~(\ref{eq:spin}) and (\ref{eq:charge}).

The $\widetilde{\rm SU}(2)$ transformation to the rotating reference form also changes the operator correspondences discussed in the main text. The $f_1$ fermions are no longer just electrons $c$, but related by
\begin{equation}
c_{\vi \alpha} \sim \Phi_{\vi \alpha \beta} f_{1\vi \beta}\,.
\label{eq:cPf}
\end{equation}
Similarly, the generalization of Eq.~(\ref{eq:CBF}) is
\begin{equation}
c_{\vi \alpha} \sim \Phi_{\vi \alpha \beta} \tilde{f}_{\vi \beta} \mbox{~where~}  \left( \begin{array}{cc} \tilde{f}_{\vi \uparrow} & - \tilde{f}_{\vi \downarrow} \\ \tilde{f}_{\vi \downarrow}^\dagger  & \tilde{f}_{\vi \uparrow}^\dagger \end{array} \right)
 \sim \mathcal{B}^\dagger \mathcal{F}_\vi \,.
\end{equation}
For the spin operator of the spin liquid, Eq.~(\ref{NambuST}) is replaced by
\begin{equation}
{S}_{\vi \ell} \sim  -\frac{1}{4}  \mbox{Tr} ( \tilde{\Phi}_\vi^\dagger \sigma^\ell \tilde{\Phi}_\vi \sigma^m ) \, \mbox{Tr} ( \mathcal{F}_{\vi}^{\phantom\dagger} \sigma^z {\sigma}^{m T} \sigma^z \mathcal{F}_{\vi}^{\dagger} ) \,, \label{NambuST2}
\end{equation}
where $\ell,m = x,y,z$, and $\tilde{\Phi}$ is the $2 \times 2$ matrix of the complex numbers $\Phi_{\alpha\beta}$. It can be checked that all three expressions in Eqs.~(\ref{eq:cPf}-\ref{NambuST2}) are invariant under all gauge transformations. In establishing Eq.~(\ref{NambuST2}), we use the following identity which lifts spinor, $V$, to vector, $O$, representations of SU(2)
\begin{equation}
V^\dagger V = 1, \quad V^{\dagger }\sigma^\ell V = O_{\ell m} \sigma^m \quad \Rightarrow \quad O^T O = 1\,.
\end{equation}

Now we can employ the above gauge and global symmetries to obtain the complete Hamiltonian of the ALM in terms of the $U_{\vi \vj}$, $\widetilde{U}_{\vi \vj}$, and $a_{\vi \vj}$ gauge fields, the bosonic Higgs fields $\Phi$ and $B$, and the fermionic matter fields $c$, $f_1$, and $f$. 
Alternatively, the Hamiltonian can also be derived by the decoupling of the interlayer exchange interactions \cite{YaHui-ancilla1,Christos:2023oru}, as discussed earlier in this appendix when we introduced the matter fields.
The Hamiltonian of Eqs.~(\ref{eq:WS}), (\ref{HKLmf}), (\ref{eq:fermionhop2}), (\ref{Yukawa}), and (\ref{Bfunctional}) is replaced by
\begin{eqnarray}
\mathcal{H}_{\rm ALM} &=& \mathcal{H}_{\rm KLg} + \mathcal{H}_{SLf} + \mathcal{H}_Y + \mathcal{E}_2 [B,U] + \mathcal{E}_4 [B,U] \nonumber \\
\mathcal{H}_{\rm KLg} &=& \sum_{\vi,\vj} \left[- t_{\vi\vj} c^\dagger_{\vi\alpha} c_{\vj\alpha} -   t_{1,\vi\vj} f^\dagger_{1\vi\alpha} e^{i a_{\vi\vj}} \widetilde{U}_{ij,\alpha\beta} f_{1\vj\beta} \right] \nonumber \\
&~&~~~~~~ -  \sum_{\vi}(  c^\dagger_{\vi\alpha} \Phi_{\vi\alpha\beta} f_{1\vi\beta}+  f^\dagger_{1\vi\alpha} \Phi^\ast_{\vi \beta \alpha} c_{\vi\beta}) \nonumber
 \\
 \mathcal{H}_{SLf} & = & \frac{iJ}{2} \sum_{\langle \vi \vj \rangle} e_{\vi \vj} \left[  \mbox{Tr} \left( \mathcal{F}_{\vi}^{\dagger}  U_{\vi \vj}^{\phantom\dagger}  \mathcal{F}_{\vj }^{\phantom\dagger} \sigma^z \widetilde{U}_{\vi\vj}^{T} \sigma^z \right) -  \mbox{Tr} \left( \mathcal{F}_{\vj }^{\dagger} U_{\vj \vi}^{\phantom\dagger} \mathcal{F}_{\vi }^{\phantom\dagger} \sigma^z \widetilde{U}_{\vj\vi}^{T} \sigma^z  \right) \right] \nonumber \\
  \mathcal{H}_Y  & =& \sum_\vi \left[  i \left( B_{1\vi}^{\vphantom\dagger} f_{\vi\alpha}^\dagger f_{1\vi \alpha}^{\vphantom\dagger} - B_{2 \vi}^{\vphantom\dagger} \varepsilon_{\alpha\beta}^{\vphantom\dagger} f_{\vi \alpha}^{\vphantom\dagger} f_{1\vi \beta}^{\vphantom\dagger} \right)
   + \mbox{H.c.}\right. \nonumber \\
   &~& ~~~~\left.   + i \bar{g} \left( B_{1\vi}^{\vphantom\dagger} f_{\vi\alpha}^\dagger  \Phi^\ast_{\vi \beta \alpha} c_{\vi\beta}^{\vphantom\dagger} - B_{2 \vi}^{\vphantom\dagger} \varepsilon_{\alpha\beta}^{\vphantom\dagger} f_{\vi \alpha}^{\vphantom\dagger} \Phi^\ast_{\vi \gamma \beta} c_{\vi\gamma}^{\vphantom\dagger} \right)
   + \mbox{H.c.}\right] \,. \label{eq:complete}
\end{eqnarray}
The terms in $\mathcal{E}_2 [B,U] + \mathcal{E}_4 [B,U] $ remain unchanged from those in Eq.~(\ref{Bfunctional}). We can also add Maxwell terms for the U(1) gauge field $a_{\vi \vj}$, Yang-Mills terms for the $\widetilde{\rm SU}(2)$ gauge field $\widetilde{U}_{\vi \vj}$ and a Higgs potential for $\Phi$ and $\widetilde{U}_{\vi \vj}$.

In the pseudogap regime studied in Sections~\ref{sec:spinliquids}-\ref{sec:dwave}, we condensed the Higgs field $\Phi_{\vi\alpha\beta} = \Phi \delta_{\alpha\beta}$, and ignored its flucutations. The coupling constant in $\mathcal{H}_Y$ in Eq.~(\ref{eq:complete}) is related to that in 
Eq.~(\ref{Yukawa}) by $g = \bar{g} \Phi^\ast$. The constant $\Phi$ produces the hole pocket and anti-nodal pseudogap in the spectrum of $c$ and $f_1$. The constant $\Phi$ also allowed us to set the $\widetilde{\rm SU}(2)$ gauge field to $\widetilde{U}_{\vi\vj} = 1$, and the U(1) gauge field $a_{\vi \vj}$ to zero.

In the strange metal regime studied in Section~\ref{sec:strange}, we worked in the approximation of ignoring the coupling to the bottom layer of ${\bm S}_2$ spins. This can be achieved in Eq.~(\ref{eq:complete}) by setting $B=0$. We further neglected the U(1) gauge field $a_{\vi \vj}$ and the $\widetilde{\rm SU} (2)$ gauge field $\widetilde{U}_{\vi \vj}$, and only included the spacetime fluctuations of the diagonal component $\Phi$ coupled to the fermions $c$ and $f_1$.

\section*{References}
\bibliography{qpm_ropp}

\end{document}